\documentclass[11pt,nofootinbib]{article}

\usepackage{amssymb}

\usepackage{amsmath}

\usepackage{bm}

\usepackage{jcappubn}

\usepackage{longtable}

\usepackage[utf8]{inputenc}

\usepackage{tikz}
\usetikzlibrary{arrows,shapes}
\usetikzlibrary{trees}
\usetikzlibrary{matrix,arrows} 				
\usetikzlibrary{positioning}				
\usetikzlibrary{calc,through}				
\usetikzlibrary{decorations.pathreplacing}  
\usepackage{pgffor}							

\usetikzlibrary{decorations.pathmorphing}	
\usetikzlibrary{decorations.markings}
\tikzset{
    photon/.style={decorate,line width=0.7pt, decoration={snake,amplitude=1.0pt}, draw},
    electron/.style={solid,line width=0.7pt,draw=black},
    proton/.style={double,line width=0.7pt,draw=black},
    hydro1/.style={dashed,line width=0.7pt,draw=black},
    hydro2/.style={dotted,line width=0.7pt,draw=black},
}

\allowdisplaybreaks

\newcommand{\beq}{\begin{equation}}
\newcommand{\eeq}{\end{equation}}
\newcommand{\bea}{\begin{eqnarray}}
\newcommand{\eea}{\end{eqnarray}}
\newcommand{\bel}{\begin{align}}
\newcommand{\eel}{\end{align}}
\newcommand{\ben}{\begin{enumerate}}
\newcommand{\een}{\end{enumerate}}

\newcommand{\pa}{\partial}

\newcommand{\na}{\nabla}

\newcommand{\ed}{{\rm d}}

\newcommand{\we}{\wedge}
\newcommand{\para}{\parallel}
\newcommand{\Lie}{{\cal L}}
\newcommand{\Ord}{{\cal O}}

\newcommand{\Tr}{{\rm Tr}}
\newcommand{\ti}{\tilde}

\newcommand{\Rs}{\mathbb{R}}
\newcommand{\Cs}{\mathbb{C}}

\newcommand{\Ss}{\mathbb{S}}
\newcommand{\Is}{\mathbb{I}}

\renewcommand\({\left(}
\renewcommand\){\right)}
\renewcommand\[{\left[}
\renewcommand\]{\right]}
\newcommand{\bra}{\langle}
\newcommand{\ket}{\rangle}

\newcommand{\ub}{\underbrace}
\newcommand{\os}{\overset}
\newcommand{\us}{\underset}

\newcommand{\nn}{\nonumber}

\newcommand{\al}{\alpha}
\newcommand{\be}{\beta}
\newcommand{\ga}{\gamma}
\newcommand{\Ga}{\Gamma}
\newcommand{\de}{\delta}
\newcommand{\De}{\Delta}
\newcommand{\ep}{\epsilon}
\newcommand{\vep}{\varepsilon}
\newcommand{\ze}{\zeta}
\newcommand{\et}{\eta}
\newcommand{\te}{\theta}
\newcommand{\vte}{\vartheta}
\newcommand{\Te}{\Theta}

\newcommand{\ka}{\kappa}
\newcommand{\la}{\lambda}
\newcommand{\La}{\Lambda}
\newcommand{\ro}{\rho}
\newcommand{\si}{\sigma}
\newcommand{\Si}{\Sigma}
\newcommand{\ta}{\tau}
\newcommand{\ph}{\phi}
\newcommand{\vph}{\varphi}
\newcommand{\ch}{\chi}
\newcommand{\om}{\omega}
\newcommand{\Om}{\Omega}

\newcommand{\cL}{{\cal L}}
\newcommand{\cV}{{\cal V}}

\newcommand{\cA}{{\cal A}}

\newcommand{\cP}{{\cal P}}

\newcommand{\cM}{{\cal M}}

\newcommand{\cC}{{\cal C}}

\title{Tetrad formalism for exact cosmological observables}

\author{Ermis Mitsou$^1$,} 
\emailAdd{ermitsou@physik.uzh.ch}
\author{Jaiyul Yoo$^{1,2}$}
\emailAdd{jyoo@physik.uzh.ch}

\affiliation{$^1$Center for Theoretical Astrophysics and Cosmology, Institute for Computational Science, University of Z\"urich, CH--8057 Z\"urich, Switzerland}
\affiliation{$^2$Physics Institute, University of Z\"urich, Winterthurerstrasse 190, CH-8057, Z\"urich, Switzerland}

\abstract{The standard description of cosmological observables is incomplete, because it does not take into account the correct angular parametrization of the sky, i.e. the one determined by the observer frame. The corresponding corrections must be taken into account for reliable results at non-linear orders. This can be accomplished by introducing an orthonormal basis, or ``tetrad", at the observer point, representing the frame with respect to which observations are performed. In this work we consider the tetrad formulation of General Relativity, thus associating tetrads to sources as well, and develop a new formalism for describing cosmological observables associated with localized sources. It is based on a manifold which we call the ``observer space-time", whose coordinates are the proper time, redshift and angles an observer uses to parametrize measurements, and on which the rest of the observables are defined. This manifold does not have to be diffeomorphic to the true space-time and allows us to resolve caustics in the latter, in contrast to similar coordinate-based formalisms. As a concrete example, we work out the definitions and equations for the angular diameter distance, weak lensing and number count observables. As for the observables associated to the CMB, they lie inside the phase space distribution of the photon fluid, so we also revisit the construction of general-relativistic matrix kinetic theory from the tetrad formalism viewpoint. Here too the latter appears as the natural approach for relating the macroscopic dynamics to the microscopic quantum field theory, and therefore for constructing the matrix Boltzmann equations, without any approximation on the gravitational side. We provide detailed discussions over some subtle issues, that are not considered in the literature, and an alternative construction of the collision term, which deviates from the standard one at higher order in the interactions. As a concrete example, we derive the lowest-order collision term for a fluid of photons, electrons and protons that includes the polarization/spin information. Finally, all our equations are derived at the fully non-linear level, i.e. without any reference to some background space-time, and in a completely coordinate-independent fashion. Thus, the present work contains all the required structures for computations in cosmology with exact and model-independent cosmological observables.}

\begin{document}

\maketitle

\flushbottom

\section{Introduction \& summary}

In the last few decades several impressive observational achievements (see e.g. \cite{KOSMET11,PLANCKcos18,PLANCKover18,Tegmark:2006az}) allowed the development of a ``concordance" model of cosmology, the so-called ``$\La$CDM" model, whose parameters are now determined with below percent accuracy. Remarkably, the physics involved in this model is mostly conservative, as it essentially relies on the theory of General Relativity (GR) and the Standard Model (SM) of particle physics. Both of them are mature modelizations of nature, in that they have been tested extensively and in various contexts for several decades, if not a century. Nevertheless, there are also important ``black boxes" in this construction, the most prominent ones arguably being the dark energy and dark matter components of the universe (or effects).\footnote{Other important open questions include the physics of inflation and the generation of baryon asymmetry in the early universe, or the microscopic physics behind the neutrino mass.} Understanding this dark sector is one of the most profound challenges of modern physics, with several future surveys being devoted to this task, be it fully or partially \cite{DESI13,LSST04,WFIRST12,SKA09,EUCLID11}. These advances will increase the precision that is required from the theoretical predictions in order to correctly interpreted the data. The case of particular interest in this work is the definition and computation of cosmological observables, i.e. the reconstruction of the physical quantities measured by the observer out of the information on the latter's light-cone. The linear order perturbation theory around the homogeneous and isotropic solution is well understood and documented, but is often insufficient for matching the aforementioned precision requirements. This is why, in the last decade, the community has been actively investigating the impact of second-order effects in the CMB lensing \cite{Challinor:2005jy,Lewis:2006fu,Hanson:2009kr,Lewis:2011fk,Pettinari:2014iha,Bonvin:2015uha,Marozzi:2016uob,Marozzi:2016und,Marozzi:2016qxl,Pratten:2016dsm,Lewis:2017ans,DiDio:2019rfy}, in galaxy number counts \cite{DiDio:2014lka,Bertacca:2014dra,Bertacca:2014wga,DiDio:2015bua,Nielsen:2016ldx,DiDio:2016gpd,Umeh:2016nuh,Jolicoeur:2017nyt,Jolicoeur:2017eyi,Jolicoeur:2018blf,Koyama:2018ttg,DiDio:2018unb,Clarkson:2018dwn,Jalilvand:2019brk,Fuentes:2019nel} and cosmological distances and weak lensing
\cite{Bernardeau:2009bm,Bernardeau:2011tc,Umeh:2012pn,Umeh:2014ana,Andrianomena:2014sya,Marozzi:2014kua,Bonvin:2015kea,Fanizza:2018qux,Gressel:2019jxw}.

Despite this important literature on the subject, it turns out that the approaches employed so far contain an approximation that is no longer justified at non-linear orders, thus potentially invalidating several results. This is a bold claim, so it is worth laying down some supporting material in order to make our point. First, we note that what one really measures in cosmology is the functional relation between observables, e.g. the relation between the temperature of the photon fluid and the angular direction in which it is observed on the sky $T(\vte)$, or the average relation between the luminosity distance and the redshift of some set of sources $\bra D_L(z) \ket$. One then immediately recognizes that there exists a subset of observables which appear as more ``fundamental", because they are able to parametrize the rest of the observables, and also because they are model-independent quantities, thus leading to the aforementioned observable functions. This fundamental set is composed of the redshift $z$ associated with some source, the observed angular parametrization of the sky $\{ \vte, \vph \}$ and the observed frequency parametrization of light spectra $\hat{\om}$. The $\{ \hat{\om}, z \}$ observables depend on the information of the observer and source 4-velocities. The $\{ \hat{\om}, z \}$ quantities are therefore uniquely defined for a given observer, but change if one alters her 4-velocity, i.e. they are defined up to a boost of the observer. As for the angular parametrization $\{ \vte, \vph \}$, it is defined only up to a global rotation of the sky, since one needs to pick a definite spatial reference frame in order to associate $\{ \vte, \vph \}$ numbers to sources. From the viewpoint of the 4-dimensional space-time manifold, this spatial frame corresponds to three space-like orthonormal vectors in the tangent space of the observer position, which are normal to the 4-velocity of the observer, i.e. they generate her ``rest-frame". Together, these four vectors therefore form an orthonormal basis of the observer's tangent space, i.e. a ``tetrad" or ``vierbein". Such a basis represents the ``observer frame" with respect to which she measures tensorial components. The prototypical example in this case is the incoming photon 4-momentum $\hat{k}$, whose components in the observer frame $e_a$ provide the observables $\{ \hat{\om}, \vte, \vph \}$
\beq \label{eq:kaparam}
(\hat{k}^a) \equiv \hat{\om} \( 1, -\,\hat{n}(\vte,\vph) \) \, , \hspace{1cm} \hat{n} \equiv \( \sin \vte \cos \vph, \sin \vte \sin \vph, \cos \vte \) \, .
\eeq 
Our key observation is that the approaches employed so far for the computation of non-linear effects take into account the 4-velocity of the observer, but not the spatial part of her frame, so that $\hat{\om}$ and $z$ are well-defined, but not $\{ \vte, \vph \}$. Rather, the only available spatial reference vectors are the ones induced by the spatial coordinate system $\{ \pa_x, \pa_y, \pa_z \}$ under consideration and these induce a different parametrization $\{ \te, \ph \} \neq \{ \vte, \vph \}$ of the observer sky manifold, through
\beq \label{eq:tephdef}
( \hat{k}^x, \hat{k}^y, \hat{k}^z ) \sim \( \sin \te \cos \ph, \sin \te \sin \ph, \cos \te \) \, .
\eeq
Within perturbation theory, the latter is usually described through the ``background" or ``unlensed" angles $\{ \bar{\te}, \bar{\ph} \}$ plus an angular deflection field $\{ \de \te\( \bar{\te}, \bar{\ph} \), \de \ph\( \bar{\te}, \bar{\ph} \) \}$. Importantly, the $\{ \pa_x, \pa_y, \pa_z \}$ vectors at the observer position are neither orthonormal, nor normal to the 4-velocity, in the coordinate systems of practical convenience in cosmological perturbation theory (synchronous, longitudinal, etc.). Consequently, the $\{ \te, \ph \}$ parametrization is {\it not} the one an observer actually uses to map the sky $\{ \vte, \vph \}$, but some diffeomorphism of the latter, with a typical amplitude proportional to the gravitational potentials at the observer. The mismatch between $\{ \te, \ph \}$ and $\{ \vte, \vph \}$ therefore leads to an artificial ``lensing" effect that has not been accounted for in the quantitative studies of the literature yet.\footnote{To avoid confusion for the reader who is specialized in the field, let us stress that the usual introduction of a Sachs basis does not resolve the issue we are pointing out here. An orthonormal basis in the tangent space of {\it a given point} on the sky manifold allows one to obtain the observed components of tensors at that point. It does not, however, provide the global observer parametrization of the sky manifold itself $\{ \vte, \vph \}$, which is necessary when computing angular correlation functions or spectra. In particular, note that the parametrization $\{ \vte, \vph \}$ induces a privileged Sachs basis to each point on the sky $\{ \pa_{\vte}, \pa_{\vph}/\sin\vte \}$, thus forming the two sky vector {\it fields} that the observer implicitly uses in practice for decomposing tensors on the sky.}  

At a more qualitative level, the necessity of introducing a full observer frame in order to obtain the correct angular parametrization of observables is a recognized fact \cite{Challinor:2000as,Challinor:2002zh,Perlick:2010zh,Pitrou:2012ge,Yoo:2014kpa,Yoo:2014sfa,Su:2014mga,Pitrou:2015iya,Yoo:2017svj,Hellaby:2017soj,Korzynski:2017nas,Yoo:2018qba,Grimm:2018nto,Grasso:2018mei,Marcori:2018cwn,Mitsou:2019ocs}. Since the effect of this frame is to correct a parametrization of observables in the observer's tangent space, its consideration leads to extra terms at the observer position in the expressions for the cosmological observables, at linear order in perturbation theory. These ``observer terms" can therefore only affect the first few multipoles of the corresponding angular spectra, so their effect is irrelevant in a multipole analysis at linear order, which is why they have been generically neglected in the literature. However, they can no longer be ignored at the non-linear level, since they couple with source and line-of-sight terms, thus affecting all multipoles.\footnote{See \cite{Mitsou:2019ocs} for a detailed discussion of this issue and, in particular, the impact on the statistics of observables.} Let us also stress that, already at the linear level, these observer terms are clearly relevant conceptually, since they are necessary for the full expression of the observable to be gauge-invariant and free of infrared divergences \cite{Yoo:2009au,Yoo:2010ni,Yoo:2014kpa,Yoo:2014sfa,Biern:2016kys,Biern:2017bzo,Yoo:2017svj,Scaccabarozzi:2017ncm,Scaccabarozzi:2018vux,Yoo:2018qba,Grimm:2018nto}. Thus, in general, as one delves into the non-linear regime, it is important that the observer frame is properly taken into account in order to avoid miscalculations and misinterpretations. 

From the viewpoint of the tetrad formulation of GR \cite{Yepez:2011bw}, where the metric information is contained and generalized in a tetrad {\it field}, the introduction of a tetrad at a single (observer) point begs for a generalization to all the points of the manifold. Here we will therefore reconsider the issue of cosmological observables with the tetrad description of space-time as our starting point. This leads to several conceptual and practical advantages at a remarkably negligible price: the introduction of three extra non-dynamical fields. For instance, the tetrads at other points than the observer one can now be interpreted as the frames of sources, thus unifying all reference frames involved in cosmological observables in a single space-time field. The extra six components that the tetrad field has with respect to the metric can therefore be interpreted as the information of 4-velocities and rest-frame orientations of an observer/source family. Since one already considers velocity fields in cosmology, the truly new information one has to keep track of here are the three fields determining the orientation of the spatial frames. Moreover, with this viewpoint the local Lorentz symmetry of the tetrad formalism is now interpreted as frame transformations, thus allowing us to access all the possible observer/source families. One can therefore select frames that are convenient for performing computations and then have access to the ones that are physically relevant by using these transformations. This freedom comes on top of the freedom to choose the coordinate system of the space-time manifold, which is also intact thanks to the fact that the observer/source frame information is now encoded in tensors: the four vectors of a tetrad. 

The original and central content of this work is the use of the tetrad field to develop a new formalism for defining and computing the aforementioned cosmological observable functions. We will focus in particular on the most important ones: the angular diameter distance, weak lensing and galaxy number counts (associated with a given source 4-velocity field) and the cosmic microwave background (CMB). Being sky maps, all these observables will be fields on the ``observer sky" $\Ss$ parametrized by the two angles $\{ \vte, \vph \}$ that the actual observer uses in practice. All the considered definitions and equations will be given at the fully non-linear level and without any reference to some ``background" (homogeneous and isotropic) space-time, so that coordinate and model-independence are manifest. In the case of localized sources, our formalism is based on the introduction of a new manifold, the ``observer space" $\cC$, which is parametrized by the fundamental observables $\{ z, \vte, \vph \}$ and has the topology of a 3-cylinder $\cC \simeq \Rs_+ \times \Ss$. It is then mapped to the observer light-cone in the space-time manifold $\cM$ through the bundle of light-like geodesics emanating from the observer position. All other observables are directly defined as functions on that observer space $\cC$, thus achieving a fully coordinate and model-independent definition of the relations between physical quantities indeed. In particular, caustics of light rays now correspond to the map $\cC \to \cM$ being non-injective, not to singularities, so our observable maps are definable and computable in the presence of strong lensing as well. This absence of obstruction in resolving caustics is another important feature of our formalism and stands in contrast to the observational coordinate \cite{ObsCoord, Nugier:2013tca} and geodesic light-cone coordinate \cite{Gasperini:2011us, BenDayan:2012pp, BenDayan:2012wi, BenDayan:2012ct, BenDayan:2013gc, Fanizza:2013doa, Nugier:2013tca, Marozzi:2014kua, DiDio:2014lka, Fanizza:2014baa, Fanizza:2015swa, Fleury:2016htl, Scaccabarozzi:2017ncm, Mitsou:2017ynv, Fanizza:2018tzp} approaches for cosmological observables. At the practical level, one no longer needs to compute redshift fluctuations and angular deflections with respect to some reference parametrization (e.g. affine parameter and unlensed angles), since now $\{ z, \vte, \vph \}$ {\it are} the parameters with respect to which our equations are defined. Thus, the operator controlling evolution down the light-cone will be the derivative with respect to $z$, while the geodesic deviation operator leading to the Jacobi map will be the derivative with respect to $\{ \vte, \vph \}$.\footnote{Higher-order angular derivatives would then allow one to go beyond the infinitesimal beam approximation and thus consider finite shapes on the sky (see \cite{Clarkson:2016zzi,Clarkson:2016ccm,Fleury:2017owg,Fleury:2018cro,Fleury:2018odh} for works on finite beams).} In order to also describe the drift of observables with respect to a given observer world-line and transport of her frame, one simply considers a specific integral line of the observer 4-velocity field and repeats the observer space construction at each point. This therefore leads to the ``observer space-time" $\Ord := \Rs \times \cC$ parametrized by $\{ \ta, z, \vte, \vph \}$, where $\ta$ is the observer proper time. By construction, the image of $\Ord$ in the true space-time $\cM$ is then the observable universe of the observer under consideration. Finally, each space $\Ss$, $\cC$ or $\Ord$ can be promoted to its ``spectral" analogue by including the observed frequency $\hat{\om}$ dimension. We also pay special attention to the effect of local Lorentz transformations, i.e. the change of observer/source family, on the spectral observer space. In particular, this induces a non-trivial coordinate transformation 
\beq
\{ z, \vte, \vph, \hat{\om} \} \to \{ \ti{z}, \ti{\vte}, \ti{\vph}, \ti{\hat{\om}} \} \, , 
\eeq
so that this space is actually endowed with an atlas. The different observable functions associated with each chart correspond to the observations of all possible observers and all possible sources in a given space-time geometry. 

In the case of the CMB, which is a diffuse source, the corresponding observables are obtained by evaluating the photon phase space distribution at the observer position and pulling-back its tangent space dependence on the spectral observer sky $\Ss_{\rm spec} \simeq \Ss \times \Rs_+$. If one is only interested in relating these observables to the corresponding ``source", which would be the last scattering surface, then one can use the geometrical optics approach (eikonal approximation), whereby the photon intensity tensor is parallely-transported from that surface along light-like geodesics up to the observer point. For this task, one can again use the observer space-time formalism described above, now applied to the case of a continuous collection of sources situated at $z = z_{\rm last.\,scat.}(\vte,\vph)$. However, this is only an approximative approach, neglecting for instance non-gravitational interactions, or the fact that the photon decoupling process is not instantaneous. In order to take into account all possible effects, one needs to consider the full dynamics of the photon phase space distribution, and of the ones it couples to, which is required anyways for the cosmological evolution of matter in general. The second important part of this work therefore contains a detailed description of matrix kinetic theory on curved space-time.\footnote{Here by ``matrix" is meant the fact that the phase space distributions take into account the possibility of quantum superposition of particle polarizations and are therefore hermitian matrix functions of phase space, instead of scalars.} As also recognized in \cite{Pitrou:2008ut,Pitrou:2008hy,Beneke:2010eg,Naruko:2013aaa,Su:2014mga,Fidler:2017pkg,Pitrou:2019hqg}, the tetrad formalism appears as the natural language for this task, especially for relating the particle phase space distribution to the underlying quantum field theory (QFT) quantities, which is required in order to obtain the collision term of the Boltzmann equation for matrix distributions. We revisit the construction of this theory, including in particular detailed discussions over subtle issues that are not addressed in the literature, and also deviate from the latter in our definition of the collision term, which leads to different results at higher order corrections in the coupling constants. Finally, another original output of our treatment is that, since we have access to the spin polarization of fermionic matter through the associated matrix distribution, we can compute the intrinsic magnetization of the corresponding fluid. In particular, this leads to an extra magnetic moment that sources the cosmic magnetic fields and might therefore be relevant in studying their generation.

In summary, in this work we propose an ab initio derivation of the essential equations regarding cosmological observables employing only exact definitions and relations. In order to achieve this without specifying a coordinate system, one needs to introduce tetrads, which is the core concept of our formalism and its distinguishing feature with respect to the standard practice. The content is organized as follows. In section \ref{sec:motivation} we motivate the use of the tetrad formalism of GR in cosmology. In section \ref{sec:mathframe} we introduce the mathematical structures upon which our formalism is built. These can be organized in three categories, each one of them being associated with a given manifold. More precisely, we discuss the fields, equations and symmetries associated with the space-time manifold in subsection \ref{sec:M}, those associated with the line manifold on which geodesics are defined in subsection \ref{sec:L} and those associated with the phase space on which particle distributions are defined in subsection \ref{sec:TM}. In section \ref{sec:geo} we then present the observer space-time formalism for observables from localized sources, which is constructed using the structures introduced in subsection \ref{sec:L}, and conclude by building the distance, weak lensing and number count maps in subsection \ref{sec:locsourceobs}. Finally, in section \ref{sec:Boltzmann} we provide the relevant derivations and discussions for general-relativistic matrix kinetic theory, based on the structures introduced in subsection \ref{sec:TM}. In subsection \ref{sec:Compton} we provide, as a concrete example, the lowest-order collision term for a fluid of photons, electrons and protons, including the full spin/polarization information, and in subsection \ref{sec:CMB} we define the CMB observable maps using the spectral observer sky construction of subsection \ref{sec:obssky}. We use natural units $8 \pi G = c = \hbar = 1$ and a space-time signature of mostly pluses.

\clearpage

\section{Motivation} \label{sec:motivation}

\subsection{Measurement and observer frame} \label{sec:mesobs}

In cosmological observations we measure the light of remote sources, both localized and diffuse. Denoting by $\cM$ the space-time manifold, the measurement event takes place in a neighborhood $\hat{\cal I} \subset \cM$ of the observer's position $\hat{P} \in \cM$. We will generically use a hat to denote evaluation at that point, or to denote quantities that are only defined there.\footnote{This notation is chosen such that it does not clog too much the equations and is inspired by the fact that $\hat{P}$ is the tip of the observer's light-cone.} For all practical purposes in cosmology, $\hat{\cal I}$ can be considered to be of infinitesimal extent, i.e. just enough to give us access to the tangent space $T_{\hat{P}} {\cal M}$. The information of cosmological observables is ultimately contained in the momentum and polarization of the incoming photons, so what we measure are the components of some tensors at $\hat{P}$, a statement which only makes sense with respect to some basis of the tangent space $T_{\hat{P}} {\cal M}$. As shown in appendix \ref{app:obsfrortho}, for a basis of $T_{\hat{P}} \cM$ to correspond to the frame with respect to which an observer is making measurements at $\hat{P}$, it must be orthonormal with respect to the metric tensor $g$ at $\hat{P}$. This basis is then referred to as the ``observer frame''. 

So let us consider a set of four vectors $\hat{e}_a \in T_{\hat{P}} \cM$ labeled by $a \in \{ 0,1,2,3 \}$ and satisfying the orthonormality condition
\beq \label{eq:Ptetdef}
\hat{g} \( \hat{e}_a, \hat{e}_b \) \equiv \et_{ab} \, ,
\eeq
where $\et$ denotes the Minkowski metric. Such a basis $\hat{e}_a$ is known as a ``tetrad" or ``vierbein". When interpreting it as an observer frame, the time-like element $\hat{e}_0$ represents the 4-velocity of the observer, while the space-like elements $\{ \hat{e}_i \}_{i=1}^3$ provide a Cartesian basis of the observer rest-frame. The condition \eqref{eq:Ptetdef} does not determine the $\hat{e}_a$ entirely, as it is invariant under internal Lorentz transformations of that basis
\beq \label{eq:heaLLT}
\hat{e}_a \to \La_a^{\,\,\,b} \hat{e}_b \, .
\eeq
The tetrad indices $a,b,c,\dots$ should therefore be understood as forming the vector representation of the Lorentz group and we displace them using the Minkowski metric $\et_{ab}$. Given the observer frame interpretation of $\hat{e}_a$ this freedom should be expected. Indeed, the angular parametrization of the observed sky is defined only up to a rotation, which explains the ${\rm SO}(3) \subset {\rm SO}(1,3)$ subgroup acting on the spatial frame $\hat{e}_i \to R_i^{\,\,\,j} \hat{e}_j$. As for the boosts, they alter the observer's 4-velocity $\hat{e}_0 \to \La_0^{\,\,\,a} \hat{e}_a$, thus allowing us to relate different observers. In particular, note that boosts span the full interior of the tangent light-cone in $T_{\hat{P}} \cM$, so we have access to all possible observers, with all possible spatial frame orientations.

Let us now consider an arbitrary coordinate system $x^{\mu} \in \{ t,x,y,z \}$ around $\hat{P}$, so that we can decompose $\hat{e}_a \equiv \hat{e}_a^{\mu} \pa_{\mu}$ and \eqref{eq:Ptetdef} becomes $\hat{g}_{\mu\nu} \hat{e}_a^{\mu} \hat{e}_b^{\nu} \equiv \et_{ab}$. Denoting by $\hat{e}^a_{\mu}$ the coefficients of the inverse matrix of $\hat{e}_a^{\mu}$, we then have that $\hat{e}^a_{\mu} \equiv \et^{ab} \hat{g}_{\mu\nu} \hat{e}_b^{\nu}$ and that $\hat{e}^a_{\mu} \ed x^{\mu} \in T^*_{\hat{P}} \cM$ is the corresponding tetrad basis of the cotangent space. The components of a tensor $T$ that are measured by the observer $\hat{e}_a$ at $\hat{P}$ are then given by the projection on that basis
\beq \label{eq:atoIO}
\hat{T}_{a_1 \dots a_m}^{b_1 \dots b_n} := \hat{e}_{a_1}^{\mu_1} \dots \hat{e}_{a_m}^{\mu_m} \, \hat{e}^{b_1}_{\nu_1} \dots \hat{e}^{b_n}_{\nu_n} \, \hat{T}_{\mu_1 \dots \mu_m}^{\nu_1 \dots \nu_n} \, .
\eeq
Under a Lorentz transformation \eqref{eq:heaLLT}, which physically modifies the observer, the components \eqref{eq:atoIO} mix accordingly in the corresponding tensor representation of the Lorentz group. Thus, exactly as in special relativity, but now locally at $\hat{P}$ (i.e. in $T_{\hat{P}} \cM$), measurement is observer-dependent. On the other hand, the quantities \eqref{eq:atoIO} are invariant under coordinate transformations, since all coordinate-induced indices are fully contracted and the tensor field is evaluated at a definite point $\hat{P}$. Thus, as one should demand in generally-covariant theories, physical observables are independent of the way we choose to parametrize space-time. 

The prototypical example of observables are those lying in the photon 4-momentum $\hat{k}$ at $\hat{P}$. Expressing it in the tetrad basis $\hat{k}^a := \hat{e}^a_{\mu} k^{\mu}$, the light-like condition becomes $\et_{ab} \hat{k}^a \hat{k}^b \equiv 0$, so the $\hat{k}^a$ numbers can be parametrized as in \eqref{eq:kaparam}. We have that $\hat{\om} := \hat{k}^0$ is the photon frequency, while $(\vte,\vph)$ is the angular position in the sky of the correspond light source, as measured by the observer $\hat{e}_a$. Thus, the $\hat{k}$ information, expressed in the basis $\hat{e}_a$, provides the numbers that the corresponding observer uses to parametrize light spectra ($\hat{\om}$) and the night sky ($\vte,\vph$). In particular, the derivative with respect to $\vte, \vph$ will lead to the construction of deviation quantities such as the Jacobi map. 

Now that we have explained these geometrical aspects of observation, let us see what happens if one does not introduce the tetrad vectors $\hat{e}_a$ at $\hat{P}$. In that case, the only available basis of $T_{\hat{P}} \cM$ is $\pa_{\mu}$, i.e. the one induced by some coordinate system $x^{\mu}$ around $\hat{P}$. In general, this basis $\pa_{\mu}$ is not orthonormal at $\hat{P}$, since, by definition 
\beq
g \( \pa_{\mu}, \pa_{\nu} \) \equiv g_{\mu\nu}  \, .
\eeq
Thus, if one wishes to interpret $\pa_{\mu}|_{\hat{P}}$ as the frame of some observer at $\hat{P}$, then the coordinate system must satisfy
\beq \label{eq:observercoord}
\hat{g}_{\mu\nu} = \et_{\mu\nu} \, .
\eeq
Put differently, \eqref{eq:observercoord} is required for the coordinate-induced components $\hat{T}_{\mu_1 \dots \mu_m}^{\nu_1 \dots \nu_n}$ of some tensor $T$ at $\hat{P}$ to be the ones measured by some observer.\footnote{Note that one can reach the present approach by starting with a general $\hat{e}_a$ and then choosing the coordinate system such that $\hat{e}_a^{\mu} \to \de_a^{\mu}$, thus effectively identifying the $a$ and $\mu$ indices.} The condition \eqref{eq:observercoord} seems pretty mild a priori, as it only constrains the coordinate system in the infinitesimal vicinity $\hat{\cal I}$ of $\hat{P}$. In fact, it is not even as constraining as a system of normal coordinates around $\hat{P}$, since the latter further requires $\pa_{\mu} g_{\nu\ro}|_{\hat{P}} = 0$. Nevertheless, as innocent as this condition may seem, it is not satisfied in the coordinate systems that are usually employed in cosmology, and the associated perturbation theory, such as the longitudinal and synchronous gauges. In particular, the practical gauges are usually defined through global conditions, in contrast to the condition \eqref{eq:observercoord} which is local. 

We conclude that, in the typical coordinate systems employed in cosmology, the tensor components cannot be interpreted as observables. An important exception are time-like components, for which one usually explicitly invokes the observer 4-velocity $\hat{u}$, i.e. the analogue of $\hat{e}_0$, which therefore allows one to construct observables such as $\hat{\om} \equiv -\hat{u}_{\mu} \hat{k}^{\mu}$. In the normal subspace to $\hat{u}$, however, there are no reference vectors to project onto, so we cannot extract the spatial components of tensors that are actually being observed. In particular, this means that we do not have access to the observed angles associated with incoming photons. The only available angles are the ones extracted from the coordinate-induced components \eqref{eq:tephdef}. Since $\hat{g}_{\mu\nu} \neq \et_{\mu\nu}$, the basis $\{ \pa_x, \pa_y, \pa_z \}|_{\hat{P}}$ is neither orthonormal, nor normal to $\hat{u}$, so $(\te,\ph)$ are not the angles an observer actually uses to parametrize the sky. 

In the standard approaches to cosmological observables this problem is not resolved, but rather hidden under the carpet of cosmological perturbation theory. One starts by considering the homogeneous and isotropic space-time that appears as the zeroth-order approximation of $g$. In the Friedmann-Lema\^itre-Robertson-Walker (FLRW) coordinates of that ``background" universe, the spatial metric at the observation point is flat, so the tensor components are indeed the ones measured by some observer, the one at rest in these coordinates. One then parametrizes the observed sky with the angles used by this observer $\bar{\te}, \bar{\ph}$. However, when fluctuations are introduced, their effect is taken into account through deviation angles $\de \te(\bar{\te}, \bar{\ph})$ and $\de \ph(\bar{\te}, \bar{\ph})$ relating the ``background" (or ``unlensed'') $(\bar{\te}, \bar{\ph})$ and ``deflected" (or ``lensed") $(\te,\ph)$ angular positions in the sky. As we just saw, however, the $(\te, \ph)$ are not the observed angles of some observer, because now the metric at the observer also includes fluctuations, so the deflections $(\de \te, \de \ph)$ are ambiguous.

\subsection{Advantages of the tetrad formulation of differential geometry} \label{sec:advtet}

Although the introduction of a tetrad $\hat{e}_a$ at the observer point is technically sufficient to define all the desired observables in a coordinate and background-independent way, it has the inelegant aspect of arising as some extra manipulation, i.e. the translation between the $T_{\hat{P}} \cM$ bases $\hat{e}_a$ and $\pa_{\mu}|_{\hat{P}}$. A simple way around the problem is to consider the tetrad formulation of differential geometry, that is, to replace the metric $g_{\mu\nu}(x)$ by a tetrad {\it field} $e_a^{\mu}(x)$, i.e. a set of four vector fields forming an orthonormal basis at every point of $\cal M$
\beq
g_{\mu\nu}(x)\, e_a^{\mu}(x)\, e_b^{\nu}(x) = \et_{ab} \, .
\eeq
This means that they carry all of the metric information since, in terms of the inverse matrices $e_{\mu}^a(x)$,
\beq \label{eq:gofe}
g_{\mu\nu}(x) = \et_{ab} \, e^a_{\mu}(x) \, e^b_{\nu}(x) \, ,
\eeq
and actually even more, because they have six more components. This extra number of fields is the only downside of the formalism, which is quickly dwarfed by its many advantages, both at the conceptual and computational level:

\begin{itemize}

\item
Unifying the observer and source frames into an ``observer family".

The time-like element $e_0(x)$ is now interpreted as the 4-velocity field of a family of ``observers", of which $\hat{e}_0 := e_0(\hat{P})$ is the true observer, while the one at any other point $P \in \cM$ is associated to some source. The space-like elements $e_i(x)$ then correspond to a spatial frame carried by these observers and it conveniently probes the rest-frame subspaces of the observer at $\hat{P}$ and of the source at $P$, by definition.

\item
Working directly with $T_{a_1 \dots a_m}^{b_1 \dots b_n}(x)$ all over $\cal M$.

Now that we have a tetrad at each point on space-time, we can consider the tensor components in that basis all over $\cal M$, i.e. Eq. (\ref{eq:atoIO}) becomes
\beq \label{eq:atoI}
T_{a_1 \dots a_m}^{b_1 \dots b_n}(x) := e_{a_1}^{\mu_1}(x) \dots e_{a_m}^{\mu_m}(x) \, e^{b_1}_{\nu_1}(x) \dots e^{b_n}_{\nu_n}(x) \, T_{\mu_1 \dots \mu_m}^{\nu_1 \dots \nu_n}(x) \, ,
\eeq
which are therefore scalar fields with respect to diffeomorphisms. For instance, we can work directly with the observed momentum components $\hat{k}^a := \hat{e}^a_{\mu} \hat{k}^{\mu}$ and the emitted ones $k_P^a := e^a_{\mu}(P)\, k^{\mu}(P)$. The invariance of these quantities under coordinate transformations is what will allow us to define cosmological observables without requiring the specification of a coordinate system.

\item
The local symmetry of observer transformations.

From Eq. (\ref{eq:gofe}) we see that the metric is invariant under a {\it local} Lorentz transformation (LLT) of the $a$ index,
\beq \label{eq:LLT}
\ti{e}_{\mu}^a(x) = \La^a_{\,\,\,b}(x) \, e_{\mu}^b(x) \, , \hspace{1cm} \et_{cd} \La^c_{\,\,\,a} \La^d_{\,\,\,b} \equiv \et_{ab} \, ,
\eeq
which therefore corresponds to another observer family, i.e. with different 4-velocities $\ti{e}_0^{\mu}$ and spatial frames $\ti{e}_i^{\mu}$. The corresponding tensor components transform accordingly
\beq
\ti{T}_{a_1 \dots a_m}^{b_1 \dots b_n}(x) = \La_{a_1}^{\,\,\,c_1}(x) \dots \La_{a_m}^{c_m}(x)\, \La^{b_1}_{\,\,\,d_1}(x) \dots \La^{b_n}_{\,\,\,d_n}(x) \, T_{b_1 \dots b_m}^{d_1 \dots d_n}(x)  \, ,
\eeq
so they are diffeomorphism scalars, but they are Lorentz tensors. They correspond to the components that are measured/emitted by the new observer family $\ti{e}_a^{\mu}(x)$. The LLTs are therefore ``observer transformations". Since the metric is invariant, this is a symmetry of the equations when everything is expressed in terms of $e^a_{\mu}(x)$. We can understand this by noting that a 6-dimensional gauge symmetry is required in order to render the extra six components of the tetrad non-physical. These extra components therefore contain the information of the tetrad orientation, i.e. the part that is not captured by the internal scalar product in Eq. (\ref{eq:gofe}). We thus see that the tetrad formulation provides an elegant unification of the gravitational and observer information, i.e. the tetrad internal product and the tetrad orientation, respectively, in a single mathematical object. Note also that the Lorentz group allows us to reach all possible observer families, because the local boosts probe the full interior of each light-cone, while the local rotations probe all possible spatial frames.

Finally, at the conceptual level, the LLT symmetry provides a coordinate-independent manifestation of the notion of relativity. Indeed, the fact that LLTs are a symmetry of the action means that the physics is observer-{\it independent}, e.g. whether two particles scattered or not is independent of the choice of $e^a_{\mu}(x)$, given some $g_{\mu\nu}(x)$. However, their recorded initial and final momenta do depend on the observer and change under an LLT. Consequently, the physics is observer-independent, but measurement is observer-{\it dependent}. Note that this has nothing to do with the choice of coordinates, so the relativity of measurement is related to the LLTs, not the coordinate transformations.\footnote{Of course one can always associate an observer family 4-velocity to some coordinate system, i.e. the family with $u^{\mu} = (1,0,0,0)$ in that system, but it cannot also have a trivial spatial frame, because this would lead to a trivial tetrad field $e_a^{\mu}(x) = \de_a^{\mu}$ and thus no curvature.} Rather, the symmetry under coordinate transformations reflects the fact that the physics is independent of the parametrization of space-time, which is independent of the notion of observer.

\item
Full covariance: freedom of gauge choice and control.

As already stressed in the case of the tetrad at the observer point, the fact that we do not have to privilege a particular coordinate system means that we are free to choose whichever diffeomorphism gauge we wish in cosmological perturbation theory, say the Newtonian one. On the other hand, in the case of LLTs, a choice of gauge amounts to a choice of observer family, and in particular its dynamics. For instance, we will see that we can choose a gauge where all observers are in free-fall and their spatial frames are parallely transported along $e_0$. Although this is certainly a valid assumption at large enough scales, we will nevertheless choose to work in a generic gauge in order to maintain generality. In particular, we will be able to extend the diffeomorphism-invariant combinations of the metric components in linear perturbation theory, the ``Bardeen variables", to the ones that are invariant under both symmetries. Finally, the more the local symmetries of our equations, the more ways we have to perform consistency checks, a tool that becomes increasingly relevant given the complexity of higher-order perturbation theory.

\item
Symmetry-based definition of cosmological observables from localized sources.

Just as a tensor is defined by its transformation properties, so can we now define a cosmological observable associated with some localized source through its transformation properties under the present symmetries. Denoting by $P$ the position of the source, which is linked by a light-like geodesic to $\hat{P}$, an associated cosmological observable $C(P, \hat{P})$ is a function of $P$ and $\hat{P}$, with the following properties:

\begin{itemize}

\item
Under a coordinate transformation the observable transforms as a bi-scalar. This means that $C(P, \hat{P})$ is invariant, because it is expressed as a function of points. If we express it as a function of the coordinates of these points we rather have
\beq
\ti{C}(\ti{x}, \ti{\hat{x}}) = C(x, \hat{x}) \, .
\eeq
A measurement cannot depend on the parametrization of space-time. 

\item
Under a LLT, the variation of $C(P, \hat{P})$ depends {\it only} on $\La^a_{\,\,\,b}(P)$ and $\La^a_{\,\,\,b}(\hat{P})$, because the only observers involved in the process are $e_a(P)$ and $e_a(\hat{P})$. The geodesic path between them solely depends on the metric information $g_{\mu\nu}(x)$, not on the choice of intermediate observers $e_a^{\mu}(x)$. This will not always be explicit in our formalism, because we will have $e_{\mu}^a(x)$ appearing in the observables all along the line of sight, but its presence will be such that only $\La^a_{\,\,\,b}(P)$ and $\La^a_{\,\,\,b}(\hat{P})$ will end up appearing under an LLT.
\\
\\
The fact that $e_0(P)$ is identified with a physical quantity, that is the source's 4-velocity, means that a boost at $P$ changes the source to the one with a different 4-velocity $\ti{e}_0(P)$, thus also modifying the corresponding observables. In the case where the source has some non-uniform shape, one can also associate to it a privileged spatial basis $e_i(P)$, say by picking three reference points that are distinguished by the shape. Then, performing a local rotation at $P$ would mean that we change the source to one that is directed differently $\ti{e}_i(P)$, thus modifying again the corresponding observables. We therefore conclude that, in the absence of such a privileged way of associating some $e_i(P)$ to the source, the corresponding observables should also be invariant under local rotations at $P$. Indeed, if they are not, it would mean that their value is ambiguous, as it depends on a choice of $e_i(P)$ that has no physical interpretation.

\end{itemize}

\item
A natural framework for kinetic theory with QFT amplitudes.

In kinetic theory, the QFT scattering amplitudes in momentum space $\cA(p_1, \dots p_n \to q_1, \dots q_m)$ carry the microscopic physics information of the collision terms for phase space particle distributions $f(x,p)$. These functions $\cA$ are computed through QFT on Minkowski space-time, so they are made of contractions of the momenta with $\et$. In the metric formalism of GR, where the Lorentz group no longer appears and the momenta are diffeomorphism vectors $p^{\mu}$, one can make the amplitude generally-covariant by simply replacing $\et \to g(x)$. The (on-shell) momentum integration too can be expressed in a covariant way by considering a 4-dimensional integral with measure $\de(g_{\mu\nu}(x) \, p^{\mu} p^{\nu} + m^2)\,\ed^4 p/\sqrt{-g(x)}$, which can then be reduced to a 3-dimensional one at the price of breaking manifest covariance. In most of the literature, which employs the metric formalism, one either works with the above prescription (see for instance \cite{Bernstein:1988bw, Kosowsky:1994cy, Ma:1995ey, Dodelson:2003ft, Durrer:2008eom, Weinberg:2008zzc}), or one invokes a specific tetrad field as an intermediate step to perform the computation of the collision term, but then turns back to the metric description as soon as possible (see for instance \cite{Durrer:1993db, Pitrou:2008ut, Pitrou:2008hy, Beneke:2010eg, Fidler:2017pkg,Pitrou:2019hqg}). In both cases the final result is the same: the collision term depends on the gravitational fields. 

Here we want to highlight that, in the tetrad formalism, and in particular in its present use where we switch to Lorentz indices as soon as possible, the collision term is much simpler. Indeed, by working with the Lorentz-indexed momenta $p^a$ the QFT amplitudes can be used as they are, since we are contracting the momenta with $\et$ in the general-relativistic case too. In particular, the momenta all transform with the same Lorentz matrix $p^a \to \La^a_{\,\,\,b}(x) \, p^b$, so the $\cA$ functions are invariant, since this is a symmetry of the $S$-matrix. The integration measure is simpler too, it is the Lorentz-invariant combination $\ed^3 p/\sqrt{m_s^2 + p^2}$ which also appears in QFT. Therefore, the collision term is much simpler, as it is completely independent of the gravitational field $e^a_{\mu}(x)$, and its only dependence on $x^{\mu}$ comes from the distributions $f_s(x,p)$ and their associated wave-functions in the matrix case. Thus, by working with Lorentz-indexed quantities one not only considers directly the relevant parametrization for observables, but one also obtains simpler collision integrals to compute in kinetic theory.

\end{itemize}

\subsection{A word on the other common interpretation of $\hat{e}_a$}  \label{sec:noJac}

Finally, let us make contact with another, quite widespread interpretation of a tetrad, when it is invoked only at a specific point, say $\hat{P}$. We note that \eqref{eq:Ptetdef}, or its coframe analogue
\beq \label{eq:gofeO}
\hat{g}_{\mu\nu} = \et_{ab}\, \hat{e}_{\mu}^a \hat{e}_{\nu}^b \, ,
\eeq
take the form of a coordinate transformation from some arbitrary coordinate system $x^{\mu}$, with metric components $g_{\mu\nu}$, to some other system $x_{\hat{P}}^a$ in which $g_{ab}$ is Minkowski at $\hat{P}$, i.e. the system of an observer at $\hat{P}$. The tetrad matrix then appears as the Jacobian of that coordinate transformation $x^a_{\hat{P}}(x)$ evaluated at $\hat{P}$
\beq \label{eq:eJac}
\hat{e}_{\mu}^a \equiv \frac{\pa x_{\hat{P}}^a}{\pa x^{\mu}}(\hat{P}) \, .
\eeq
Going from that arbitrary system $x^{\mu}$ to some other arbitrary system $\ti{x}^{\mu}$ then reproduces the tensorial transformation rule of the $\mu$ index
\beq
\ti{\hat{e}}^a_{\mu} \equiv \frac{\pa x_{\hat{P}}^a}{\pa \ti{x}^{\mu}}(\hat{P}) = \frac{\pa x_{\hat{P}}^a}{\pa x^{\nu}}(\hat{P})\, \frac{\pa x^{\nu}}{\pa \ti{x}^{\mu}}(\hat{P}) \equiv \hat{e}^a_{\nu}\, \frac{\pa x^{\nu}}{\pa \ti{x}^{\mu}}(\hat{P}) \, ,
\eeq
and so on for the tetrad vectors $\hat{e}_a^{\mu}$. In this process, the observer coordinates $x_{\hat{P}}^a$ are fixed, hence the consistency with their Lorentz index. Indeed, these coordinates can only be transformed to the ones of some other observer at $\hat{P}$, by definition, since we must maintain $\hat{g}_{ab} \equiv \et_{ab}$. The allowed coordinate transformations are the ones preserving (\ref{eq:gofeO}), i.e. those that reduce to Poincar\'e around $\hat{P}$ 
\beq
\ti{x}_{\hat{P}}^a = q^a + \La^a_{\,\,\,b} x_{\hat{P}}^b + \Ord((x_{\hat{P}} - \hat{x}_{\hat{P}})^2) \, .
\eeq
The corresponding frame would then only be sensitive to the Lorentz transformation (i.e. not the translation)
\beq
\ti{\hat{\pa}}_{a,\hat{P}} = \La_a^{\,\,\,b} \hat{\pa}_{b,\hat{P}} \, , \hspace{1cm} \Rightarrow \hspace{1cm} \ti{\hat{e}}_a = \La_a^{\,\,b} \hat{e}_b \, , 
\eeq
and therefore reproduces the Lorentz transformation of the $a$ index. Here we wish to stress that the interpretation of the tetrad as a Jacobian of some coordinate transformation $x^a_{\hat{P}}(x)$ is relevant only in the case where it is used at a single point $\hat{P} \in \cal M$. This is because a tetrad at some other point $\hat{P}'$ will correspond to the Jacobian of some other transformation $x_{\hat{P}'}^a(x)$, i.e. the one trivializing the metric at $\hat{P}'$, not $\hat{P}$. Indeed, if the tetrad were the Jacobian of a single coordinate transformation all over $\cal M$, then space-time would be flat, as one could perform that transformation to get $g_{\mu\nu} \to \et_{\mu\nu}$ everywhere. 

Thus, in the presence of a tetrad {\it field} the underlying observer coordinates associated with each point lose their relevance and one only retains the necessary information for observations, i.e. the basis at each tangent space $T_P \cM$. We therefore believe that the ``Jacobian matrix" interpretation of the tetrad loses its appeal in this context and can even become misleading. We prefer the more gauge-theoretical viewpoint where the Lorentz indices ``$a$" simply correspond to some internal gauge symmetry, just as in Yang-Mills theory, with no reference whatsoever to any particular coordinate system $x_P^a$. The $e_a^{\mu}(x)$ are therefore simply a set of four vector fields and we privilege no coordinate system in describing their dynamics. The only aspects that one can retain from the Jacobian matrix picture is that, for a given $P \in \cM$, there always exists a coordinate system in which $e_a^{\mu}(P) = \de_a^{\mu}$.

\clearpage

\section{Mathematical framework} \label{sec:mathframe}

\subsection{General preliminaries} \label{sec:genprel}

The mathematical framework we are going to discuss in this section mainly involves three distinct manifolds:
\begin{itemize}

\item
$\cM$: the 4-dimensional space-time manifold on which space-time fields are based,

\item
$\Lie$: the 1-dimensional line manifold on which the world-line fields are based, 

\item
$T^* \cM$: the 8-dimensional cotangent bundle manifold on which the Boltzmann distributions are based.

\end{itemize}
These are respectively presented in each one of the following subsections, along with their associated symmetries and the equations of motion for the fields they host. In particular, in each case, we provide the description in the tetrad formalism.

Before we jump into the details, let us discuss an important point that has to do with the mathematical description of the involved physical degrees of freedom and leads in particular to two complementary perspectives to electromagnetism. In the cosmological setting, the space-time fields (defined on $\cM$) correspond to the coherent ``long" wave-length fluctuations of the underlying quantum fields, which are therefore well described classically.\footnote{We adopt the effective field theory approach to gravity, so that it makes sense to talk about a corresponding quantum field, even if it is not fundamental.} Here we will consider the gravitational and electromagnetic contributions, but one could also include extra degrees of freedom that appear for instance in theories of inflation or dark energy. For coherent long wave-length and low frequency excitations to arise the corresponding field must be light, weakly interacting and bosonic.\footnote{Fermionic fields can only give rise to elementary quantum excitations (particles) and bound states thereof. Indeed, because of Pauli's exclusion principle, the large occupation number configurations that are required in order to reach classical behavior do not exist.} Gravity survives because it is a long-range force and also universally attractive, i.e. the field is excitable at all scales and there are no opposite charges that could cancel out the effect. As for electromagnetism, it is also long-range, but the universe is electrically neutral on average, so the long wave-length electric field modes are suppressed. Nevertheless, we do infer the existence of magnetic fields on cosmological scales through observations (see \cite{Durrer:2013pga} and references therein), the origin of which is still a subject of speculation to this date, which is why we include the electromagnetic space-time field. 

On the other hand, we also have particle-like excitations which are very well localized with respect to cosmological scales. If the corresponding particle fluids are dilute enough, their dynamics can be described statistically through 1-particle phase space distributions $f(x,p)$ on $T^* \cM$. As in the case of the space-time fields, the $x^{\mu}$-dependence captures only long wave-lengths, i.e. the distribution fluctuates mildly in space-time, while the ``short" length scales associated with the individual particle dynamics are captured by the momentum dependence $p_{\mu}$. For instance, the typical wave-length of a CMB photon today is of the order of the millimeter. Note also that, although such photons are produced abundantly, this is not the case for short wave-length gravitons, because their coupling is so weak that their production requires energy densities that are not resolved at cosmological scales. Thus, in cosmology all gravitational effects can be taken into account through the classical space-time field alone. 

The bottom-line is that we will take into account both ``manifestations" of electromagnetism, i.e. the phase space distribution of photons {\it as well as} the coherent long wave-length space-time field, the two being effectively treated as different non-interacting components of the universe. This is consistent because photons have no self-interactions. As for the world-line fields, we will only focus on the light-like geodesics that connect a localized source to the observer, along with related objects. We choose to describe them in field-theoretical language as well, because this will make transparent a lot of manipulations and facilitate the relation to the other two manifolds. 

Finally, since respecting symmetries is a central aspect of this work, in appendix \ref{app:diffsym} we propose for the interested reader a comprehensive discussion about the diffeomorphism symmetry and, in particular, the distinction between its ``active" (pullbacks) and ``passive" (coordinate transformations) versions. The difference is purely conceptual at the level of local equations, but has practical implications when integrals are involved. This is the case here since cosmological observables are defined through integrals over $\Lie$. In the appendix, however, we choose to illustrate our discussion with the space-time manifold $\cM$ instead, in order to avoid specificities of the 1-dimensional case. For the uninterested reader, we directly mention our terminology. We will use the acronyms P$\cal M$D (resp. A$\cal M$D) for the passive (resp. active) space-time diffeomorphisms, P$\Lie$D (resp. A$\Lie$D) for the passive (resp. active) world-line diffeomorphisms and P$T^*\cM$D (resp. A$T^*\cM$D) for the passive (resp. active) cotangent bundle diffeomorphisms. When the passive/active distinction is irrelevant, the first letter will be dropped.

\subsection{Space-time fields}  \label{sec:M}

In this section we introduce the tetrad field and the local Lorentz symmetry that is associated with it, and express the Einstein-Maxwell equations in this language. We then discuss the fact that the tetrad formalism is the only way to couple spinors to gravity and we finally present a privileged choice of observers, i.e. the free-falling ones, providing the corresponding gauge-fixing condition of the local Lorentz symmetry.

\subsubsection{Tetrad field and local Lorentz transformations} \label{sec:tetandtrans}

We choose four vector fields $e_a^{\mu}(x)$, indexed by $a \in \{ 0,1,2,3 \}$, such that they are orthonormal
\beq \label{eq:ge2et}
g_{\mu\nu} e_a^{\mu} e_b^{\nu} \equiv \et_{ab} \, , \hspace{1cm} \et_{ab} := {\rm diag} \( -1, 1, 1, 1 \) \, .
\eeq
We will be using lower-case Latin letters from the beginning of the alphabet $a,b,c, \dots$ to denote these indices, and those starting at $i,j,k, \dots$ to denote the spatial part, i.e. $i \in \{ 1,2,3 \}$. We thus demand that $e_0^{\mu}$ is time-like, while the $e_i^{\mu}$ are space-like. The four vectors $e_a := e_a^{\mu} \pa_{\mu}$ form a basis of the tangent space at each space-time point, an alternative to the coordinate-induced basis $\pa_{\mu}$. Denoting by $e_{\mu}^a$ the inverse matrix of $e_a^{\mu}$, i.e.
\beq
e_{\mu}^a e_a^{\nu} \equiv \de_{\mu}^{\nu} \, , \hspace{1cm} e_{\mu}^a e^{\mu}_b \equiv \de^a_b \, ,
\eeq
we see that the $e_{\mu}^a$ transform as a set of four covectors under space-time diffeomorphisms. The fields $e_a^{\mu}$ or $e^a_{\mu}$ are known as a ``tetrad", a ``vierbein", or simply a ``frame".\footnote{Usually these names refer to the vectors $e_a^{\mu}$, so the covectors $e_{\mu}^a$ are then the ``coframe", but we will not make this distinction here.} The coordinate-induced basis $\pa_{\mu}$ has the advantage of commuting, but is not orthonormal
\beq
\[ \pa_{\mu}, \pa_{\nu} \] \equiv 0 \, , \hspace{1cm} g \( \pa_{\mu}, \pa_{\nu} \) \equiv g_{\mu\nu} \, ,
\eeq
while the tetrad basis is orthonormal, but does not commute
\beq
\[ e_a, e_b \] \equiv C_{ab}^{\,\,\,\,\,\,c} e_c \, , \hspace{1cm} g \( e_a, e_b \) \equiv \et_{ab} \, , 
\eeq
where the 
\beq \label{eq:Cdef}
C_{ab}^{\,\,\,\,\,\,c} := -2 e_a^{\mu} e_b^{\nu} \pa_{[\mu} e^c_{\nu]} 
\eeq
are the ``structure coefficients". One can now express Eq. (\ref{eq:ge2et}) as a decomposition of the metric field
\beq \label{eq:gofe}
g_{\mu\nu} = \et_{ab}\, e_{\mu}^a e_{\nu}^b  \, .
\eeq
In matrix notation this would read
\beq
\bm{g} = \bm{e}^T \cdot \bm{\et} \cdot \bm{e} \, ,
\eeq
so the determinants are related by
\beq
e := \det e_{\mu}^a = \sqrt{-g} \, .
\eeq
From Eq. (\ref{eq:gofe}) one sees that the metric is invariant under a local Lorentz transformation (LLT) of the $a$ index,
\beq \label{eq:LLT}
\ti{e}_{\mu}^a(x) = \La^a_{\,\,\,b}(x) \, e_{\mu}^b(x) \, , \hspace{1cm} \et_{cd}\, \La^c_{\,\,\,a} \La^d_{\,\,\,b} = \et_{ab} \, ,
\eeq
so it makes sense to call the $a,b,c, \dots$ ``Lorentz" indices and displace them using $\et_{ab}$. Thus, the $a = 0$ index is displaced with $-1$ and the $a = i$ index is displaced using $\de_{ij}$. We will then refer to the $\mu,\nu,\ro, \dots$ indices of tensors as ``diffeomorphism" indices,\footnote{In the literature one also finds the terminology ``holonomic" for the diffeomorphism indices, while those defined with respect to some general basis that is not induced by a coordinate system, i.e. with non-vanishing structure coefficients, are called ``anholonomic".} since they mix under $\cal M$Ds. One can then check that $e_{\mu}^a$ is nothing but $e^{\mu}_a$ with its indices displaced by the appropriate metrics
\beq
e_{\mu}^a \equiv g_{\mu\nu} \et^{ab} e_b^{\nu} \, ,
\eeq
and vice-versa. With $e_a^{\mu}$ and $e_{\mu}^a$ we can express diffeomorphism tensors as diffeomorphism scalars, but Lorentz tensors
\beq \label{eq:atoI}
T_{a_1 \dots a_n}^{b_1 \dots b_m}(x) := e_{a_1}^{\mu_1}(x) \dots e_{a_n}^{\mu_n}(x) \, e^{b_1}_{\nu_1}(x) \dots e^{b_m}_{\nu_m}(x) \, T_{\mu_1 \dots \mu_n}^{\nu_1 \dots \nu_m}(x) \, ,
\eeq
and vice-versa. The Lorentz indices are {\it internal} indices, in total analogy with the Yang-Mills indices of the Standard Model. A P$\cal M$D will only change the way the $e_a := e_a^{\mu} \pa_{\mu}$ vector is represented in the coordinate-induced basis $\pa_{\mu}$, but it will not mix it with the $e_{b \ne a}$, the LLTs will. 

An interesting conceptual difference with respect to the metric formalism is the way in which the Lorentzian signature condition is imposed. The signature of the metric is the set of signs of its eigenvalues. In the metric formalism one has to restrict the set of considered metrics $g_{\mu\nu}$ to the ones having Lorentzian signature, i.e. a condition in field space, which therefore seems somewhat inelegant. In contrast, in the tetrad formalism the metric \eqref{eq:gofe} has Lorentzian signature {\it whatever} the eigenvalues of $e_{\mu}^a$, as long as they are all non-zero, so we now only need the condition $e \neq 0$ in field space. The reason for this is that the signature information is now ``hardwired" through the choice of internal metric, or equivalently, of the internal group. Indeed, it is the fact that we choose to contract the $a$ indices with $\et_{ab}$, or equivalently, to act on them with SO(1,3), which makes $g_{\mu\nu}$ Lorentzian. Had we chosen SO$(4)$ as our local group, the corresponding invariant metric would have been $\de_{ab}$ and would have thus led to a $g_{\mu\nu}$ metric with Euclidean signature.

\subsubsection{Connections and curvatures} \label{sec:connscurv}

Given the local symmetries of the theory, i.e. $\cal M$Ds and LLTs, we introduce a gauge field (or ``connection'') for each one of them in order to form covariant derivatives. For $\cal M$Ds we have the ``affine connection'' $\Ga^{\ro}_{\,\,\,\mu\nu}$. It is a scalar under LLTs, but under a P$\cal M$D it transforms as
\beq \label{eq:Gatrans}
\ti{\Ga}^{\ro}_{\,\,\,\mu\nu}(\ti{x}) = \frac{\pa \ti{x}^{\ro}}{\pa x^{\ga}}(x) \, \frac{\pa x^{\al}}{\pa \ti{x}^{\mu}}(\ti{x}(x)) \, \frac{\pa x^{\be}}{\pa \ti{x}^{\nu}}(\ti{x}(x)) \, \Ga^{\ga}_{\,\,\,\al\be}(x) + \frac{\pa \ti{x}^{\ro}}{\pa x^{\ga}}(x) \, \frac{\pa^2 x^{\ga}}{\pa \ti{x}^{\mu} \pa \ti{x}^{\nu}}(\ti{x}(x)) \, ,
\eeq
and under an A$\cal M$D it varies by
\bea
\de_{\xi} \Ga^{\ro}_{\,\,\,\mu\nu} & = & - \pa_{\mu} \pa_{\nu} \xi^\ro - \Lie_{\xi} \Ga^{\ro}_{\,\,\,\mu\nu} + \Ord(\xi^2) \label{eq:dGapa} \\
 & \equiv &  - \na_{(\mu} \[ \na_{\nu)} \xi^\ro - T^{\ro}_{\,\,\,\nu)\si} \xi^{\si} \] + R^{\ro}_{\,\,\,(\mu\nu)\si}\, \xi^{\si} + \frac{1}{2}\, \Lie_{\xi} T^{\ro}_{\,\,\,\mu\nu} + \Ord(\xi^2)  \label{eq:dGana} \, , 
\eea
where $\na$ is the covariant derivative with respect to $\cal M$Ds, e.g.
\beq
\na_{\mu} X^{\nu} := \pa_{\mu} X^{\nu} + \Ga^{\nu}_{\,\,\,\ro\mu} X^{\ro} \, ,
\eeq
while
\beq \label{eq:TRdef}
T^{\ro}_{\,\,\,\mu\nu} := \Ga^{\ro}_{\,\,\,\nu\mu} - \Ga^{\ro}_{\,\,\,\mu\nu} \, , \hspace{1cm} R^{\ro}_{\,\,\,\si\mu\nu} := \pa_{\mu} \Ga^{\ro}_{\,\,\,\si\nu} - \pa_{\nu} \Ga^{\ro}_{\,\,\,\si\mu} + \Ga^{\ro}_{\,\,\,\al\mu} \Ga^{\al}_{\,\,\,\si\nu} - \Ga^{\ro}_{\,\,\,\al\nu} \Ga^{\al}_{\,\,\,\si\mu} \, ,
\eeq
are the corresponding torsion and the curvature tensors. In Eq. \eqref{eq:dGapa}, by ``$\Lie_{\xi} \Ga^{\ro}_{\,\,\,\mu\nu}$" we mean the action of the Lie derivative $\Lie_{\xi}$ on $\Ga^{\ro}_{\mu\nu}$ as if it were a tensor of rank three 
\beq
\Lie_{\xi} \Ga^{\ro}_{\,\,\,\mu\nu} := \xi^{\si} \pa_{\si} \Ga^{\ro}_{\,\,\,\mu\nu} - \Ga^{\si}_{\,\,\,\mu\nu} \pa_{\si} \xi^{\ro} + \Ga^{\ro}_{\,\,\,\si\nu} \pa_{\mu} \xi^{\si} + \Ga^{\ro}_{\,\,\,\mu\si} \pa_{\nu} \xi^{\si} \, .
\eeq
Eq. \eqref{eq:dGapa} shows that the non-tensorial part of $\de_{\xi} \Ga^{\ro}_{\,\,\,\mu\nu}$ is simply $-\pa_{\mu} \pa_{\nu} \xi^\ro$, i.e. the linearization of the rightmost term in Eq. \eqref{eq:Gatrans}. On the other hand, Eq. \eqref{eq:dGana} has the advantage of being explicitly covariant to lowest order, as it should, since $\de_{\xi} \Ga^{\ro}_{\,\,\,\mu\nu}$ is the difference of two infinitesimally close connections. Finally, the torsion and curvature fields defined in Eq. \eqref{eq:TRdef} transform as tensors.

For LLTs we have the connection associated with an internal group, the Lorentz group SO$(1,3)$, just as in Yang-Mills theory, although in that case the group is compact. That connection is therefore a covector under $\cal M$Ds with values in the Lorentz algebra $\Si^{ab}_{\mu} = - \Si^{ba}_{\mu}$ that transforms non-linearly under LLTs
\beq \label{eq:omtrans}
\ti{\Si}^{ab}_{\mu} = \La^a_{\,\,\,c}\, \La^b_{\,\,\,d}\, \Si^{cd}_{\mu} + \La^a_{\,\,\,c} \pa_{\mu} \La^{bc} \, ,
\eeq
known as the ``spin connection''. Alternatively, the antisymmetry in the $ab$ indices can be understood as following from the requirement that the internal Minkowski metric be compatible with the covariant derivative
\beq
0 = \na_{\mu} \et^{ab} \equiv \pa_{\mu} \et^{ab} + \Si^a_{\,\,\,c \mu} \et^{cb} + \Si^b_{\,\,\,c \mu} \et^{ac} \equiv \Si_{\mu}^{ab} + \Si_{\mu}^{ba}  \, .
\eeq
Expressing the Lorentz transformations in terms of generators $\bm{\La} = e^{-\bm{\te}}$, where $\te_{ab} \equiv - \te_{ba}$, we can write the variation of $\Si^{ab}_{\mu}$ under an LLT in Eq. (\ref{eq:omtrans}) as
\beq \label{eq:omvar}
\de_{\te} \Si^{ab}_{\mu} := \ti{\Si}^{ab}_{\mu} - \Si^{ab}_{\mu} = \pa_{\mu} \te^{ab} + \Si^a_{\,\,\,c \mu} \te^{cb} + \Si^b_{\,\,\,c \mu} \te^{ac} + \Ord(\te^2) \equiv \na_{\mu} \te^{ab} + \Ord(\te^2)  \, ,
\eeq
which is again covariant to lowest order since $\de_{\te} \Si^{ab}_{\mu}$ is the difference of two infinitesimally close connections. Since we have two connections, $\Ga$ and $\Si$, we will denote by $\na$ the fully covariant derivative, and by $\na^{\Ga}$ and $\na^{\Si}$ the ones that are covariant only with respect to the corresponding symmetries, when acting on tensors with both types of indices. Next, one can also define torsion and curvature fields associated with the spin connection
\beq \label{eq:TeOmdef}
\Te^a_{\mu\nu} := \na^{\Si}_{\mu} e_{\nu}^a - \na^{\Si}_{\nu} e_{\mu}^a \, , \hspace{1cm} \Om^{ab}_{\mu\nu} := \pa_{\mu} \Si_{\nu}^{ab} - \pa_{\nu} \Si_{\mu}^{ab} + \Si^a_{\,\,\,c\mu} \Si^{cb}_{\nu} - \Si^a_{\,\,\,c\nu} \Si^{cb}_{\mu}  \, .
\eeq
Note that the latter is completely analogous to the field strength 2-form of Yang-Mills theory, while the former depends also on the tetrad information, unlike the affine torsion which depends exclusively on the affine connection \eqref{eq:TRdef}. Both of the fields in \eqref{eq:TeOmdef} are tensors under $\cM$Ds, thanks to the antisymmetric derivatives, and transform linearly under LLTs
\beq
\ti{\Te}^a_{\mu\nu} = \La^a_{\,\,\,b} \Te^b_{\mu\nu} \, , \hspace{1cm} \ti{\Om}^{ab}_{\mu\nu} = \La^a_{\,\,\,c} \La^b_{\,\,\,d} \Om^{cd}_{\mu\nu} \, .
\eeq
In metric GR one selects a preferred affine connection known as the Levi-Civita connection, which is uniquely defined as the metric-compatible and torsion-free connection
\beq
\na_{\ro} g_{\mu\nu} = 0 \, , \hspace{1cm} T^{\ro}_{\,\,\,\mu\nu} = 0 \, ,
\eeq
respectively. The latter implies that $\Ga^{\ro}_{\,\,\,\mu\nu}$ has as many independent components as $\pa_{\ro} g_{\mu\nu}$ and therefore that it is fully determined by the equation $\na_{\ro} g_{\mu\nu} = 0$, whose solution are the Christoffel symbols
\beq
\Ga^{\ro}_{\,\,\,\mu\nu} = \frac{1}{2}\, g^{\ro\si} \( \pa_{\mu} g_{\nu\si} + \pa_{\nu} g_{\mu\si} - \pa_{\si} g_{\mu\nu} \) \, .
\eeq
In tetrad GR, one can proceed in an analogous fashion. We start by requiring that the connections are tetrad-compatible, i.e. the fully covariant equation (also known as ``the tetrad postulate")
\beq \label{eq:tetpost}
\na_{\mu} e_{\nu}^a \equiv \pa_{\mu} e_{\nu}^a - e_{\ro}^a \Ga^{\ro}_{\,\,\,\nu\mu} + \Si^a_{\,\,\,b\mu} e^b_{\nu} = 0 \, .
\eeq
In particular, this implies that we can switch index types through the covariant derivative, e.g.
\beq
\na_{\mu} X^a \equiv \na_{\mu} \( e^a_{\nu} X^{\nu} \) = e^a_{\nu} \na_{\mu} X^{\nu} \, .  
\eeq
By contracting Eq. \eqref{eq:tetpost} with tetrads we derive a relation between the two connections, i.e. the affine connection is
\beq \label{eq:Gaofom}
\Ga^{\mu}_{\,\,\,\nu\ro} = e^{\mu}_a \na^{\Si}_{\ro} e_{\nu}^a \, ,
\eeq
while the spin connection is
\beq \label{eq:omofGa}
\Si^a_{\,\,\,b\mu} = e^a_{\nu} \na_{\mu}^{\Ga} e^{\nu}_b \, .
\eeq
Note that both sides of \eqref{eq:Gaofom} are consistently covariant under LLTs, but not $\cal M$Ds, while both sides of \eqref{eq:omofGa} are consistently covariant under $\cal M$Ds, but not LLTs. In particular, the last equation shows that now the spin connection basically amounts to the information of the parallel transport of the tetrad vectors along themselves. 

Now since the equation of tetrad compatibility \eqref{eq:tetpost} relates uniquely the two connections $\Ga$ and $\Si$, it also relates uniquely the corresponding torsion and curvature tensors defined in Eqs. \eqref{eq:TRdef} and \eqref{eq:TeOmdef}, respectively. Indeed, taking the antisymmetric part of Eq. \eqref{eq:Gaofom} we find that the affine and spin torsions are the same, but just expressed in different bases
\beq \label{eq:torsionrel}
T^{\ro}_{\,\,\,\mu\nu} = e^{\ro}_a \Te^a_{\mu\nu} \, .
\eeq 
On the other hand, plugging Eq. \eqref{eq:omofGa} in Eq. \eqref{eq:TeOmdef} one finds
\beq \label{eq:curvrel}
\Om^{ab}_{\mu\nu} = e_{\ro}^a e^{b\si} R^{\ro}_{\,\,\,\si\mu\nu} \, .
\eeq
In analogy with the metric case, further demanding zero torsion fully determines the connections in terms of the tetrad. The affine connection $\Ga^{\ro}_{\,\,\,\mu\nu}$ is again the Christoffel symbols of the metric, because we have the same conditions
\beq
\na_{\ro} g_{\mu\nu} \equiv e_{a\mu} \na_{\ro} e_{\nu}^a + e_{a\nu} \na_{\ro} e_{\mu}^a = 0 \, , \hspace{1cm} T^{\ro}_{\,\,\,\mu\nu} = 0 \, .
\eeq 
As for the spin connection, the condition $\Te^a_{\mu\nu} = 0$ alone fully determines it in terms of $e^a_{\mu}$, because these are as many equations as the number of components in $\Si^{ab}_{\mu}$, i.e. independently of whether tetrad compatibility is imposed or not. One thus finds the torsion-free spin connection
\beq \label{eq:LVsc}
\Si^{ab}_{\mu} = e^{a\nu} \pa_{[\mu} e_{\nu]}^b - e^{b\nu} \pa_{[\mu} e_{\nu]}^a - e_{c \mu} e^{a\nu} e^{b\ro} \pa_{[\nu} e_{\ro]}^c   \, ,
\eeq
which can also be expressed as a linear combination of the structure coefficients of Eq. (\ref{eq:Cdef})
\beq
\Si_{ab\mu} \equiv \frac{1}{2} \[ C_{abc} - C_{bca} - C_{cab} \] e_{\mu}^c  \, .
\eeq
With the condition of zero torsion $R_{\mu\nu\ro\si}$ becomes the Riemann tensor of $g_{\mu\nu}$ and, given Eq. \eqref{eq:curvrel}, $\Om^{ab}_{\mu\nu}$ is nothing but the Riemann tensor partially expressed in the tetrad basis, so from now on we write $R^{ab}_{\mu\nu}$ instead. From this relation one sees trivially why the first pair of indices of the Riemann tensor is antisymmetric as well, because in the tetrad viewpoint these index pair parametrizes the Lorentz algebra. One can then define the usual curvature tensors with only Lorentz indices
\beq \label{eq:fullLorentzR}
R_{abcd} := R_{ab\mu\nu} e^{\mu}_c e^{\nu}_d \, , \hspace{1cm} R_{ab} := R^c_{\,\,\,acb} \equiv e_a^{\mu} e_b^{\nu} R_{\mu\nu}[g] \, , \hspace{1cm} R := R^a_{\,\,\,a} \equiv R[g] \, , 
\eeq
and also the Weyl tensor
\beq \label{eq:Weyldef}
W_{abcd} := R_{abcd} - \et_{a[c} R_{d]b} + \et_{b[c} R_{d]a} + \frac{1}{3}\, \et_{a[c} \et_{d]b} R \equiv e_a^{\mu} e_b^{\nu} e_c^{\ro} e_d^{\si} W_{\mu\nu\ro\si}[g] \, .
\eeq
Because the latter is fully traceless and shares all the symmetries of the Riemann tensor, all of its information lies in two symmetric traceless spatial tensors, the so-called ``electric'' and ``magnetic'' components associated with the observer $e_a^{\mu}$
\beq \label{eq:EBWeyldef}
E_{ij} := W_{0i0j} \, , \hspace{1cm} B_{ij} := - \frac{1}{2}\, \vep_{ikl} W_{0jkl} \, ,
\eeq
since we then have
\beq
W_{0ijk} = - B_{il} \vep_{ljk} \, , \hspace{1cm} W_{ijkl} = \de_{ik} E_{jl} - \de_{il} E_{jk} - \de_{jk} E_{il} + \de_{jl} E_{ik} \, .
\eeq
As for the electromagnetic field $A_{\mu}$, it is a $\cM$D covector and LLT scalar, but varies under a U(1) gauge transformation (U(1)GT)
\beq \label{eq:U1gt}
\ti{A}_{\mu} = A_{\mu} + \pa_{\mu} \te \, ,
\eeq
so that the invariant curvature is the Maxwell tensor
\beq
F_{\mu\nu} := \pa_{\mu} A_{\nu} - \pa_{\nu} A_{\mu} \, , \hspace{1cm} F_{ab} := e_a^{\mu} e_b^{\nu} F_{\mu\nu} \, .
\eeq
The electric and magnetic fields measured by the observer family $e_a^{\mu}$ are then 
\beq
E^i := F^{0i} \, , \hspace{1cm} B^i := \frac{1}{2}\, \vep^{ijk} F^{jk} \, ,
\eeq
respectively. Finally, it is also useful to define
\beq \label{eq:fullLorentz}
\Si_{abc} := \Si_{ab\mu} e^{\mu}_c \, , \hspace{1cm} A_a := e_a^{\mu} A_{\mu} \, , \hspace{1cm}  \pa_a := e_a^{\mu} \pa_{\mu} \, ,  \hspace{1cm} \na_a := e_a^{\mu} \na_{\mu} \, ,
\eeq
where the $\Si_{abc}$ are also known as the ``Ricci rotation coefficients". Rearranging \eqref{eq:omofGa} we find
\beq \label{eq:RicciRot}
\na_a^{\Ga} e^{\mu}_b \equiv \Si^c_{\,\,\,ba} e_c^{\mu}  \, ,
\eeq
so these coefficients control the parallel transport of the tetrad vectors along themselves. In particular, since $e_0^{\mu}$ is the 4-velocity, the 4-acceleration vector $a^{\mu} := \na_0^{\Ga} e_0^{\mu}$ in the tetrad basis is nothing but
\beq \label{eq:omi00acc}
a^0 \equiv 0 \, , \hspace{1cm} a^i \equiv \Si^i_{\,\,\,00} \, .
\eeq
We can also compute the Fermi-Walker derivative of the spatial frame along the 4-velocity field
\beq \label{eq:omij0prec}
\na^{\rm FW}_0 e_i^{\mu} := \na_0^{\Ga} e_i^{\mu} - g \( e_i, a \) e_0^{\mu} + g \( e_i, e_0 \) a^{\mu} \equiv -\, \Si_{ij0} e_j^{\mu} - \Si_{i00} e_0^{\mu} \, .
\eeq
We see that the time components of the spin connection $\Si_{i00}$ and $\Si_{ij0}$ control the 4-acceleration of the observer family and the precession of its spatial frame, respectively. 
 
To conclude this subsection, we stress again that, although the tetrad has both types of indices, it only requires the notion of spin connection to form fully covariant objects. This is because the condition of zero spin torsion $\Te^a_{\mu\nu} = 0$ amounts to as many equations as the number of components in $\Si^{ab}_{\mu}$ and thus determines the latter uniquely in terms of the tetrad (Eq. \eqref{eq:LVsc}). One can then form full scalars through the curvature of $\Si^{ab}_{\mu}$. The tetrad-compatibility condition $\na_{\mu} e_{\nu}^a = 0$ is therefore superfluous and can be alternatively understood as a way of constructing an affine connection $\Ga^{\ro}_{\,\,\,\mu\nu}$ out of the tetrad, i.e. Eq. \eqref{eq:Gaofom}, if one wants to introduce one. As for the matter sector, note that there too $\Ga^{\ro}_{\,\,\,\mu\nu}$ is not required because all derivatives can be expressed as antisymmetric (``exterior'') derivatives of differential forms (at least in the Standard Model). On the other hand, a tetrad and a spin connection {\it are} required in order to couple to spinors, as we will see in subsection \ref{sec:spinors}.

\subsubsection{Tetrad General Relativity} \label{sec:TGR}

The action we will consider is made of three terms
\beq
S = S_{\rm EH} + S_{\rm EM} + S_{\rm m} \, ,
\eeq
where 
\beq \label{eq:SEH}
S_{\rm EH} = \frac{1}{16\pi G} \int \ed^4 x \, e \[ e_a^{\mu} e_b^{\nu} R_{\mu\nu}^{ab} - 2 \La \] \, ,
\eeq
is the Einstein-Hilbert action in the tetrad language,
\beq
S_{\rm EM} = - \frac{1}{4} \int \ed^4 x\, e F_{\mu\nu} F^{\mu\nu} \, , 
\eeq
is the Maxwell action of electrodynamics and $S_{\rm m}$ is the matter action. However, since matter will be described through the Boltzmann formalism, we will not specify $S_{\rm m}$, but will inlcude the matter content directly at the level of the equations of motion. Setting to zero the variation of $S$ with respect to $e_a^{\mu}$ we find 
\beq \label{eq:EOM}
R^{ab}_{\mu\nu} e_b^{\nu} - \frac{1}{2}\, R^{bc}_{\nu\ro} e_b^{\nu} e_c^{\ro} e_{\mu}^a  + \La e_{\mu}^a =  8\pi G\, T_{\mu}^a \, ,
\eeq
where we defined
\beq
T_{\mu}^a := - \frac{1}{e} \frac{\de}{\de e_a^{\mu}} \[ S_{\rm EM} + S_{\rm m} \] \, .
\eeq
If $S_{\rm m}$ depends on $e_{\mu}^a$ only through the metric combination in Eq. (\ref{eq:gofe}), we find that $T_{\mu}^a$ is indeed the energy-momentum tensor with one index in the tetrad basis
\bea
T_{\mu}^a & := & - \frac{1}{e} \frac{\de }{\de e_a^{\mu}} \[ S_{\rm EM} + S_{\rm m} \]  \equiv - \frac{1}{\sqrt{-g}}\, \frac{\pa g^{\nu\ro}}{\pa e_a^{\mu}}\, \frac{\de }{\de g^{\nu\ro}} \[ S_{\rm EM} + S_{\rm m} \]  \nn \\
 & \equiv & - \frac{2}{\sqrt{-g}}\, e^{a \nu} \frac{\de }{\de g^{\mu\nu}} \[ S_{\rm EM} + S_{\rm m} \]  \equiv e^{a \nu} T_{\mu\nu} \, .
\eea
Thus, if we express the equations of motion with only diffeomorphism indices, i.e. contracting Eq. (\ref{eq:EOM}) with $e_{a\nu}$, we recover the standard Einstein equations of the metric $g_{\mu\nu}$
\beq \label{eq:EOME}
R_{\mu\nu} - \frac{1}{2}\, g_{\mu\nu} R + \La g_{\mu\nu} =  8 \pi G T_{\mu\nu} \, , \hspace{1cm} T_{\mu\nu} := e^a_{\mu} T_{a\nu}  \, .
\eeq
As a last alternative, one can consider the equation with only Lorentz indices, i.e. contracting Eq. (\ref{eq:EOM}) with $e_b^{\mu}$
\beq \label{eq:EIJ}
R_{ab} - \frac{1}{2}\, \et_{ab} R + \La \et_{ab} = 8\pi G\, T_{ab}  \, , \hspace{1cm} T_{ab} := T_{a\mu} e^{\mu}_b \, .
\eeq
All tensors in Eq. (\ref{eq:EIJ}) are also symmetric in $ab$. To understand this, note that the presence of an $N$-dimensional gauge symmetry reflects itself as $N$ undetermined field combinations and thus as $N$ identities satisfied by the equations of motion. The $\cal M$D symmetry manifests itself as the Bianchi identity $\na_a R^{ab} \equiv \na^b R/2$ and thus energy-momentum conservation
\beq \label{eq:EMcons}
\na_a T^{ab} = 0  \, ,
\eeq
when the equations of motion are satisfied, while the LLT symmetry manifests itself as the vanishing of the antisymmetric part $R_{[ab]} \equiv 0$ and thus leads to
\beq
T_{[ab]} = 0 \, , 
\eeq
again when the equations of motion are satisfied.

Given the interpretation of the tetrad, the $T^{ab}$ components are the energy density $T^{00}$, momentum density $T^{0i}$, pressure $T^{ii}/3$ and anisotropic stress $T^{ij} - \de^{ij} T^{kk}/3$ measured by the observer family. The electromagnetic contribution being
\beq \label{eq:EMTEM}
T_{ab}^{\rm EM} = F_{ac} F_b^{\,\,\,c} - \frac{1}{4} \, \et_{ab} \, F_{cd} F^{cd} \, ,
\eeq 
we have
\bea
T_{\rm EM}^{00} & \equiv & \frac{1}{2} \( E^i E^i + B^i B^i \) \, , \\
T_{\rm EM}^{0i} & \equiv & \vep^{ijk} E^j B^k  \, , \\
\frac{1}{3}\, T_{\rm EM}^{ii} & \equiv & \frac{1}{3}\, T_{\rm EM}^{00}  \, , \\
T_{\rm EM}^{ij} - \frac{1}{3}\, \de^{ij} T_{\rm EM}^{kk} & \equiv & - E^i E^j - B^i B^j + \frac{1}{3}\, \de^{ij} \( E^k E^k + B^k B^k \) \, .
\eea
As for the matter contribution $T^{\rm m}_{ab}$, it makes sense to decompose it in its own rest-frame. The fluid 4-velocity $V^a$ with respect to the observer family $e_a$ is defined as the unit-normed time-like eigenvector 
\beq
T^{\rm m}_{ab} V^b = - \ro V_a \, , \hspace{1cm} V_a V^a \equiv -1 \, ,
\eeq
with the eigenvalue $\ro$ being the rest-frame energy density, and thus   
\beq
T^{\rm m}_{ab} = \( \ro + p \) V_a V_b + p\, \et_{ab} + \Pi_{ab} \, ,
\eeq
where $p$ is the pressure and $\Pi_{ab}$ is the anisotropic stress tensor, obeying
\beq
\Pi_a^a \equiv 0 \, , \hspace{1cm} V^a \Pi_{ab} \equiv 0 \, .
\eeq
The energy density, momentum density, pressure and anisotropic stress measured by the observer in terms of the ones in the fluid's rest-frame are then simply
\bea
T_{\rm m}^{00} & \equiv & \ro \( 1 + V^i V^i \) + p V^i V^i + \Pi^{ii} \, , \\
T_{\rm m}^{0i} & \equiv & \[ \( \ro + p \) \( 1 + V^k V^k \) \de^{ij} + \Pi^{ij} \] \frac{V^j}{\sqrt{1 + V^l V^l}}  \, , \\
\frac{1}{3}\, T_{\rm m}^{ii} & \equiv & p + \frac{1}{3} \[ \( \ro + p \) V^i V^i + \Pi^{ii} \] \, , \\
T_{\rm m}^{ij} - \frac{1}{3}\, \de^{ij} T_{\rm m}^{kk} & \equiv & \( \ro + p \) V^i V^j + \Pi^{ij} - \frac{1}{3}\, \de^{ij} \[ \( \ro + p \) V^k V^k + \Pi^{kk} \] \, ,
\eea
respectively, and we have used the algebraic constraints of $V_a$ and $\Pi_{ab}$. Finally, setting to zero the variation of $S$ with respect to $A_{\mu}$ and contracting with a tetrad we find the Maxwell equation in the tetrad basis
\beq \label{eq:Maxwell}
\na_a F^{ab} = J^b \, ,
\eeq
where
\beq \label{eq:Jcurrentdef}
J^a := -e^a_{\mu}\, \frac{\de S_{\rm m}}{\de A_{\mu}} \, ,
\eeq
is the total electric current measured by the observer family. As a consequence of U(1)GT invariance, it is conserved
\beq  \label{eq:Jelcons}
\na_a J^a = 0 \, ,
\eeq 
when the equations of motion are satisfied, as is required for the consistency of Eq. \eqref{eq:Maxwell}.

\subsubsection{Spinors and gravity} \label{sec:spinors}

Let us now consider the description of spinors in the presence of gravity, which will be relevant when dealing with the Boltzmann equation of spin-1/2 particles. It is another important advantage of the tetrad formalism that it arises as the only way to incorporate spinor fields in the presence of a non-trivial geometry. To understand this, note first that spinors are defined as half-integer spin representations of the universal cover of the Lorentz group SO$(1,3)$ that is SL$(2,\Cs)$. In the metric description one usually interprets the diffeomorphism group as a generalization of the Poincar\'e transformations of Minkowski space-time, i.e.
\beq
\ti{x}^{\mu} = \La^{\mu}_{\,\,\,\nu} x^{\nu} + a^{\mu} \to f^{\mu}(x) \, ,
\eeq
where $\bm{\La}$ and $a$ are a constant Lorentz matrix and vector. This then implies a simple generalization for the transformation of tensor indices, i.e. integer spin representations, that is given by the replacement (here for a vector)
\beq
\ti{V}^{\mu}(\ti{x}) = \La^{\mu}_{\,\,\,\nu} V^{\nu}(x) \to \frac{\pa f^{\mu}}{\pa x^{\nu}}\, V^{\nu}(x) \, ,
\eeq
in the transformation rules of flat space-time. But the possibility of performing this replacement relies crucially upon the fact that the indices of the Lorentz matrix are space-time indices, i.e. that SO$(1,3)$ can be viewed as a subgroup of the diffeomorphism group. This is not the case of SL$(2,\Cs)$, so there is no such generalization for the half-integer representations where the corresponding transformation matrix $U(\La)$ has spinor indices. Spinors therefore require the action of the Lorentz group, be it global or local, and thus the presence of tetrads. 

Here we will work for definiteness with Dirac spinors, because all types of spin-1/2 particles can be expressed in this representation. We will consider a Dirac mass in concrete examples, but the Majorana and massless cases can be easily obtained with minor modifications. We will denote the Dirac indices by $\frak{a}, \frak{b}, \frak{c}, \dots$, but most of the time these will be kept implicit for simplicity, as is customary. Thus, a Dirac spinor $\psi^{\frak{a}}(x)$ is a set of scalars under $\cM$Ds, while under an LLT with parameter $\te_{ab}(x)$
\beq
\ti{\psi}(x) = U(x)\,\psi(x) \, , \hspace{1cm} U(x) := \exp \[ - \frac{1}{4}\, \te_{ab}(x)\, \ga^a \ga^b \] \, , 
\eeq
where the $\ga^a$ are the gamma matrices obey the Clifford algebra
\beq \label{eq:Clifford}
\{ \ga^a, \ga^b \} = -2 \et^{ab} \, .
\eeq
Now that the spinor indices are ``internal" from the viewpoint of $\cM$Ds, the local action of the Lorentz group on $\psi$ is qualitatively the same as the one of the SU$(N)$ group in Yang-Mills theory. The only difference is that the group dimension and signature are related to the ones of the space-time manifold $\cM$. 

The need for a tetrad becomes also obvious when trying to construct the kinetic part of the general-relativistic Dirac Lagrangian. One must turn the diffeomorphism index of the derivative $\pa_{\mu}$ into a Lorentz index in order to contract with $\ga^a$, i.e. the combination $\ga^a e_a^{\mu} \pa_{\mu} \equiv \ga^a \pa_a$.\footnote{Another common choice is to define instead the Dirac matrix fields $\ga^{\mu}(x) := \ga^a e_a^{\mu}(x)$, which therefore obey the modified Clifford algebra $\{ \ga^{\mu}, \ga^{\nu} \} \equiv -2g^{\mu\nu}$, so that $\ga^a e_a^{\mu} \pa_{\mu} \equiv \ga^{\mu} \pa_{\mu}$, but all this is only a matter of interpretation.} Moreover, since the Lorentz group now acts locally, one must consider the covariant derivative in the Dirac representation
\beq
\na_{\mu} \psi := \[ \pa_{\mu} + \frac{1}{4}\, \Si_{ab \mu} \ga^a \ga^b \] \psi \, . 
\eeq
The Dirac Lagrangian reads
\bea
L_{\rm D} & := & \frac{i}{2} \[ \bar{\psi} \ga^a \na_a \psi - \na_a \bar{\psi} \ga^a \psi \] - m \bar{\psi} \psi \nn \\
 & \equiv & \frac{i}{2} \[ \bar{\psi}\, \ga^a \pa_a \psi - \pa_a \bar{\psi}\, \ga^a \psi  \] - m \bar{\psi} \psi - \frac{1}{2}\, \vep_{abcd}\, \Si^{abc} S^d \, , \label{eq:LDirac}
\eea
where we have used the identity 
\beq \label{eq:3gammaiden}
\ga_a \ga_b \ga_c \equiv -\et_{ab} \ga_c + \et_{ca} \ga_b - \et_{bc} \ga_a + i \vep_{abcd} \ga^d \ga^5   \, , \hspace{1cm} \ga^5 := i \ga^0 \ga^1 \ga^2 \ga^3 \, ,
\eeq
and
\beq  \label{eq:JJtcurrents}
S^a := \frac{1}{2}\, \bar{\psi} \ga^a \ga^5 \psi \, ,
\eeq
is the spin pseudo-current. Perhaps a more familiar expression for the latter is found when its spatial part is expressed in terms of left/right-handed Weyl spinors
\beq
S^i \equiv \frac{1}{2} \[ \psi^{\dagger}_L \si^i \psi_L + \psi^{\dagger}_R \si^i \psi_R \] \, ,
\eeq 
where the $\si^i$ are the Pauli matrices. Just as the electric current is defined through Eq. \eqref{eq:Jcurrentdef}, the spin current can be defined by varying the Dirac action with respect to an independent spin connection
\beq \label{eq:spincurrent}
S_{abc} := - 2e_{c\mu} \frac{\de S_{\rm D}}{\de \Si^{ab}_{\mu}} = \vep_{abcd} S^d \, .
\eeq
As for the energy-momentum tensor
\beq \label{eq:Tabpsi}
T_{ab} := - \frac{1}{e} \, e_b^{\mu} \, \frac{\de S_{\rm D}}{\de e^{a\mu}} \equiv - \frac{i}{2} \[ \bar{\psi} \ga_a \na_b \psi - \na_b \bar{\psi} \ga_a \psi  \] - \frac{1}{2}\,\na_c S_{ab}^{\,\,\,\,\,\,c} + \et_{ab} L_{\rm D} \, ,
\eeq
it is indeed symmetric when the equation of motion of $\psi$ 
\beq
\[ i\ga^a \na_a - m \] \psi = 0 \, , 
\eeq
is satisfied. To see this, multiply the above equation with $\ga^{[b} \ga^{c]}$ from the left and use \eqref{eq:3gammaiden} to find the relation
\beq
\ga^{[a} \na^{b]} \psi = \frac{i}{2} \[ m \ga^{[a} \ga^{b]} - \vep^{abcd} \ga_c \ga^5 \na_d \] \psi \, ,
\eeq 
with which it is trivial to show that the antisymmetric part of \eqref{eq:Tabpsi} vanishes. 

Finally, let us come back to the connection with the action of the Poincar\'e group in Minkowski space-time field theory. In that case there exists a gauge where $e_{\mu}^a = \de_{\mu}^a$, thus effectively identifying the two types of indices. This gauge fixing is then preserved under a subgroup of combined $\cal M$Ds and LLTs (which we express in terms of the passive transformation for the former)
\beq
\ti{x}^{\mu} = \La^{\mu}_{\,\,\,\nu} x^{\nu} + a^{\mu} \, , \hspace{1cm} \La^a_{\,\,\,b} = \de^a_{\mu} \La^{\mu}_{\,\,\,\nu} \de_b^{\nu}  \, .
\eeq 
The necessity of combining the two symmetries in order to preserve $e_{\mu}^a = \de_{\mu}^a$ is what gives to the Dirac spinor on flat space-time its usual transformation under the Poincar\'e group, i.e. the Lorentz transformation acting on the coordinate-dependence and the one mixing the spinor indices are the same. If we now compute the Noether current associated with the global Lorentz symmetry, we find $J_{abc} = L_{abc} + S_{abc}$, where $L_{abc}$ is the ``orbital" part, while $S_{abc}$ is the intrinsic spin part given in Eq. \eqref{eq:spincurrent}. On the other hand, if we compute the Noether current associated with space-time translations we find Eq. \eqref{eq:Tabpsi}, but without the $\sim \na_c S_{ab}^{\,\,\,\,\,\,c}$ term, which is therefore not symmetric. This is a well-known feature and, as with any Noether current, one has the freedom to add an independently conserved term to obtain some desired property.\footnote{Indeed, the divergence of the extra term is $\sim \na^b \na_c S_{ab}^{\,\,\,\,\,\,c} \equiv \vep_{abcd} \na^b \na^c \ti{S}^d \equiv - \vep_{abcd} R^{ebcd} \ti{S}_e \equiv 0$.} Here we see that this ``corrective" additional term $\sim \na_c S_{ab}^{\,\,\,\,\,\,c}$ is automatically obtained in the definition \eqref{eq:Tabpsi} of $T_{ab}$ thanks to the fact that we considered the torsion-free spin connection. The resulting symmetric tensor is known as the ``Belinfante-Rosenfeld" energy-momentum tensor.

\subsubsection{Parallel-transported tetrads} \label{sec:paralltrans}

Let us now discuss a specific class of observers that appears as a natural choice in cosmology. Indeed, as a first approximation on cosmological scales, observers and/or sources are in free-fall, i.e. their 4-velocity obeys the geodesic equation with respect to the weak gravitational fields involved in cosmological perturbation theory. From Eq. \eqref{eq:omi00acc} we see that this corresponds to setting
\beq \label{eq:e0fix0}
\Si_{i00} = 0 \, .
\eeq
This fixes the local boost symmetry, but not the local rotation one, since the latter acts linearly on $\Si_{i00}$. We can thus still locally rotate the spatial frames with an arbitrary rotation matrix field 
\beq \label{eq:resrot}
e_0^{\mu}(x) \to e_0^{\mu}(x) \, , \hspace{1cm} e_i^{\mu}(x) \to R_i^{\,\,j}(x) \, e_j^{\mu}(x) \, , \hspace{1cm} R_i^{\,\,k} R_j^{\,\,k} = \de_{ij} \, .
\eeq
Unlike the case of $e_0^{\mu}$, however, the motion of the observers alone does not provide a privileged $e_i^{\mu}$. Therefore, in the absence of any more input, the simplest motion along a geodesic flow, which can be expressed through a $\cal M$D-covariant condition, is the parallel transport along $e_0^{\mu}$, just as it is the case for $e_0^{\mu}$ itself
\beq \label{eq:eifix0}
\na^{\Ga}_0 e_i^{\mu} = 0 \, , \hspace{1cm} \Leftrightarrow \hspace{1cm} \Si_{ij0} = 0 \, ,
\eeq
which, according to Eq. \eqref{eq:omij0prec}, amounts to choosing non-precessing frames. Along with Eq. (\ref{eq:eifix0}), we thus derive
\beq \label{eq:PTT}
\na^{\Ga}_0 e_a^{\mu} = 0 \, , \hspace{1cm} \Leftrightarrow \hspace{1cm} \Si_{ab0} = 0 \, ,
\eeq
and we can therefore refer to it as the ``parallel-transported tetrad gauge'' (PTT). The fact that we can express this gauge in terms of the spin connection is convenient, because we directly see that it corresponds to exactly as many conditions as the number of dimensions of the Lorentz group (six). This, however, does not correspond to a complete gauge fixing. Indeed, the condition (\ref{eq:PTT}) is similar to the Weyl gauge of electrodynamics, where the time component of the vector potential $A_{\mu}$ is set to zero
\beq
A_t = 0 \, .
\eeq
The subtlety here is that we actually have the Lorentz time-component $\Si_{ab0}$ instead of the coordinate one $\Si_{ab t}$. In the Weyl gauge of electrodynamics one has a residual gauge symmetry that are the time-independent transformations, i.e. Eq. \eqref{eq:U1gt} with $\te = \te(\vec{x})$. Here, using the transformation of $e_{\mu}^a$ and $\Si^{ab}_{\mu}$ under a LLT, we find that the condition (\ref{eq:PTT}) is maintained by a further LLT if the latter satisfies
\beq \label{eq:LLTres}
\La_0^{\,\,\,c} \pa_c \La^{ab} + \La_0^{\,\,\,i} \La^a_{\,\,\,c} \Si^{bc}_{\,\,\,\,\,\,i} = 0 \, .
\eeq
To get more insight into this equation we can express the Lorentz matrix in terms of the generators $\bm{\La} = e^{-\bm{\te}}$ and consider the transformation to linear order in $\te_{ab}$ 
\beq \label{eq:LLTreste}
\pa_0 \te_{ab} - \te_{0i} \, \Si_{abi} + \Ord(\te^2) = 0 \, .
\eeq
Since $e_0^{\mu}$ is time-like, $\pa_0 \equiv e_0^{\mu} \pa_{\mu}$ takes the form of a convective derivative, up to some multiplicative factor, so we know that this type of equation admits solutions, at least locally. We thus have a residual gauge symmetry of the same kind as in electrodynamics. 

We can now understand this situation as follows. The fact that the PTT gauge (\ref{eq:PTT}) involves only a time-like derivation of the tetrad means that it only determines its evolution in time. The residual LLTs then correspond to the freedom to choose the frame arbitrarily on some time-like hypersurface, say the initial data surface. Once this is done, the full tetrad field is uniquely determined by Eq. (\ref{eq:PTT}). In the case of boosts, the residual gauge freedom amounts to the freedom of choosing among all possible free-falling observer 4-velocity fields, while for rotations, it amounts to the freedom of choosing among all possible parallel-transported spatial frames along $e_0^{\mu}$. Although the PTT gauge is physically suitable in many cases, here we will not fix the LLT symmetry to maintain generality.

\subsection{World-line fields}  \label{sec:L}

In this section we consider the dynamics of a free point-particle of mass $m$ and charge $q$ from a field-theoretical viewpoint. The particle position now appears as four scalar fields on a 1-dimensional manifold $\Lie$, with values in the space-time manifold $\cM$, and we consider the associated action. We also discuss the geometry and symmetries associated with $\cL$, especially in the $m,q = 0$ case, a feature which will be useful in the construction of the observer space. Finally, we derive the particle equations of motion in the tetrad formalism.

\subsubsection{Geometrical considerations} \label{sec:geodefprop}

Geometrically speaking, a path of a point-particle in $\cal M$ is a map from a one-dimensional manifold $\Lie$ to the space-time manifold $\cal M$ 
\beq
\ga : \Lie \to {\cal M} \, , 
\eeq
and its image $\ga(\Lie) \subset \cM$ is the particle's world-line. Given a coordinatization $\la : \Lie \to \Rs$ and the space-time coordinates $x^{\mu} : \cM \to \Rs^4$, we can describe the $\ga$ map through four functions $\ga^{\mu}(\la)$. The latter is a set of four scalars under $\Lie$-diffeomorphisms ($\Lie$D). In the passive version (P$\Lie$D) that are the $\la$-reparametrizations, we have
\beq
\ti{\la} = \ti{\la}(\la) \, , \hspace{1cm} \Rightarrow \hspace{1cm} \ti{\ga}^{\mu}(\ti{\la}) = \ga^{\mu}(\la) \, ,
\eeq
while in the active version (A$\Lie$D) we have the action of the Lie derivative with respect to a vector field $\ka$ on $\Lie$
\beq
\de_{\ka} \ga^{\mu} = - \Lie_{\ka} \ga^{\mu} + \Ord(\ka^2) \equiv - \ka \pa_{\la} \ga^{\mu} + \Ord(\ka^2) \, .
\eeq
In the case of P$\cal M$Ds, the $\ga^{\mu}$ transform as coordinates
\beq \label{eq:PMDga}
\ti{x}^{\mu} = \ti{x}^{\mu}(x) \, , \hspace{1cm} \Rightarrow \hspace{1cm} \ti{\ga}^{\mu}(\la) = \ti{x}^{\mu}(\ga(\la)) \, .
\eeq
In the case of A$\cal M$Ds, however, there is a subtlety. If we interpret $\ga$ as simply some continuous collection of points in $\cal M$ with no relation to the fields whatsoever, then our pullback definition of A$\cal M$Ds would suggest
\beq
\de_{\xi} \ga^{\mu} = 0 \, ,
\eeq
since we move the fields while keeping the points fixed. However, here the $\ga^{\mu}$ we are interested in is not any path, but one that must ultimately obey the geodesic equation, which therefore relates it to the space-time fields. This means that, if we pullback the fields, then we must also move the geodesic in order to obtain the same physical configuration and thus a symmetry of the geodesic equation. Therefore, by treating $\ga$ as a ``dynamical" object, i.e. one that has an action coupled to $g_{\mu\nu}$, we force it to transform as all other dynamical objects (the space-time fields) in order to maintain a symmetry. The A$\cal M$D transformation of a geodesic is therefore simply given by the expression of the passive one in Eq. (\ref{eq:PMDga}) in terms of the generating vector field
\beq \label{eq:dexigamu}
\de_{\xi} \ga^{\mu} = \xi^{\mu}(\ga) + \Ord(\xi^2) \, .
\eeq
If we do not transform $\ga^{\mu}$ along with the fields, then we effectively obtain a configuration corresponding to a different path than the one we started with, so the geodesic equation is not invariant. As a result, the P$\cal M$Ds and A$\cal M$Ds have basically no conceptual difference for a geodesic and thus also for the (scalar) space-time fields $X$ evaluated on it, i.e. we have respectively
\beq
\ti{X}(\ti{\ga}) = X(\ga) \, , \hspace{1cm} \de_{\xi}[X(\ga)] = [\de_{\xi} X](\ga) + \de_{\xi} \ga^{\mu} [\pa_{\mu} X](\ga) + \Ord((\de_{\xi}\ga)^2) = 0 \, .
\eeq 
For this reason, the ``passive'' and ``active'' prefixes will only refer to the way that same transformation is usually expressed, i.e. in terms of a coordinate transformation, or in terms of the generating vector field, respectively. This situation is ultimately due to the fact that, for $\ga$, the $\cM$D symmetries are internal symmetries, and there is thus no distinction bewteen passive and active versions, as it is for instance also the case for LLTs. This is also why we only referred to P$\cM$Ds in the transformation rules of a cosmological observable in section \ref{sec:advtet}, because the A$\cM$Ds are basically the same transformation.\footnote{In fact, in the case of cosmological observables the $\de_{\xi} \ga^{\mu} = 0$ choice, i.e. not moving the geodesic along, would lead to yet another problem. If the space-time fields move while the points are held fixed, then it is not guaranteed that the resulting $\ti{C}(P, \hat{P})$ will still be connecting two points that are linked by a light-like geodesic. This means that only a subgroup of such active diffeomorphisms is actually defined on a cosmological observable linking $\hat{P}$ and $P$.} In contrast to the above remarks about $\cM$Ds, the distinction between P$\Lie$Ds and A$\Lie$Ds is important, because the solutions of observables will involve integrals over $\Lie$ with non-trivial boundaries (see appendix \ref{app:diffsym}).

\subsubsection{Action for point-particles} \label{eq:geoaction}

We start by considering the following action
\beq \label{eq:SPm}
S = \int \ed \la \[ \frac{1}{2} \, \ep^{-1} g_{\mu\nu}(\ga)\, \pa_{\la} \ga^{\mu} \pa_{\la} \ga^{\nu} - q A_{\mu}(\ga)\, \pa_{\la} \ga^{\mu} - \frac{1}{2}\,\ep m^2 \] \, ,
\eeq
where $\ep$ is the analogue of a tetrad covector on $\Lie$, i.e. a ``monad'' (or ``einbein''). Here we will adopt the convention of $\la$ having length dimensions, which means that $\ep$ has dimensions of length too. The monad is invariant under $\cM$Ds and LLTs, since it knows nothing about $\cM$, while under a P$\Lie$D it transforms as a covector
\beq \label{eq:epPLDtrans}
\ti{\ep}(\ti{\la}) = \frac{\pa \la}{\pa \ti{\la}} (\ti{\la}(\la))\, \ep(\la) \, ,
\eeq
and under an A$\Lie$D
\beq
\de_{\ka} \ep = - \Lie_{\ka} \ep + \Ord(\ka^2) \equiv - \ka \pa_{\la} \ep - \ep \pa_{\la} \ka + \Ord(\ka^2) \equiv - \pa_{\la} (\ka \ep) + \Ord(\ka^2) \, .
\eeq
The action (\ref{eq:SPm}) is then invariant under $\cal M$Ds and P$\Lie$Ds and varies by a boundary term under A$\Lie$Ds and U(1)GTs, so all these transformations are symmetries. In particular, we recognize that $-\ep^2$ is the metric and $\ep$ the corresponding volume form on $\Lie$. Here there is no non-trivial analogue of the LLT symmetry because the corresponding group is trivial in one dimension. Thus, (\ref{eq:SPm}) is the action of four scalar fields $\ga^{\mu}$ on a manifold $\Lie$ with monad $\ep$ and ``cosmological" constant $m^2$. We can then consider the Legendre transform of this action with respect to $\pa_{\la} \ga^{\mu}$, where the conjugate momentum is
\beq \label{eq:kofgadot}
K_{\mu} := \frac{\pa L}{\pa (\pa_{\la} \ga^{\mu})} = g_{\mu\nu}(\ga) \, \ep^{-1} \pa_{\la} \ga^{\nu} - q A_{\mu}(\ga) \, ,
\eeq
to obtain the canonical action
\beq \label{eq:Sgeocan}
S = \int \ed \la \[ K_{\mu} \pa_{\la} \ga^{\mu} - \ep H \] \, , \hspace{1cm} H := \frac{1}{2} \[ m^2 + g^{\mu\nu}(\ga) \( K_{\mu} + q A_{\mu}(\ga) \) \( K_{\nu} + q A_{\nu}(\ga) \) \] \, .
\eeq
As expected, $\ep$ plays the same role as the lapse function in the ADM formulation of GR, so a choice of $\ep$ amounts to a choice of $\la$-parametrization. The constraint it imposes $H = 0$ is nothing but the dispersion relation for a particle of mass $m$ and charge $q$. Note also that $K_{\mu}$ transforms like a covector under $\cM$Ds, but it transforms non-linearly under U(1)GTs \eqref{eq:U1gt}
\beq \label{eq:U1gtk}
\ti{K}_{\mu} = K_{\mu} - q (\pa_{\mu} \te)(\ga) \, ,
\eeq
so that the combination $K_{\mu} + q A_{\mu}(\ga)$ appearing in $H$ is consistently invariant. Going back to the Lagrangian description \eqref{eq:SPm}, if $m \neq 0$, we can integrate out $\ep$, i.e. replace it with the solution of its own equation of motion
\beq \label{eq:epsol}
\ep := \frac{1}{m} \, \sqrt{- g_{\mu\nu}(\ga)\, \pa_{\la} \ga^{\mu} \pa_{\la} \ga^{\nu}} \, ,
\eeq
to derive the well-known action\footnote{This is the one-dimensional analogue of the relation between the Polyakov and  Nambu-Goto actions, including the Kalb-Ramond term, in string theory \cite{Becker:2007zj}.}
\beq \label{eq:Sm}
S = - \int \ed \la \[ m \sqrt{- g_{\mu\nu}(\ga)\, \pa_{\la} \ga^{\mu} \pa_{\la} \ga^{\nu}} + q A_{\mu}(\ga)\, \pa_{\la} \ga^{\mu} \] \, ,
\eeq
governing the dynamics of a point-particle of mass $m$ and charge $q$. Note that, although $\ep$ is totally determined in terms of the rest of the fields through Eq. (\ref{eq:epsol}), it can still take any value depending on the $\la$-parametrization we choose because $\sqrt{- g_{\mu\nu}(\ga)\, \pa_{\la} \ga^{\mu} \pa_{\la} \ga^{\nu}}$ transforms as a covector on $\Lie$. This freedom is in agreement with $\ep$ being the ``lapse function'' on $\Lie$ from the canonical viewpoint. 

In the $m \neq 0$ case, a standard choice of parametrization is proper time, i.e. the momentum/velocity relation \eqref{eq:kofgadot} becomes
\beq
K_{\mu}|_{\rm proper\,\,time} = m g_{\mu\nu}(\ga)\, \pa_{\la} \ga^{\mu} - q A_{\mu}(\ga) \, , \hspace{1cm} \Rightarrow \hspace{1cm} \ep_{\rm proper\,\,time} = \frac{1}{m} \, .
\eeq
This is obviously not defined in the case of massless particles such as photons $m = 0$. Nevertheless, one usually works with the closest analogue that is the affine parametrization $\ep = {\rm const.}$, which is implicitly chosen when setting $K^{\mu} = {\rm const.} \times \pa_{\la} \ga^{\mu}$. We will later exploit this freedom to pick a parametrization that is well-suited for observational cosmology.

We thus conclude that the actions in (\ref{eq:SPm}) and (\ref{eq:Sm}) share the same classical dynamics, so the former is a legitimate description of point-particle dynamics. The advantage of the action (\ref{eq:SPm}), however, is that it has a non-singular $m \to 0$ limit, which is the case of interest for cosmological observables, i.e. the case of light-like geodesics. For $m = 0$ one encounters new features. First, note that $\ep$ is no longer determined by the equations of motion already in the Lagrangian formalism, i.e. in Eq. (\ref{eq:SPm}), because it is not present in its own equation of motion. This means that it is free to choose, i.e. it is not related to the rest of the fields as in the massive case \eqref{eq:epsol}, and can therefore be used to neutralize one more degree of freedom in $\ga^{\mu}$. This reflects the fact that a light-like path is constrained to lie in a submanifold of $\cM$ with one less dimension than in the time-like case, i.e. a light-cone. Another manifestation of this property is the presence of an additional internal symmetry of the action (\ref{eq:SPm}) when $m=0$ that is a combined conformal transformation of the space-time and world-line geometries
\beq \label{eq:lgeoconf}
g_{\mu\nu}(x) \to C(x)\, g_{\mu\nu}(x) \, , \hspace{1cm} A_{\mu}(x) \to A_{\mu}(x) \, , \hspace{1cm} \ep(\la) \to C(\ga(\la))\, \ep(\la) \, .
\eeq

\subsubsection{Equations of motion}  \label{sec:goingtet}

Let us first note that, given the transformation properties of $\ep$, the operator 
\beq
\pa_{\ep} := \ep^{-1} \pa_{\la}
\eeq
is invariant under $\Lie$Ds and therefore qualifies as a ``physical'' derivative (it is the analogue of $\pa_a := e_a^{\mu}\pa_{\mu}$ on $\cM$). Moreover, it is convenient to define the U(1)GT-invariant 4-momentum in the tetrad basis
\beq \label{eq:kaofKmu}
k_a := e_a^{\mu}(\ga) \[ K_{\mu} + q A_{\mu}(\ga) \] \, . 
\eeq
Just as its diffeomorphism-indexed counterpart, the $k^a$ are four $\Lie$D scalars, but they transform as a vector under LLTs
\beq \label{eq:LLTkada}
\ti{k}^a = \La^a_{\,\,\,b}(\ga)\, k^b \, ,
\eeq
and are invariant under $\cal M$Ds.\footnote{As a concrete example in the A$\cM$D case, we first note that 
\beq
\de_{\xi} e_{\mu}^a = -\Lie_{\xi} e_{\mu}^a + \Ord(\xi^2) \, , \hspace{1cm} \de_{\xi} \ep = 0 \, ,
\eeq
so, dropping $\Ord(\xi^2)$ terms,
\bea
\de_{\xi} k^a & = & \de_{\xi} \[ e^a_{\mu}(\ga)\, \pa_{\ep} \ga^{\mu} \] = \de_{\xi} \[ e^a_{\mu}(\ga) \] \pa_{\ep} \ga^{\mu} + e^a_{\mu}(\ga)\, \pa_{\ep} \de_{\xi} \ga^{\mu} \nn \\
 & = & \[ \( \de_{\xi} e^a_{\mu}\)(\ga) + \(\pa_{\nu} e^a_{\mu}\)(\ga)\, \de_{\xi} \ga^{\nu} \] \pa_{\ep} \ga^{\mu} + e^a_{\mu}(\ga)\, \pa_{\ep} \de_{\xi} \ga^{\mu} = 0 \, . \nn 
\eea} We can now express the equations of motion of the action \eqref{eq:Sgeocan} in a neat way. First, the variation with respect to $k_{\mu}$ yields the velocity/momentum relation \eqref{eq:kofgadot}, now reading
\beq \label{eq:kofgtet}
\pa_{\ep} \ga^{\mu} = e^{\mu}_a(\ga) \, k^a \, .
\eeq
Next, the variation with respect to $\ga^{\mu}$ leads to
\beq \label{eq:geotet}
\na_{\ep} k^a = q F^a_{\,\,\,b}(\ga)\, k^b \, ,
\eeq
where we have defined the covariant derivative $\na_{\ep}$ with respect to LLTs on $\Lie$
\beq \label{eq:naep}
\na_{\ep} X^a := \pa_{\ep} X^a + \Si^a_{\,\,\,bc}(\ga)\, k^c X^b \, ,
\eeq
which therefore commutes with $e^a_{\mu}(\ga)$. Eqs. \eqref{eq:kofgtet} and \eqref{eq:geotet} with $q = 0$ are nothing but the geodesic equation, in first-order form, and the $\sim q$ term is nothing but the Lorentz force. Finally, varying the action \eqref{eq:Sgeocan} with respect to $\ep$ on finds the mass-shell condition for a free point-particle 
\beq \label{eq:lltet}
m^2 + k_a k^a = 0 \, .
\eeq

\subsection{Phase space fields} \label{sec:TM}

In this section we consider the dynamics of phase space distributions for free point-particles. Starting from the cotangent bundle $T^*\cM$ of the space-time manifold $\cM$, we build the corresponding Lorentz bundle $L\cM$ using the tetrad field and then the on-shell phase space $\cP \cM$. We pay special attention to the symmetries associated with each case. We then derive the Liouville equation for the distributions on $\cP \cM$, generalize to the case of tensor/spinor valued ones and also define their moments and corresponding conservation equations.

\subsubsection{Geometrical considerations} \label{sec:geomconsBoltz}

We start by considering the off-shell phase space of 1-particle dynamics (i.e. without imposing any mass condition) that is the cotangent bundle manifold $T^* {\cal M}$. This is an 8-dimensional manifold conventionally parametrized by the pair $x^{\mu}$ and $P_{\mu}$. Incidentally, this is the space in which the canonical line fields $(\ga^{\mu}(\la), K_{\mu}(\la))$ introduced in subsection \ref{eq:geoaction} take their values. As a manifold on its own right, one can consider the corresponding group of diffeomorphisms $T^*\cM$D, whose passive version relates all possible coordinate systems 
\beq \label{eq:CTTsM}
\ti{x}^{\mu} = \ti{x}^{\mu} (x,P) \, , \hspace{1cm} \ti{P}_{\mu} = \ti{P}_{\mu} (x,P) \, .
\eeq
However, $T^* {\cal M}$ is not any manifold, but inherits its structure from $\cal M$, so this constrains the set of admissible coordinate systems and thus the diffeomorphisms that relate them. Indeed, the coordinate systems on $T^*{\cal M}$  (its ``atlas") are the ones related only by the subgroup of the transformations (\ref{eq:CTTsM}) that corresponds to the following representation of P$\cal M$Ds 
\beq
\ti{x}^{\mu} = \ti{x}^{\mu}(x) \, , \hspace{1cm} \ti{P}_{\mu} = \frac{\pa x^{\nu}}{\pa \ti{x}^{\mu}}(\ti{x}(x))\, P_{\nu} \, ,
\eeq
and is therefore only 4-dimensional instead of the 8-dimensional $T^*\cM$D. For the active version, i.e. acting on fields on $T^*{\cal M}$, the corresponding 8-dimensional generating vector field $\Xi \equiv \( \xi^{\mu}, \pi_{\mu}[\xi] \)$ satisfies
\beq \label{eq:U1gtTsM}
\xi^{\mu} = \xi^{\mu}(x) \, , \hspace{1cm} \pi_{\mu} = -\pa_{\mu} \xi^{\nu}(x)\, P_{\nu} \, ,
\eeq
and we have again the corresponding Lie derivative that generates the transformation. For instance, a scalar field $f_*(x,P)$ on $T^*{\cal M}$ transforms as
\beq
\de_{\xi} f_* = - \Lie_{\Xi} f_* + \Ord(\xi^2) = - \xi^{\mu} \pa_{\mu} f_* + \pa_{\mu} \xi^{\nu} P_{\nu} \, \frac{\pa f_*}{\pa P_{\mu}} + \Ord(\xi^2) \, .
\eeq
Another special case of P$T^*{\cal M}$Ds are the ones which correspond to the U(1)GTs and which we deduce from Eq. \eqref{eq:U1gtk}
\beq
\ti{x}^{\mu} = x^{\mu} \, , \hspace{1cm} \ti{P}_{\mu} = P_{\mu} - q \pa_{\mu} \te(x) \, .
\eeq
Now considering $T^*{\cal M}$ as the (off-shell) covariant phase space of a particle endows it with more structure, because one has access to a larger subgroup of $T^*{\cal M}$Ds, the ``canonical transformations''. These can be defined as the $T^*{\cal M}$Ds that preserve the canonical 1-form on $T^*{\cal M}$
\beq
C := P_{\mu} \ed x^{\mu} \, ,
\eeq
{\it up to a total derivative}. One can then check that $C$ is invariant under P$\cal M$Ds and varies by a total derivative under A$\cM$Ds and U(1)GTs, so these symmetries are particular cases of canonical transformations. To make contact with the canonical action-based geodesic formalism discussed in subsection \ref{eq:geoaction}, if we consider a definite path in phase space 
\beq
(x^{\mu}, P_{\mu}) = (\ga^{\mu}(\la), K_{\mu}(\la)) \, ,
\eeq
then evaluating the canonical 1-form on that path (i.e. its pullback to $\Lie$) gives the world-line 1-form
\beq
C(\ga,K) = K_{\mu} \ed \ga^{\mu} \equiv K_{\mu} \pa_{\la} \ga^{\mu} \ed \la \, .
\eeq
This is the combination that enters the canonical action \eqref{eq:Sgeocan}, by definition, and thus allows one to identify the rest as (minus) the ``Hamiltonian". The canonical transformations leave this combination invariant, up to a boundary term in the canonical action (\ref{eq:Sgeocan}), so the latter remains in canonical form. Another property of the canonical transformations is that they preserve the canonical volume form on $T^*{\cal M}$
\beq
{\rm vol}_* := \frac{1}{(2\pi)^4} \, \ed^4 x \we \ed^4 P \, ,
\eeq
where
\beq
\ed^4 x := \frac{1}{4!}\, \vep_{\mu\nu\ro\si} \, \ed x^{\mu} \we \ed x^{\nu} \we \ed x^{\ro} \we \ed x^{\si} \, , \hspace{1cm} \ed^4 P := \frac{1}{4!}\, \vep^{\mu\nu\ro\si} \, \ed P_{\mu} \we \ed P_{\nu} \we \ed P_{\ro} \we \ed P_{\si} \, .
\eeq
The $(2\pi)^{-4}$ normalization comes from the fact that the elementary phase space volume is the Planck constant $h \equiv 2\pi \hbar$, for each space-time dimension, and we use the $\hbar = 1$ normalization. It is then conventional, in particle physics and cosmology, to associate this normalization with the momentum coordinates and this is reflected in the definition of the Fourier transform. 

Let us now switch to the tetrad formalism, where the momentum components of interest are the ones in the tetrad basis, and in particular the U(1)GT-invariant ones
\beq
p^a := e^{a\mu}(x) \[ P_{\mu} + q A_{\mu}(x) \] \, ,
\eeq
in analogy with Eq. \eqref{eq:kaofKmu}. From the viewpoint of $T^*{\cal M}$, the change of coordinates
\beq \label{eq:Padef}
x^{\mu} \to x^{\mu} \, , \hspace{1cm} P_{\mu} \to p^a
\eeq
corresponds to a passive $T^*\cM$D that is {\it not} a canonical transformation, so the volume form transforms non-trivially to
\beq
{\rm vol}_* \to {\rm vol}_L = \frac{1}{(2\pi)^4} \, (e \, \ed^4 x) \we \[ \frac{1}{4!}\, \vep_{abcd} \, \ed p^a \we \ed p^b \we \ed p^c \we \ed p^d \] \, , \hspace{1cm} e := \det e_{\mu}^a \equiv \sqrt{-g} \, ,
\eeq
where in the first round bracket we recognize the volume form on $\cal M$.\footnote{This is obtained by noting that, whenever the exterior derivative $\ed$ acts on either $e_{\mu}^a(x)$ or $A_{\mu}(x)$ the corresponding terms vanish because they are proportional to the exterior product of five $\ed x^{\mu}$.} By performing a non-canonical coordinate transformation, the corresponding space is technically no longer $T^*{\cal M}$, because we have changed the transformations that act on it and thus its geometric structure. Indeed, now the P$\cal M$Ds are represented as
\beq
\ti{x}^{\mu} = \ti{x}^{\mu}(x) \, , \hspace{1cm} \ti{p}^a = p^a \, ,
\eeq 
and we also have LLT coordinate transformations
\beq \label{eq:LLTcoortrans}
\ti{x}^{\mu} = x^{\mu} \, , \hspace{1cm} \ti{p}^a = \La^a_{\,\,\,b}(x) \, p^b \, .
\eeq 
We have thus traded the cotangent bundle $T^*{\cal M}$ for a vector bundle based on $\cal M$ with structure group SO$(1,3)$. We will refer to it as the ``Lorentz" bundle and denote it by $L{\cal M}$. In order to get the action of the A$\cal M$Ds, we first transform our scalar to the new coordinates
\beq \label{eq:ftrans}
f_L( x, p ) := f_*(x, P(p)) \equiv f_*(x, e_{\mu a}(x) \,p^a - q A_{\mu}(x)) \, ,
\eeq
and obtain 
\beq
\de_{\xi} f_L = - \xi^{\mu} \pa_{\mu} f_L + \Ord(\xi^2) \, .
\eeq
Observe that, since now the LLTs are coordinate transformations on $L{\cal M}$, i.e. a passive transformation, they also have their active counterpart on fields over that manifold, generated by the Lie derivative with respect to some vector field. We proceed as in the $\cal M$D case, i.e. we express \eqref{eq:LLTcoortrans} in terms of the generators
\beq
\ti{x}^{\mu} = x^{\mu} \, , \hspace{1cm} \ti{p}^a = \La^a_{\,\,\,b}(x)\, p^b = p^a - \te^a_{\,\,\,b}(x) \, p^b + \Ord(\te^2) \, ,
\eeq
and identify the generating vector field
\beq
\Te = \( 0, \te^a \) \, , \hspace{1cm} \te^a := -\te^a_{\,\,\,b}(x) \, p^b \, .
\eeq
Thus, writing down the coordinate transformation \eqref{eq:LLTcoortrans} for the scalar
\beq
\ti{f}_L(x, \bm{\La}(x) \, p) = f_L(x,p) \, ,
\eeq
we can relabel the variables to derive the active LLT 
\beq
\ti{f}_L(x,p) = f_L(x,\bm{\La}^{-1}(x)\, p) = e^{-\Lie_{\Te}} f_L(x,p) \equiv \exp \[ -\te_a^{\,\,\,b}(x)\, p^a \frac{\pa}{\pa p^b} \] f_L(x,p) \, ,
\eeq
and the variation is
\beq
\de_{\te} f_L = - \Lie_{\Te} f_L + \Ord(\te^2) = - \te_a^{\,\,\,b} p^a \frac{\pa f_L}{\pa p^b}  + \Ord(\te^2) \, .
\eeq
Finally, note that we can also define a covariant Fourier transformation for scalars on $L \cM$
\beq \label{eq:FLdef}
F_L(x, X) := \int \frac{\ed^4 p}{(2\pi)^4} \, e^{i p_a X^a} f_L(x,p) \, ,
\eeq
and thus a dual space parametrized by $\( x^{\mu}, X^a \)$, where the ``internal" coordinates $X^a$ transform as
\beq \label{eq:XaLLT}
\ti{X}^a = \La^a_{\,\,\,b}(x)\, X^b \, ,
\eeq
under passive LLTs. We will see how to interpret these $X^a$ coordinates in section \ref{sec:Boltzmann}.

\subsubsection{On-shell phase space and Liouville operator} \label{sec:Liou}

The (off-shell) 1-particle density distribution of kinetic theory is a scalar field $f_*$ on $T^* \cM$, or alternatively, a scalar field $f_L$ on $L\cM$. In the absence of particle interactions, i.e. for the ``free'' theory described by the canonical action (\ref{eq:Sgeocan}), the Liouville theorem states that $f_L$ is conserved when evaluated on a solution $( \ga^{\mu}(\la), k^a(\la))$
\beq 
\pa_{\la} f_L( \ga(\la), k(\la) ) = 0 \, .
\eeq
Distributing $\pa_{\la}$, using the equations of motion of $\ga^{\mu}$ and $K_{\mu}$, i.e. Eqs. \eqref{eq:kofgtet} and \eqref{eq:geotet}, and demanding that the result holds for all solutions, we find the Liouville equation on $L\cM$
\beq \label{eq:Liouvilleoff}
p^a \[ \na^L_a + q F_a^{\,\,\,b}(x) \, \frac{\pa}{\pa p^b} \] f_L = 0  \, , 
\eeq 
where we have defined the covariant derivative on $L\cM$
\beq \label{eq:Liouom}
\na^L_a \equiv \pa_a - \Si^b_{\,\,\,ca}(x)\, p^c \frac{\pa}{\pa p^b} \, .
\eeq
Until now we have only used the geodesic equation, i.e. the equations of motion of $\ga^{\mu}$ and $K_{\mu}$, but we still have the constraint imposed by $\ep$, i.e. the mass-shell condition \eqref{eq:lltet}. The latter implies that $f_L$ actually lives on a 7-dimensional submanifold $L_m{\cal M} \subset L{\cal M}$, the on-shell covariant phase space defined by
\beq \label{eq:onshell}
p_a p^a + m^2 = 0 \, ,
\eeq
which is consistently preserved under evolution
\beq \label{eq:onshellflow}
p^b \[ \na^L_b + q F_b^{\,\,\,c}(x) \, \frac{\pa}{\pa p^c} \] \( p_a p^a + m^2 \) \equiv 0 \, .
\eeq
A first advantage of working on $L{\cal M}$, instead of $T^* \cM$, is that the on-shell condition (\ref{eq:onshell}) depends on the momentum coordinates $p^a$ alone and has a simple solution
\beq \label{eq:p0ofpi}
p^0 = E_p := + \sqrt{m^2 + \vec{p}^2} \, ,
\eeq
contrary to the condition we would have had in $T^*\cM$
\beq \label{eq:onshellpg}
g^{\mu\nu}(x) \[ p_{\mu} + q A_{\mu}(x) \] \[ p_{\nu} + q A_{\nu}(x) \] + m^2 = 0 \, .
\eeq
The $L_m \cM$ manifold is therefore parametrized by $\( x^{\mu}, p^i \)$ and the on-shell density distribution is defined as
\beq \label{eq:f7def}
f( x, \vec{p}) := f_L(x, E_p, \vec{p})  \, .
\eeq 
From now on all $p^a$ occurrences are implicitly considered on-shell, i.e. $p^0 \equiv E_p$. The $\cal M$Ds still act as usual on $L{\cal M}$, but the LLTs now act in a mass-dependent non-linear way. The passive version is
\beq \label{eq:PLLT3dfiber}
\ti{x}^{\mu} = x^{\mu} \, , \hspace{1cm} \ti{p}^i = \La^i_{\,\,j}(x) \, p^j + \La^i_{\,\,0}(x) \, E_p  \, ,
\eeq 
while the corresponding active one is
\bea 
 & & \ti{f}(x,p^i) = f(x, (\La^{-1})^i_{\,\,j}(x) \, p^j + (\La^{-1})^i_{\,\,0}(x) \, E_p) \nn \\
 & \Rightarrow & \de_{\te} f = \( \te^{0j}(x)\, E_p - \te^{ij}(x)\, p^i \) \frac{\pa f}{\pa p^j} + \Ord(\te^2) \, . \label{eq:ALLT3dfiber}
\eea 
One can next define a volume form on $L_m {\cal M}$, that is invariant under passive $\cal M$Ds and LLTs, by integrating over the one of $L{\cal M}$ with a Dirac delta imposing the on-shell constraint
\beq \label{eq:voldef}
{\rm vol}_m := \int_{p^0 > 0} {\rm vol}_L \, 2\pi \de \( p_a p^a + m^2 \) = (e\, \ed^4 x) \we \frac{\ed^3 p}{(2\pi)^3 2E_p} \, ,
\eeq
where
\beq
\ed^3 p := \frac{1}{3!} \, \vep_{ijk} \, \ed p^i \we \ed p^j \we \ed p^k \, .
\eeq
Note that now the two factors inside ${\rm vol}_m$ are separately invariant under both symmetries. Repeating Liouville's theorem on $f(x,\vec{p})$, i.e. $\pa_{\la} f(\ga^{\mu}(\la),k^i(\la)) = 0$ for all geodesic solutions we obtain the Liouville equation on $L_m \cM$
\beq \label{eq:Lioueq}
{\rm L} f = 0  \, ,
\eeq
where ${\rm L}$ is the covariant Liouville operator in the tetrad basis
\bea
{\rm L} & := & p^a \[ \pa_a - \( \Si_{iba} p^b + q F_{ia} \) \frac{\pa}{\pa p^i} \] \label{eq:Liouop}  \\
 & \equiv & E_p \pa_0 + p^i \pa_i + \[ \( \Si_{0i0} E_p - q E_i \) E_p + \( \( \Si_{0ij} - \Si_{ij0} \) E_p - q \vep_{ijk} B_k \) p^j - \Si_{ijk} p^j p^k \] \frac{\pa}{\pa p^i} \, , \nn
\eea
where $p^a := ( E_p, \vec{p} )$ is the on-shell 4-momentum. This expression is not explicitly Lorentz-invariant, which is unavoidable because $\pa_{p^0}$ is not defined on $f(x,\vec{p})$. To see that it is indeed Eq. \eqref{eq:Liouvilleoff} constrained on the mass shell $m$, we can use Eq. \eqref{eq:f7def} and the antisymmetry of $\Si$ and $F$ to find
\bea
{\rm L} f & \equiv & p^a \[ \pa_a f_L - \( \Si_{iba} p^b + q F_{ia} \) \( \frac{p^i}{E_p} \frac{\pa}{\pa p^0} + \frac{\pa}{\pa p^i} \) f_L \]_{p^0 = E_p} \nn \\
 & \equiv & p^a \[ \pa_a f_L - \( \Si^0_{\,\,\,ba} p^b + q F^0_{\,\,\,a} \) \frac{\pa}{\pa p^0} \, f_L - \( \Si_{iba} p^b + q F_{ia} \) \frac{\pa}{\pa p^i} \, f_L \]_{p^0 = E_p} \nn \\
 & \equiv & p^a \[ \pa_a f_L - \( \Si^c_{\,\,\,ba} p^b + q F^c_{\,\,\,a} \) \frac{\pa}{\pa p^c} \, f_L \]_{p^0 = E_p} \nn \\
 & \equiv & \[ p^a \na_a^L f_L \]_{p^0 = E_p} \, .  
\eea
Finally, note that the more usual definition of the Liouville operator is rather $E_p^{-1} {\rm L} = \pa_0 + \dots$, which has the dimensions of a time-derivative, but this is not invariant under local boosts, so here we prefer the Lorentz-invariant definition. 

The above construction can be generalized straightforwardly to 4-dimensional tensor fields on $\cP \cM$. These are fields with Lorentz indices $f_{a_1 \dots a_n}(x,\vec{p})$ and possibly an extra Dirac index $f^{\frak{a}}_{a_1 \dots a_n}(x,\vec{p})$ which we keep again implicit.\footnote{A single such index is enough, since any even set of spinor indices can be turned into Lorentz indices.} Note that all the components of such a field must lie on the same mass shell $m$ for the LLTs to be well-defined, e.g. in the passive case 
\beq \label{eq:ftenstrans}
\ti{f}_{a_1 \dots a_n}(\ti{x}, \ti{\vec{p}}) = \La_{a_1}^{\,\,\,b_1}(x) \dots \La_{a_n}^{\,\,\,b_n}(x)\, U(x)\, f_{b_1 \dots b_n}(x,\vec{p}) \, ,
\eeq
where the coordinates are related by the LLT (\ref{eq:PLLT3dfiber}), and must also have the same charge $q$ for the U(1)GTs to be well-defined
\beq
\ti{f}_{a_1 \dots a_n}(x, \vec{p}) = e^{iq \te(x)} f_{a_1 \dots a_n}(x,\vec{p}) \, .
\eeq
Evaluating such a tensor on a given geodesic solution $(\ga(\la), \vec{k}(\la))$ one obtains a Lorentz tensor on $\Lie$, so that one must use $\na_{\la}$ instead of $\pa_{\la}$ to get a covariant generalization of the Liouville theorem. The resulting covariant Liouville operator then reads
\bea
{\rm L} f_{a_1 \dots a_n} & := & p^a \[ \pa_a - \( \Si_{iba} p^b + q F_{ia} \) \frac{\pa}{\pa p^i} \] f_{a_1 \dots a_n} \label{eq:Ltens} \\
 & & +\, p^c \[ \sum_{k=1}^n \Si_{a_k\,\,c}^{\,\,\,\,\,b} f_{a_1 \dots a_{k-1}b a_{k+1}\dots a_n} + \frac{1}{4}\, \Si_{abc} \, \ga^a \ga^b f_{a_1 \dots a_n} \] \, , \nn
\eea
i.e. it simply takes into account the mixing of the indices by involving the corresponding spin connection factors. In the absence of the Dirac index there is no $U(x)$ matrix and no $\sim \ga^a \ga^b$ term in Eqs. \eqref{eq:ftenstrans} and \eqref{eq:Ltens}, respectively.

Finally, in the presence of more than one particle species $f_s(x,\vec{p})$, the passive version of LLTs is no longer defined because there is no unique mass $m$ to consider inside $E_p$ in Eq. (\ref{eq:PLLT3dfiber}). This is due to the fact that each $f_s$ is defined on a different mass shell $L_m \cM \subset L \cM$. The active LLTs in Eq. (\ref{eq:ALLT3dfiber}), however, still work perfectly well, since they act on fields 
\beq \label{eq:ALLT3dfibers}
\ti{f}_s(x,p^i) = f_s(x,(\La^{-1})^i_{\,\,j}(x) \, p^j + (\La^{-1})^i_{\,\,0}(x) \, E_{p,s}) \, ,
\eeq
where now 
\beq
E_{p,s} := \sqrt{m^2_s + \vec{p}^2} \, . 
\eeq
One can therefore adopt the following geometric viewpoint. The $x^{\mu}$ and $p^i$ coordinates parametrize a single space that we denote by $\cP \cM$, the on-shell (Lorentz) phase space, on which all $f_s(x,\vec{p})$ are defined. The $\cM$Ds can still act in both their passive and active version on $\cP \cM$, but the LLTs are only defined as active transformations on the fields $f_s(x,\vec{p})$. Note also that now each $f_s$ comes with its own volume form
\beq \label{eq:voldefs}
{\rm vol}_s := (e\, \ed^4 x) \we \frac{\ed^3 p}{(2\pi)^3 2E_{p,s}} \, ,
\eeq
and its own $(m_s, q_s)$-dependent Liouville operator.

\subsubsection{Distribution moments}

Given the volume form \eqref{eq:voldefs}, we can use its Lorentz-invariant momentum factor to define the moments of $f_s$, i.e. the space-time Lorentz tensors
\beq \label{eq:momentdef}
T_s^{a_1 \dots a_n}(x) := \int \frac{\ed^3 p}{(2\pi)^3 E_{p,s}} \, f_s( x, \vec{p} ) \, p_s^{a_1} \dots p_s^{a_n} \, , \hspace{1cm} n>0 \, , \hspace{1cm} p_s^a := (E_{p,s}, \vec{p}) \, ,
\eeq
of which the $n=1$ and $n=2$ cases are the particle number current vector and energy momentum tensor of the $s$ species, respectively, as measured by the observer family $e_a$. The electric current vector of the $s$ species is then
\beq
J_s^a := q_s T^a_s \, .
\eeq
Note that here we focus exclusively on scalar distribution functions, a restriction that will be justified in section \ref{sec:Boltzmann}. In the absence of interactions, i.e. if ${\rm L} f_s = 0$ holds, the moments obey the following conservation equation
\beq \label{eq:momentcons}
\na_{a_1} T_s^{a_1 \dots a_n} = n q_s F_{a_1}^{\,\,\,(a_1} T_s^{a_2 \dots a_n)}  \, .
\eeq 
To see this, first express them as an integral over the 4-momenta $p^a$
\beq
T_s^{a_1 \dots a_n}(x) = 2\int_{p^0 > 0} \frac{\ed^4 p}{(2\pi)^4} \, 2\pi \de \( p_a p^a + m_s^2 \) f_{L,s}( x, p ) \,p^{a_1} \dots p^{a_n} \, ,
\eeq
and use Eqs. (\ref{eq:onshellflow}) and (\ref{eq:Liouvilleoff}) to get
\bea
\na_{a_1} T_s^{a_1 \dots a_n}(x) & = & 2\int_{p^0 > 0} \frac{\ed^4 p}{(2\pi)^4} \, 2\pi \de \( p_a p^a + m_s^2 \) \[ p^{a_1} \na^L_{a_1} f_{L,s}( x, p )  \] p^{a_2} \dots p^{a_n} \label{eq:TTEMcons} \\
 & = & -2q_s F_{a_1}^{\,\,\,b}(x)  \int_{p^0 > 0} \frac{\ed^4 p}{(2\pi)^4} \, 2\pi \de \( p_a p^a + m_s^2 \) p^{a_1} \dots p^{a_n}  \frac{\pa}{\pa p^b}\,f_{L,s}( x, p ) \nn \\
 & = & 2q_s F_{a_1}^{\,\,\,b}(x)  \int_{p^0 > 0} \frac{\ed^4 p}{(2\pi)^4} \, f_{L,s}( x, p ) \, \frac{\pa}{\pa p^b} \[ 2\pi \de \( p_a p^a + m_s^2 \) p^{a_1} \dots p^{a_n} \] \nn \\
 & = & 2n q_s F_{a_1}^{\,\,\,(a_1}(x)  \int_{p^0 > 0} \frac{\ed^4 p}{(2\pi)^4} \, f_{L,s}( x, p ) \, 2\pi \de \( p_a p^a + m_s^2 \) p^{a_2} \dots p^{a_n)}  \nn \\
 & \equiv & n q_s F_{a_1}^{\,\,\,(a_1}(x) \, T_s^{a_2 \dots a_n)}(x) \nn \, .
\eea
In the case of the energy momentum tensor, the total one is given by
\beq
T^{ab} := \sum_s T^{ab}_s + T^{ab}_{\rm EM} \, ,
\eeq
where the last term corresponds to the contribution of the electromagnetic field given in Eq. \eqref{eq:EMTEM}. The conservation equation \eqref{eq:momentcons}, along with the Maxwell equation \eqref{eq:Maxwell}, then imply the usual conservation of $T^{ab}$ 
\bea
\na^a T_{ab} & \equiv & \sum_s \na^a T_{ab}^s + \na^a T_{ab}^{\rm EM} \nn \\ 
 & = & \sum_s q_s F^a_{\,\,\,b} T^s_a + \na_a F^{ac} F_{bc} + \frac{3}{2}\,F^{ac} \na_{[a} F_{bc]} \nn \\
 & = & \sum_s q_s F^a_{\,\,\,b} T^s_a + \sum_s J_s^c F_{bc} + \frac{3}{2}\, F^{ac} \na_{[a} F_{bc]} \equiv 0 \, , \label{eq:Tconsnocol}
\eea
where the first two terms cancel each other out and the last term vanishes through the Bianchi identity in the tetrad basis.

\clearpage

\section{Observer space-time formalism} \label{sec:geo}

In this section we start by defining the fundamental observables and express the equation of motion of photons in terms of them. The boundary value of the photon 4-momentum at the observer, and in the $\hat{e}_a$ frame, provides the observed frequency and angle parametrization, and we fix the reparametrization symmetry of the geodesic map to the (log-)redshift parametrization. We then successively define the observer sky, observer space, their spectral generalizations and their mapping to the observer's light-cone, paying attention to the involved symmetries, and also discuss the issue of caustics. The construction is completed by introducing a Sachs basis along each photon path and fixing its rotational freedom by matching it to the natural dyad on the observer sky, thus obtaining the Sachs basis that is actually used to build sky maps of tensorial quantities. With this we can then describe the Jacobi and volume maps within this formalism to obtain the fundamental observables associated with localized sources. Finally, we also consider the possibility of describing the drift of cosmological observables by discussing the extension of the observer space to the observer space-time and the relation of the latter to the observational coordinates.

\subsection{Fundamental observables}

We start by considering two events $\hat{P}$ and $P$ corresponding to reception and emission of a light signal, respectively, or the ``observer" and ``source" points. In the eikonal approximation of light propagation, the two events are therefore connected by a light-like geodesic $\ga \subset \cM$, i.e. $\hat{P}, P \in \ga$. The tetrad field $e_a$ describes an observer family with four-velocities $e_0$ and spatial frames $e_i$. In particular, $e_0(\hat{P})$ corresponds to the observer 4-velocity, $e_i(\hat{P})$ to the spatial frame that is used to measure spatial tensor components, while $e_0(P)$ corresponds to the source 4-velocity. 

We start by considering the point-particle equations of motion in the tetrad basis \eqref{eq:kofgtet}, \eqref{eq:geotet} and \eqref{eq:lltet} for the photon case $m,q = 0$. We first use \eqref{eq:lltet} to express $k^a$ in terms of its independent components 
\beq \label{eq:kaofomandni}
k^a = \om \( 1, -n^i \) \, , \hspace{1cm} n^i n^i \equiv 1 \, .
\eeq
Note that $\om(\la)$ and $-n^i(\la)$ are the frequency and propagation direction as measured by $e_a(\ga(\la))$. Denoting by $\hat{\la}$ the parameter value corresponding to the observer 
\beq
\ga(\hat{\la}) \equiv \hat{P} \, ,
\eeq
we have that 
\beq
\hat{\om} := \om(\hat{\la}) \, , \hspace{1cm} \hat{n}^i := n^i(\hat{\la}) \, , 
\eeq
are the observed frequency of the signal and its position in the sky, while the observed redshift from a source at $\la$ is given by
\beq
z(\la) := \hat{\om}^{-1} \om(\la) - 1 \, .
\eeq 
It will be convenient to work instead with the ``log-redshift" variable
\beq
\ze := \log \( 1 + z \) \equiv \log \frac{\om}{\hat{\om}} \, ,
\eeq
which coincides with $z$ only if $z \ll 1$. The transformations under LLTs are
\bea 
\ti{\om} & = & \La_{\para} \om \, , \label{eq:otrans} \\
\ti{\ze} & = & \ze + \log \frac{\La_{\para}}{\hat{\La}_{\para}}  \, ,  \label{eq:zetrans}\\
\ti{n}^i & = & \La_{\para}^{-1} \( \La^i_{\,\,j} n^j - \La^i_{\,\,0} \)  \, , \label{eq:ntrans}
\eea
where we have defined 
\beq \label{eq:LaLdef}
\La_{\para} := \La^0_{\,\,\,0} - \La^0_{\,\,\,i} n^i \, ,
\eeq
and from now on it is understood that space-time fields, such as $\La^a_{\,\,\,b}(x)$, are implicitly evaluated on the light-like geodesic $\ga$. Another combination that will appear often is
\beq \label{eq:LaTdef}
(\La_{\perp})^i_{\,\,\,j} := \La^i_{\,\,j} + \La_{\para}^{-1} \( \La^i_{\,\,k} n^k - \La^i_{\,\,0} \) \La^0_{\,\,\,j} \, ,
\eeq
which maps vectors that are normal to $n^i$ to vectors that are normal to the new one $\ti{n}^i$ given in \eqref{eq:ntrans}. For any two vectors $X^i$ and $Y^i$ that are normal to $n^i$, we then have the identity
\beq
(\La_{\perp})^i_{\,\,\,j} X^j \, (\La_{\perp})^i_{\,\,\,k} Y^k \equiv X^i Y^i \, ,
\eeq
i.e. $\La_{\perp}$ preserves the Euclidean norm, because the Lorentz contraction effect is absorbed in the transformation of $n^i$. It will also be convenient to have the corresponding variations in terms of the generators  
\bea
\de_{\te} \om & = & \om \[ n^i \te^{0i} + \Ord(\te^2) \] \, , \label{eq:ovar} \\
\de_{\te} \ze & = & n^i \te^{0i} - \hat{n}^i \hat{\te}^{0i} + \Ord(\te^2)  \, , \label{eq:zvar} \\
\de_{\te} n^i & = & n^{ij} \te^{0j} - \te^{ij} n^j + \Ord(\te^2) \, , \label{eq:nvar}
\eea
where
\beq
n^{ij} := \de^{ij} - n^i n^j \, , \hspace{1cm} n^i n^{ij} \equiv 0 \, ,
\eeq
is the projector to the normal subspace to $n^i$. Note, in particular, that the transformation rules of $\ze$ and $n^i$ conform to the ones of a cosmological observable, i.e. these quantities are $\cM$D-invariant and only depend on $\La^a_{\,\,\,b}(\hat{P})$ and $\La^a_{\,\,\,b}(P)$ under LLTs.

\subsection{Redshift parametrization} \label{sec:redpara}

Let us now write down the evolution equations \eqref{eq:kofgtet} and \eqref{eq:geotet} for the case of interest $q = 0$ in terms of $\ze$ and $n^i$
\bea
\pa_{\la} \ga^{\mu} & = & \ep \om \[ e_0^{\mu} - n^i e_i^{\mu} \] \, , \label{eq:gadotep} \\
\pa_{\la} \ze & = & \ep \om \Si_{0\para\para} \, , \label{eq:zdotep} \\ 
\pa_{\la} n^i & = & \ep \om n^{ij} \Si_{j \para\para} \, , \label{eq:ndotep} 
\eea
and we will also use the notation
\beq
X_{\para} := \om^{-1} k^a X_a \equiv \[ e_0^{\mu} - n^i e_i^{\mu} \] X_{\mu} \, .
\eeq
Given Eq. \eqref{eq:zdotep}, it is convenient to consider $\la$ as dimensionless, so that now $\ep$ has dimensions of area, and we now fix the $\cL$D gauge by the following condition
\beq \label{eq:gfLD}
\ep = \frac{1}{\om \Si_{0\para\para}} \, ,
\eeq
in which case the log-redshift equation becomes trivial
\beq
\pa_{\la} \ze = 1 \, ,  
\eeq
so that
\beq \label{eq:laofz}
\la = \hat{\la} + \ze  \, .
\eeq
Setting for definiteness $\hat{\la} \equiv 0$, we get that in this gauge the log-redshift $\ze$ {\it is} the parametrization of the geodesic. This is very convenient since this is the parametrization the actual observer uses in practice. More precisely, the observer uses $z \equiv e^{\ze}-1$, but any differential equation given in terms of $\ze$ can be written in terms of $z$ straightforwardly. This gauge is of course not defined for space-times or regions where $\Si_{0\para\para}$ can go through zero, so from now on we are constraining our field of applications to cosmology with mild inhomogeneity and anisotropy.\footnote{For generic space-times, a computationally convenient parametrization would rather be $\ep = \om^{-1}$, because then Eqs. \eqref{eq:gadotep}, \eqref{eq:zdotep} and \eqref{eq:ndotep} become independent of $\om$.\label{ft:nonsingep}}

The evolution equations (\ref{eq:gadotep}) and (\ref{eq:ndotep}) now become
\bea
\pa_{\ze} \ga^{\mu} & = & \frac{e_0^{\mu} - n^i e_i^{\mu}}{\Si_{0\para\para}} \, , \label{eq:gadot} \\
\pa_{\ze} n^i & = & n^{ij} \, \frac{\Si_{j \para\para}}{\Si_{0\para\para}} \, , \label{eq:ndot} 
\eea
or, in terms of $k^a$
\beq \label{eq:geogaka}
\pa_{\ze} \ga^{\mu} = \frac{e_a^{\mu} k^a}{\om \Si_{0\para\para}} \, , \hspace{1cm} \pa_{\ze} k^a = - \frac{\om \Si^a_{\,\,\,\para\para}}{\Si_{0\para\para}} \, .
\eeq
Note that Eq. (\ref{eq:gfLD}) and the resulting (\ref{eq:laofz}) are invariant under $\cM$Ds and local rotations, but not under local boosts, because the right-hand sides transform non-trivially. Thus, every local boost must now be compensated by some $\Lie$D in order to preserve the gauge condition (\ref{eq:gfLD}), i.e. the latter actually breaks both symmetries down to a combination of the two. Another way to see this is that, now that we use an observer-dependent parameter $\ze$, $\Lie$ gets reparametrized under local boosts.

For the passive case, the compensating P$\Lie$D is simply the one maintaining the relation in Eq. (\ref{eq:laofz}), i.e. $\la$ must transform as in Eq. (\ref{eq:zetrans}). To check this at the level of the gauge condition (\ref{eq:gfLD}), we apply both a P$\Lie$D and a LLT on the inverse quantities for convenience 
\bea
\frac{\pa \ti{\la}}{\pa \la}\, \ep^{-1} & = & \frac{1}{\La_{\para} \om}\, \La_0^{\,\,\,a} \[ \Si_{abc} + \La^d_{\,\,\,b} \pa_c \La_{da} \] k^b k^c \, .
\eea
Simplifying the derivative term
\bea
k^b k^c \La_0^{\,\,\,a} \La^d_{\,\,\,b} \pa_c \La_{da} & = & - k^b k^c \La^d_{\,\,\,b} \La_{da}  \pa_c \La_0^{\,\,\,a} \equiv - k_a k^c \pa_c \La_0^{\,\,\,a} \os{(\ref{eq:kofgtet})}{=} - k_a \pa_{\ep} \La_0^{\,\,\,a}  \nn \\
& = & - \pa_{\ep} \( \La_0^{\,\,\,a} k_a \) + \La_0^{\,\,\,a} \pa_{\ep} k_a \os{(\ref{eq:geotet})}{=} \pa_{\ep} \( \La_{\para}\om \) - \La_0^{\,\,\,a} \Si_{abc} k^b k^c
\eea
and setting again $\ze = \la$, one gets
\beq
\frac{\pa \ti{\la}}{\pa \la} = \pa_{\la} \log \[ \La_{\para} e^{\la} \]  \, ,
\eeq
which can be solved with the boundary condition $\hat{\ti{\la}} = 0$ to yield
\beq \label{eq:tilaofz}
\ti{\la}(\la) = \la + \log \frac{\La_{\para}}{\hat{\La}_{\para}}  \, ,
\eeq
and is indeed the transformation of $\ze$ in Eq. (\ref{eq:zetrans}). In the active compensation case, one can look at the variation of $\ze$ in Eq. (\ref{eq:zvar}) to get the generating vector $\ka$ to linear order
\beq \label{eq:kaofte}
\ka = \te^{0i} n^i - \hat{\te}^{0i} \hat{n}^i + \Ord(\te^2) \, .
\eeq
For the remaining variables, under a P$\Lie$D-compensated LLT we have 
\beq \label{eq:ntranscPLD}
\ti{\ga}^{\mu}(\ti{\ze}) = \ga^{\mu}(\ze) \, , \hspace{1cm} \ti{n}^i(\ti{\ze}) =  \[ \La_{\para}^{-1} \( \La^i_{\,\,j} n^j - \La^i_{\,\,0} \) \](\ze)  \, , 
\eeq
since these quantities are $\Lie$D scalars. Under an A$\Lie$D-compensated LLT, we have
\beq \label{eq:nvarcALD}
\de_{\te} \ga^{\mu} = - \ka\, \pa_{\ze} \ga^{\mu} + \Ord(\te^2) \, , \hspace{1cm} \de_{\te} n^i = -\ka \, \pa_{\ze} n^i + n^{ij} \te^{0j} - \te^{ij} n^j + \Ord(\te^2) \, . 
\eeq
Now whether we use a P$\Lie$D or a A$\Lie$D compensation is irrelevant at the level of the resulting equations of motion, they will be invariant under the combined transformation either way by construction. However, at the level of the cosmological observables, which involve integrals over $\Lie$, it is important that we use a P$\Lie$D, otherwise we will have extra terms due to the boundary of the integral (see the end of appendix \ref{app:diffsym}). Finally, now that we have fixed $\ep$, it is more convenient to use $\na_{\ze}$ instead of $\na_{\ep}$ (see \eqref{eq:naep}), which therefore reads
\beq  \label{eq:nala}
\na_{\ze} X^a \equiv \pa_{\ze} X^a + \Si^a_{\,\,\,b\mu} \pa_{\ze} \ga^{\mu} X^b = \pa_{\ze} X^a + \frac{\Si^a_{\,\,\,b\para}}{\Si_{0\para\para}}\, X^b \, ,
\eeq
and we have used \eqref{eq:gadot} in the last step. In particular, the second equation of \eqref{eq:geogaka} reads $\na_{\ze} k^a = 0$.

\subsection{Observer sky}  \label{sec:obssky}

We now take full advantage of the fact that the boundary data $\hat{k}^a$ provide the parametrization of the observables that the actual observer uses in practice. The observed position vector $\hat{n}^i$ is an over-parametrization of the unit-sphere $\Ss$ that is the ``observer sky" (or ``celestial sphere''). The observed frequency $\hat{\om} \in \Rs_+$ is the parameter with respect to which the observed spectrum of some source's light is given. Together, $\hat{n}^i$ and $\hat{\om}$ thus parametrize what is topologically a half-infinite 3-cylinder 
\beq
\Ss_{\rm spec} := \Ss \times  \Rs_+ \, , 
\eeq
that we will call the ``spectral observer sky". As we will see, however, there is no physical distance associated with $\hat{\om}$, which is why we will not refer to this as a ``cylinder". 

Let us now describe the geometry of that space in a manner that will be convenient later on. Since $\hat{n}^i$ and $\hat{\om}$ are $\cM$D-invariant, the only relevant transformations here are the LLTs. We start by expressing $\hat{n}^i$ in terms of two angles $\vte^{\hat{A}} \in \{ \vte, \vph \}$, choosing the $i = 3$ direction as the zenith one
\beq \label{eq:tevphobs}
\hat{n}(\vte) = \( \sin \vte \cos \vph, \, \sin \vte \sin \vph, \, \cos \vte \) \, . 
\eeq
We want the relation in Eq. (\ref{eq:tevphobs}) to hold for all observers, i.e. under a LLT
\beq \label{eq:niOinv}
\ti{\hat{n}}(\ti{\vte}) = \( \sin \ti{\vte} \cos \ti{\vph}, \, \sin \ti{\vte} \sin \ti{\vph}, \, \cos \ti{\vte} \) \equiv \hat{n}(\ti{\vte}) \, ,
\eeq
meaning that $\hat{n}^i$ and $\vte^{\hat{A}}$ are alternative parametrizations of $\Ss$ with a fixed functional relation. In particular, note that the above relation is {\it not} the transformation of a scalar on $\Ss$, which would rather read $\ti{\hat{n}}(\ti{\vte}) = \hat{n}(\vte)$ and thus change the functional dependence of $\hat{n}$ on the angles. Evaluating (\ref{eq:ntranscPLD}) at $\ti{\ze} = \ze = 0$ in order to get the relation between $\ti{\vte}^{\hat{A}}$ and $\vte^{\hat{A}}$
\beq \label{eq:niOvtetrans}
\ti{\hat{n}}^i(\ti{\vte}) = \hat{\La}_{\para}^{-1}(\vte) \[ \hat{\La}^i_{\,\,\,j} \hat{n}^j(\vte) - \hat{\La}^i_{\,\,\,0}\] \, , \hspace{1cm} \hat{\La}_{\para}(\vte) \equiv \hat{\La}^0_{\,\,\,0} - \hat{\La}^0_{\,\,\,j} \hat{n}^j(\vte) \, ,
\eeq
we have that the LLT at $\hat{P}$ in terms of $\vte^{\hat{A}}$ takes the form of a coordinate transformation on $\Ss$
\beq \label{eq:SOLLT}
\ti{\vte}(\vte) = \arccos \[ \frac{\hat{\La}^3_{\,\,\,i} \hat{n}^i(\vte) - \hat{\La}^3_{\,\,\,0}}{\hat{\La}^0_{\,\,\,0} - \hat{\La}^0_{\,\,\,j} \hat{n}^j(\vte)} \] \, , \hspace{1cm} \ti{\vph}(\vte) = \arctan \[ \frac{\hat{\La}^2_{\,\,\,i} \hat{n}^i(\vte) - \hat{\La}^2_{\,\,\,0}}{\hat{\La}^1_{\,\,\,j} \hat{n}^j(\vte) - \hat{\La}^1_{\,\,\,0}} \] \, .
\eeq
In particular, the coordinate transformation induced by a local rotation at $\hat{P}$ is itself a rotation. As for the observed frequency parameter $\hat{\om}$, its transformation in Eq. \eqref{eq:otrans} now means that it mixes with the $\vte^{\hat{A}}$ coordinates
\beq \label{eq:hatomLLT}
\ti{\hat{\om}} = \hat{\La}_{\para}(\vte)\, \hat{\om} \, .
\eeq
The Jacobian matrix of the $\{ \vte, \vph, \hat{\om} \}$ coordinate transformation is therefore not in block-diagonal form and the partial derivatives on $\Ss_{\rm spec}$ transform as follows
\beq  \label{eq:paC3LLT} 
\ti{\pa}_{\hat{A}} = \frac{\pa \vte^{\hat{B}}}{\pa \ti{\vte}^{\hat{A}}} \[ \pa_{\hat{B}} + \hat{\La}_{\para}^{-1}(\vte)\, \hat{\La}^0_{\,\,\,i} \pa_{\hat{B}} \hat{n}^i(\vte)\, \hat{\om} \pa_{\hat{\om}} \] \, , \hspace{1cm} \ti{\pa}_{\hat{\om}} = \hat{\La}_{\para}^{-1}(\vte) \, \pa_{\hat{\om}} \, .
\eeq
The fact that the transformation of $\pa_{\hat{\om}}$ does not depend on $\pa_{\hat{A}}$ means that fields on $\Ss_{\rm spec}$ with no $\hat{\om}$ dependence have that property for all observers. Thus, if one is not interested in spectral distributions on the sky $f(\vte,\hat{\om})$, but rather fields on $\Ss$ alone $f(\vte)$, then one simply sets $\pa_{\hat{\om}} f = 0$ and this condition is conserved under LLTs. 

Now given this $\{ \vte, \vph, \hat{\om} \}$ parametrization and Eq. \eqref{eq:kaofomandni}, the LLT-invariant line-element on $\Ss_{\rm spec}$ is 
\bea
\ed^2 l_{\Ss_{\rm spec}} & := & \et_{ab} \, \ed \hat{k}^a \ed \hat{k}^b \nn \\
 & = & \et_{ab} \[  \pa_{\hat{A}} \hat{k}^a \pa_{\hat{B}} \hat{k}^b \, \ed \vte^{\hat{A}} \ed \vte^{\hat{B}} + 2 \pa_{\hat{\om}} \hat{k}^a \pa_{\hat{A}} \hat{k}^b \, \ed \hat{\om} \, \ed \vte^{\hat{A}} + \pa_{\hat{\om}} \hat{k}^a \pa_{\hat{\om}} \hat{k}^b \, \ed \hat{\om}^2 \] \nn \\
 & = & \et_{ab} \, \pa_{\hat{A}} \hat{k}^a \pa_{\hat{B}} \hat{k}^b \, \ed \vte^{\hat{A}} \ed \vte^{\hat{B}}  \nn \\
 & = & \hat{\om}^2 \pa_{\hat{A}} \hat{n}^i \pa_{\hat{B}} \hat{n}^i \, \ed \vte^{\hat{A}} \ed \vte^{\hat{B}}  \nn \\
 & = & \hat{\om}^2 \ed l^2_{\Ss} \, , \label{eq:dl2spec}
\eea
where
\beq \label{eq:dOmdef}
\ed l^2_{\Ss} := \ed \vte^2 + \sin^2 \vte \, \ed \vph^2 \, ,
\eeq
is the line-element of the observer sky $\Ss$. In particular, we can express it in terms of a metric
\beq \label{eq:SABdef}
\ed^2 l_{\Ss} \equiv S_{\hat{A}\hat{B}}(\vte)\, \ed \vte^{\hat{A}} \, \ed \vte^{\hat{B}} \, , \hspace{1cm} S_{\hat{A}\hat{B}}(\vte) := \pa_{\hat{A}} \hat{n}^i \pa_{\hat{B}} \hat{n}^i = \( \begin{array}{cc} 1 & 0 \\ 0 & \sin^2 \vte \end{array} \)_{\hat{A}\hat{B}} \, ,
\eeq
which is the one that is used in actual observations. The fact that there is no physical distance associated to a $\ed \hat{\om}$ displacement in \eqref{eq:dl2spec} means that, geometrically speaking, $\Ss_{\rm spec}$ is a stack of superimposed regular spheres $\Ss$ parametrized by $\hat{\om}$. Under LLTs, the invariance of $\ed^2 l_{\Ss_{\rm spec}}$, along with Eqs. \eqref{eq:hatomLLT} and \eqref{eq:dl2spec}, implies the following transformation for the metric
\beq \label{eq:SABtrans}
\ti{S}_{\hat{A}\hat{B}}(\ti{\vte}) = \hat{\La}_{\para}^{-2}(\vte) \, \frac{\pa \vte^{\hat{C}}}{\pa \ti{\vte}^{\hat{A}}}(\ti{\vte}(\vte)) \, \frac{\pa \vte^{\hat{D}}}{\pa \ti{\vte}^{\hat{B}}}(\ti{\vte}(\vte)) \, S_{\hat{C}\hat{D}}(\vte) \, .
\eeq
Thus, one needs to work with the full $\Ss_{\rm spec}$ in order to interpret the effect of LLTs as a coordinate transformation, given by \eqref{eq:SOLLT} and \eqref{eq:hatomLLT}. If instead one restricts to the subspace $\Ss$, then LLTs induce both a coordinate and a conformal transformation \eqref{eq:SABtrans}. To understand the presence of the conformal factor $\hat{\La}_{\para}^{-2}(\vte)$ in \eqref{eq:SABtrans}, note that rotations are an isometry of the metric, i.e. they preserve the functional relation $\ti{S}_{\hat{A}\hat{B}}(\vte) \os{\rm rot.}{=} S_{\hat{A}\hat{B}}(\vte)$, but boost-induced coordinate transformations \eqref{eq:SOLLT} are not.  
However, the function $S_{\hat{A}\hat{B}}(\vte)$ given in \eqref{eq:SABdef} is the same for all observers, since we obtained it without specifying the latter. Indeed, using the transformation property \eqref{eq:niOinv}, we find that under an LLT
\beq
\ti{S}_{\hat{A}\hat{B}}(\ti{\vte}) := \ti{\pa}_{\hat{A}} \ti{\hat{n}}^i(\ti{\vte})\, \ti{\pa}_{\hat{B}} \ti{\hat{n}}^i(\ti{\vte}) = \ti{\pa}_{\hat{A}} \hat{n}^i(\ti{\vte})\, \ti{\pa}_{\hat{B}} \hat{n}^i(\ti{\vte}) \equiv S_{\hat{A}\hat{B}}(\ti{\vte}) \, ,
\eeq
just like the $\hat{n}^i(\vte)$. Thus, we can understand the conformal factor in \eqref{eq:SABtrans} as a compensator in order to make $S_{\hat{A}\hat{B}}(\vte)$ invariant under boosts too. One can actually check this explicitly, by computing the transformation of the line-element \eqref{eq:dOmdef} under \eqref{eq:SOLLT}
\beq \label{eq:dl2Strans}
\ed \ti{l}^2_{\Ss} := \ed \ti{\vte}^2 + \sin^2 \ti{\vte} \, \ed \ti{\vph}^2 = \hat{\La}^{-2}_{\para}(\vte) \( \ed \vte^2 + \sin^2 \vte \, \ed \vph^2 \) \equiv \hat{\La}_{\para}^{-2}(\vte)\, \ed l^2_{\Ss} \, .
\eeq 
so by pulling out a $\hat{\La}_{\para}^{-2}$ factor in \eqref{eq:SABtrans} the $S_{\hat{A}\hat{B}}(\vte)$ functions remain the same indeed. Finally, the transformation \eqref{eq:dl2Strans} provides a clear interpretation of the effect of boosts on the observer sky. Since $\hat{\La}_{\para}(\vte)$ appears as a ``radius" in the line-element $\hat{\La}_{\para}^2 \ed l^2_{\Ss}$, the latter describes the geometry of an ellipsoid directed along $\sim \hat{\La}^0_{\,\,\,i}$. This conformal factor therefore accounts for the stretch/compression of angular distances on the sky under a boost. 

Let us next define an orthonormal basis $S_A^{\hat{A}}(\vte)$, with internal indices $A \in \{ 1, 2 \}$, i.e. a ``dyad" (or ``zweibein"), associated with the metric $S_{AB}$ 
\beq \label{eq:SbAAdef}
S_{\hat{A} \hat{B}} \, S^{\hat{A}}_A S^{\hat{B}}_B = \de_{AB} \, ,
\eeq
with inverse $S_{\hat{A}}^A$, which therefore transforms in the vector analogue of Eq. \eqref{eq:SABtrans} under LLTs at $\hat{P}$
\beq 
\ti{S}^{\hat{A}}_A(\ti{\vte}) = \hat{\La}_{\para}(\vte) \, \frac{\pa \ti{\vte}^{\hat{A}}}{\pa \vte^{\hat{B}}}(\vte) \, S^{\hat{B}}_A(\vte) \, .
\eeq
As for the internal (unhatted) indices, they can mix under a {\it local} rotation symmetry on $\Ss$ without altering the defining equation \eqref{eq:SbAAdef} 
\beq \label{eq:locrotS}
\ti{S}_A^{\hat{A}}(\vte) = R_A^{\,\,\,B}(\vte)\, S_B^{\hat{A}}(\vte) \, , \hspace{1cm} R^{AB}(\vte) = \exp \[ - \al(\vte) \, \vep^{AB} \] = \de^{AB} \cos \al(\vte) - \vep^{AB} \sin \al(\vte)  \, ,
\eeq
and are therefore displaced with $\de^{AB}$. Just as in the case of the $\hat{n}^i(\vte)$ and $S_{\hat{A}\hat{B}}(\vte)$ functions, we would also like to have fixed $S^A_{\hat{A}}(\vte)$ functions for all observers, which we choose to be
\beq \label{eq:dyad}
S^A_{\hat{A}}(\vte) = \( \begin{array}{cc} 1 & 0 \\ 0 & \sin \vte \end{array} \)^A_{\hat{A}}  \, .
\eeq
This choice is not invariant under neither LLTs nor local rotations, because both lead to a non-diagonal matrix in general. We have again broken two symmetries down to a combination of them that preserves the gauge condition. Moreover, here too we still have the full LLT freedom, as any such transformation can be compensated by some local rotation. The transformation which preserves \eqref{eq:dyad} is therefore of the form
\beq \label{eq:sbAAtrans}
\ti{S}^{\hat{A}}_A(\ti{\vte}) = R_A^{\,\,\,B}(\vte) \, \hat{\La}_{\para}(\vte)\, \frac{\pa \ti{\vte}^{\hat{A}}}{\pa \vte^{\hat{B}}}(\vte) \, S^{\hat{B}}_B(\vte)   \, ,
\eeq
where now the the compensating angle $\al(\vte)$ in $R^A_{\,\,\,B}(\vte)$ is entirely determined by the Lorentz generator $\hat{\te}_{ab}$, such that
\beq
\ti{S}_{\hat{A}}^A(\ti{\vte}) = S_{\hat{A}}^A(\ti{\vte}) \, .
\eeq
To show that such a compensating local rotation exists, we just need to isolating $R_A^{\,\,\,B}$ in Eq. \eqref{eq:sbAAtrans}
\beq
R^A_{\,\,\,B}(\vte) =  \hat{\La}^{-1}_{\para}(\vte)\, \ti{S}^{\hat{A}}_A(\ti{\vte}) \, \frac{\pa \ti{\vte}^{\hat{A}}}{\pa \vte^{\hat{B}}}(\ti{\vte}) \, S_{\hat{B}}^B(\vte) \, ,
\eeq 
and show that it is a rotation matrix indeed, i.e. $R R^T = {\rm id}$. Using \eqref{eq:SABtrans}, we get
\bea
R_A^{\,\,\,C}(\vte)\, R_B^{\,\,\,C}(\vte) & = & \hat{\La}_{\para}^{-2}(\vte) \, \ti{S}_A^{\hat{A}}(\ti{\vte})\, \ti{S}_B^{\hat{B}}(\ti{\vte}) \, \frac{\pa \vte^{\hat{C}}}{\pa \ti{\vte}^{\hat{A}}}(\ti{\vte})\, \frac{\pa \vte^{\hat{D}}}{\pa \ti{\vte}^{\hat{B}}}(\ti{\vte}) \, S^C_{\hat{C}}(\vte)\, S^C_{\hat{D}}(\vte) \nn \\
 & \equiv & \hat{\La}_{\para}^{-2}(\vte) \, \ti{S}_A^{\hat{A}}(\ti{\vte})\, \ti{S}_B^{\hat{B}}(\ti{\vte}) \, \frac{\pa \vte^{\hat{C}}}{\pa \ti{\vte}^{\hat{A}}}(\ti{\vte})\, \frac{\pa \vte^{\hat{D}}}{\pa \ti{\vte}^{\hat{B}}}(\ti{\vte}) \, S_{\hat{C}\hat{D}}(\vte) \nn \\
 & \equiv & \hat{\La}_{\para}^{-2}(\vte) \, \ti{S}_A^{\hat{A}}(\ti{\vte})\, \ti{S}_B^{\hat{B}}(\ti{\vte}) \, \ti{S}_{\hat{A}\hat{B}}(\ti{\vte}) \equiv \de_{AB} \, .
\eea 
Thus, we still have the LLT freedom at $\hat{P}$, we must only be aware that this symmetry now induces a compensating local rotation on the tangent space of $\Ss$, just as it induces a compensating $\Lie$D on $\Lie$, and both transformations are controlled by $\hat{\La}^a_{\,\,\,b}$ and $\hat{n}^i(\vte)$.

\subsection{Observer space}

\subsubsection{Bundle of geodesics} \label{sec:vteparam}

From \eqref{eq:gadot} and \eqref{eq:ndot} we see that $\ga^{\mu}(\ze)$ is uniquely determined by the boundary data $\hat{n}^i$ and $\hat{\ga} = \hat{P}$, i.e. it is independent of $\hat{\om}$. This means that the points of $\Ss$ are in a one-to-one correspondence with the light-like geodesic paths $\ga \subset \cM$ going through $\hat{P}$.\footnote{This is not the case for time-like geodesics, where one needs all of the three $\hat{k}^i$ numbers to distinguish among all possible geodesics at $\hat{P}$. Indeed, in the light-like case the geodesics are constrained to lie on the light-cone, whereas in the time-like case they probe its interior, which has one more dimension.} We can therefore label these geodesics by $\vte^{\hat{A}}$, leading to a bundle of paths $\ga(\ze,\vte)$ generating the light-cone of $\hat{P}$. Simply put, we are parametrizing the geodesics by their boundary data $\hat{n}^i$ in the $\hat{e}_a$ frame, which are directly the observed angles. Along with the redshift parameter $z$ (or $\ze$), these are the fundamental observables the actual observer uses to parametrize events on the light-cone in practice. As a consequence, any $\Lie$-field $X(\ze)$ acquires a $\vte^{\hat{A}}$-dependence $X(\ze, \vte)$ which, just as $\ga^{\mu}(\ze,\vte)$, effectively makes it a field on 
\beq \label{eq:3Cdef}
\cC := \Lie \times \Ss \, .
\eeq
We will refer to this space as the ``observer space". If we are also interested in spectral distributions, then these are fields of the form $X(\ze,\vte,\hat{\om})$, thus living on the ``spectral observer space" 
\beq
\cC_{\rm spec} := \Lie \times \Ss_{\rm spec} \equiv \cC \times \Rs_+ \, .
\eeq
Let us now pay closer attention to the boundary conditions at the observer position $\hat{P}$. By construction, the image of $\cC$ under the $\ga$ map is the observer light-cone $\ga(\cC) \subset \cM$, i.e. the subspace of $\cM$ spanned by all the light-like geodesics attached to $\hat{P}$, so in particular  
\beq \label{eq:gluingcond}
\lim_{\ze \to 0} \ga(\ze, \vte) \equiv \hat{P} \, , \hspace{1cm} \forall \, \vte^{\hat{A}} \, . 
\eeq
Consequently, the space-time fields evaluated on $\cC$ obey
\beq
\lim_{\ze \to 0} X(\ga(\ze, \vte)) \equiv X(\hat{P}) \, , \hspace{1cm} \forall \, \vte^{\hat{A}} \, . 
\eeq
Since the image $\ga(\cC)$ has conical topology, the $\ga$ map is continuous at $\ze = 0$, but not differentiable there. More precisely, its angular derivative is well-defined 
\beq \label{eq:BCpaAga}
\lim_{\ze \to 0} \pa_{\hat{A}}\ga^{\mu}(\ze,\vte) \equiv 0 \, ,  \hspace{1cm} \forall \, \vte^{\hat{A}} \, , 
\eeq
because of (\ref{eq:gluingcond}) and the fact that the cone is smooth in these directions. However, $\pa_{\ze} \ga^{\mu}$ is multivalued at $\ze = 0$
\beq
\lim_{\ze \to 0} \pa_{\ze} \ga^{\mu}(\ze,\vte) = \pa_{\ze} \ga^{\mu}(0, \vte) \, , 
\eeq
since $\pa_{\ze} \ga^{\mu}(0,\vte)$ is precisely the boundary information $\hat{n}^i(\vte) \sim \hat{e}^i_{\mu} \pa_{\ze} \ga^{\mu}(0,\vte)$ by construction, and also
\beq
\lim_{\ze \to 0} n^i(\ze,\vte) = \hat{n}^i(\vte) \, .
\eeq
Consequently, for space-time fields evaluated on the geodesic
\beq
\lim_{\ze \to 0} \pa_{\hat{A}} X(\ga(\ze, \vte)) = \lim_{\ze \to 0} \pa_{\hat{A}} \ga^{\mu} [\pa_{\mu} X](\ga(\ze, \vte)) \equiv 0 \, ,  \hspace{1cm} \forall \, \vte^{\hat{A}} \, , 
\eeq
but, given \eqref{eq:geogaka},
\beq
\lim_{\ze \to 0} \pa_{\ze} X(\ga(\ze, \vte)) = \lim_{\ze \to 0} \pa_{\ze} \ga^{\mu} [\pa_{\mu} X](\ga(\ze, \vte)) = \Si_{0\para\para}^{-1} [\pa_{\para} X](\ga(\ze, \vte)) \neq 0  \, . 
\eeq
Note that these conditions are not inconsistent, because the $\ga^{\mu}(\ze,\vte)$ are fields on $\cC$, which is a 3-cylinder \eqref{eq:3Cdef}, {\it not} a 3-cone, so they can have an angular dependence for all $\ze$, i.e. including at the $\ze = 0$ value. It is only the image $\ga(\cC)$ that has a conical topology in $\cM$ because of the boundary condition in Eq. (\ref{eq:gluingcond}). The latter then only affects the way in which geodesics are glued together, i.e. their $\vte^{\hat{A}}$-dependence given in Eq. (\ref{eq:BCpaAga}).  A crucial requirement for this construction is that the angular parameters $\vte^{\hat{A}}$ are {\it not} part of some coordinate system $x^{\mu}$ on $\cM$, but a parametrization of $\pa_{\ze} \ga^{\mu}$ data in $T_{\hat{P}} \cM$. Indeed, if the $\vte^{\hat{A}}$ were angular coordinates on $\cM$, i.e. parametrizing the cone $\ga(\cC)$ instead of the cylinder $\cC$, then they would necessarily be ill-defined at $\hat{P}$. 

Finally, note the following important property, which actually holds for all $\la$-parametrizations of the light-cone $(\la, \vte^{\hat{A}})$, i.e. not only the log-redshift one $\la = \ze$ corresponding to Eq. \eqref{eq:gfLD}. Using the geodesic light-like deviation equations
\beq
\pa_{\ep} \ga^{\mu} = e^{\mu}_a k^a \, , \hspace{1cm} \na_{\ep} k^a \equiv \pa_{\ep} k^a + \Si^a_{\,\,\,b\mu} \pa_{\ep} \ga^{\mu} k^b = 0 \, ,
\eeq
one obtains straightforwardly
\beq
\pa_{\ep} \( k_a e^a_{\mu} \pa_{\hat{A}} \ga^a \) = 0 \, .  
\eeq
The boundary condition in Eq. \eqref{eq:BCpaAga} then implies that this quantity is zero everywhere, and therefore that $\pa_{\ze} \ga^{\mu}$ and $\pa_{\hat{A}} \ga^{\mu}$ are orthonormal
\beq  \label{eq:pazepaAnorm}
g_{\mu\nu} \pa_{\ze} \ga^{\mu} \pa_{\hat{A}} \ga^{\nu} = 0 \, .
\eeq

\subsubsection{Caustic resolution} \label{sec:caustics}

From the previous paragraph we understand that the 3-cylinder construction is clearly distinct from the observational coordinate \cite{ObsCoord, Nugier:2013tca} and geodesic light-cone coordinate \cite{Gasperini:2011us, BenDayan:2012pp, BenDayan:2012wi, BenDayan:2012ct, BenDayan:2013gc, Fanizza:2013doa, Nugier:2013tca, Marozzi:2014kua, DiDio:2014lka, Fanizza:2014baa, Fanizza:2015swa, Fleury:2016htl, Scaccabarozzi:2017ncm, Mitsou:2017ynv, Fanizza:2018tzp} formalisms, where the angles associated with incoming light-rays are part of a specific coordinate system on $\cM$. Although these angles are not defined at the observer, they are an unambiguous parametrization of the incoming light-like geodesics because they are constant along these paths, by construction. There is, however, an important disadvantage in this approach. The fact that the angles are part of a coordinate system on $\cM$ implies that this formalism cannot handle caustic singularities, i.e. the case where two light-like geodesics based at the observer position cross each other at some other point down the past light-cone. Indeed, being parametrized by the coordinate angles, their crossing implies a coordinate singularity and thus the breakdown of the coordinate system. 

In contrast, with the 3-cylinder parametrization one works directly with the observed angles, by construction, and the presence of caustics does not lead to singularities. Consider for instance two light rays with observed angles $\vte^{\hat{A}}$ and $\vte'^{\hat{A}}$ crossing each other at respective log-redshifts $\ze$ and $\ze'$, i.e. the case
\beq \label{eq:gamunotinj}
\ga^{\mu}(\ze, \vte) = \ga^{\mu}(\ze', \vte') \, .
\eeq
In particular, note that this is already the case for $\ze = 0$, where all angles are sent to the same space-time point $\hat{P}$. The possibility of having \eqref{eq:gamunotinj} simply means that the $\ga$ map is not injective, not that it is singular. The important difference is that the space-time coordinates, where the singularity occurs, are not the parameters with respect to which we solve our equations, but rather the parameters of the target space. One must therefore simply keep in mind that two different points on $\cC$ may correspond to the same point on the light-cone $\ga(\cC) \subset \cM$. As already discussed in subsection \ref{sec:redpara}, the only limitation for the applicability of our formalism comes from the choice of the redshift parametrization, which is ill-defined at low $\ze$, where the Hubble expansion and observer velocities are comparable. Indeed, in the specific case $\Si_{0\para\para} = 0$, the geodesic equations \eqref{eq:gadot} and \eqref{eq:ndot} do become singular, and thus so does the $\ga^{\mu}(\ze,\vte)$ function. In these cases, one should choose another $\la$-parametrization (see for instance footnote \ref{ft:nonsingep}).

We therefore stress again that it is crucial not to mistake the space $\cC$ for a submanifold of $\cM$, because this would mean that $\ga$ is a coordinate transformation on that submanifold, which must therefore be bijective. Rather, $\cC$ is a distinct space, parametrized by the coordinates $\ze$ and $\vte^{\hat{A}}$, that is mapped to the topologically different $\ga(\cC) \subset \cM$ through a non-injective map $\ga$.

\subsubsection{Induced coordinate transformations on $\cC$ from LLTs}

Now a LLT induces the coordinate transformation in Eqs. (\ref{eq:zetrans}), (\ref{eq:SOLLT}) and (\ref{eq:hatomLLT}) on $\cC_{\rm spec}$, which we repeat here in order to stress the fact that these coordinates mix under local boosts
\bea
\ti{\ze}(\ze,\vte,\hat{\om}) & = & \ze + \log \frac{\La^0_{\,\,\,0}(\ze,\vte) - \La^0_{\,\,\,i}(\ze,\vte)\, n^i(\ze,\vte)}{\hat{\La}^0_{\,\,\,0} - \hat{\La}^0_{\,\,\,j} \hat{n}^j(\vte)} \, , \nn \\
\ti{\vte}^{\hat{A}}(\ze,\vte,\hat{\om}) & = & \ti{\vte}^{\hat{A}}(\vte) \, , \label{eq:zevteomLLT} \\
\ti{\hat{\om}}(\ze,\vte,\hat{\om}) & = & \[ \hat{\La}^0_{\,\,\,0} - \hat{\La}^0_{\,\,\,i} \hat{n}^i(\vte) \] \hat{\om} \, . \nn
\eea
The corresponding Jacobian matrix is therefore not in block-diagonal form and the partial derivatives on $\cC_{\rm spec}$ transform as follows
\bea 
\ti{\pa}_{\ze} & = & \frac{\La_{\para} \Si_{0\para\para}}{\pa_{\para} \La^0_{\,\,\,0} - \pa_{\para} \La^0_{\,\,\,j} n^j + \La^0_{\,\,\,0} \Si_{0\para\para}  - \La^0_{\,\,\,j} \Si_{j\para\para}}\, \pa_{\ze}   \, ,  \\
\ti{\pa}_{\hat{A}} & = & \frac{\pa \vte^{\hat{B}}}{\pa \ti{\vte}^{\hat{A}}} \[ \pa_{\hat{B}} - \Si_{0\para\para}\, \frac{\pa_{\hat{B}} \La^0_{\,\,\,0} - \pa_{\hat{B}} \La^0_{\,\,\,i} n^i - \La^0_{\,\,\,i} \pa_{\hat{B}} n^i + \La_{\para} \hat{\La}^{-1}_{\para} \hat{\La}^0_{\,\,\,i} \pa_{\hat{B}} \hat{n}^i}{\pa_{\para} \La^0_{\,\,\,0} - \pa_{\para} \La^0_{\,\,\,j} n^j + \La^0_{\,\,\,0} \Si_{0\para\para}  - \La^0_{\,\,\,j} \Si_{j\para\para}} \, \pa_{\ze} + \hat{\La}_{\para}^{-1} \hat{\La}^0_{\,\,\,i} \pa_{\hat{A}} \hat{n}^i \, \hat{\om} \pa_{\hat{\om}} \] \, , \nn \\
 \label{eq:paALLT} \\
\ti{\pa}_{\hat{\om}} & = & \hat{\La}_{\para}^{-1} \pa_{\hat{\om}} \, , 
\eea
where $\pa_{\para} := \pa_0 - n^i \pa_i$ and we have used Eq. \eqref{eq:ndot}. The fact that the $\pa_{\ze}$ and $\pa_{\hat{\om}}$ derivatives are only rescaled implies that the fields that are independent of $\ze$ and/or $\hat{\om}$ remain so for all observers. For instance, $\ga^{\mu}(\ze,\vte)$ and $n^i(\ze,\vte)$ are consistently independent of $\hat{\om}$. Also, this transformation of $\pa_{\ze}$ and $\pa_{\hat{\om}}$ implies that the corresponding invariant derivatives are simply
\beq
\pa_{\ep} = \hat{\om} \Si_{0\para\para} e^{\ze} \pa_{\ze}  \hspace{1cm} {\rm and} \hspace{1cm} \hat{\om} \pa_{\hat{\om}} \, .
\eeq
As for $\pa_{\hat{A}}$, it does not mix with the other two derivatives only at $\ze, \hat{\om} = 0$, so we will have to be careful about that. We nevertheless define the corresponding derivative in the dyad basis
\beq
\pa_A := S_A^{\hat{A}} \pa_{\hat{A}} \, ,
\eeq
which also transforms non-linearly, i.e. it mixes with $\pa_{\ze}$ and $\pa_{\hat{\om}}$ 
\beq
\ti{\pa}_A = \hat{\La}_{\para}(\vte) \, R_A^{\,\,\,B}(\vte)\, \pa_B + \dots \, ,
\eeq
where we have used \eqref{eq:sbAAtrans}. Next, we already know that $\ga^{\mu}(\ze,\vte)$ transforms as a set of four scalars under the LLT-induced reparametrization of $\ze$. As for the $\vte^{\hat{A}}$ dependence, we note that a LLT changes the tetrad basis $e_a$, and thus the $\hat{n}^i$ boundary data, but not the $\ga$ map. We thus still have the same geodesic and it therefore transforms as a scalar with respect the LLT-induced $\vte^{\hat{A}}$ reparametrization as well
\beq
\ti{\ga}^{\mu}(\ti{\ze},\ti{\vte}) = \ga^{\mu}(\ze,\vte) \, .
\eeq
Similarly, for $k^a$ we have
\beq
\ti{k}^a(\ti{\ze},\ti{\vte}) = \La^a_{\,\,\,b}(\ze,\vte)\, k^b(\ze,\vte) \, .
\eeq

\subsection{Full-sky Sachs basis} \label{sec:Sachs}

We now wish to complement the $n^i(\ze,\vte)$ direction with a 2-dimensional basis of its normal subspace that is parallel-transported along each geodesic. By decomposing vectors that are normal to $n^i$ in that basis we will therefore ``factor out" the trivial part of their dynamics which is due to parallel transport. However, parallel transport is only defined in terms of full Lorentz vectors, whereas the discussed basis will be purely spatial. We can therefore start by considering two dimensionless Lorentz vectors $k_A^a(\ze,\vte)$ on $\cC$, with $A \in \{ 1,2 \}$, i.e. transforming as
\beq
\ti{k}_A^a(\ti{\ze},\ti{\vte}) = \La^a_{\,\,\,b}(\ze,\vte) \, k_A^b(\ze,\vte) \, ,
\eeq
under LLTs and invariant under $\cM$Ds, and also satisfying
\beq \label{eq:SachsPT}
\na_{\ze} k_A^a = 0 \, , \hspace{1cm} k_a k_A^a = 0 \, ,  \hspace{1cm} k_{Aa} k^a_B = \de_{AB} \, ,
\eeq
where the last two equations are consistently preserved under parallel transport. We will refer to $k_A^a$ as a ``Sachs basis" and to the corresponding indices $A, B, C, \dots$ as ``Sachs indices''. The choice of the latter is not a coincidence, as we will see below. Observe that Eq. \eqref{eq:SachsPT} determines that basis only up to a shift of the form
\beq \label{eq:Sachsshift}
\ti{k}_A^a(\ze,\vte) = k_A^a(\ze,\vte) + \al_A(\vte)\, k^a(\ze,\vte) \, , 
\eeq
and a rotation of the $A$ indices
\beq \label{eq:globrotsAi}
\ti{k}_A^a(\ze,\vte) = R_A^{\,\,\,B}(\vte)\, k^a_B(\ze,\vte) \, , \hspace{0.5cm} R^{AB}(\vte) = \exp \[ -\al \, \vep^{AB}(\vte) \] = \de^{AB} \cos \al(\vte) - \vep^{AB} \sin \al(\vte)  \, ,
\eeq
where $\al_A$ and $\al$ are independent of $\ze$ in order to maintain parallel transport, so these are fields on $\Ss$. We will refer to these transformations as ``Sachs shifts" and ``Sachs rotations'', respectively. From the geometrical viewpoint, the shift freedom corresponds to the 2-parameter family of possible of 2-dimensional space-like normal subspaces to $k^a$, while the rotational freedom corresponds to a choice of orthonormal basis within that subspace. 

Now note that  
\beq \label{eq:nAdef}
n_A := - k_A^0 \, ,
\eeq
is generically non-zero, which means that $k_A^{\mu} := e_a^{\mu} k^a_A$ is not normal to the 4-velocity of the observer family
\beq
n_A \equiv e_0^{\mu} k_{A\mu} \neq 0 \, ,
\eeq
because $e_0^{\mu}$ is not parallel-transported along $\ga$ in general. The Sachs shifts allow us to set $n_A(\ze,\vte) = 0$ for a given value of $\ze$, i.e. as a boundary condition, but then the parallel transport of $k_A^a$ will generically induce $n_A \neq 0$ at other values. Solving the algebraic constraints \eqref{eq:SachsPT} on $k^a_A$, we readily derive that it takes the following form
\beq \label{eq:kAadecomp}
k_A^a =  - n_A\, \frac{k^a}{\om} + n_A^a \, , \hspace{1cm} n_A^a := \( 0, n_A^i \) \, , \hspace{1cm} n^i n_A^i \equiv 0 \, , \hspace{1cm} n_A^i n^i_B \equiv \de_{AB} \, .
\eeq
In particular, under a Sachs shift \eqref{eq:Sachsshift}
\beq \label{eq:Sachsshiftn}
\ti{n}_A(\ze,\vte) = n_A(\ze,\vte) - \al_A(\vte)\, \om(\ze) \, , \hspace{1cm} \ti{n}_A^i(\ze,\vte) = n_A^i(\ze,\vte) \, .
\eeq
From the first equation of \eqref{eq:kAadecomp} we see that $n_A$ actually controls the longitudinal direction $\sim k^a$, which is why it will end up dropping from the quantities of interest. As for the $n_A^i$, we see that they form an orthonormal basis of the subspace normal to $n^i$ and thus satisfy the extra identities
\beq \label{eq:sprop}
\de^{AB} n_A^i n_B^j \equiv n^{ij} \, , \hspace{1cm} \vep^{AB} n^i_A n^j_B \equiv \vep^{ijk} n^k \, .
\eeq
Moreover, the corresponding 4-dimensional diffeomorphism vectors $n_A^{\mu} := e^{\mu}_a n_A^a$ are both normal to $k^{\mu} := e^{\mu}_a k^a$ and the 4-velocity of the observer family $e_0^{\mu}$ 
\beq
k_{\mu} n_A^{\mu} \equiv 0 \, ,  \hspace{1cm} e_{0\mu} n^{\mu}_A \equiv 0 \, .
\eeq
The $n_A^{\mu}$ are therefore what one usually refers to as the ``Sachs basis" and 
\beq
n^{\mu\nu} := n_A^{\mu} n_A^{\nu} \equiv n^{ij} e_i^{\mu} e_j^{\nu} \, ,
\eeq
is the so-called ``screen projector". However, the $n_A^a$ are not Lorentz vectors, because $n_A^0 = 0$ is not a Lorentz-invariant condition, but they are $\cM$D scalars and thus the $n_A^{\mu}$ are $\cM$D vectors. Since here we privilege the Lorentz basis, we will work with $k_A^a$ in order to preserve Lorentz covariance. In terms of $n_A$ and $n^i_A$, transformations under LLTs read
\beq  \label{eq:sAsAiLLT}
\ti{n}_A(\ti{\ze},\ti{\vte}) = \[ \La_{\para} n_A - \La^0_{\,\,\,i} n^i_A \](\ze,\vte) \, , \hspace{1cm} \ti{n}_A^i(\ti{\ze},\ti{\vte}) = (\La_{\perp})^i_{\,\,\,j}(\ze,\vte)\, n_A^j(\ze,\vte) \, ,
\eeq
in the P$\Lie$D-compensated case and
\bea
\de_{\te} n_A & := & - \ka\, \pa_{\ze} n_A + n^i \te^{0i} n_A + \te^{0i} n^i_A + \Ord(\te^2) \, , \\
\de_{\te} n_A^i & := & - \ka\, \pa_{\ze} n_A^i - \( n^i \te^{0j} + \te^{ij} \) n_A^j + \Ord(\te^2) \, ,
\eea
in the A$\Lie$D-compensated case, while the evolution equation in \eqref{eq:SachsPT} becomes
\bea
\pa_{\ze} n_A & = & n_A - \frac{\Si_{0i \para}}{\Si_{0\para\para}} \, n^i_A \, , \label{eq:sAdot} \\ 
\pa_{\ze} n_A^i & = & \frac{n^i \Si_{0j\para} - \Si_{ij\para}}{\Si_{0\para\para}}\, n_A^j \, .  \label{eq:sAidot}
\eea
This shows again that, even if we set $n_A = 0$ at some $\ze$ value, this quantity will be turned on by parallel transport. Moreover, note that both the LLT and evolution equation of $n_A^i$ are independent of $n_A$, which means that we do not need to keep track of the latter, as only the $n_A^i$ information will turn out to be relevant. 

Finally, we must select a definite basis $k_A^a(\ze,\vte)$ by providing boundary conditions for it at the observer sky $\ze = 0$, and in doing so fixing the shift \eqref{eq:Sachsshift} and rotation \eqref{eq:globrotsAi} ambiguities. To that end, we note that we have already privileged a basis in the tangent space of $\Ss$, the one given by the dyad \eqref{eq:dyad}, which is nothing but the unit-normalization of the angular directions $\pa_{\hat{A}}$ on the sky. For the spatial part $n_A^i$ we thus pick
\beq  \label{eq:sAOidef}
\hat{n}_A^i(\vte) := n_A^i(0,\vte) = S_A^{\hat{A}}(\vte) \, \pa_{\hat{A}} \hat{n}^i(\vte) \equiv \pa_A \hat{n}^i(\vte) \, ,
\eeq
so that, using \eqref{eq:dyad} and \eqref{eq:tevphobs}
\beq \label{eq:sAiO}
\hat{n}_1(\vte) = \( \cos \vte \cos \vph, \cos \vte \sin \vph, -\sin \vte \)  \, , \hspace{1cm} \hat{n}_2(\vte) = \( -\sin \vph, \cos \vph, 0 \) \, . 
\eeq
These are the unit-normed longitudinal and latitudinal vectors fields on $\Ss$, just like the dyad, but when $\Ss$ is seen as a subset of $\Rs^3$. We first note that Eq. \eqref{eq:sAOidef} is consistent because it has the required properties
\beq \label{eq:sAOiprop}
\hat{n}^i \hat{n}_A^i \equiv 0 \, , \hspace{1cm} \hat{n}_A^i \hat{n}_B^i \equiv \de_{AB} \, , \hspace{1cm} \de^{AB} \hat{n}_A^i \hat{n}_B^j \equiv \hat{n}^{ij} \hspace{1cm} \vep^{AB} \hat{n}_A^i \hat{n}_B^j \equiv \vep^{ijk} \hat{n}^k \, .
\eeq
We must now also make the equality \eqref{eq:sAOidef} invariant under LLTs. Put differently, as in the case of $\hat{n}^i(\vte)$, $S_{\hat{A}\hat{B}}(\vte)$ and $S_{\hat{A}}^A(\vte)$, the $\hat{n}_A^i(\vte)$ given by Eq. \eqref{eq:sAiO} must now be the same functions of $\vte^{\hat{A}}$ for all observers
\beq
\ti{\hat{n}}_A^i(\ti{\vte}) = \hat{n}_A^i(\ti{\vte}) \, .
\eeq
To achieve this, we first compute the variation of the right-hand side using Eqs. \eqref{eq:niOvtetrans}, \eqref{eq:sbAAtrans} and \eqref{eq:paALLT} 
\beq
\ti{\pa}_A \ti{\hat{n}}^i(\ti{\vte}) = \hat{\La}_{\para}(\vte)\, R_A^{\,\,\,B}(\vte)\, \pa_B \[ \hat{\La}^{-1}_{\para}(\vte) \( \hat{\La}^i_{\,\,\,j} \hat{n}^j(\vte) - \hat{\La}^i_{\,\,\,0} \) \] = R_A^{\,\,\,B}(\vte) \[ (\hat{\La}_{\perp})^i_{\,\,\,j}(\vte)\, \pa_B \hat{n}^j(\vte) \] \, ,
\eeq
and see that it reproduces the transformation \eqref{eq:sAsAiLLT} of $n_A^i$, but along with a Sachs rotation \eqref{eq:globrotsAi} given by the LLT-compensating rotation $R^A_{\,\,\,B}(\vte)$ of the dyad basis
\beq \label{eq:sAiOLLT}
\ti{\hat{n}}_A^i(\ti{\vte}) = R_A^{\,\,\,B}(\vte)\, (\hat{\La}_{\perp})^i_{\,\,\,j}(\vte) \, \hat{n}^j_B (\vte) \, . 
\eeq
Thus, with this choice of boundary conditions \eqref{eq:sAOidef} we identify the local rotation symmetry on $\Ss$ (see Eq. \eqref{eq:locrotS}) with the global Sachs rotations along each geodesic $\vte^{\hat{A}}$ (see Eq. \eqref{eq:globrotsAi}), so that the $A$ indices of $n_A^i$ and of $S_A^{\hat{A}}$ are now indeed the same. Remember that the Sachs rotations are independent of $\ze$, in order to preserve parallel transport, but they can depend on $\vte^{\hat{A}}$, i.e. they can be different for each light-like geodesic reaching the observer. By fixing the boundary condition \eqref{eq:sAOidef} we are fixing the freedom of independent Sachs rotations by matching them to the local rotations on $\Ss$, thus making them act solely as compensators of LLTs at the observer point $\hat{P}$. For a given observer $\hat{e}_a$, we have thus fully determined an orthonormal basis $\{ n^i, n_A^i \}$ for the spatial part of Lorentz vectors on all of $\cC$. This basis is adapted to the factorization of that space $\Lie \times \Ss$ in that $n^i$ generates the propagation direction, while the $n_A^i$ generate the transverse directions.  

Let us now consider the boundary conditions for $\hat{n}_A(\vte) := n_A(0,\vte)$. We first remind that, since $n_A$ will not appear in any of the physical observables, its value is actually irrelevant. Nevertheless, as a matter of aesthetics, one might be interested in considering the privileged choice $\hat{n}_A = 0$, meaning that $\hat{k}_A^{\mu}$ is normal to the observer velocity $\hat{e}_0^{\mu}$. However, this condition is not preserved under local boosts at $\hat{P}$, because the LLT of $\hat{n}_A$ is not linear (see Eq. \eqref{eq:sAsAiLLT}). With the compensating Sachs rotation that is now required it reads
\beq \label{eq:sAOLLT}
\ti{\hat{n}}_A(\ti{\vte}) = R_A^{\,\,\,B}(\vte) \[ \hat{\La}_{\para}(\vte)\, \hat{n}_B(\vte) - \hat{\La}^0_{\,\,\,i} \hat{n}_B^i(\vte) \] \, .
\eeq 
Nevertheless, remember that $n_A$ also transforms non-linearly under the Sachs shift transformations \eqref{eq:Sachsshiftn} which, just like the Sachs rotations, can now depend on the observed angles $\vte^{\hat{A}}$. We can therefore use these shifts to compensate the non-linear part of \eqref{eq:sAOLLT} and the required parameter is
\beq
\al_A(\vte) = - \hat{\om}^{-1} \hat{\La}^0_{\,\,\,i} R_A^{\,\,\,B}(\vte) \, \hat{n}_B^i(\vte) \, .
\eeq
With this, the boundary condition that makes the Sachs basis $k_A^a$ purely spatial at the observer
\beq \label{eq:Sshiftgf}
\hat{n}_A(\vte) = 0 \, ,
\eeq
is preserved under LLTs. Thus, just as for the Sachs rotations, the Sachs shift freedom can be completely fixed by acting as a compensator of LLTs in order to preserve some condition. Nevertheless, when defining observables, it is still useful to check invariance under these shifts to make sure that they do not depend on $n_A$.

\subsection{Transformation rule summary}

To conclude this subsection, let us summarize the LLT transformations of the fields on $\cC$. We remind that our construction has fixed the $\Lie$D, Sachs rotation and Sachs shift symmetries, with these transformations now being induced by LLTs in order to preserve the conditions \eqref{eq:gfLD}, \eqref{eq:dyad} and \eqref{eq:Sshiftgf}, respectively. In the P$\Lie$D-compensated case we have
\bea
\ti{\ga}^{\mu}(\ti{\ze},\ti{\vte}) & = & \ga^{\mu}(\ze,\vte) \, , \label{eq:gaPLLT} \\
\ti{n}^i(\ti{\ze},\ti{\vte}) & = & \[ \La_{\para}^{-1} \( \La^i_{\,\,j} n^j - \La^i_{\,\,0} \) \](\ze, \vte) \, , \label{eq:nPLLT} \\
\ti{n}_A(\ti{\ze},\ti{\vte}) & = & R_A^{\,\,\,B}(\vte) \[ \La_{\para} n_B - \( \La^0_{\,\,\,i} - \hat{\La}^0_{\,\,\,i} \)  n^i_B \](\ze,\vte) \, , \label{eq:sAPLLT} \\
\ti{n}_A^i(\ti{\ze},\ti{\vte}) & = & R_A^{\,\,\,B}(\vte) \, (\La_{\perp})^i_{\,\,\,j}(\ze,\vte)\, n_B^j(\ze,\vte) \, , \label{eq:sAiPLLT} 
\eea
In the A$\Lie$D-compensated case, which is expressed in terms of variations, we must be careful because each side is evaluated at different angles, thus bringing two possibilities, as in the case of the $\cM$D symmetry for instance (see appendix \ref{app:diffsym}). In the case of the $\ze$ parameter, this is a genuine coordinate system on the manifold $\Lie$, so an active diffeomorphism corresponds to a variation at fixed $\ze$, as we have implicitly done so far. The $\vte^{\hat{A}}$ parameters, however, are rigidly related to the boundary values of the $n^i$ field, i.e. we have $\ti{\hat{n}}^i(\ti{\vte}) = \hat{n}^i(\ti{\vte})$, instead of a scalar field transformation $\ti{\hat{n}}^i(\ti{\vte}) = \hat{n}^i(\vte)$. It therefore makes no sense to compute the variation at fixed $\vte^{\hat{A}}$, since we are precisely looking at how $n^i$ varies. We must thus use
\beq
\de_{\te} X(\ze,\vte) := \ti{X}(\ze, \ti{\vte}) - X(\ze,\vte) \, , \hspace{1cm} {\it not} \hspace{1cm} \de_{\te} X(\ze,\vte) := \ti{X}(\ze, \vte) - X(\ze,\vte) \, ,
\eeq
since the latter would give zero for $n^i$ at $\ze = 0$. With this definition of $\de_{\te}$ we thus get
\bea
\de_{\te} \ga^{\mu} & = & -\ka\, \pa_{\ze} \ga^{\mu} + \Ord(\te^2) \, , \label{eq:gaALLT} \\
\de_{\te} n^i & = & -\ka\, \pa_{\ze} n^i + n^{ij} \te^{0j} - \te^{ij} n^j + \Ord(\te^2) \, , \label{eq:nALLT} \\
\de_{\te} n_A & = & -\ka\, \pa_{\ze} n_A - \al\, \vep_{AB}\,  n_B + n^i \te^{0i} n_A + \( \te^{0i} - \hat{\te}^{0i} \) n^i_A + \Ord(\te^2) \, , \label{eq:sAALLT}  \\
\de_{\te} n_A^i & = & -\ka\, \pa_{\ze} n_A^i - \al\, \vep_{AB}\,  n_B^i - \( n^i \te^{0j} + \te^{ij} \) n_A^j + \Ord(\te^2) \, . \label{eq:sAiALLT} 
\eea

\subsection{Observables from localized sources} \label{sec:locsourceobs}

We can now consider the cosmological observables that are associated to localized sources. These are typically frequency-independent and are therefore defined on the observer space $\cC$.

\subsubsection{Distances and weak lensing}

We start by noting that $\pa_{\hat{A}} \ga^{\mu}(\ze,\vte)$ is the linear map relating the difference in space-time position between two infinitesimally close geodesics at $( \ze, \vte )$ to the corresponding observed angular deviation $\ed \vte^{\hat{A}}$
\beq
\ed_{\vte} \ga^{\mu} := \ed \ga^{\mu}|_{\ed \ze = 0} \equiv \pa_{\hat{A}} \ga^{\mu} \, \ed \vte^{\hat{A}} \, .
\eeq
Just as $\pa_{\ze} \ga^{\mu}$, this is a vector under $\cM$Ds, i.e. in the passive and active versions we have
\beq
\pa_{\hat{A}} \ti{\ga}^{\mu} = \frac{\pa \ti{x}^{\mu}}{\pa x^{\nu}}(\ga) \, \pa_{\hat{A}} \ga^{\nu} \, , \hspace{1cm} \de_{\xi} \pa_{\hat{A}} \ga^{\mu} = [\pa_{\nu} \xi^{\mu}](\ga)\, \pa_{\hat{A}} \ga^{\nu} + \Ord(\xi^2) \, ,
\eeq
respectively. Therefore, reminding that $\pa_A := S_A^{\hat{A}} \pa_{\hat{A}}$, the following quantity
\beq \label{eq:gaaAdef}
J^A_{\,\,\,B} := k_a^A e^a_{\mu} \pa_B \ga^{\mu}  \, ,
\eeq
is invariant under $\cM$Ds. It is also invariant under Sachs shifts, thanks to Eq. \eqref{eq:pazepaAnorm}, which here translates into $k_{\mu} \pa_{\hat{A}} \ga^{\mu} = 0$, so it is consistently independent of $n_A$
\beq
J^A_{\,\,\,B} \equiv n^A_i e^i_{\mu} \pa_B \ga^{\mu}  \, .
\eeq
Finally, it transforms linearly under LLTs, which is non-trivial because the angular derivative $\pa_{\hat{A}}$ mixes with $\pa_{\ze}$ and $\pa_{\hat{\om}}$ according to Eq. \eqref{eq:paALLT}. In this case, however, the non-angular derivatives cancel out because $k_a^A e^a_{\mu} \pa_{\ze} \ga^{\mu} \sim k_a^A k^a \equiv 0$ and $\pa_{\hat{\om}} \ga^{\mu} \equiv 0$. Thus, in the P$\Lie$D-compensated case we have
\beq \label{eq:gaABPLLT}
\ti{J}_{AB}(\ti{\ze},\ti{\vte}) = R_A^{\,\,\,C}(\vte)\, R_B^{\,\,\,D}(\vte) \, \hat{\La}_{\para}(\vte)\, J_{CD}(\ze,\vte) \, ,  
\eeq
while in the A$\Lie$D-compensated case we have
\beq \label{eq:gaABALLT} 
\de_{\te} J_{AB} = -\ka \, \pa_{\ze} J_{AB} + \hat{n}^i \hat{\te}^{0i} J_{AB} - \al \[ \vep_{AC} J_{CB} + \vep_{BC} J_{AC} \] + \Ord(\te^2) \, . 
\eeq
In particular, note that $J_{AB}$ is only sensitive to LLTs at $\hat{P}$. The boundary condition \eqref{eq:BCpaAga} implies
\beq \label{eq:BCJAB}
\hat{J}_{AB}(\vte) := J_{AB}(0,\vte) \equiv 0 \, , 
\eeq
It will be useful to use a matrix notation $\bm{J}$ for $J_{AB}$ in what follows. This $2\times 2$ matrix is the ``Jacobi map" \cite{Schneider:1992,Seitz:1994xf,Bonvin:2005ps,Lewis:2006fu,Bonvin:2008ni,Perlick:2010zh,Fleury:2013sna,Fanizza:2013doa,Fleury:2014rea,Pitrou:2015iya,Fleury:2015hgz,Grimm:2018nto} which relates the physical observed angular deviation on the sky $S^A_{\hat{A}}\, \ed \vte^{\hat{A}}$ to the corresponding physical vector normal to $n^i$ in the source's rest-frame
\beq
n^A_i e^i_{\mu} \ed_{\vte} \ga^{\mu} \equiv J^A_{\,\,\,B} \[ S^B_{\hat{A}}\, \ed \vte^{\hat{A}} \] \, .
\eeq
For instance, the physical area at $\ga(\ze,\vte)$ normal to $n^i(\ze,\vte)$ corresponding to the observed solid angle element
\beq
\ed \Om := \frac{1}{2}\,\vep_{AB} \[ S^A_{\hat{A}} \, \ed \vte^{\hat{A}} \] \we \[ S^B_{\hat{B}} \, \ed \vte^{\hat{B}} \] \equiv \sin \vte\, \ed \vte \we \ed \vph \, ,
\eeq
is given by
\beq \label{eq:dAdOm}
\ed A :=\frac{1}{2}\, \vep_{AB} \[ n^A_i e^i_{\mu} \ed_{\vte} \ga^{\mu} \] \we \[ n^B_j e^j_{\nu} \ed_{\vte} \ga^{\nu} \] = \[ \det \bm{J} \] \ed \Om \, .
\eeq
Thus, from (\ref{eq:dAdOm}) we have that
\beq
D(\ze,\vte) := \sqrt{\det \bm{J}(\ze,\vte)} \, ,
\eeq
is the angular diameter distance at $(\ze,\vte)$. Let us now define
\beq \label{eq:muABdef}
K^A_{\,\,\,B} := - \hat{\om}^{-1} k^A_a \na_B k^a \equiv e^{\ze} n^A_i \[  \pa_B n^i - \Si^i_{\,\,\,\para\mu} \pa_B \ga^{\mu} \] \, ,
\eeq
where 
\beq \label{eq:naAXa}
\na_A X^a := \pa_A X^a + \Si^a_{\,\,\,b \mu} \pa_A \ga^{\mu} X^b  \, , 
\eeq
is formally the covariant derivative along the angular directions for Lorentz vectors, in total analogy with the one in the longitudinal direction (\ref{eq:nala})
\beq
\na_{\ze} X^a := \pa_{\ze} X^a + \Si^a_{\,\,\,b \mu} \pa_{\ze} \ga^{\mu} X^b \, ,
\eeq
and both are consistent with the covariant derivative on space-time fields evaluated on the geodesic, e.g.
\beq
\na_{\ze,A} [X^a(\ga)] \equiv \pa_{\ze,A} \ga^{\mu} [\na_{\mu} X^a](\ga) \, . 
\eeq
The subtlety, however, is again that the angular derivative $\pa_{\hat{A}}$ mixes with $\pa_{\ze}$ and $\pa_{\hat{\om}}$ according to Eq. \eqref{eq:paALLT}, so $\na_A$ is truly a covariant derivative only if it enters in some specific combination. This is the case in Eq. \eqref{eq:muABdef} because the $\sim \pa_{\ze}$ terms form $\na_{\ze}$ and thus drop by the geodesic equation
\beq
\sim \pa_{\ze} k^a + \om^a_{\,\,\,b\mu} \pa_{\ze} \ga^{\mu} k^b \equiv \na_{\ze} k^a = 0 \, ,
\eeq
while the $\sim \pa_{\hat{\om}}$ term vanishes because
\beq
\sim k_a^A \pa_{\hat{\om}} k^a \equiv k_a^A \pa_{\hat{\om}} \[ \hat{\om} e^{\ze} \( 1, -n^i \) \] = k_a^A \hat{\om}^{-1} k^a \equiv 0 \, .
\eeq
Thus, just as $J_{AB}$, the $K_{AB}$ combination transforms linearly under LLTs. In the P$\Lie$D-compensated case we have
\beq \label{eq:muABPLLT}
\ti{K}_{AB}(\ti{\ze}, \ti{\vte}) = R_A^{\,\,\,C}(\vte) \, R_B^{\,\,\,D}(\vte)\, K_{CD}(\ze,\vte)  \, ,
\eeq
while for the A$\Lie$D-compensated case
\beq \label{eq:muABALLT}
\de_{\te} K_{AB} = -\ka\, \pa_{\ze} K_{AB} - \al \[ \vep_{AC} K_{CB} + \vep_{BC} K_{AC} \]  + \Ord(\te^2)  \, .
\eeq
Also just like $J_{AB}$, the transformation of $K_{AB}$ solely depends on LLTs at the observer point $\hat{P}$. As for the boundary conditions, they are simply
\beq \label{eq:BCmuAB}
\hat{K}_{AB}(\vte) := K_{AB}(0,\vte) = \hat{n}_A^i \hat{\pa}_B \hat{n}^i \equiv \hat{n}_A^i \hat{n}_B^i \equiv \de_{AB} \, .
\eeq
Using Eqs. \eqref{eq:gaaAdef}, \eqref{eq:muABdef} and $\na_{\ze} k^a = \na_{\ze} k_A^a = 0$, we then have that
\beq \label{eq:gaABdot}
\pa_{\ze} J_{AB} = - \frac{K_{AB}}{e^{\ze} \Si_{0\para\para}} \, ,
\eeq
and
\beq \label{eq:KABdotint}
\pa_{\ze} K_{AB} = - \hat{\om}^{-1} k_A^a \na_{\ze} \na_B k_a = - \hat{\om}^{-1} k_A^a \[ \na_{\ze}, \na_B \] k_a = \frac{e^{\ze}}{\Si_{0\para\para}} \, R_{a\para b \para} \, k_A^a e^b_{\mu} \pa_B \ga^{\mu}  \, .
\eeq 
Now note that, because of \eqref{eq:pazepaAnorm}, the combination $e^a_{\mu} \pa_A \ga^{\mu}$ obeys $k_a e^a_{\mu} \pa_A \ga^{\mu} = 0$ and therefore decomposes into a purely spatial part and a $\sim k^a$ part. From the definition \eqref{eq:gaaAdef} of $J_{AB}$ we get 
\beq \label{eq:invertJacrel}
e^a_{\mu} \pa_A \ga^{\mu} = \( 0, n^i_B J^B_{\,\,\,A} \) + c_A k^a \, ,
\eeq
for some $c_A$. Plugging this in \eqref{eq:KABdotint}, the antisymmetry of $R_{abcd}$ in its last two indices eliminates the $\sim c_A$ term and, using also \eqref{eq:kAadecomp}, we finally get
\beq \label{eq:KABdot}
\pa_{\ze} K_{AB} = \frac{e^{\ze}}{\Si_{0\para\para}} \, R_{A\para C \para} J_{CB} \, , 
\eeq
where we have introduced the following notation for $\cM$-tensor fields evaluated on $\ga$
\beq
X_{n \dots} := n^i X_{i \dots} \, , \hspace{1cm} X_{A \dots} := n^i_A X_{i \dots} \, .
\eeq
Eqs. \eqref{eq:gaABdot} and \eqref{eq:KABdot} are essentially the projection on the Sachs basis of the geodesic deviation equation, in first-order form, with $\pa_A$ playing the role of the deviation operator. Along with the boundary conditions \eqref{eq:BCJAB} and \eqref{eq:BCmuAB}, these equations completely determine the Jacobi map $\bm{J}$. In the literature these equations are usually given in terms of the affine parameter $\la$, i.e. the one defined by the gauge $\ep = {\rm const.}$, so the novel aspect here is that we have expressed these equations directly in terms the observed (log-)redshift parametrization $\ze$.

It is now convenient to decompose the Riemann tensor in \eqref{eq:KABdot} into its Ricci and Weyl parts, and in particular the electric and magnetic components of the latter, to get, in matrix notation
\beq \label{eq:JKpaze}
\pa_{\ze} \bm{J} = - \frac{1}{e^{\ze} \om_{0\para\para}}\, \bm{K} \, , \hspace{1cm} \pa_{\ze} \bm{K} = \frac{e^{\ze}}{\om_{0\para\para}} \[ \frac{1}{2}\, R_{\para \para} + W_+ \bm{\si}_+ + W_{\times} \bm{\si}_{\times} \] \bm{J} \, ,
\eeq
where 
\beq \label{eq:Paulipcdef}
\bm{\si}_+ := \( \begin{array}{cc} 1 & 0 \\ 0 & -1 \end{array} \) \, , \hspace{1cm} \bm{\si}_{\times} := \( \begin{array}{cc} 0 & 1 \\ 1 & 0 \end{array} \) \, ,
\eeq
\beq
W_+ := E_+ + B_{\times} \, ,   \hspace{1cm} W_{\times} := E_{\times} - B_+ \, , 
\eeq
and
\beq
E_{+,\times} := \si_{+,\times}^{AB} E_{AB} \, ,   \hspace{1cm} B_{+,\times} := \si_{+,\times}^{AB} B_{AB} \, .
\eeq
We next express these equations in terms of physically-interpretable quantities. We first split $\bm{J}$ into its determinant and conformal parts
\beq \label{eq:JofDJc}
\bm{J} \equiv D \bm{J}_c \, , \hspace{1cm} \det \bm{J}_c \equiv 1 \, ,  
\eeq
and then express the latter through $\frak{sl}(2,\Rs)$ generators
\beq  \label{eq:JcofSi}
\bm{J}_c := \exp \[ s_{\circ} \bm{\vep} + s_+ \bm{\si}_+ + s_{\times} \bm{\si}_{\times} \] \equiv \bm{1} \cosh S + \( s_{\circ} \bm{\vep} + s_+ \bm{\si}_+ + s_{\times} \bm{\si}_{\times} \) \frac{\sinh S}{S} \, ,
\eeq
where
\beq \label{eq:vepmat}
\bm{\vep}_{AB} \equiv \vep_{AB} \, , \hspace{1cm} S := \sqrt{- s_{\circ}^2 + s_+^2 + s_{\times}^2} \, . 
\eeq
In particular, the eigenvalues of $\bm{J}$ now read 
\beq
\la_{\pm}[\bm{J}] = D e^{\pm S} \, ,
\eeq
and we also define the complex combination
\beq
s := s_+ + i s_{\times}  \, , \hspace{1cm} S \equiv \sqrt{- s_{\circ}^2 + |s|^2}  \, .
\eeq
Thus, for a source located at $\ga(\ze,\vte)$ with 4-velocity $e_0^{\mu}(\ga(\ze,\vte))$, $D(\ze,\vte)$ is the angular diameter distance to it, $s_{\circ}(\ze,\vte)$ is the angle by which the observed image has been rotated with respect to the Sachs basis, while the $s(\ze,\vte)$ parametrizes the shear deformation of that image with respect to the Sachs basis. Compared to the usual parametrizations of $\bm{J}_c$ (see e.g. \cite{Seitz:1994xf,Perlick:2010zh,Fleury:2014rea}), the advantage of the group-theoretically motivated one we chose in Eq. \eqref{eq:JcofSi} is that it is independent of the order in which the rotation and shear effects are considered, since they are both described through generators of the corresponding Lie algebra.

Let us now decompose $\bm{K}$. The fact that $\bm{J}_c \in {\rm SL}(2,\Rs)$, implies that 
\beq
\bm{Q}_c := (\pa_{\ze} \bm{J}_c) \bm{J}_c^{-1} \, , 
\eeq
is an element of $\frak{sl}(2,\Rs)$
\beq
\Tr \, \bm{Q}_c \equiv 0 \, .
\eeq
From the first equation of \eqref{eq:JKpaze} we then see that it is convenient to parametrize $\bm{K}$ as 
\beq
\bm{K} = \( \te \bm{1} + \bm{K}_c \) \bm{J}_c \, , \hspace{1cm} \bm{K}_c := \si_{\circ} \bm{\vep} + \si_+ \bm{\si}_+ + \si_{\times} \bm{\si}_{\times} \, , \hspace{1cm} \si := \si_+ + i \si_{\times}   \, .
\eeq
In terms of our new variables $\{ D, s_{\circ}, s, \te, \si_{\circ}, \si \}$, the boundary conditions \eqref{eq:BCJAB} and \eqref{eq:BCmuAB} translate into
\beq \label{eq:BCDfX}
D(0,\vte) = s_{\circ}(0,\vte) = s(0,\vte) = \si_{\circ}(0,\vte) = \si(0,\vte) = 0 \, , \hspace{1cm} \te(0,\vte) = 1 \, .
\eeq
Let us now express \eqref{eq:JKpaze} in terms of these variables. Multiplying the two equations by $\bm{J}_c^{-1}$ from the right and splitting the result into pure-trace and traceless parts, we obtain
\beq \label{eq:Jeqdec}
\pa_{\ze} D = - \frac{\te}{e^{\ze} \Si_{0\para\para}} \, , \hspace{1cm} \bm{Q}_c = - \frac{1}{e^{\ze} D \Si_{0\para\para}}\, \bm{K}_c \, ,
\eeq
and
\beq \label{eq:ydot}
\pa_{\ze} \te = \frac{1}{2 \Si_{0\para\para}} \[ \frac{1}{e^{\ze} D} \, \Tr\, ( \bm{K}_c^2 ) + e^{\ze} D R_{\para\para} \] \, ,
\eeq
\beq \label{eq:Kcdot}
\pa_{\ze} \bm{K}_c = \frac{1}{\Si_{0\para\para}} \[ \frac{1}{e^{\ze} D} \, \te \bm{K}_c + e^{\ze} D \( W_+ \bm{\si}_+ + W_{\times} \bm{\si}_{\times} \) \] \, ,
\eeq
where we have used  the second equation of \eqref{eq:Jeqdec} to simplify the last two and the fact that $\bm{K}_c^2 \sim \bm{1}$. Before we proceed further, we note that the $\sim \bm{\vep}$ component of \eqref{eq:Kcdot}, i.e. the evolution equation for $\si_{\circ}$, is a first-order linear differential equation for $\si_{\circ}$ with no source. Given the boundary condition \eqref{eq:BCDfX} we therefore have
\beq
\si_{\circ} = 0 \, ,
\eeq 
i.e. $\bm{K}$ is symmetric, everywhere on $\cC$. We next compute the only non-trivial term
\bea
\bm{Q}_c & \equiv & \[ \frac{\sinh 2 S}{2S}\, \pa_{\ze} s_{\circ} + \( 1 - \frac{\sinh 2 S}{2S} \) \frac{\pa_{\ze} S}{S}\, s_{\circ}  + \frac{\sinh^2 S}{S^2} \( s_+ \pa_{\ze} s_{\times} - s_{\times} \pa_{\ze} s_+ \) \] \bm{\ep} \nn \\
 & & + \[ \frac{\sinh 2 S}{2S}\, \pa_{\ze} s_+ + \( 1 - \frac{\sinh 2 S}{2S} \) \frac{\pa_{\ze} S}{S}\, s_+  + \frac{\sinh^2 S}{S^2} \( s_{\circ} \pa_{\ze} s_{\times} - s_{\times} \pa_{\ze} s_{\circ} \) \] \bm{\si}_+ \nn \\
 & & + \[ \frac{\sinh 2 S}{2S}\, \pa_{\ze} s_{\times} + \( 1 - \frac{\sinh 2 S}{2S} \) \frac{\pa_{\ze} S}{S}\, s_{\times}  + \frac{\sinh^2 S}{S^2} \( s_+ \pa_{\ze} s_{\circ} - s_{\circ} \pa_{\ze} s_+ \) \] \bm{\si}_{\times} \, , \nn \\
\eea
and, isolating the $\sim \pa_{\ze}$ terms in \eqref{eq:Jeqdec}, \eqref{eq:ydot} and \eqref{eq:Kcdot}, we finally find the system
\bea
\pa_{\ze} D & = & - \frac{1}{e^{\ze} \Si_{0\para\para}}\, \te \, , \label{eq:dotD} \\
\pa_{\ze} s_{\circ} & = & - \frac{1}{e^{\ze} D \Si_{0\para\para}} \[ \frac{1}{S^2} \( 1 - S \coth S \) {\rm Re} \( \bar{\si} s \) s_{\circ} + {\rm Im} \( \bar{\si} s \) \] \, , \label{eq:dotSio} \\
\pa_{\ze} s & = & - \frac{1}{e^{\ze} D \Si_{0\para\para}} \[ \( S \coth S + i s_{\circ} \) \si + \frac{1}{S^2} \( 1 - S \coth S \) {\rm Re} \( \bar{\si} s \) s \] \, , \label{eq:dotSi}  \\
\pa_{\ze} \te & = & \frac{1}{\Si_{0\para\para}} \[ \frac{1}{e^{\ze} D}\, |\si|^2 + \frac{1}{2}\, e^{\ze} D R_{\para\para} \] \, , \label{eq:dotchD} \\
\pa_{\ze} \si & = & \frac{1}{\Si_{0\para\para}} \[ \frac{1}{e^{\ze} D} \, \te \si + e^{\ze} D \( W_+ + i W_{\times} \) \] \, . \label{eq:dotch}
\eea
Note that the closed subsystem of Eqs. \eqref{eq:dotD}, \eqref{eq:dotchD} and \eqref{eq:dotch} for the set $\{ D, \te, \si \}$ is the analogue of the Sachs equations in our formalism. In particular, $\te$ and $\si$ are known as the optical expansion and shear, respectively. Also, observe that the right-hand side of \eqref{eq:dotSio} is made entirely of second-order terms in perturbation theory around the FLRW space-time, so that, given the boundary conditions \eqref{eq:BCDfX}, $s_{\circ}$ vanishes at first order, in agreement with \cite{Fleury:2015hgz, Yoo:2018qba, DiDio:2019rfy}.\footnote{In \cite{DiDio:2019rfy} the computation is technically the same, in that one compares the rotation angle between a parallely-transported vector (here $k_A^{\mu}$) and a vector transported according to the geodesic deviation equation (here $\pa_A \ga^{\mu}$), but these vectors have different physical interpretations than the ones considered here.} It is also interesting that, using \eqref{eq:dotSio} and \eqref{eq:dotSio}, one obtains
\beq
\pa_{\ze} S = - \frac{{\rm Re} \( \bar{\si} s \)}{D e^{\ze} \Si_{0\para\para} S} \, .
\eeq
This is a much simpler equation than \eqref{eq:dotSio}, so one could consider working with $S$ instead and recovering the angle, up to a sign, through 
\beq
s_{\circ} \equiv \sqrt{|s|^2 - S^2} \, .
\eeq
Finally, the P$\Lie$D-compensated LLT rules \eqref{eq:gaABPLLT} and \eqref{eq:muABPLLT} translate into
\bea
\ti{D}(\ti{\ze},\ti{\vte}) & = & \hat{\La}_{\para}(\vte)\, D(\ze,\vte) \, ,  \\
\ti{s}_{\circ}(\ti{\ze},\ti{\vte}) & = & s_{\circ}(\ze,\vte) \, ,  \\
\ti{s}(\ti{\ze},\ti{\vte}) & = & e^{-2i\al(\vte)} s(\ze,\vte) \, , \\
\ti{\te}(\ti{\ze}, \ti{\vte}) & = & \te(\ze,\vte) \, , \\
\ti{\si}(\ti{\ze}, \ti{\vte}) & = & e^{-2i\al(\vte)} \si(\ze,\vte) \, ,
\eea
while the A$\Lie$D-compensated ones \eqref{eq:gaABALLT} and \eqref{eq:muABALLT} give 
\bea
\de_{\te} D & = & - \ka \pa_{\ze} D + \hat{n}^i \hat{\te}^{0i} D + \Ord(\te^2) \, , \\
\de_{\te} s_{\circ} & = & - \ka \pa_{\ze} s_{\circ} + \Ord(\te^2) \, , \\
\de_{\te} s & = & - \ka \pa_{\ze} s - 2 i \al s + \Ord(\te^2)  \, , \\
\de_{\te} \te & = & - \ka \pa_{\ze} \te + \Ord(\te^2) \, , \\
\de_{\te} \si & = & - \ka \pa_{\ze} \si - 2 i \al \si + \Ord(\te^2) \, .
\eea
Note that these transformations abide to the criteria of cosmological observables, i.e. they solely depend on the LLTs at the observer and source positions. In fact, in the present case there is actually no dependence on the LLT parameter at the source position, implying in particular that these quantities do not depend on the source's 4-velocity. In contrast, this is not the case of the luminosity distance, given by Etherington's distance-duality equation
\beq \label{eq:DLdef}
D_L := \( 1 + z \)^2 D \equiv e^{2\ze} D \, ,
\eeq 
because the redshift is sensitive to LLTs at the source.

\subsubsection{Volume and source number density}

We now wish to find the relation between the physical volume occupied by some source $\ed V(\ga(\ze,\vte))$ in its rest-frame $e_0(\ga(\ze,\vte))$ and the corresponding observed solid angle $\ed \Om(\vte)$ and observed redshift interval $\ed z(\ze,\vte)$. This information is needed in order to infer a number density from number counts of localized sources, e.g. galaxies. We first split $\ed V$ into an area element $\ed A$ that is normal to $n^i(\ze,\vte)$ and a length element $\ed L$ that goes along that direction
\beq
\ed V := \ed A \, \ed L \, . 
\eeq
From the previous subsection we already have the relation between the are and the solid angle $\ed A = D^2 \ed \Om$, so we look for the $\ed L \sim \ed z$ relation. The length element $\ed L$ can be defined implicitly through the corresponding derivative operator acting on space-time fields $X$ evaluated on the geodesic
\beq
\pa_L X := n^i \pa_i X \, ,
\eeq
which is (minus) the spatial part of $\pa_{\para} := \pa_0 - n^i \pa_i$. To relate $\pa_L$ to $\pa_z$, we need to go back to a generic $\la$-parametrization of the line manifold $\Lie$ and note that equation (\ref{eq:gadotep}) gives 
\beq
\pa_{\para} = (\ep \om)^{-1} \pa_{\la} \ga^{\mu} \pa_{\mu}  \, ,
\eeq
so for fields on $\Lie$ this reduces to
\beq
\pa_{\para} =  (\ep \om)^{-1} \pa_{\la} \, . 
\eeq
The operator $\pa_L$ can then be consistently defined on such a field $Y$, only if it obeys an equation of the form
\beq
(\ep \om)^{-1} \pa_{\la} Y = Z_{\para} \equiv Z_0 - n^i Z_i \, ,
\eeq
for some Lorentz vector $Z_a$, because then we can unambiguously infer the spatial part of the variation
\beq
\pa_L Y := n^i Z_i \, .
\eeq
The redshift observable $z$ obeys such an equation (\ref{eq:zdotep}), so we can extract the spatial variation
\beq
\pa_L z = \( 1 + z \) \Si_{0\para i} n^i \equiv \( 1 + z \) \Si^0_{\,\,\,ij} n^i n^j \equiv \( 1 + z \) \Si^0_{\,\,\,nn}  \, ,
\eeq
thus leading to
\beq
\ed L = \( \pa_L z \)^{-1} \ed z = \frac{\ed z}{\( 1 + z \) \Si^0_{\,\,\,nn}}  \, .
\eeq
We conclude that the desired ratio
\beq
\ed V \equiv \cV \, \ed z \, \ed \Om \, ,
\eeq
is given by
\beq
\cV := \frac{D^2}{e^{\ze} \Si^0_{\,\,\,nn}} \, .
\eeq
Had we chosen to define the ratio $\cV$ in terms of the log-redshift interval $\ed V \equiv \cV \, \ed \ze \, \ed \Om$, the result would have been even simpler $\cV = D^2/\Si^0_{\,\,\,nn}$. Note that $\cV$ is not a cosmological observable, according to our definition, because its LLT depends on $\pa_{\mu} \te_{ab}$ through $\Si_{0ij}$, and thus to the Lorentz matrices $\La^a_{\,\,\,b}$ at points around the source position $\ga(\ze,\vte)$. This is ultimately due to the presence of $\ed z$, whose transformation also depends on $\pa_{\mu} \te_{ab}$ because it is the differential of the cosmological observable $z$. Finally, given the observed number of sources $\ed N(\ze, \vte)$ in the interval $\ed z\, \ed \Om$, we can infer the corresponding number density at $\ga(\ze,\vte)$
\beq \label{eq:nobs}
n := \frac{\ed N}{\ed V} = \frac{1}{\cV} \frac{\ed N}{\ed z\, \ed \Om} \, .
\eeq

\subsection{Observer space-time} \label{sec:observerspace}

The observer space $\cC$ covers the full light-cone of the observer at the point $\hat{P}$. We can, however, consider also the possibility of performing observations at a later time in order to measure the ``drift" of observables \cite{Korzynski:2017nas,Marcori:2018cwn}, and therefore from some other point $\hat{P}'$ in the future light-cone of $\hat{P}$. This means that, instead of a single observer point $\hat{P}$, we must select a specific observer world-line $\hat{P}(\ta)$, where $\ta \in \Rs$ can be conveniently chosen to be the proper time of the observer. To each point $\hat{P}(\ta)$ of this world-line we can then associate a corresponding observer space $\cC(\ta)$ and its spectral extension $\cC_{\rm spec}(\ta)$. This amounts to considering a continuous family of $\ga^{\mu}(\ze,\vte)$ maps parametrized by $\ta$, i.e. $\ga^{\mu}(\ta,\ze,\vte)$, so that all $\cC$-fields acquire a $\ta$ dependence, just as we did for the $\vte^{\hat{A}}$ parametrization in subsection \ref{sec:vteparam}. The observer world-line is then given by 
\beq
\hat{\ga}^{\mu}(\ta) := \ga^{\mu}(\ta,0,\vte) \, ,
\eeq
since there is no $\vte^{\hat{A}}$ dependence at $\ze = 0$. In the single observer point case we considered $\hat{e}_0$ as the observer 4-velocity, so in the observer world-line case the path $\hat{\ga}^{\mu}(\ta)$ must be an integral line of the 4-velocity field $e_0^{\mu}(x)$. If $\ta$ is to denote the proper time of the observer, then the relation is simply
\beq \label{eq:hatgaeq}
\pa_{\ta} \hat{\ga}^{\mu}(\ta) = e_0^{\mu}(\hat{\ga}(\ta)) \, .
\eeq
Given some reference point $\hat{P}$, this equation completely determines the world-line and therefore the corresponding function $\ga^{\mu}(\ta,\ze,\vte)$. Therefore, in general, if $\ze \neq 0$
\beq 
\pa_{\ta} \ga^{\mu}(\ta,\ze,\vte) \neq e_0^{\mu}(\ga(\ta,\ze,\vte)) \, .
\eeq
Now the space on which $\ga^{\mu}(\ta,\ze,\vte)$ is defined is 
\beq
\Ord := \Rs \times \cC
\eeq
and we will refer to it as the ``observer space-time". Again, if some fields on that space also depend on the observed frequency $\hat{\om}$, then they are actually defined on the ``spectral observer space-time" 
\beq
\Ord_{\rm spec} := \Ord \times \Rs_+ \, . 
\eeq
We thus have that $\ga^{\mu}(\ta,\ze,\vte)$ probes all the points of space-time that are connected to the observer world-line by some light-like geodesic, i.e. the image $\ga(\Ord)$ is the {\it observable universe} of that observer, by definition. The space $\Ord$ is parametrized by the proper time $\ta$, log-redshift $\ze$ and angles $\vte^{\hat{A}}$ at which the corresponding signal was observed, i.e. exactly the parameters the observer has access to in practice. It is important, however, to notice that $\{ \ta, \ze, \vte \}$ is {\it not} a coordinate system on $\cM$ in general. Indeed, as we have already discussed in the case of $\cC$ in subsection \ref{sec:caustics}, the map
\beq
\ga : \Ord \to \cM \, ,
\eeq
is not injective, firstly because it maps the full observer sky to a single point 
\beq
\ga(\ta,0,\vte) = \hat{P}(\ta) \, , 
\eeq
and secondly because, in the presence of caustics on the light-cone $\ta$, we have
\beq
\ga(\ta,\ze,\vte) = \ga(\ta,\ze',\vte') \, , \hspace{1cm} \( \ze, \vte \) \neq \( \ze', \vte' \) \, .
\eeq
Moreover, in the presence of strong gravitational fields, the $\ga$ map can also be non-surjective, i.e. there may very well be points of $\cM$ that are not connected by any light-like geodesic to the observer world-line $\hat{P}(\ta)$. Thus, $\{ \ta, \ze, \vte \}$ can be interpreted only as a {\it local} coordinate system in patches where the gravitational field is such that $\ga^{\mu}\( \ta, \ze, \vte \)$ is invertible. At the level of the observables, they too acquire a unique dependence on $\ta$, determined by the unique $\ga^{\mu}(\ta,\ze,\vte)$ map, and their drift is now simply obtained by taking the derivative with respect to $\ta$. Importantly, this drift will depend on the dynamics of the observer under consideration and therefore on the choice of LLT gauge in the vicinity of the world-line. As already discussed in subsection \ref{sec:paralltrans}, the natural choice in cosmology is the choice of free-falling non-precessing observers. 

It is interesting to consider a patch where the $\ga$ map is indeed a diffeomorphism, so that the $\{ \ta, \ze, \vte \}$ can be interpreted as a set of local space-time coordinates. Using $\ch^{\hat{\mu}}$ to collectively denote these coordinates, the corresponding metric is obtained by performing the coordinate transformation $x^{\mu} \to \ch^{\hat{\mu}}$, i.e. pulling back $g$ along the $\ga$ map
\beq
g_{\hat{\mu}\hat{\nu}}(\chi) := g_{\mu\nu}(\ga(\ch))\, \pa_{\hat{\mu}} \ga^{\mu}(\ch)\, \pa_{\hat{\nu}} \ga^{\nu}(\ch) \, ,
\eeq 
and Eqs. \eqref{eq:geogaka} and \eqref{eq:pazepaAnorm} lead to a line-element of the form
\beq
\ed s^2 = \al \ed \ta^2 + 2 \be \ed \ta \ed \ze + h_{\hat{A}\hat{B}} \( \ed \vte^{\hat{A}} + v^{\hat{A}} \ed \ta \) \( \ed \vte^{\hat{B}} + v^{\hat{B}} \ed \ta \) \, ,
\eeq
which is the one of observational coordinates \cite{ObsCoord, Nugier:2013tca}. This is not surprising, since the $\ta = {\rm const.}$ hypersurfaces are light-cones and the $\vte^{\hat{A}}$ angles are constant along the light-like geodesics composing it. In particular, using Eq. \eqref{eq:invertJacrel} we find that the 2-metric is essentially the square of the Jacobi map and, with Eqs. \eqref{eq:JofDJc} and \eqref{eq:JcofSi}, 
\bea
h_{\hat{A}\hat{B}}(\ch) & \equiv & \( S_{\hat{A}}^A S_{\hat{B}}^B \)(\vte) \[ J^C_{\,\,\,A} J_{CB} \](\ta,\ze,\vte)  \\
 & \equiv & \( S_{\hat{A}}^A S_{\hat{B}}^B \)(\vte) \[ D^2 \( \bm{1} \cosh (2 S) + \( s_+ \bm{\si}_+ + s_{\times} \bm{\si}_{\times} \) \frac{\sinh (2S)}{S} \) \]_{AB}(\ta,\ze,\vte) \, . \nn
\eea
On the other hand, Eqs. \eqref{eq:geogaka}, \eqref{eq:dyad}, \eqref{eq:nAdef}, \eqref{eq:Sshiftgf} provide the following values on the observer world-line
\beq
\al(\ta,0,\vte) = -1 \, , \hspace{1cm} \be(\ta,0,\vte) = - \hat{\Si}^{-1}_{0\para\para}(\ta) \, , \hspace{1cm} v^{\hat{A}}(\ta,0,\vte) = 0 \, , 
\eeq
while \eqref{eq:BCDfX} and \eqref{eq:hatgaeq} imply $D(\ch) = -\hat{\Si}^{-1}_{0\para\para}(\ta)\, \ze + \Ord(\ze^2)$, so
\beq
\lim_{\ze \to 0} \ze^{-2} h_{\hat{A}\hat{B}}(\ch)\, \ed \vte^{\hat{A}} \ed \vte^{\hat{B}} = \hat{\Si}^{-2}_{0\para\para}(\ta) \( \ed \vte^2 + \sin^2 \vte\, \ed \vph^2 \) \, .
\eeq
These conditions are exactly the ones defining the system of observational coordinates in the redshift parametrization \cite{ObsCoord, Nugier:2013tca}, which is the same as the log-redshift one close to the observer since $z = \ze + \Ord(\ze^2)$. This exercise therefore provides a nice consistency check of our formalism. We wish, however, to remind one last time that working with arbitrary coordinates and the $\ga$ map allows one to resolve caustics, which is not the case when using observational coordinates. 

Finally, we should also discuss the effect of LLTs on $\Ord$, as we did for the observer sky $\Ss$ and then the observer space $\cC$. As for the $\ze$, $\vte^{\hat{A}}$ and $\hat{\om}$ parameters, $\ta$ is trivially invariant under $\cM$Ds, since it serves as an internal parameter of a $\cM$D-covariant equation \eqref{eq:hatgaeq}. Under an LLT, however, the new 4-velocity field $\ti{e}_0 = \La_0^{\,\,\,a} e_a$ implies a different $\ti{\hat{\ga}}^{\mu}$ solution. In particular, the $\ti{\hat{\ga}}$ and $\hat{\ga}$ solutions can share at most one point $\hat{P}$ in general, which is the point we were considering implicitly when working with $\cC$ alone. For the $\ta$ value corresponding to $\hat{\ga}(\ta) = \hat{P}$, the transformation will therefore be the one given in Eq. \eqref{eq:zevteomLLT}. However, for the other $\ta$ values the transformation will be much more complicated. Indeed, we must transform all constituents of Eq. \eqref{eq:hatgaeq} and then find the corresponding transformation $(\ta,\ze,\vte) \to (\ti{\ta}, \ti{\ze}, \ti{\vte})$, which therefore requires solving differential equations in $\ta$ and also $\ze$ (the light-like geodesic equation). For this reason, we will not discuss further the issue of observer transformations on $\Ord$, i.e. our observable drift results will hold for generic observers, but we will not provide the map relating the ones of two different observer world-lines.

\clearpage

\section{General-relativistic matrix kinetic theory} \label{sec:Boltzmann}

\subsection{Collisions and the microscopic space-time} \label{sec:microst}

In section \ref{sec:Liou} we have derived the evolution equation for the distribution functions $f_s(x,\vec{p})$ for a gas of ``free" particles, i.e. particles whose trajectories are solely altered by the non-trivial space-time geometry and electromagnetic field. Let us now consider the presence of interparticle forces, i.e. ``collisions". In this case, one must distinguish between two important space/time scales, namely, the typical separation between two successive collisions $L_{\rm free}$ and the typical space/time extent of the collision event itself $L_{\rm coll}$. 

In the case where the gas is ``dilute" enough $L_{\rm free} \gg L_{\rm coll}$, the particles spend most of their time in free motion, so one can describe the effect of collisions as merely changing a given free state to some other free state. In particular, this means that the degrees of freedom of the gas are the ones of a collection of free particles. This is what we have already implicitly assumed in the construction of $f_s(x, \vec{p})$ and the Liouville equation, since these are defined on the mass shells $E_{p,s} = \sqrt{m_s^2 + \vec{p}^2}$ of the free particles. Thus, $f_s(x,\vec{p})$ typically varies in $x^{\mu}$ over scales $\sim L_{\rm free}$ and obeys an equation of the form
\beq \label{eq:kineq}
{\rm L} f_s(x,\vec{p}) = C_s(x,\vec{p}) \, ,
\eeq
where $C_s$ is the ``collision term'' of the $s$ species capturing deviations from free motion.

In the opposite case of a ``dense" gas $L_{\rm free} \sim L_{\rm coll}$, the interactions are an integral part of the dynamics and therefore alter its description qualitatively. In particular, the degrees of freedom are no longer the ones of free particles, but rather collective excitations, whose precise structure is in general hard to obtain. More precisely, the spectral distribution of the system is not of the form $\sim \de^{(4)}(p_a p^a + m^2)$, as we have used until now (implicitly or explicitly), but rather a generic function of the 4-momentum norm $A(p^2)$. Moreover, since free motion is no longer the typical behavior of the particles, we cannot simply replace $\de^{(4)}(p^2 + m^2) \to A(p^2)$ in our equations at the level of the 8-dimensional phase space $L \cM$, i.e. we cannot treat the gas as a collection of particle species with a continuous mass spectrum.\footnote{There is an exception to this conclusion, i.e. there are cases where interactions are frequent but where the dilute gas machinery can still be applied. This occurs when $A(p^2)$ exhibits sharp enough maxima around some $p^2 = -m^2$ value, in which case the degrees of freedom are effective particles (``quasi-particles") with effective mass $m$. More specifically, we need the width of $A(p^2)$ around $-m^2$ to be small compared to both $L^{-2}_{\rm coll}$ and $L^{-2}_{\rm free}$.} Thus, the case of dense gases, such as in the very early universe, cannot be modeled using some $f_s(x,\vec{p})$ and Eq. \eqref{eq:kineq}, i.e. as free motion that is perturbed by sporadic collisions. Instead, one must consider a more fundamental non-equilibrium QFT description \cite{Kadanoff:1989, Berges:2004yj, Buss:2011mx}. 

Here we therefore focus on the case of ``dilute" gases, which is a valid assumption in cosmology way after the reheating era. We will refer to $L_{\rm free}$ and $L_{\rm coll}$ as the ``macroscopic" and ``microscopic" scales, respectively. In the absence of unstable particles, the collision term is dominated by $2 \leftrightarrow 2$ scattering and is given by the ``Boltzmann-Uehling-Uhlenbeck equation'' (BUU)
\bea
C_s(x,\vec{p}) & = & \frac{1}{2} \int \frac{\ed^3 p_1}{(2\pi)^3 2 E_{p_1,1}} \frac{\ed^3 p_2}{(2\pi)^3 2 E_{p_2,2}} \frac{\ed^3 p_3}{(2\pi)^3 2 E_{p_3,3}} \, (2\pi)^4 \de^{(4)} ( p_1 + p_2 - p_3 - p_s ) \label{eq:BUU}   \\
 & & \hspace{1cm} \times \, |\cA|^2(\vec{p}_1, \vec{p}_2 \to \vec{p}_3, \vec{p})  \[ f_1 f_2 \( 1 \pm f_3 \) \( 1 \pm f_s \) - \( 1 \pm f_1 \) \( 1 \pm f_2 \) f_3 f_s \] \nn \, ,
\eea
where $f_k := f_k(x,\vec{p}_k)$. Here $\cA$ is the ``matrix element", or ``amplitude", associated with the scattering event\footnote{The matrix element is usually denoted by ``$\cM$", but here this already denotes the space-time manifold.} and it is related to the $S$-matrix of the QFT through
\beq
\bra \vec{p}_3, \vec{p} | S - \Is | \vec{p}_1, \vec{p}_2 \ket \equiv (2\pi)^4 \de^{(4)} ( p_1 + p_2 - p_3 - p_s ) \, i \cA (\vec{p}_1, \vec{p}_2 \to \vec{p}_3, \vec{p}) \, .
\eeq
The $\pm$ distinguishes between bosons ($+$) and fermions ($-$) and these $\Ord(f^3)$ terms implement the ``Bose enhancement'' and ``Pauli blocking'' effects. The fact that $C_s(x,\vec{p})$ is independent of $\pa^{n>0}_{\mu} f_s(x, \vec{p})$ reflects the separation of scales, i.e. that $f_s(x,\vec{p})$ typically varies over macroscopic space-time scales, whereas the collision term only captures the microscopic ones. Indeed, when taking the dilute limit from non-equilibrium QFT \cite{Kadanoff:1989, Berges:2004yj, Buss:2011mx}, the collision term of the BUU equation appears as the zeroth order result in a derivative expansion probing the inhomogeneities in $x^{\mu}$. 

Let us now discuss some important structural aspects of Eq. \eqref{eq:BUU}. First, from the mathematical viewpoint, the fact that we are considering a full scattering process at every $x^{\mu}$ means that the ``space-time" in which the QFT is defined is not $\cM$. Rather, by definition, it is the space-time that is Fourier dual to the $p^a$ coordinates, i.e. the one which is parametrized by the $X^a$ coordinates defined at the end of subsection \ref{sec:geomconsBoltz} and appearing in Eq. \eqref{eq:FLdef} in particular. The fact that the $p^a$ data coordinatize the tangent spaces means that the $X^a$ can be thought of as parametrizing an infinitesimal space-time in the tetrad basis of the tangent space around each $x^{\mu}$. In the usual derivations of the Boltzmann equation, there is a {\it single} space-time coordinate $x^{\mu}$ and the separation between the macroscopic and microscopic scales is performed by simplifying/neglecting terms depending on their behaviour with respect to $x^{\mu}$. 

The present mathematical framework provides a radically different structure for implementing this idea. Instead of separating scales with respect to a single space-time coordinate $x^{\mu}$, we literally have two such coordinates that already represent ``macroscopic" and ``microscopic" spaces-times. The former is the $x^{\mu}$ coordinate on $\cM$, the ``macroscopic'' space-time capturing the variation of the distributions. The latter is the $X^a$ coordinate of the ``microscopic'' space-time, the one on which we compute the scattering matrix $S$ through the Fourier dual parameters $p^a$. Note that this is a Minkowski space-time, because the $X^a$ only mix under Lorentz transformations (\ref{eq:XaLLT}). The $x^{\mu}$ dependence of the Lorentz matrix $\La^a_{\,\,\,b}(x)$ now reflects the fact that one can choose different frames at each point $x^{\mu}$, i.e. different observers for each scattering event. From the viewpoint of $X^a$, however, these are global Lorentz transformations, i.e. the usual symmetry of QFT. 

Thus, an important property of this construction is that it allows one to match the symmetries of GR to the ones of QFT on flat space-time, without compromising the former. One can therefore directly plug the QFT amplitudes in the collision term without performing any kind of approximation. The only approximation here is the extreme separation between macro and micro scales and the classical treatment of gravity, since the latter is by construction a ``macro" entity. The disadvantage of having this separation of scales ``hardwired'' into the mathematical structure is that we do not have access to effects of intermediate scale, as one could recover perturbatively in the usual approach. However, in the case of cosmology this is not really a problem, as the separation between ``macro'' and ``micro'' scales is huge. On the other hand, the advantages of this structure are important, especially for deriving the desired generalization of (\ref{eq:BUU}), as we will soon discuss.

Finally, observe that $C_s$ depends on the momenta $\vec{p}$, not $X^a$, meaning that the involved (statistical) states in the microscopic QFT are invariant under translations $X^a \to X^a + c^a$. Thus, every microscopic space must host itself a dilute, but very large number of particles, so that we can reach statistical homogeneity. We are therefore treating the cosmic fluids as a collection of infinitesimal homogeneous thermodynamic systems, one at each $x^{\mu}$, in which scattering events take place. In this refined description, the microscopic spaces at each $x^{\mu}$ will be renamed ``mesoscopic", since each one of them hosts a full thermodynamic system, as opposed to the individual particle interactions that occur at the truly ``microscopic" scale $L_{\rm coll}$.

\subsection{Quantum superposition and matrix distributions} \label{sec:Lioulim}

In the cases of physical interest, on top of 3-momenta $\vec{p}$, there are extra discrete labels $s_k$ determining the 1-particle states $| \vec{p}, \{ s_k \} \ket$ in the QFT of the microscopic space-time. These distinguish among different particle species, spin, polarization, flavor, etc. Also, since this is the particle content of an interacting theory, it contains bound states of the fundamental particles, as well as unstable particles whose life-time is long enough ($\gg L_{\rm coll}$) to be considered as part of the spectrum. Here we choose to store all of these indices $\{ s_k \}$ inside a single index $s$ for notational simplicity, so the 1-particle states of the QFT read $| \vec{p}, s \ket$. For instance, if we only consider photons and electrons, then we have that $s$ takes four values. For example, ``1"  could denote a photon state with positive (circular) polarization, ``2" a photon state with negative polarization, ``3" an electron with spin ``up" and ``4" an electron with spin ``down". 

Now one has to take into account the fact that quantum states can be superposed, meaning that one cannot simply generalize $f(x,\vec{p}) \to f_s(x,\vec{p})$, as we did in section \ref{sec:TM} for the species indexation for instance. Indeed, this privileges some basis $| \vec{p}, s \ket$ in the underlying Hilbert space, thus neglecting all the possible state superpositions of the form $\sum_s \al_s | \vec{p}, s \ket$. If we chose to work with a different polarization basis $| \vec{p}, s \ket \to U_{ss'} |\vec{p},s' \ket$, where $U_{ss'}$ is a unitary matrix, then the corresponding distribution $f_s(x,\vec{p})$ would either not be real or would no longer be an array but a matrix.

As we will see in subsection \ref{sec:reltoQFT}, the solution is to consider a hermitian matrix in the discrete index $f_{ss'}(x,\vec{p})$ that arises naturally in the quantum context and serves as a two-point correlation function, thus capturing the information of superposed states. In particular, under a change of basis, one would now get (in matrix notation)
\beq \label{eq:fssptransHilb}
\ti{\bm{f}} = \bm{U} \bm{f} \bm{U}^{\dagger} \, ,
\eeq
which therefore remains consistently hermitian. Note that the $s$ parametrization we use here is non-redundant, i.e. each $s$ value corresponds to a physical state, and is obtained after decomposing the microscopic quantum fields $\ph_{\dots}(X)$ in some basis of wave-functions, to be discussed in subsection \ref{sec:wavefunctions}. In particular, $f_{ss'}$ is a set of scalars both under $\cM$Ds and LLTs.

One must also pay attention to the fact that some superpositions are forbidden by (super-) selection rules \cite{Streater:1989vi}. For instance, one cannot have a superposition of bosonic and fermionic 1-particle states, meaning 
\beq \label{eq:Gcond}
|s| \neq |s'| \hspace{1cm} \Rightarrow \hspace{1cm} f_{ss'}|_{s\neq s'} \equiv 0 \, ,
\eeq  
where $|s| \in \{0,1\}$ denotes the Grassmann parity\footnote{That is, $|s| = 0$ for bosons and $|s| = 1$ for fermions.} of the particle $s$, or of states with different charge
\beq \label{eq:qcond}
q_s \neq q_{s'} \hspace{1cm} \Rightarrow \hspace{1cm} f_{ss'}|_{s \neq s'} \equiv 0 \, .
\eeq  
Because of these rules, the $f_{ss'}$ matrix will generically be in block diagonal form, with each block corresponding to a subspace of superposable 1-particle states. Another natural question is whether one should also double the $\vec{p}$ entries in $f_{ss'}(x,\vec{p})$, since $\vec{p}$ labels quantum oscillators just as $s$. As we will shortly see, the reason only one $\vec{p}$ dependence remains is nothing but the translational invariance in $\vec{X}$ that we require at the mesoscopic level. 

To our knowledge, the corresponding matrix kinetic theory for $f_{ss'}$ has been initially developed on flat space-time in the context of neutrino flavor mixing \cite{Dolgov:1980cq, Barbieri:1990vx, Sigl:1992fn, Raffelt:1992uj, Vlasenko:2013fja} and has also been applied to curved space-time for the case of CMB polarization \cite{Kosowsky:1994cy, Pitrou:2008ut, Pitrou:2008hy, Beneke:2010eg} and the fermionic case \cite{Fidler:2017pkg,Pitrou:2019hqg}, using either the background space-time approach or the intermediate tetrad field approach (see also \cite{Volpe:2015rla} for a summary of matrix kinetic theory techniques and further references.). In the following subsections, we will derive a generalization of the BUU equation for matrix distributions that includes all the possible microscopic QFT processes. However, before we proceed, we must first discuss some limitations of the formalism that are due to the classical nature of the Liouville operator. 

Remember that the Liouville operator (\ref{eq:Liouop}) depends on a mass $m$, through $E_p$, and even the distribution $f(x,\vec{p}) := f_L(x,E_p,\vec{p})$ implicitly depends on a mass. Thus, in the presence of two $s$ indices in the distribution $f_{ss'}$, there is an ambiguity in the choice of mass parameter, since the available ones form an array $m_s$, not a matrix $m_{ss'}$. If, in any set of superposable 1-particle states, all particles have the same mass, then there is no ambiguity, but what if this is not the case? One could a priori think that the QFT would then have a selection rule forbidding the superposition of such particles, thus leading to a consistent $f_{ss'}|_{s \neq s'} = 0$ throughout evolution, if the initial conditions satisfy that condition. However, it turns out that nature provides us with at least one counter-example, which is relevant at cosmological scales: neutrinos. Indeed, the mass eigenstates of neutrinos $| \vec{p}, s \ket$, i.e. those that do not mix under free evolution, {\it are} mixed by the weak interactions. Thus, starting with $f_{ss'}|_{s \neq s'} = 0$ at some time, we will have $f_{ss'}|_{s \neq s'} \neq 0$ at latter times. The question therefore remains: what mass should one associate to the matrix element $f_{ss'}$ given the array $m_s$? 

A first guiding remark is that LLT covariance forces us to consider a definite mass $m_{ss'}$ for every $f_{ss'}$ component. To see this, note that active LLTs (\ref{eq:ALLT3dfiber}) bring in a mass dependence through $E_p$ and, for this to be a representation of the Lorentz symmetry, that energy must be of the form $E_p = \sqrt{m^2 + p^2}$ for some mass $m$, i.e. to derive from the Lorentz-invariant condition $p_a p^a + m^2 = 0$. Even if we chose to work with the off-shell distribution $f_{L,ss'}(x,p)$, Liouville's theorem guarantees that evolution will not mix different mass shells, so we would just be working with a continuous family of $f_{ss'}(x,\vec{p})$ distributions that cannot be ``superposed" in some quantum sense. Thus, the only generalization of the Liouville operator that is LLT-invariant is the trivial generalization of Eq. \eqref{eq:Liouop}
\bea  
{\rm L} f_{ss'} & := & \[ E_{p,ss'} \pa_0 + p^i \pa_i \] f_{ss'} \label{eq:Lioupssp} \\
 & & + \[ \( \Si_{0i0} E_{p,ss'} + q_s E_i \) E_{p,ss'} + \[ \( \Si_{0ij} - \Si_{ij0} \) E_{p,ss'} + q_s \vep_{ijk} B_k \] p^j - \Si_{ijk} p^j p^k \] \frac{\pa}{\pa p^i} \, f_{ss'}  \, ,  \nn
\eea
where the energies $E_{p,ss'}$ are of the usual form
\beq \label{eq:Epssofmss}
E_{p,ss'} := \sqrt{m_{ss'}^2 + p^2} \, ,
\eeq
for some set of masses $m_{ss'}$. Note that the charge array $q_s$ is unambiguous here, because in that case we do have a selection rule \eqref{eq:qcond}. Eq. \eqref{eq:Lioupssp} is therefore simply the standard Liouville operator for each individual component $f_{ss'}$. Unfortunately, however, the form (\ref{eq:Lioupssp}) treats the off-diagonal terms $f_{s\neq s'}$ as distribution functions associated with some effective particle of mass $m_{ss'}$, {\it not} as a distribution measuring the quantum superposition of particles $s$ and $s'$. The classical aspect of the involved physics in this formalism is therefore at odds with the quantum interpretation of $f_{ss'}$. The fact that the latter is associated with propagation along two different kinds of paths $\ga_s$ and $\ga_{s'}$ because of $m_s \neq m_{s'}$ seems to require the notion of quantum superposition to be somehow present already at the geometric level and therefore goes beyond the present formalism. Consequently, the Liouville operator (\ref{eq:Lioupssp}) is bound to miss quantum effects of order $\Ord((m_s-m_{s'})/E)$ in this case. In particular, having lost the quantum nature of $f_{ss'}$ in that respect, there is no privileged way of determining the $m_{ss'}$ numbers out of $m_s$.

Nevertheless, one could still hope for some guidance from QFT by noting that $m_{ss'}$ is reminiscent of the mass matrix in flavor space. In that case, the new 1-particle states, i.e. the ``flavor eigenstates'', are related by a unitary matrix
\beq
| \vec{p}, s \ket \to U^*_{ss'} | \vec{p}, s' \ket \, , 
\eeq
so the corresponding annihilation operators are related by
\beq
a_s \to U_{ss'} a_{s'} \, ,
\eeq
and the free Hamiltonian becomes
\beq \label{eq:Hnondiag}
H_0 = \int \frac{\ed^3 p}{(2\pi)^3}\, E_{p,s} a^{\dagger}_{\vec{p},s} a_{\vec{p},s} \to  \int \frac{\ed^3 p}{(2\pi)^3}\, E_{p,ss'} a^{\dagger}_{\vec{p},s} a_{\vec{p},s'} \, ,
\eeq
i.e. one involving a hermitian matrix of energies
\beq
E_{p,ss'} := U^*_{rs} E_{p,r} U_{rs'} \, .
\eeq
The flavor eigenstates therefore mix under free evolution, leading in particular to a forward scattering term \cite{Sigl:1992fn,Fidler:2017pkg}  in the evolution of $\bm{f}$ (in matrix notation)
\beq
\sim i \[ \bm{E}_p, \bm{f}(x,\vec{p}) \] \, .
\eeq
This could therefore seem as a natural candidate for the energy matrix entering (\ref{eq:Lioupssp}). There are, however, two reasons why this cannot be the case. First, in the limit case of a diagonal $\bm{E}_p$ matrix we have that the $f_{s\neq s'}$ components propagate at the speed of light, independently of the mass scales on the diagonal. This situation is clearly unphysical, as one would expect the off-diagonal terms to have a qualitatively intermediate behavior, not a completely different one. 

Second, this kind of energy matrix is not related to a mass matrix through some on-shell condition holding individually for each component (\ref{eq:Epssofmss}). Instead, it is related to the corresponding mass matrix
\beq
(m^2)_{ss'} := U^*_{rs} m^2_r U_{rs'} \, ,
\eeq
through a matrix relation
\beq \label{eq:Epssofmss2}
{\bm E}_p = \sqrt{{\bm m}^2 + \vec{p}^2} \, ,
\eeq
i.e. which mixes the matrix components, as opposed to Eq. \eqref{eq:Epssofmss}. In particular, this equation is not Lorentz-invariant, because such a transformation would lead to a matrix of momenta as well.\footnote{The Hamiltonian operator \eqref{eq:Hnondiag} of course still transforms as the time-component of a Lorentz vector, thanks to a non-trivial transformation of the ladder operators, but the $E_{p,ss'}$ components do not.} In fact, choosing another basis than the mass eigenstates is not a ``Lorentz-friendly" operation to begin with, because under a Lorentz transformation the flavor eigenstates become a superposition of kets with {\it different} momenta
\beq
U^*_{ss'} | \vec{p}, s' \ket \to U^*_{ss'} | \La^i_{\,\,j} p^j + \La^i_{\,\,0} E_{p,s'}, s' \ket \, . 
\eeq 
In retrospect, we can now understand the problem as the impossibility of building a Liouville operator that is both LLT-invariant and covariant under unitary transformations of the $s$ index (change of Hilbert space basis)
\beq
\bm{f} \to \bm{U} \bm{f} \bm{U}^{\dagger} \, ,
\eeq
{\it if} the components that are mixed belong to different mass shells. LLT-invariance requires the form (\ref{eq:Lioupssp}) and the relation (\ref{eq:Epssofmss}), whereas the Hilbert basis-independence requires matrix-type multiplications of $f_{ss'}$ and $E_{p,ss'}$ and the matrix relation (\ref{eq:Epssofmss2}). For this reason, one can only obtain a consistent generalized BUU equation in the case where there are selection rules forbidding the superposition of mass eigenstates with different mass, i.e.
\beq \label{eq:mcond}
m_s \neq m_{s'} \hspace{1cm} \Rightarrow \hspace{1cm} f_{ss'}|_{s\neq s'} \equiv 0 \, ,
\eeq  
just as the selection rules we already have for Grassmann parity \eqref{eq:Gcond} and charge \eqref{eq:qcond}. Thus, the matrix distribution obeys a Boltzmann equation of the form
\beq \label{eq:Boltzssp}
{\rm L} f_{ss'} = C_{ss'} \, ,
\eeq
where
\bea  
{\rm L} f_{ss'} & := & \[ E_{p,s} \pa_0 + p^i \pa_i \] f_{ss'} \label{eq:Lioupsspunamb} \\
 & & + \[ \( \om_{0i0} E_{p,s} - q_s E_i \) E_{p,s} + \[ \( \om_{0ij} - \om_{ij0} \) E_{p,s} - q_s \vep_{ijk} B_k \] p^j - \om_{ijk} p^j p^k \] \frac{\pa}{\pa p^i} \, f_{ss'} \, ,  \nn
\eea
is the unambiguous scalar Liouville operator. The moments of the distribution (\ref{eq:momentdef}) are also straightforwardly generalized to the trace over each non-trivial diagonal block $B$ of $\bm{f}$
\beq \label{eq:momentsdefss}
T_B^{a_1 \dots a_n}(x) := \sum_{s \in B} \int \frac{\ed^3 p}{(2\pi)^3 E_{p,B}} \, f_{ss}( x, \vec{p} ) \, p_s^{a_1} \dots p_s^{a_n} \, , \hspace{1cm} n>0 \, , \hspace{1cm} p_s^a := ( E_{p,s}, \vec{p} ) \, ,
\eeq
so that the energy $E_{p,s}$ entering the denominator and the $p^a_s$ are not ambiguous as they contain the single mass parameter $m_s$ associated with all the particles in $B$. With this definition we still have that these moments are conserved in the absence of collisions, only now Eq. \eqref{eq:momentcons} generalizes to
\beq \label{eq:momentcons2}
\na_{a_1} T_B^{a_1 \dots a_n} = n q_B F_{a_1}^{\,\,\,(a_1} T_B^{a_2 \dots a_n)} \, .
\eeq
Fortunately, this issue with the Liouville generalization to neutrino matrix distributions is irrelevant in practice thanks to the extreme separation between the neutrino decoupling scale $\La_{\rm dec} \sim 10^6 \, {\rm eV}$ and the neutrino mass scale $m_{\nu} \sim 0.1\,{\rm eV}$. With these we can split the universal neutrino time-line into three phases, namely, the one where temperature is $\La_{\rm dec} \gg T \gg m_{\nu}$, which we will refer to as the ``middle" era, and the two neighboring periods, which we respectfully refer to as the ``early" and ``late" eras. During the early era, the neutrino distribution is overwhelmingly supported on ultra-relativistic $\vec{p}$ values, so one can safely set the masses to zero.\footnote{See \cite{Cardall:2007zw} for an analogous limit using the flavor eigenstates, in which case it is the mass in the Liouville operator only that is set to zero.} The neutrinos are interacting through the weak force, meaning that the mass eigenstates are mixed and therefore that we have a non-diagonal neutrino block in $\bm{f}$. This is consistent with Eq. \eqref{eq:mcond} since the masses are all effectively zero. We next arrive in the middle era, where the zero mass approximation still holds, but now the interactions are negligible as well, so the neutrino distribution effectively obeys the massless Liouville equation, again in agreement with Eq. \eqref{eq:mcond}. The problem arises in the late era, because now the masses are no longer negligible, but we must still evolve a non-diagonal neutrino block in $\bm{f}$, in contradiction with Eq. \eqref{eq:mcond}. In practice, however, not all of this information is needed in order to evolve the quantities of interest and the cosmological observables are only sensitive to the trace of $\bm{f}$ since we cannot directly measure the cosmic neutrino background polarization. In the absence of neutrino interactions, other than gravitational, the only quantities that are required for closing the evolution equations are the neutrino moments \eqref{eq:momentsdefss}, which obey themselves a closed set of hierarchical evolution equations. The $f_{s \neq s'}$ information thus effectively drops out. We can therefore accurately describe neutrinos in this formalism if we evolve all of $\bm{f}$ up to the middle era and then simply retain $\Tr \, \bm{f}$ for the rest of the integration.

\subsection{Matrix distribution from a QFT density matrix} \label{sec:reltoQFT}

We now provide the concrete relation between the matrix distribution $f_{ss'}$ and the mesoscopic QFT. We consider a given macroscopic space-time point $x^{\mu}$ and focus on the QFT that lives in the corresponding mesoscopic space-time with coordinates $X^a$. We will therefore omit the $x^{\mu}$ dependencies in what follows, although one should keep in mind that all of the objects that we are about to define and use do depend on that variable. This is simply because the LLTs are by definition Lorentz transformations that can be different at every $x^{\mu}$, so their action through unitary transformations on quantum states or operators also depends on $x^{\mu}$. Thus, these objects will depend on $x^{\mu}$ for a generic observer family $e_a(x)$.   

In the statistical context, the state of the system is described by a density matrix $\ro \equiv \ro(x)$, which can be formally written as
\beq \label{eq:rodecomp}
\ro = \sum_{\psi} p_{\psi} | \psi \ket \bra \psi | \, ,
\eeq
for some orthonormal basis $| \psi \ket$. The $p_{\psi}$ correspond to the statistical probability of the system being in the quantum state $| \psi \ket$, so it is constant in mesoscopic time $T := X^0$. Consequently, $\ro$ is $T$-dependent in the Schr\"odinger picture and $T$-independent is the Heisenberg one, i.e. contrary to usual operators. Since the $p_{\psi}$ are probabilities we have
\beq \label{eq:roprop}
\Tr\, \ro \equiv \sum_{\psi} p_{\psi} \equiv 1 \, , \hspace{1cm} \ro^{\dagger} \equiv \ro \, , \hspace{1cm} \bra \psi | \ro | \psi \ket \equiv p_{\psi} \geq 0 \, , \,\,\, \forall \psi \, ,
\eeq
and the quantum statistical expectation value of some observable $\Ord$ is thus given by
\beq
\bra \Ord \ket_{\ro} := \Tr \[ \ro \, \Ord \] \equiv \sum_{\psi} p_{\psi} \bra \psi | \Ord | \psi \ket \, .
\eeq
Here we choose to work in the interaction picture, so $\ro$ evolves as
\beq
\ro(T') = U(T',T)\, \ro(T)\, U^{\dagger}(T',T) \, ,
\eeq
where 
\beq \label{eq:Uint}
U(T',T) \equiv {\cal T} \exp \[ - i \int_T^{T'} \ed \ti{T}\, H_{\rm int.}(\ti{T})  \] \, ,
\eeq
${\cal T}$ is the time-ordering operator and $H_{\rm int.}$ is the interaction Hamiltonian in the interaction picture, thus made of freely evolving fields. 

Remember that we assume each mesoscopic space-time to consist of a homogeneous thermodynamic system, thus harboring a large number of particles, but that is dilute enough so that collisions are rare. Put differently, the typical time spent in free motion $L_{\rm free}$ is much larger than the typical time scale of the collision event $L_{\rm coll}$. Here this translates in the fact that, between two consecutive collisions, the states have the time to reach asymptotic states, i.e. states that behave as free states. Each collision can therefore be treated as a standard scattering process, i.e. from asymptotic ``in" state to asymptotic ``out" state
\beq
\ro_{\rm in} := \lim_{T \to -\infty} \ro(T) \hspace{1cm} \ro_{\rm out} := \lim_{T \to \infty} \ro(T) \, .
\eeq
The states $| \psi \ket$ involved in $\ro_{\rm in, out}$ are superpositions of Fock states $| \vec{p}_1, s_1, \dots, \vec{p}_n, s_n \ket$ defined through creation operators $a^{\dagger}_{\vec{p},s}$ acting on a vacuum state $| 0 \ket$
\beq \label{eq:Fock}
| \vec{p}_1, s_1, \dots, \vec{p}_n, s_n \ket := \sqrt{2 E_{p_n,s_n}} \, a^{\dagger}_{\vec{p}_n, s_n} \dots \sqrt{2 E_{p_1,s_1}} \, a^{\dagger}_{\vec{p}_1, s_1} | 0 \ket \, , \hspace{1cm} a_{\vec{p},s} | 0 \ket \equiv 0 \, .
\eeq
Here $s \in \{ 1, \dots, D \}$ is the aforementioned discrete index collectively parametrizing spin states, flavor, species, particle/anti-particle pairs, etc. The ladder operators obey canonical (anti-)commutation relations
\beq \label{eq:CCR}
[ a_{\vec{p},s}, a^{\dagger}_{\vec{p}',s'} ]_{|s||s'|} = (2\pi)^3 \de^{(3)}(\vec{p} - \vec{p}')\, \de_{ss'}  \, , \hspace{1cm} [ a_{\vec{p},s}, a_{\vec{p}',s'} ]_{|s||s'|} = 0 \, ,
\eeq
where
\beq
\[ A, B \]_n := AB - (-1)^n BA \, ,
\eeq
which then imply the following symmetries
\beq \label{eq:comrelFock}
| \vec{p}_1, s_1, \dots, \vec{p}_k, s_k, \dots, \vec{p}_l, s_l, \dots, \vec{p}_n, s_n \ket = (-1)^{|s_k||s_l|} | \vec{p}_1, s_1, \dots, \vec{p}_l, s_l, \dots, \vec{p}_k, s_k, \dots, \vec{p}_n, s_n \ket \, .
\eeq
The asymptotic states are eigenstates of the ``asymptotic" Hamiltonian
\beq \label{eq:freeHam}
H_{\rm asy.} := \sum_s \int \frac{\ed^3 p}{(2\pi)^3}\, E_{p,s} N_{\vec{p},s} \, , \hspace{1cm} E_{p,s} := \sqrt{m_s^2 + \vec{p}^2} \, ,
\eeq
where
\beq
N_{\vec{p},s} := a^{\dagger}_{\vec{p},s} a_{\vec{p},s} \, , 
\eeq
are the number operators. The total Hamiltonian is then $H = H_{\rm asy.} + H_{\rm int.}$, while the total momentum, number and charge operators are given by
\beq \label{eq:totmom}
\vec{P} := \sum_s \int \frac{\ed^3 p}{(2\pi)^3} \, \vec{p} \, N_{\vec{p},s} \, , \hspace{1cm} N := \sum_s \int \frac{\ed^3 p}{(2\pi)^3} \, N_{\vec{p},s} \hspace{1cm} Q := \sum_s \int \frac{\ed^3 p}{(2\pi)^3} \, q_s N_{\vec{p},s}  \, ,
\eeq
respectively. 

Let us now briefly discuss one of the ``standard subtleties" of QFT about the particle content and the operators given above (see \cite{Weinberg:1995mt, Peskin:1995ev} for clarifications and more details on this). To that end, we denote collectively by ``$\la$" the coupling constants in the QFT action, such that for $\la = 0$ the theory is a collection of free harmonic oscillators. The $| \vec{p}, s \ket$ states are the 1-particle states of the interacting theory, i.e. eigenstates of $H$ and $\vec{P}$ that are irreducible representations of the Poincar\'e group. Because of this, the $| \vec{p}, s \ket$ set contains fundamental stable particles, i.e. particles corresponding to some field operator in $H$ (e.g. the electron), but also bound states (e.g. the hydrogen atom). On the other hand, it does not contain 1-particle states of the $\la = 0$ theory which, once $\la \neq 0$, become either strongly coupled at the energies of interest (e.g. quarks) or unstable (e.g. the Higgs or excited bound states). In the former case, such particles would only appear through bound states (e.g. pions) but, precisely because the coupling is strong, a perturbative approach is only possible through some effective action where the bound state is treated as a fundamental particle. In the case of unstable particles, if they are long-lived ($\gg L_{\rm coll}$) then it makes sense to consider them as part of the spectrum instead of as a resonance, i.e. as quasi-eigenstates of $H$. 

Thus, the ladder operators introduced above are generally non-trivial functions of the ones corresponding to the field operators. These combinations take into account the virtual particles that ``dress" the fundamental particles and sustain the bound states. In particular, in the case of bound states these are effective ladder operators obeying approximate canonical (anti-)commutation relations only at scales that do not resolve the internal structure of the bound system. The ``asymptotic" Hamiltonian $H_{\rm asy.}$ is therefore {\it not} the same as the Hamiltonian of the $\la = 0$ theory, i.e. the ``free" Hamiltonian $H_0$, because it needs $\la \neq 0$ to have bound particle eigenstates and it also contains the renormalized masses and couplings. This operator therefore matches $H$ on 1-particle states $H | \vec{p}, s \ket = H_{\rm asy.} | \vec{p}, s \ket = E_{p,s} | \vec{p}, s \ket$, but not on the multi-particle states of Eq. \eqref{eq:Fock}. This only works if one takes superpositions of them which localize and separate in space the individual particles well enough, so that they evolve as approximately free, i.e. asymptotic states. Keeping in mind this subtlety, one can then work with these Fock states in the asymptotic regions.

We now have everything we need to express the phase space distribution in terms of mesoscopic QFT operators. The expectation value of the $s$-particle number density in the asymptotic regions is
\beq \label{eq:fQ}
f^{\rm in,out}_s(\vec{p}) := V^{-1} \bra N_{\vec{p},s} \ket_{\ro_{\rm in,out}} \equiv V^{-1} \Tr \[ \ro_{\rm in,out} N_{\vec{p},s} \] \, ,
\eeq 
where 
\beq \label{eq:QFTvoldef}
V := (2\pi)^3 \de^{(3)}(\vec{p} = 0) \equiv \frac{(2\pi)^3}{\ed^3 p} \, ,
\eeq
is the ``volume" of the mesoscopic space-time, a singular constant that drops out of the physical quantities. Making explicit the $x^{\mu}$ dependence of $f^{\rm in,out}_s$ through the one of $\ro_{\rm in,out}$, we get that $f^{\rm in,out}_s(x,\vec{p})$ has the interpretation of the average number density of $s$-particles in phase space at $T \to \pm \infty$, i.e. it {\it is} the Boltzmann distribution $f_s(x,\vec{p})$ before and after scattering. Now note that (\ref{eq:fQ}) appears as the diagonal of the ``correlation" function
\beq
f^{\rm in,out}_{ss'}(\vec{p}, \vec{p}') := V^{-1} \bra a^{\dagger}_{\vec{p}',s'} a_{\vec{p},s} \ket_{\ro_{\rm in,out}} \, .
\eeq
Remember, however, that only homogeneous states should be considered in the mesoscopic space-time, meaning that
\beq \label{eq:homro}
[ \vec{P}, \ro ] = 0 \, ,
\eeq
where $\vec{P}$ is the momentum operator, a condition that is consistently preserved under evolution in $T$, because of the Jacobi identity of the commutator and the conservation of $\vec{P}$
\beq
\frac{\pa}{\pa T} \, [ \vec{P}, \ro ] = -i [ \vec{P}, [ H_{\rm int.}, \ro ]] \equiv - [ H_{\rm int.}, [\ro, \vec{P}] ] - [ \ro, [ \vec{P}, H_{\rm int.} ]] = - [ H, [\ro, \vec{P}] ] - [ \ro, [ \vec{P}, H ]] = 0 \, .
\eeq
Equation \eqref{eq:homro} then implies that the $| \psi \ket$ states appearing in Eq. \eqref{eq:rodecomp} are eigenstates of $\vec{P}$, so
\beq \label{eq:rosspp0}
f^{\rm in,out}_{ss'}(\vec{p}, \vec{p}') \sim \de^{(3)}(\vec{p} - \vec{p}') \, .
\eeq
In contrast, nothing keeps the $s$ index from mixing, so we must consider the hermitian matrix distribution
\beq \label{eq:fsspp}
f^{\rm in,out}_{ss'}(\vec{p}) := V^{-1} \bra N_{\vec{p},s's} \ket_{\ro_{\rm in,out}} \, ,
\eeq
where we have defined
\beq
N_{\vec{p},ss'} := a^{\dagger}_{\vec{p},s} a_{\vec{p},s'} \, ,
\eeq
and we now understand the singular normalization $V$ as canceling the one coming from the Dirac delta in Eq. \eqref{eq:rosspp0} evaluated at $\vec{p} = \vec{p}'$. Note that, contrary to $N_{\vec{p},s}$, these operators do not necessarily commute among themselves
\beq
\[ N_{\vec{p},ss'}, N_{\vec{q},rr'} \] = (2\pi)^3 \de^{(3)} (\vec{p} - \vec{q}) \[ N_{\vec{p},sr'} \de_{rs'} - N_{\vec{p},rs'} \de_{sr'} \] \, .
\eeq
Moreover, their time-evolution in the interaction picture is
\beq
N_{\vec{p},ss'}(T') = e^{i \( E_{p,s} - E_{p,s'} \) \( T' - T \)} N_{\vec{p},ss'}(T) \, .
\eeq
However, as we saw in subsection \ref{sec:Lioulim}, we will only consider the cases where $m_s = m_{s'}$ in any non-trivial block of $f_{ss'}$, so in what follows $N_{\vec{p},ss'}$ will be constant in $T$. 

Following the standard argumentation for deriving the Boltzmann equation, the statistical state in the ``in" region can be assumed to be minimally correlated, i.e. it is entirely determined by the corresponding 1-particle distribution $f^{\rm in}_{ss'}(\vec{p})$ instead of a full BBGKY-like hierarchy. More precisely, the higher order moments
\beq \label{eq:correlfunc}
\sim \bra a^{\dagger}_{\vec{p}_1,s_1} \dots a^{\dagger}_{\vec{p}_n,s_n} a_{\vec{q}_1, r_1} \dots a_{\vec{q}_m, r_m} \ket_{\ro_{\rm in}} \, ,
\eeq
factorize into products of the two-point functions
\bea  
\langle a^{\dagger}_{\vec{p}',s'} a_{\vec{p},s} \rangle_{\ro_{\rm in}} & \equiv & (2\pi)^3 \de^{(3)}(\vec{p} - \vec{p}') \, f^{\rm in}_{ss'}(\vec{p}) \, , \label{eq:simptraces1} \\
\langle a_{\vec{p},s} a^{\dagger}_{\vec{p}',s'} \rangle_{\ro_{\rm in}} & \equiv & (2\pi)^3 \de^{(3)}(\vec{p} - \vec{p}') \[ \de_{ss'} + (-1)^{|s||s'|} f^{\rm in}_{ss'}(\vec{p}) \] \, , \label{eq:simptraces2}
\eea
where the first equation is the definition of $f^{\rm in}_{ss'}(\vec{p})$, while the second one is obtained by using Eq. \eqref{eq:CCR}. This is the assumption of ``molecular chaos" in the quantum context, by which the particle momenta are uncorrelated {\it before} scattering and thus $f^{\rm in}_{ss'}(\vec{p})$ is a complete enough description of the state. This will not hold in general for the ``out" state $\ro_{\rm out}$, i.e. it will not be expressible solely in terms of its 1-particle distribution $f^{\rm out}_{ss'}(\vec{p})$, because the collision will correlate the outcoming states. Nevertheless, since the gas is dilute, the macroscopic free evolution between two successive scattering events is long enough to make the higher-order correlation functions decay, thus leading again to an uncorrelated ``in" statistical state for the next scattering event. This unequal treatment of the ``in" and ``out" regions breaks the time-reversal symmetry and thus generates the ``arrow of time" at the mesoscopic and macroscopic levels. Thus, the distribution $f_{ss'}(x,\vec{p})$ that will ultimately obey the Boltzmann equation is $f^{\rm in}_{ss'}(x,\vec{p})$. For this reason, from now on we focus on the ``in" region and simplify the notation to $\ro := \ro_{\rm in}$ and $f_{ss'}(\vec{p}) := f^{\rm in}_{ss'}(\vec{p})$.

\subsection{Expressing $\ro$ and the entropy current in terms of $f_{ss'}$} \label{sec:entropy}

Since $\ro(x)$ is entirely determined by $f_{ss'}(x,\vec{p})$, the relation in Eq. \eqref{eq:fsspp} can be inverted, subject to the conditions \eqref{eq:roprop} and \eqref{eq:homro}. In particular, this will allow us to express the entropy density
\beq \label{eq:entropydef}
s(x) := - V^{-1} \Tr \[ \ro(x) \log \ro(x) \] \, ,
\eeq
in terms of $f_{ss'}(x,\vec{p})$. Given that the only non-trivial operator involved in Eq. \eqref{eq:fsspp} is $N_{\vec{p},ss'}$, the inversion must take the form
\beq \label{eq:rotherdev}
\ro(x) = Z^{-1}(x)\, F\[ -\int \frac{\ed^3 p}{(2\pi)^3} \, w_{ss'}(x,\vec{p}) \, N_{\vec{p},ss'} \] \, ,
\eeq
where $F$ is some monotonic function determined by its Taylor series, the normalization factor
\beq
Z(x) := \Tr \, F\[ -\int \frac{\ed^3 p}{(2\pi)^3} \, w_{ss'}(x,\vec{p}) \, N_{\vec{p},ss'} \] \, ,
\eeq
gives $\Tr \, \ro_{\rm in} \equiv 1$ and $w_{ss'}$ is a hermitian matrix
\beq
w^*_{ss'}(x,\vec{p}) = w_{s's}(x,\vec{p}) \, ,
\eeq
so that $\ro$ is a hermitian operator. We thus have that $w_{ss'}$ has as many independent components as $f_{ss'}$ so that we can relate the two in a bijective way. But we also have the undetermined $F$ function, so there is still some ambiguity in inverting \eqref{eq:fsspp}. To fix $F$, we can be guided by the special case of thermal and chemical equilibrium
\beq \label{eq:roeq}
\ro_{\rm eq.}(x) = \frac{\exp \[ -\be(x) \( u_a(x) \, P^a_{\rm asy.} - \mu(x)\, N \) \]}{{\rm Tr}\, \exp \[ -\be(x) \( u_a(x)\, P^a_{\rm asy.} - \mu(x)\, N \) \]}  \, ,
\eeq
where 
\beq
P^a_{\rm asy.} := \( H_{\rm asy.}, \vec{P} \) \, , \hspace{1cm} u_a u^a \equiv -1 \, ,
\eeq
and $H_{\rm int.}$ can be neglected to a first approximation in the dilute gas case. Here the ``mesoscopic" functions $\be(x)$, $\mu(x)$ and $u^a(x)$, i.e. that are independent of the microscopic state $( \vec{p}, s )$, are the inverse temperature, the chemical potential and the fluid's 4-velocity with respect to the observer family $e_a$ at $x^{\mu}$, respectively. For \eqref{eq:roeq} to hold we thus need
\beq \label{eq:thereq}
F_{\rm eq.} = \exp \, , \hspace{1cm} w_{ss', {\rm eq.}}(x,\vec{p}) = \be(x) \[ u_a(x)\, p_s^a - \mu(x) \] \de_{ss'} \, , \hspace{1cm} p_s^a := \( E_{p,s}, \vec{p} \) \, .
\eeq
To lowest order in the deviations from equilibrium, we can therefore consider the fixed operatorial dependence $F = \exp$, thus reducing the problem to expressing $w_{ss'}$ in terms of $f_{ss'}$. Independently of the proximity to equilibrium, however, this choice of $F$ is also motivated by the fact that it maximizes the entropy density \eqref{eq:entropydef} when seen as a functional of $w_{ss'}(x,\vec{p})$, with $x^{\mu}$ considered as an external set of fixed parameters. To see this, note that $F \equiv \exp$ implies
\beq \label{eq:sLeglogZ}
s(x) \equiv \int \frac{\ed^3 p}{(2\pi)^3}\, w_{ss'}(x,\vec{p})\, f_{ss'}(x,\vec{p}) + V^{-1} \log Z(x) \, ,
\eeq
and
\beq \label{eq:fdewZ}
f_{ss'}(x,\vec{p}) \equiv \frac{\de}{\de w_{ss'}(x,\vec{p})} \[ -V^{-1} \log Z(x) \] \, .
\eeq 
Therefore, $f_{ss'}$ and $w_{ss'}$ become Legendre-conjugate variables with respect to the functional $-V^{-1} \log Z(x)$, while the entropy becomes the Legendre transform of that functional. It is therefore independent of $w_{ss'}$
\beq
\frac{\de s(x)}{\de w_{ss'}(x,\vec{p})} \equiv 0 \, ,
\eeq
which is precisely the statement of entropy maximization. With this educated guess for the $F$ function, we can now compute the relation between $f_{ss'}$ and $w_{ss'}$ in order to invert Eq. \eqref{eq:fsspp}. We first note that $N_{\vec{p},ss'}$ can be expressed as a tensor product of operators defined on each $\vec{p}$ oscillator Hilbert space
\beq
N_{\vec{p},ss'} = V \[ \dots \otimes \mathbb{I} \otimes \dots \otimes \mathbb{I} \otimes \us{\vec{p}}{\ub{N_{ss'}}} \otimes \mathbb{I} \otimes \dots \otimes \mathbb{I} \otimes \dots \] \, , \hspace{1cm} N_{ss'} := a^{\dagger}_s a_{s'} \, ,
\eeq
where the $a_s$ are the unit-normalized ladder operators of a set of $D$ oscillators, i.e.
\beq
\[ a_s, a_{s'}^{\dagger} \]_{\pm} = \de_{ss'}  \, , \hspace{1cm} \[ a_s, a_{s'} \]_{\pm} = \[ a_s^{\dagger}, a_{s'}^{\dagger} \]_{\pm} = 0 \, .
\eeq
We can thus formally factorize
\beq \label{eq:roinofrop}
\ro(x) = \bigotimes_{\vec{p} \, \in \, \Rs^3} \ro_{\vec{p}}(x) \, ,  \hspace{1cm}  V^{-1} \log Z(x) = \int \frac{\ed^3 p}{(2\pi)^3} \log Z_{\vec{p}}(x) 
\eeq
where
\beq \label{eq:roZp}
\ro_{\vec{p}}(x) := Z^{-1}_{\vec{p}}(x) \, \exp \[ - w_{ss'}(x,\vec{p}) \, N_{ss'} \] \, , \hspace{1cm} Z_{\vec{p}}(x) := \Tr \exp \[ - w_{ss'}(x,\vec{p}) \, N_{ss'} \] \, ,
\eeq
are the density matrices and partition functions of each $\vec{p}$ factor, and we have used 
\beq \label{eq:tracefact}
\Tr \[ A \otimes B \] \equiv \Tr A \times \Tr B \, .
\eeq
We next note that, since $w_{ss'}(x,\vec{p})$ is a hermitian matrix, it can be diagonalized using a unitary matrix $U_{\ti{s}s}(x,\vec{p})$
\beq
\ti{w}_{\ti{s}\ti{s}'} = U_{\ti{s}s} U^*_{\ti{s}'s'} w_{ss'}  = {\rm diag}(\ti{w}_{\ti{s}})_{\ti{s}\ti{s}'} \, , \hspace{1cm} U_{\ti{s} s} U_{\ti{s}' s}^* = \de_{\ti{s}\ti{s}'} \, , \hspace{1cm} U_{\ti{s}s} U_{\ti{s}s'}^* = \de_{ss'} \, .
\eeq
Defining the linear combinations for each $(x,\vec{p})$ value
\beq \label{eq:tildebasis}
\ti{a}_{\ti{s}} := U_{\ti{s}s}\, a_s \, ,  
\eeq 
we get that they also obey canonical commutation relations 
\beq
\[ \ti{a}_{\ti{s}}, \ti{a}_{\ti{s}'}^{\dagger} \]_{\pm} = \de_{\ti{s}\ti{s}'}  \, , \hspace{1cm} \[ \ti{a}_{\ti{s}}, \ti{a}_{\ti{s}'} \]_{\pm} = \[ \ti{a}_{\ti{s}}^{\dagger}, \ti{a}_{\ti{s}'}^{\dagger} \]_{\pm} = 0 \, ,
\eeq
and therefore simply correspond to the ladder operators associated with particles in a different polarization basis. Note that, since $w_{ss'} \equiv 0$ if $s$ and $s'$ are not superposable, we have that $U_{\ti{s}s}$ is in block-diagonal form and, in particular, $U_{\ti{s}s} = 0$ if $|\ti{s}| \neq |s|$. With this we now get an expression involving the standard number operators $\ti{N}_{\ti{s}} := \ti{a}^{\dagger}_{\ti{s}} \ti{a}_{\ti{s}}$
\beq
Z_{\vec{p}}(x) \equiv \Tr \exp \[ - \ti{w}_{\ti{s}}(x,\vec{p}) \, \ti{N}_{\ti{s}} \]  \, .
\eeq
To compute the trace, we consider the orthonormal occupation number basis
\beq \label{eq:occnumbasis}
| n_1, \dots, n_D \ket := \frac{(\ti{a}_D^{\dagger})^{n_D}}{\sqrt{n_D!}} \dots \frac{(\ti{a}_1^{\dagger})^{n_1}}{\sqrt{n_1!}} | 0 \ket \, , \hspace{1cm} \ti{a}_s |0 \ket \equiv 0 \, ,
\eeq
so that $n_{\ti{s}} \in \mathbb{N}$ in the bosonic case $|\ti{s}| = 0$ and $n_{\ti{s}} \in \{ 0,1 \}$ in the fermionic case $|\ti{s}| = 1$. Using $\ti{N}_{\ti{s}} | n_1, \dots, n_D \ket \equiv n_{\ti{s}} | n_1, \dots, n_D \ket$ and $[ \ti{N}_{\ti{s}}, \ti{N}_{\ti{s}'} ] \equiv 0$, we can thus write
\beq
Z_{\vec{p}}(x) \equiv \sum_{n_1, \dots, n_D} \bra n_1, \dots, n_D | \exp \[ - \ti{w}_{\ti{s}}(x,\vec{p}) \, \ti{N}_{\ti{s}} \] | n_1, \dots, n_D \ket \equiv \prod_{{\ti{s}} = 1}^D \sum_n \exp \[ - \ti{w}_{\ti{s}}(x, \vec{p})\, n \] \, ,
\eeq
and then
\bea
\log Z_{\vec{p}}(x) & = & \log \prod_{{\ti{s}} = 1}^D \[ 1 - (-1)^{|\ti{s}|} e^{-\ti{w}_{\ti{s}}(x,\vec{p})} \]^{(-1)^{|\ti{s}|+1}} = - \sum_{\ti{s} = 1}^D (-1)^{|\ti{s}|} \log \[ 1 - (-1)^{|\ti{s}|} e^{-\ti{w}_{\ti{s}}(x,\vec{p})} \]  \nn \\
 & \equiv & - \Tr \[ \bm{1}_{\circ} \log \( \bm{1} - \bm{1}_{\circ} e^{-\ti{\bm{w}}(x,\vec{p})} \) \]  = - \Tr \[ \bm{1}_{\circ} \log \( \bm{1} - \bm{1}_{\circ} e^{-\bm{w}(x,\vec{p})} \) \]   \, ,
\eea
where we switched to matrix notation in the second line, we defined
\beq
[\bm{1}_{\circ}]_{ss'} := (-1)^{|s||s'|} \de_{ss'}  \, ,
\eeq
and the trace appearing here is over the $s$ indices. Thus,
\beq
-V^{-1} \log Z(x) = \int \frac{\ed^3 p}{(2\pi)^3}\, \Tr \[ \bm{1}_{\circ} \log \( \bm{1} - \bm{1}_{\circ} e^{-\bm{w}(x,\vec{p})} \) \] \, ,
\eeq
and therefore, using \eqref{eq:fdewZ},
\beq
\bm{f} = \[ e^{\bm{w}} - \bm{1}_{\circ} \]^{-1} \, , \hspace{1cm} \bm{w} = \log \[ \bm{f}^{-1} + \bm{1}_{\circ} \]  \, ,
\eeq
where it is understood that these matrix functions are defined by their Taylor series. In the case of thermal and chemical equilibrium (\ref{eq:thereq}), we recover the well-known Bose-Einstein and Fermi-Dirac distributions
\beq
\bm{f}_{\rm eq.}(x,\vec{p}) = {\rm diag}\[ e^{\be(x) \[ u_a(x)\, p_s^a - \mu(x) \]} - (-1)^{|s|}\]^{-1} \, .
\eeq
Defining
\beq
\bm{f}_{\circ}(x,\vec{p}) := \bm{1} + \bm{1}_{\circ} \bm{f}(x,\vec{p}) \, ,
\eeq
we the density matrix \eqref{eq:rotherdev} becomes
\beq \label{eq:rotherdev2}
\ro(x) = Z^{-1}(x) \, \exp \int \frac{\ed^3 p}{(2\pi)^3} \[ \log \[  \frac{\bm{f}(x,\vec{p})}{\bm{f}_{\circ}(x,\vec{p})} \]_{ss'} N_{\vec{p},ss'} \] \, ,
\eeq
and
\beq \label{eq:Zfin}
Z(x) = \exp \[ V \int \frac{\ed^3 p}{(2\pi)^3} \, \Tr \[ \bm{1}_{\circ} \log \bm{f}_{\circ}(x,\vec{p}) \] \] \, ,
\eeq
while the entropy density \eqref{eq:sLeglogZ} takes the form
\beq \label{eq:sfin}
s(x) = \int \frac{\ed^3 p}{(2\pi)^3} \, \Tr \[ - \bm{f} \log \bm{f} + \bm{1}_{\circ} \bm{f}_{\circ} \log \bm{f}_{\circ} \](x,\vec{p}) \, .
\eeq 
We can also eliminate the singular $V$ factor that appears in \eqref{eq:rotherdev2} through $Z(x)$. We first write 
\beq
\ro(x) = \exp \int \frac{\ed^3 p}{(2\pi)^3} \[ \log \[  \frac{\bm{f}(x,\vec{p})}{\bm{f}_{\circ}(x,\vec{p})} \]_{ss'} N_{\vec{p},ss'} - \[ \bm{1}_{\circ} \log \bm{f}_{\circ}(x,\vec{p}) \]_{ss'} V \de_{ss'} \Is \] \, ,
\eeq
then use
\beq
V \de_{ss'} \Is \equiv \[ a_{\vec{p},s'}, a^{\dagger}_{\vec{p},s} \]_{|s||s'|} \equiv a_{\vec{p},s'} a^{\dagger}_{\vec{p},s} - (-1)^{|s||s'|} a^{\dagger}_{\vec{p},s} a_{\vec{p},s'}  \, ,
\eeq
and thus find
\beq \label{eq:rotherdev3} 
\ro(x) = \exp \int \frac{\ed^3 p}{(2\pi)^3} \[ \[  \log  \bm{f}(x,\vec{p}) \]_{ss'} a^{\dagger}_{\vec{p},s} a_{\vec{p},s'} - \[ \bm{1}_{\circ} \log \bm{f}_{\circ}(x,\vec{p}) \]_{ss'} a_{\vec{p},s'} a^{\dagger}_{\vec{p},s} \]  \, .
\eeq
Note that we have used $\log \bm{A} + \log \bm{B} = \log (\bm{A}\bm{B})$, which does not hold for generic matrices, but here does because both can be simultaneously diagonalized. 

Finally, from Eq. \eqref{eq:sfin} we note that $s(x)$ is not a Lorentz scalar, because the measure $\ed^3 p$ is not invariant. This fact could have also been inferred already from the lonesome volume factor in Eq. \eqref{eq:entropydef}. Rather, it is the time-component of a Lorentz vector, the entropy current
\beq \label{eq:scurrent}
s^a(x) := \sum_s \int \frac{\ed^3 p}{(2\pi)^3 E_{p,s}} \, p_s^a \[ - \bm{f} \log \bm{f} + \bm{1}_{\circ} \bm{f}_{\circ} \log \bm{f}_{\circ} \]_{ss}(x,\vec{p}) \, , \hspace{1cm} s^0(x) \equiv s(x) \, ,
\eeq
where we have recovered the Lorentz-invariant measure $\ed^3 p/E_{p,s}$. As in the case of the moments \eqref{eq:momentsdefss}, here too we can define the entropy current associated with some block $B$ of $\bm{f}$
\beq
s_B^a(x) := \sum_{s \in B} \int \frac{\ed^3 p}{(2\pi)^3 E_{p,s}} \, p_s^a \[ - \bm{f} \log \bm{f} + \bm{1}_{\circ} \bm{f}_{\circ} \log \bm{f}_{\circ} \]_{ss}(x,\vec{p}) \, .
\eeq
Taking the divergence of this quantity and proceeding as in Eqs. \eqref{eq:TTEMcons} and \eqref{eq:Tconsnocol}, i.e. expressing the integral as a 4-dimensional one along with a Dirac delta imposing the dispersion relation, we find that it is proportional to $p^a \na_a^L f_L$. Thus, in the absence of collisions, entropy is conserved in the covariant sense
\beq \label{eq:divs0nocol}
\na_a s_B^a = 0 \, ,
\eeq
i.e. any local variation in entropy must compensated by some variation in a nearby region through some entropy current.

\subsection{The collision term} \label{sec:collision}

We now want to determine the collision term, i.e. the right-hand side of Eq. \eqref{eq:Boltzssp}. We observe that the action of the Liouville operator on $f_{ss'}(x,\vec{p})$ corresponds to a time-like derivation in macroscopic space-time, up to an $E_{p,s}$ factor. Since $f_{ss'}(\vec{p})$ describes the ``in" state $\ro$ and we want a time-step in macroscopic time to correspond to a full mesoscopic scattering process, we equate $\rm L$ to the finite $T$-derivative, i.e.
\beq \label{eq:mMmatch}
C_{ss'}(\vec{p}) \os{!}{=} \frac{E_{p,s}}{T} \[ f^{\rm out}_{ss'}(\vec{p}) - f^{\rm in}_{ss'}(\vec{p}) \] \equiv \frac{E_{p,s}}{T} \[ f^{\rm out}_{ss'}(\vec{p}) - f_{ss'}(\vec{p}) \] \, ,
\eeq
where here the ``in" and ``out" regions are defined at $\mp\, T/2$, respectively, and the $T \to \infty$ limit is understood. Again, we only consider QFTs for which the right-hand side of \eqref{eq:mMmatch} is identically zero if $|s| \neq |s'|$, $q_s \neq q_{s'}$ or $m_s \neq m_{s'}$, so that the $E_{p,s}$ factor in particular is not ambiguous. Let us also stress that Eq. (\ref{eq:mMmatch}) is a matching condition between the macroscopic and microscopic dynamics that we impose {\it by hand}, i.e. it is not derivable from more fundamental equations in the present framework. Nevertheless, this type of finite $T$-derivative with the $T \to \infty$ limit is what one technically does when computing cross-sections and decay rates in QFT. We thus have
\bea
C_{ss'}(\vec{p}) & = & \frac{E_{p,s}}{VT} \[ \bra N_{\vec{p},s's} \ket_{\ro_{\rm out}} - \bra N_{\vec{p},s's} \ket_{\ro} \] \nn \\
 & \equiv & \frac{E_{p,s}}{VT}\, \Tr \[ \ro_{\rm out} N_{\vec{p},s's} - \ro N_{\vec{p},s's} \] \nn \\
 & = & \frac{E_{p,s}}{VT}\, \Tr \[ S \ro\, S^{\dagger} N_{\vec{p},s's} - \ro N_{\vec{p},s's} \] \nn \\
 & \equiv & \frac{E_{p,s}}{VT}\, \langle S^{\dagger} N_{\vec{p},s's} S - N_{\vec{p},s's} \rangle_{\ro} \nn \\
 & \equiv & \frac{E_{p,s}}{VT}\, \langle S^{\dagger} \[ N_{\vec{p},s's}, S \] \rangle_{\ro} \, , \label{eq:Col1}
\eea 
where
\beq
S := \lim_{\ep \to 0^+} \lim_{T \to \infty(1 - i\ep)} U(T/2,-T/2) \, ,
\eeq
is the $S$-matrix and $U(T,T')$ is given in Eq. \eqref{eq:Uint}. The $\ep > 0$ regularization guarantees convergence and ends up producing the $i\ep$ prescription of the Feynman propagator in perturbation theory \cite{Peskin:1995ev}. Note that both sides of Eq. \eqref{eq:Col1} are consistently hermitian matrices, although this property is no longer explicit in the last line.

To express this in terms of scattering amplitudes, we then consider the deviation from the identity $i \De := S - \mathbb{I}$, so that
\beq \label{eq:C1}
C_{ss'}(\vec{p}) \equiv \frac{E_{p,s}}{VT} \, \langle \De^{\dagger} \[ N_{\vec{p},s's}, \De \] \rangle_{\ro} \, . 
\eeq
In deriving this expression, we have used the fact that the term linear in $\De$ vanishes
\beq
\langle \[ N_{\vec{p},s's}, \De \] \rangle_{\ro} \equiv \Tr \( \ro \[ N_{\vec{p},s's}, \De \] \) \equiv \Tr \( \[ \ro, N_{\vec{p},s's} \] \De \) = 0 \, ,
\eeq
which is found by going to the tilded basis defined in subsection \ref{sec:entropy} where $\ti{\bm{f}}$ is diagonal so that only the commuting number operators appear in the expression. The scattering amplitudes $\cA$ are implicitly defined using the Fock states of the ``in" region
\bea
 & & \bra \vec{p}_1, s_1, \dots, \vec{p}_n, s_n | \De | \vec{q}_1, r_1, \dots, \vec{q}_m, r_m \ket \label{eq:TMrel} \\
 & & =:  (2\pi)^4 \de^{(4)} \( \sum_{k=1}^n p_k - \sum_{l=1}^m q_l \) \cA \( \vec{q}_1, r_1, \dots, \vec{q}_m, r_m \to \vec{p}_1, s_1, \dots, \vec{p}_n, s_n \) \, , \nn
\eea
and are computed using the connected-amputated Feynman diagrams. The $\cA$ functions inherit the (anti-)commutation symmetries of the Fock basis \eqref{eq:comrelFock} and obey
\beq \label{eq:vactoan}
\cA \( {\rm vacuum} \to \dots \) \equiv \cA \( \dots \to {\rm vacuum} \) \equiv 0  \, ,
\eeq
and also
\beq \label{eq:1ptoan}
\cA \( \vec{p}, s \to \dots \) \equiv \cA \( \dots \to \vec{p}, s \) \equiv 0 \, ,
\eeq
if the $s$ particle is stable. This is because the corresponding kets are time-translation invariant $S | 0 \ket = | 0 \ket$ and $S |\vec{p}, s \ket = |\vec{p}, s \ket$, so $\De$ sends them to zero. The vacuum case can also be simply understood from the viewpoint of energy conservation. Knowing the components of the operator $T$ in the full ``in" Fock basis then allows us to express it in terms of the ``in" ladder operators, i.e. its cluster decomposition \cite{Weinberg:1995mt}
\bea 
\De & = & \sum_{n,m=0}^{\infty} \frac{1}{n! m!} \int \( \prod_{k=1}^n \frac{\ed^3 p_k}{(2\pi)^3 \sqrt{2 E_{p_k, s_k}}} \) \( \prod_{l=1}^m \frac{\ed^3 q_l}{(2\pi)^3 \sqrt{2 E_{q_l, r_l}}} \) (2\pi)^4 \de^{(4)} \( \sum_{k=1}^n p_k - \sum_{l=1}^m q_l \) \nn \\
 & & \times \cA_c \( \vec{q}_1, r_1, \dots, \vec{q}_m, r_m \to \vec{p}_1, s_1, \dots, \vec{p}_n, s_n \) a^{\dagger}_{\vec{p}_n, s_n} \dots a^{\dagger}_{\vec{p}_1, s_1} a_{\vec{q}_1, r_1} \dots a_{\vec{q}_m, r_m}  \, , \label{eq:clustdec} 
\eea
where $\cA_c$ are the ``fully" connected scattering amplitudes, i.e. those corresponding to fully connected Feynman diagrams. Indeed, the Feynman diagrams contributing to $\cA$ must have every external line connected to some vertex, but they can have several disconnected components.\footnote{This subtlety is usually overlooked in QFT textbooks where one focuses on the simplest non-trivial amplitudes, such as $2 \to 2$, which are fully connected.} As an example where the difference is relevant, consider the $4 \to 4$ amplitude in $\la \ph^4$ theory \cite{Weinberg:1995mt}
\bea
\bra \vec{p}_1, \vec{p}_2, \vec{p}_3, \vec{p}_4 | \De | \vec{q}_1, \vec{q}_2, \vec{q}_3, \vec{q}_4 \ket & = & (2\pi)^4 \de^{(4)} \( p_1 + p_2 + p_3 + p_4 - q_1 - q_2 - q_3 - q_4 \)  \\
 & & \times \cA_c \( \vec{q}_1, \vec{q}_2, \vec{q}_3, \vec{q}_4 \to \vec{p}_1, \vec{p}_2, \vec{p}_3, \vec{p}_4 \) \nn \\
 & & + \, (2\pi)^4 \de^{(4)} \( p_1 + p_2 - q_1 - q_2 \)\, (2\pi)^4 \de^{(4)}(p_3 + p_4 - q_3 - q_4) \nn \\
 & & \times \cA_c \( \vec{q}_1, \vec{q}_2 \to \vec{p}_1, \vec{p}_2 \)  \cA_c \( \vec{q}_3, \vec{q}_4 \to \vec{p}_3, \vec{p}_4 \) + \dots   \nn
\eea
where the ellipses contain the terms that appropriately symmetrize the $\vec{p}_k, \vec{q}_l$ entries. The $\sim \cA_c(4 \to 4)$ term corresponds to the diagrams of the form
\beq
\begin{tikzpicture}[line width=1 pt, scale=2]
\begin{scope}[shift={(0.5,0)}]
\draw[electron] (-0.5,0.55) -- (0.5,-0.55);
\draw[electron] (-0.7,0.2) -- (0.7,-0.2);
\draw[electron] (-0.7,-0.2) -- (0.7,0.2);
\draw[electron] (-0.5,-0.55) -- (0.5,0.55);
\filldraw (0,0) circle (5pt);
\end{scope}
\end{tikzpicture}
\eeq
while the $\sim \cA_c(2 \to 2)\, \cA_c(2 \to 2)$ terms contain the diagrams of the form
\beq
\begin{tikzpicture}[line width=1 pt, scale=2]
\begin{scope}[shift={(0.5,0)}]
\draw[electron] (-0.5,0.8) -- (0.5,0.2);
\draw[electron] (-0.5,0.2) -- (0.5,0.8);
\filldraw (0,0.5) circle (3pt);
\draw[electron] (-0.5,-0.2) -- (0.5,-0.8);
\draw[electron] (-0.5,-0.8) -- (0.5,-0.2);
\filldraw (0,-0.5) circle (3pt);
\end{scope}
\end{tikzpicture} 
\eeq
We can now compute $C_{ss'}(x,\vec{p})$ in Eq. \eqref{eq:C1}. To find the commutator $\[ N_{\vec{p},s's}, \De \]$, where $\De$ is given by Eq. \eqref{eq:clustdec}, we use the Leibniz rule $\[ N, AB \] \equiv \[ N, A \] B + A \[ N, B \]$, thanks to which we only need the elementary commutators
\beq
[ N_{\vec{p},s's}, a^{\dagger}_{\vec{q},r} ] \equiv (2\pi)^3 \de^{(3)}(\vec{p} - \vec{q}) \, \de_{rs} a^{\dagger}_{\vec{p},s'} \, ,  \hspace{1cm}  [ N_{\vec{p},s's}, a_{\vec{q},r} ] \equiv -(2\pi)^3 \de^{(3)}(\vec{p} - \vec{q}) \, \de_{rs'} a_{\vec{p},s} \, .
\eeq
Given the symmetries of $\cA_c$, all the commutators involving a creation (resp. annihilation) operator contribute the same, thus simply giving rise to a factorial weight. The result can be written as 
\beq
C_{ss'}(\vec{p}) \equiv C^+_{ss'}(\vec{p}) - C^-_{ss'}(\vec{p}) \, ,
\eeq
where
\bea
C^+_{ss'}(\vec{p}) & := & \frac{\sqrt{E_{p,s}}}{\sqrt{2}VT}\, \sum_{n, n', m, m' = 0}^{\infty} \frac{1}{n! m! n'! m'!} \label{eq:C+1} \\
 & & \times \int \( \prod_{k=1}^n \prod_{k'=1}^{n'} \prod_{l=1}^m \prod_{l'=1}^{m'} \frac{\ed^3 p_k \, \ed^3 p'_{k'} \, \ed^3 q_l \, \ed^3 q'_{l'}}{(2\pi)^3 (2\pi)^3 (2\pi)^3 (2\pi)^3 \sqrt{2 E_{p_k, s_k} 2 E_{p'_{k'}, s'_{k'}} 2 E_{q_l, r_l} 2 E_{q'_{l'}, r'_{l'}}}} \) \nn \\
 & & \times  (2\pi)^4 \de^{(4)} \( \sum_{k=1}^n p_k - \sum_{l=1}^m q_l \)  (2\pi)^4 \de^{(4)} \( p + \sum_{k'=1}^{n'} p'_{k'} - \sum_{l'=1}^{m'} q'_{l'} \) \nn   \\
 & & \times \cA^*_c ( \vec{q}_1, r_1, \dots, \vec{q}_m, r_m \to \vec{p}_1, s_1, \dots, \vec{p}_n, s_n ) \nn \\
 & & \times \cA_c ( \vec{q}'_1, r'_1, \dots, \vec{q}'_{m'}, r'_{m'} \to \vec{p}'_1, s'_1, \dots, \vec{p}'_{n'}, s'_{n'}, \vec{p}, s) \nn \\
 & & \times \langle a^{\dagger}_{\vec{q}_m, r_m} \dots a^{\dagger}_{\vec{q}_1, r_1} a_{\vec{p}_1, s_1} \dots a_{\vec{p}_n, s_n} a^{\dagger}_{\vec{p}, s'}  a^{\dagger}_{\vec{p}'_{n'}, s'_{n'}} \dots a^{\dagger}_{\vec{p}'_1, s'_1} a_{\vec{q}'_1, r'_1} \dots a_{\vec{q}'_{m'}, r'_{m'}} \rangle_{\ro}  \, , \nn
\eea
and
\bea
C^-_{ss'}(\vec{p}) & := &  \frac{\sqrt{E_{p,s}}}{\sqrt{2}VT}\, \sum_{n, n', m, m' = 0}^{\infty} \frac{1}{n! m! n'! m'!} \label{eq:C-1} \\
 & & \times \int \( \prod_{k=1}^n \prod_{k'=1}^{n'} \prod_{l=1}^m \prod_{l'=1}^{m'} \frac{\ed^3 p_k \, \ed^3 p'_{k'} \, \ed^3 q_l \, \ed^3 q'_{l'}}{(2\pi)^3 (2\pi)^3 (2\pi)^3 (2\pi)^3 \sqrt{2 E_{p_k, s_k} 2 E_{p'_{k'}, s'_{k'}} 2 E_{q_l, r_l} 2 E_{q'_{l'}, r'_{l'}}}} \) \nn \\
 & & \times  (2\pi)^4 \de^{(4)} \( \sum_{k=1}^n p_k - \sum_{l=1}^m q_l \)  (2\pi)^4 \de^{(4)} \( \sum_{k'=1}^{n'} p'_{k'} - \sum_{l'=1}^{m'} q'_{l'} - p \) \nn   \\
 & & \times \cA^*_c ( \vec{q}_1, r_1, \dots, \vec{q}_m, r_m \to \vec{p}_1, s_1, \dots, \vec{p}_n, s_n ) \nn \\
 & & \times \cA_c ( \vec{p}, s', \vec{q}'_1, r'_1, \dots, \vec{q}'_{m'}, r'_{m'} \to \vec{p}'_1, s'_1, \dots, \vec{p}'_{n'}, s'_{n'} ) \nn \\
 & & \times \langle a^{\dagger}_{\vec{q}_m, r_m} \dots a^{\dagger}_{\vec{q}_1, r_1} a_{\vec{p}_1, s_1} \dots a_{\vec{p}_n, s_n} a^{\dagger}_{\vec{p}'_{n'}, s'_{n'}} \dots a^{\dagger}_{\vec{p}'_1, s'_1} a_{\vec{p}, s} a_{\vec{q}'_1, r'_1} \dots a_{\vec{q}'_{m'}, r'_{m'}} \rangle_{\ro}  \, . \nn
\eea
By looking at the position of the $(\vec{p},s)$ entry in the scattering amplitude $\cA_c$, we recognize in $C^{\pm}$ the ``creation" and ``annihilation" terms of the collision, respectively. Here the 4-momenta appearing in the Dirac deltas are on-shell
\beq
p_k \equiv (E_{p_k,s_k}, \vec{p}_k) \, , \hspace{1cm} p'_k \equiv (E_{p'_k,s'_k}, \vec{p}'_k) \, , \hspace{1cm} q_k \equiv (E_{q_k,r_k}, \vec{q}_k) \, , \hspace{1cm} q'_k \equiv (E_{q'_k,r'_k}, \vec{q}'_k) \, ,
\eeq
and it is understood that we sum over repeated discrete indices. 

We must now compute the quantum statistical expectation value appearing in the last lines using Eq. \eqref{eq:rotherdev3}. Expressing the trace in the occupation number basis, we see that only the terms containing an equal number of creation and annihilation operators can be non-zero. Since $\ro$ is a function of creation/annihilation pairs \eqref{eq:rotherdev3}, the non-zero terms are the ones with $n + m' = n' + m + 1$ for $C^+$ and $n + m' + 1 = n' + m$ for $C^-$. This allows us to eliminate the sum over $m'$. 

The expectation value will therefore be a sum of products of Dirac and Kronecker deltas (up to $\bm{f}$-dependent factors) that force the ladder operators to come in creation/annihilation pairs of equal momenta. Fortunately, we only need to consider the cases where all the creation/annihilation pairs have distinct momenta, because the other cases are of measure zero in the integration. Factorizing again the trace \eqref{eq:tracefact} as in subsection \ref{sec:entropy}, we are therefore only left with the simplest traces, that is, Eqs. \eqref{eq:simptraces1} and \eqref{eq:simptraces2}. When a creation/annihilation pair is converted into deltas we will say it has been ``contracted". 

Consider now the case where one of the momenta in each contracted creation/annihilation pair appears in $\cA_c$ and the other in $\cA^*_c$. For every $(n,n',m)$ value, there is only one term of this kind, because of the symmetries of $\cA_c$, so that it only picks up a combinatoric factor $N!$, where $N$ is the number of involved pairs. Let us call these terms the ``proper" collision terms and let us denote the rest by $F^{\pm}_{ss'}(\vec{p})$. Performing the proper contraction of $C^{\pm}(\vec{p})$ and eliminating some of the integrals with the resulting Dirac deltas we find (and after renaming some indices)
\bea
C^+_{ss'}(\vec{p}) & \equiv & \frac{1}{2}\, f^{\circ}_{s'' s'}(\vec{p}) \,  \sum_{n, m = 0}^{\infty} \frac{1}{n! m!} \label{eq:C+} \\
 & & \times \int \( \prod_{k=1}^n \frac{\ed^3 p_k}{(2\pi)^3 2 E_{p_k, s_k}} \prod_{l=1}^m \frac{\ed^3 q_l }{(2\pi)^3 2 E_{q_l, r_l} } \)  (2\pi)^4 \de^{(4)} \( p + \sum_{k=1}^n p_k - \sum_{l=1}^m q_l \) \nn   \\
 & & \times \, f_{r_1 r'_1}(\vec{q}_1) \dots f_{r_m r'_m}(\vec{q}_m)  \, \cA^*_c ( \vec{q}_1, r'_1, \dots, \vec{q}_m, r'_m \to \vec{p}_1, s'_1, \dots, \vec{p}_n, s'_n, \vec{p}, s'' )  \nn \\ 
 & & \times \, f^{\circ}_{s'_1 s_1}(\vec{p}_1) \dots f^{\circ}_{s'_n s_n}(\vec{p}_n) \, \cA_c ( \vec{q}_1, r_1, \dots, \vec{q}_m, r_m \to \vec{p}_1, s_1, \dots, \vec{p}_n, s_n, \vec{p}, s)  \nn \\
 & & + \, F^+_{ss'}(\vec{p})  \, , \nn
\eea
and
\bea
C^-_{ss'}(\vec{p}) & \equiv & \frac{1}{2} \, f_{s s''}(\vec{p}) \sum_{n, m = 0}^{\infty} \frac{1}{n! m!} \label{eq:C-} \\
 & & \times \int \( \prod_{k=1}^n \frac{\ed^3 p_k}{(2\pi)^3 2 E_{p_k, s_k}} \prod_{l=1}^m \frac{\ed^3 q_l }{(2\pi)^3 2 E_{q_l, r_l} } \)  (2\pi)^4 \de^{(4)} \( p + \sum_{k=1}^n p_k - \sum_{l=1}^m q_l \) \nn   \\
 & & \times \, f_{s_1 s'_1}(\vec{p}_1) \dots f_{s_n s'_n}(\vec{p}_n) \, \cA^*_c (\vec{p}, s'', \vec{p}_1, s'_1, \dots, \vec{p}_n, s'_n \to \vec{q}_1, r'_1, \dots, \vec{q}_m, r'_m)  \nn \\
 & & \times \, f^{\circ}_{r'_1 r_1}(\vec{q}_1) \dots f^{\circ}_{r'_m r_m}(\vec{q}_m)  \, \cA_c ( \vec{p}, s', \vec{p}_1, s_1, \dots, \vec{p}_n, s_n \to \vec{q}_1, r_1, \dots, \vec{q}_m, r_m)   \nn \\
 & & + \, F^-_{ss'}(\vec{p})  \, . \nn
\eea
Here we have used the fact that $f_{ss'} \equiv 0$ if $m_s \neq m_{s'}$ and the distributional identity 
\beq
(\de^{(4)}(p))^2 \equiv \de^{(4)}(0) \de^{(4)}(p) \, , \hspace{1cm} \de^{(4)}(p = 0) \equiv \frac{VT}{(2\pi)^4} \, ,
\eeq
which is how the singular $VT$ in the denominator is canceled, just as in the textbook computation of cross-sections and decay rates in QFT. In appendix \ref{sec:compactBUU} we show that the contraction pattern of discrete indices in Eqs. \eqref{eq:C+} and \eqref{eq:C-} allows for a compact formulation in terms of generalized matrix products.

Let us now consider the nature of the $\bm{F}^{\pm}$ terms more closely. By definition, the involved amplitudes contain at least one particle whose momentum is unaffected by the collision, so these are partially forward scattering processes. We must distinguish two cases: either $\vec{p}$ is among the unaffected momenta, or it is not. In the former case, a close inspection of the equations, along with the use of $f_{ss'} \equiv 0$ when $|s| \neq |s'|$ and $m_s \neq m_{s'}$, shows that any such term appearing in $\bm{C}^+$ is canceled by a term in $\bm{C}^-$, and vice-versa. This can be understood intuitively by the fact that, if the $\vec{p}$ particle scatters forward, then there is no difference between the ``creation" and ``annihilation" processes. As for the ones in which the $\vec{p}$ particle is affected, they can only enter as higher-order corrections to the proper ones. Indeed, to every proper process $m \to n$, there corresponds an infinite tower of processes involving $m + k \to n + k$ amplitudes, where the $k$ extra dummy particles have the same initial and final momenta. These are therefore of higher order in the coupling constants of the QFT and can be neglected to a first approximation.    

Thus, to lowest order in the coupling constants, the collision term is made of the $2 \to 2$ scattering term and, if there are also unstable particles, the corresponding decay/creation terms, i.e.
\beq \label{eq:BUUgen}
{\rm L} f_{ss'}(x,\vec{p}) = C_{ss'}^{2 \leftrightarrow 2}(x,\vec{p}) + \sum_{n = 2}^{\infty} \frac{1}{n!} \, C_{ss'}^{1 \leftrightarrow n}(x,\vec{p}) + \dots
\eeq
where
\bea
C_{ss'}^{2 \leftrightarrow 2}(x,\vec{p}) & = & \frac{1}{4} \int \frac{\ed^3 p_2}{(2\pi)^3 2 E_{p_2, s_2}} \frac{\ed^3 q_1 }{(2\pi)^3 2 E_{q_1, r_1} } \frac{\ed^3 q_2}{(2\pi)^3 2 E_{q_2, r_2} } \, (2\pi)^4 \de^{(4)} \( p + p_2 - q_1 - q_2 \) \nn \\
 & & \hspace{0.8cm} \times \[  f_{r_1 r'_1}(x,\vec{q}_1) \, f_{r_2 r'_2}(x,\vec{q}_2) \, f^{\circ}_{s'' s'}(x,\vec{p})\, f^{\circ}_{s'_2 s_2}(x,\vec{p}_2) \right. \label{eq:C22} \\
 & & \hspace{1.3cm} \times  \, \cA ( \vec{q}_1, r_1, \vec{q}_2, r_2 \to \vec{p}_2, s_2, \vec{p}, s ) \, \cA^* ( \vec{q}_1, r'_1, \vec{q}_2, r'_2 \to \vec{p}_2, s'_2, \vec{p}, s'') \nn \\
 & & \hspace{1.3cm} - \, f_{s s''}(x,\vec{p}) \, f_{s_2 s'_2}(x,\vec{p}_2)  \, f^{\circ}_{r'_1 r_1}(x,\vec{q}_1) \, f^{\circ}_{r'_2 r_2}(x,\vec{q}_2)  \nn \\
 & & \hspace{1.3cm} \left. \times \, \cA ( \vec{p}, s', \vec{p}_2, s_2, \to \vec{q}_1, r_1, \vec{q}_2, r_2)  \, \cA^* (\vec{p}, s'', \vec{p}_2, s'_2, \to \vec{q}_1, r'_1, \vec{q}_2, r'_2)  \]   \, ,  \nn
\eea
and
\bea
C_{ss'}^{1 \leftrightarrow n}(x,\vec{p}) & = & \frac{1}{2} \int \( \prod_{k=1}^n \frac{\ed^3 p_k}{(2\pi)^3 2 E_{p_k, s_k}} \) \, (2\pi)^4 \de^{(4)} \( p - \sum_{k=1}^n p_k \) \nn \\
 & & \hspace{0.8cm} \times \[ f_{s_1 s'_1}(x,\vec{p}_1) \dots f_{s_n s'_n}(x,\vec{p}_n) \, f^{\circ}_{s'' s'}(x,\vec{p})   \right. \label{eq:C1n} \\
 & & \hspace{1.3cm} \times \, \cA_c ( \vec{p}_1, s_1, \dots, \vec{p}_n, s_n \to \vec{p}, s) \, \cA^*_c ( \vec{p}_1, s'_1, \dots, \vec{p}_n, s'_n \to \vec{p}, s'' ) \nn \\
 & & \hspace{1.3cm} - \,f_{s s''}(x,\vec{p}) \, f^{\circ}_{s'_1 s_1}(x,\vec{p}_1) \dots f^{\circ}_{s'_n s_n}(x,\vec{p}_n) \nn \\
 & & \hspace{1.3cm} \left. \times  \, \cA_c ( \vec{p}, s' \to \vec{p}_1, s_1, \dots, \vec{p}_n, s_n) \, \cA^*_c (\vec{p}, s'' \to \vec{p}_1, s'_1, \dots, \vec{p}_n, s'_n)   \]   \, . \nn
\eea
We have thus generalized the BUU equation \eqref{eq:BUU} in two aspects: it can now handle matrix distributions, and thus non-trivial polarizations, but it also includes all possible microscopic collision processes contained in the expressions \eqref{eq:C+} and \eqref{eq:C-}.\footnote{If needed, the partially forward scattering contributions $\bm{F}^{\pm}$ can be derived from Eqs. \eqref{eq:C+1} and \eqref{eq:C-1}.} Note, however, that one would expect these higher order contributions to be relevant in regimes where the fluid is no longer dilute enough for the whole kinetic formalism to apply. Nevertheless, it is useful to have them if one is interested in next-to-leading corrections. 

Finally, observe that our expressions for the collision term are not at all explicitly hermitian. To obtain an explicitly hermitian collision term one should use the equivalent expression
\beq
C_{ss'}(\vec{p}) = \frac{E_{p,s}}{2VT}\, \langle \( S^{\dagger} \[ N_{\vec{p},s's}, S \] + \[ S^{\dagger}, N_{\vec{p},s's} \] S \) \rangle_{\ro} \, ,
\eeq
instead of the the last line in Eq. \eqref{eq:Col1}, in which case the result is simply the hermitian part of the expressions derived above. On the other hand, verifying hermiticity, instead of simply imposing it, may serve as a useful consistency check. In doing so one must note, however, that the integrand of the collision term will not be hermitian in general, only the integrated quantity will. This is because the hermiticity of the last line in Eq. \eqref{eq:Col1} relies on the the unitarity of the $S$ matrix $S S^{\dagger} \equiv \mathbb{I}$, which therefore involves a product, and in the momentum basis in which we work such products correspond to integrals over momenta.

\subsection{Comparison with the literature on the collision term} \label{sec:collcomp}

In the more standard approach to kinetic theory, employed for instance in \cite{Dolgov:1980cq, Barbieri:1990vx, Sigl:1992fn, Raffelt:1992uj, Vlasenko:2013fja, Kosowsky:1994cy, Pitrou:2008ut, Pitrou:2008hy, Beneke:2010eg, Fidler:2017pkg,Pitrou:2019hqg}, the formalism contains a single time variable and the macroscopic and mesoscopic scales are separated within that variable. In particular, this implies non-trivial extra manipulations in order to properly disentangle the two regimes. From our viewpoint, this corresponds to identifying evolution along the mesoscopic time $T$ with evolution along the world-line parameter $\la$ of the geodesics that are used in deriving the Liouville operator. Thus, instead of relating $\pa_{\la}$ to a finite difference in $T$ as in \eqref{eq:mMmatch}, one rather relates $\pa_{\la} \os{!}{=} \pa_T$. Evaluating $f_{ss'}(x,\vec{p})$ on a specific geodesic we then get
\beq \label{eq:palatopaT}
\pa_{\la} f_{ss'}(\ga(\la), \vec{k}(\la)) \os{!}{=} V^{-1} \Tr \[ \frac{\pa \ro(\ga)}{\pa T}\, N_{\vec{p},s's} \] = - i V^{-1} \Tr \[ \[ H_{\rm int.}, \ro(\ga) \] N_{\vec{p},s's} \]  \, ,
\eeq 
and, since the above equation must hold for all geodesic paths, one finally gets
\beq \label{eq:stdBoltz}
{\rm L} f_{ss'}(x,\vec{p}) = - i E_{p,s} V^{-1} \Tr \[ \[ H_{\rm int.}, \ro \] N_{\vec{p},s's} \] \equiv - i E_{p,s} V^{-1} \langle \[ N_{\vec{p},s's}, H_{\rm int.} \] \rangle_{\ro} \, .
\eeq 
We have to be careful, however, because $\pa_{\la}$ is LLT-invariant, but $\pa_T$ is not, so the above equation apparently breaks that symmetry. In the finite derivative case \eqref{eq:mMmatch} this was not a problem, because the finite difference from $T = - \infty$ to $T = \infty$ leads to the $S$ matrix, which is Lorentz invariant. To correct the situation in the present case, we note that in the above construction $\pa_{\la}$ leads to the Liouville operator ${\rm L}$, which has dimensions of mass squared, and that one can find an analogous Lorentz-invariant generalization of $\pa_T$, namely
\beq \label{eq:genpaT}
\pa_T \to p^a \frac{\pa}{\pa X^a} \, .
\eeq
With this, the combination $E_{p,s} H_{\rm int.}$ in \eqref{eq:stdBoltz} would generalize to $- p_s^a P_a^{\rm int.}$, where $P^a_{\rm int.} := (H_{\rm int.}, \vec{P})$. But, since this operator enters through a commutator with $\ro$, and the later is homogeneous \eqref{eq:homro}, the result would be again \eqref{eq:stdBoltz}. This equation is therefore LLT-invariant indeed, although not explicitly. 

Thus, with the present prescription, instead of finding the full $S$ matrix on the right-hand side as in \eqref{eq:Col1}, i.e. a full scattering event from $T = - \infty$ to $T = \infty$, one gets the interaction Hamiltonian that generates an infinitesimal increment in time. Although to lowest order in the interactions both approaches lead to the same collision term, we believe that the one we chose \eqref{eq:mMmatch} is more consistent with the assumptions behind kinetic theory (dilute gas and molecular chaos) and behind the applicability of perturbative QFT. Indeed, for the use of QFT amplitudes we need the existence of asymptotic states, meaning that the particles in our gas must be mostly free, up to sporadic interactions, i.e. the gas must be dilute. This is also necessary for the applicability of the molecular chaos assumption, i.e. one needs to have clearly separated ``in" $(T \to -\infty)$ and ``out" $(T \to \infty)$ asymptotic regions in order to justify the fact that $\ro_{\rm in} := \lim_{T \to -\infty} \ro(T)$ is completely determined by $f^{\rm in}_{ss'}(\vec{p})$, whereas $\ro_{\rm out} := \lim_{T \to \infty} \ro(T)$ requires a full tower of correlation functions $f^{\rm out}_{s_1 \dots s_n}(\vec{p}_1, \dots, \vec{p}_{n-1})$. One can then follow the evolution of $f_{ss'} := f^{\rm in}_{ss'}$, express $f^{\rm out}_{ss'}$ as the fully scattered $f_{ss'}$ information, and compute the difference, as we did in Eq. \eqref{eq:mMmatch}. In contrast, in the usual approach where $\pa_{\la} \os{!}{=} \pa_T$, one identifies the microscopic dynamics with the macroscopic ones, so that the particles are treated as being in a continuous state of interaction. There are therefore no clear ``in" and ``out" phases for the use of asymptotic QFT states to be justified and for the molecular chaos hypothesis to be implemented unambiguously. Finally, from the purely mathematical viewpoint, as already argued in section \ref{sec:microst}, the time variable $T$ of the microscopic QFT, or more generally the corresponding space-time coordinates $X^a$, are the Fourier conjugates of the momentum variables $p^a$, which are clearly independent of the macroscopic coordinates $x^{\mu}$ or the geodesic parameter $\la$, which is usually taken to be the proper time for massive particles. In light of this mathematical structure, considering the identification $\pa_{\la} \os{!}{=} \pa_T$, or its covariant generalization \eqref{eq:genpaT}, seems rather unnatural.

\subsection{Exact conservation equations of the collisional equation} \label{sec:conservcoll}

We now want to show that the total energy-momentum tensor 
\beq 
T^{ab}(x) := \int \frac{\ed^3 p}{(2\pi)^3 E_{p,s}} \, f_{ss}( x, \vec{p} ) \, p_s^a p_s^b + T^{ab}_{\rm EM} \, , 
\eeq
and the total electric current
\beq 
J^a(x) := \int \frac{\ed^3 p}{(2\pi)^3 E_{p,s}} \, q_s f_{ss}( x, \vec{p} ) \, p_s^a \, , 
\eeq
are exactly conserved 
\beq
\na_a T^{ab} = 0 \, , \hspace{1cm} \na_a J^a = 0 \, ,
\eeq
in the presence of interactions as well, as required for the consistency of the Einstein-Maxwell equations. We start with $T^{ab}$ and proceed as in Eqs. \eqref{eq:TTEMcons} and \eqref{eq:Tconsnocol}, i.e. we express the integral as a 4-dimensional one along with a Dirac delta imposing the dispersion relation and use the Maxwell equations and Bianchi identity of $F_{ab}$ for the electromagnetic part. This leaves us with an integral depending on the collision term only, which we can re-express as a 3-dimensional integral
\beq
\na_b T^{ab}(x) = \int \frac{\ed^3 p}{(2\pi)^3 E_{p,s}} \, C_{ss}(x,\vec{p}) \, p_s^a  \, .
\eeq
It is then convenient to use Eq. \eqref{eq:Col1} for the collision term, i.e.
\beq
\na_b T^{ab}(x) \sim \int \frac{\ed^3 p}{(2\pi)^3} \, p_s^a \, \langle S^{\dagger} [ N_{\vec{p},ss}, S ] \rangle_{\ro(x)} \, ,
\eeq
so that, using the definitions of the asymptotic Hamiltonian \eqref{eq:freeHam} and momentum operators \eqref{eq:totmom}, we find
\beq
\na_b T^{ab}(x) \sim \langle S^{\dagger} [ P_{\rm asy.}^a, S ] \rangle_{\ro(x)} \, , \hspace{1cm} P_{\rm asy.}^a := (H_{\rm asy.}, \vec{P}) \, .
\eeq  
For the spatial components $a = i$ we obtain trivially zero because microscopic time-evolution conserves momentum so $[\vec{P}, S] = 0$. For the time component $a = 0$ the situation is a bit less trivial, because $H_{\rm asy.}$ is not the full Hamiltonian $H$ and $[H, H_{\rm asy.}] \neq 0$. Nevertheless, $H_{\rm asy.}$ coincides with $H$ on asymptotic states by definition, so the conservation of $H$ implies
\beq
S H_{\rm asy.} | {\rm in} \ket = S H | {\rm in} \ket \equiv H S | {\rm in} \ket \equiv H | {\rm out} \ket = H_{\rm asy.} | {\rm out} \ket = H_{\rm asy.} S | {\rm in} \ket \, ,
\eeq
for all $|{\rm in}\ket$, meaning that $[S, H_{\rm asy.}] = 0$ on the states of interest. Thus, $H_{\rm asy.}$ is not necessarily constant during evolution, but the initial and final values at $T \to \mp \infty$ are the same, which concludes our proof
\beq
\na_b T^{ab} = 0 \, .
\eeq
For the electric current we proceed similarly and arrive at
\beq
\na_a J^a(x) \sim \langle S^{\dagger} [ Q, S ] \rangle_{\ro(x)} = 0 \, ,
\eeq
where $Q$ is the total charge operator \eqref{eq:totmom}, which is also exactly conserved during microscopic evolution and therefore commutes with the $S$ matrix. As for the total entropy current \eqref{eq:scurrent}, again the same procedure leads to
\beq \label{eq:Htheorem}
\na_a s^a(x) = \sum_s \int \frac{\ed^3 p}{(2\pi)^3 E_{p,s}} \[ \bm{C} \log \frac{\bm{f}_{\circ}}{\bm{f}} \]_{ss}(x,\vec{p}) \, .
\eeq
If we only consider the $2 \to 2$ scattering term, and work in the tilded basis where $\ti{f}_{\ti{s}\ti{s}'}$ is diagonal, we recover the set-up of the standard Boltzmann equation, in which case the $H$-theorem $\na_a s^a \geq 0$ is proved in the usual way. In the general case, however, it is not clear to us how to proceed, so we will not consider this issue further.

\subsection{Liouville-transported wave-functions and tensor distributions} \label{sec:wavefunctions}

The $s$ indexation of the 1-particles states $| \vec{p}, s \ket$ we have considered so far parametrizes exactly the degrees of freedom of the QFT, so it is not in a one-to-one correspondence with the quantum field components in general. The typical example is the (mesoscopic) quantum electromagnetic field $A_a(X)$, which has four components, while there are only two physical photon states. Another example is the Dirac field $\psi(X)$, which has eight real components, but corresponds to only four physical states, two for the particle and two for the anti-particle. The relation between the field indices and the $s$ indices is given by the so-called ``wave-functions". 
As we will see, these wave-functions are also required in order to express the BUU equations in terms of scattering matrix elements that depend solely on the momenta, not depend on $x^{\mu}$. Here we will discuss the two examples cited above, since these are the most relevant ones for cosmology.

\subsubsection{Photons}

We focus on the $2\times 2$ block of $\bm{f}$ corresponding to the photon distribution. For later convenience, and in order to remain close to the usual conventions, we will use $r$ to denote the indices of this matrix and $k^a$ to denote the null 4-momentum, e.g. $f_{rr'}(x,\vec{k})$. In this case the wave-functions are usually referred to as ``polarization vectors" $\ep_r^a(x,\vec{k})$, $r \in \{1,2\}$, which can be chosen such that
\beq \label{eq:algcondepsa} 
\et_{ab}\, \ep_r^a \ep_{r'}^b \equiv \de_{rr'} \, , \hspace{1cm} k_a \ep_r^a \equiv 0  \, ,
\eeq  
and arise when expressing the quantum field in terms of ladder operators in the asymptotic region
\beq \label{eq:microAaX}
A_a(X) = \int \frac{\ed^3 k}{(2\pi)^3 \sqrt{2k}}\, \ep^r_a(\vec{k}) \[ a_{\vec{k},r}\, e^{i k_b X^b} + a^{\dagger}_{\vec{k},r}\, e^{-i k_b X^b} \] \, ,
\eeq
where $k := |\vec{k}|$. With this choice the above field is completely gauge-fixed. It satisfies the Lorenz gauge
\beq
\frac{\pa A^a}{\pa X^a} = 0 \, ,
\eeq
which reduces the number of independent components from four to three. This condition is then preserved under a residual gauge transformation
\beq \label{eq:U1AX}
\ti{A}_a = A_a + \frac{\pa \te}{\pa X^a} \, ,
\eeq
where the gauge parameter obeys a free massless wave-equation
\beq \label{eq:U1AXres}
\frac{\pa^2 \te}{\pa X_a \pa X^a} = 0 \, ,
\eeq
just as the asymptotic field $A_a$ in the Lorentz gauge. This therefore allows us to eliminate one more component, a ``longitudinal" polarization, leaving us with the two physical polarizations of Eq. \eqref{eq:microAaX}. 

An important reason for invoking wave-functions at this level is that they are necessary in order to relate the photon $2 \times 2$ block $f_{rr'}$ to observations. Indeed, what the observer family $e_a(x)$ actually measures is the Lorentz tensor on $\cP\cM$
\beq \label{eq:fabdef}
f_{ab}(x,\vec{k}) := f_{rr'}(x,\vec{k})\, \ep^r_a(x,\vec{k})\, \ep^{r'}_b(x,\vec{k}) \, ,
\eeq
which therefore obeys
\beq \label{eq:fabalgprop}
f^*_{ab}(x,\vec{k}) \equiv f_{ba}(x,\vec{k}) \, , \hspace{1cm} k^a f_{ab}(x,\vec{k}) \equiv 0 \, .
\eeq
Note that $f_{ab}(x,\vec{k})$ now appears as the Fourier transform of the correlation function $\bra A_a(X)\, A^{\dagger}_b(Y) \ket_{\ro}$ for the (complexified) free quantum field. We will see that it is convenient to express the polarization-dependent BUU equation in terms of $f_{ab}$
\beq \label{eq:BUUfab}
{\rm L} f_{ab} = C_{ab} := \ep^r_a \ep^{r'}_b C_{rr'} \, ,
\eeq 
where here L is the Liouville operator for tensor distributions \eqref{eq:Ltens}. To achieve this, we can constrain the $x^{\mu}$-dependence of $\ep_r^a$ to be such that these are Liouville-transported vector distributions
\beq
{\rm L} \ep_r^a(x,\vec{k}) := k^b \[ \( \pa_b - \Si_{icb} k^c \frac{\pa}{\pa k^i} \) \ep_r^a + \Si^a_{\,\,\,cb} \ep_r^c \](x,\vec{k}) \equiv 0 \, ,
\eeq
a condition which is consistent with the algebraic relations \eqref{eq:algcondepsa}. Along with the above differential equation, these conditions imply that the $\ep_r^a$ basis is the phase space analogue of the Sachs basis $k_A^a$ associated with a particular geodesic that we built in section \ref{sec:Sachs}. In fact, by evaluating $\ep_r^a$ on a given light-like geodesic $\ep_r^a(\ga(\ze),\vec{k}(\ze))$ we obtain a Sachs basis by construction, i.e. a field on $\Lie$ satisfying \eqref{eq:SachsPT}. Thus, the polarization basis will have similar properties with the Sachs basis. For instance, the conditions \eqref{eq:algcondepsa} determine the basis $\ep_r^a$ only up to a shift of the form
\beq \label{eq:epraGT}
\ti{\ep}_r^a(x,\vec{k}) = \ep_r^a(x,\vec{k}) + \te_r(x,\vec{k})\, k^a \, ,  \hspace{1cm} {\rm L} \te_r = 0 \, ,
\eeq
and an internal rotation
\beq \label{eq:eprarot}
\ti{\ep}_r^a(x,\vec{k}) = R_{rr'}(x,\vec{k})\, \ep_{r'}^a(x,\vec{k}) \, , \hspace{1cm} {\rm L} R_{rr'} = 0 \, .
\eeq
At the level of $A_a(X)$, the shift transformation \eqref{eq:epraGT} reproduces a residual gauge transformation, i.e. \eqref{eq:U1AX} with a gauge parameter satisfying \eqref{eq:U1AXres}
\beq
\te(X) = -i \int \frac{\ed^3 k}{(2\pi)^3 \sqrt{2k}}\, \te_r(\vec{k}) \[ a_{\vec{k},r}\, e^{i k_a X^a} - a^{\dagger}_{\vec{k},r}\, e^{-i k_a X^a} \] \, .
\eeq
Therefore, the ambiguity captured by $\te_r$ corresponds to the ambiguity of how to eliminate the longitudinal polarization with the residual gauge freedom. Consequently, although we work only with the physical photon polarizations, we still have a gauge ambiguity in our choice of $\ep_r^a$ and the physics must be independent of that choice. The analogy with the Sachs basis is that the corresponding shift transformation only affects the $n_A$ component and that the physical observables are independent of that quantity. On the other hand, the internal rotation \eqref{eq:eprarot} does not lead to a gauge transformation for $A_a(X)$, because the corresponding field strength 
\beq \label{eq:FabX}
F_{ab} := \frac{\pa A_b}{\pa X^a} - \frac{\pa A_a}{\pa X^b} \, ,
\eeq
is not invariant, but gets rotated. The transformed gauge field is still a plane wave solution, but a physically distinct one, so the physical observables will generically not be invariant under \eqref{eq:eprarot}, but rather covariant. In the case of the Sachs basis too, the Sachs rotations change the angular basis on the observer sky, which is why we had to fix this freedom in order to match the basis the observer really uses. 

Let us now decompose $f_{rr'}$ as follows
\beq
\bm{f} \equiv \frac{1}{2} \( I \bm{1} + i V \bm{\vep} + \bm{P}  \) \, , 
\eeq
where $I, \bm{P}, V$ are real,
\beq
\bm{P}^T \equiv \bm{P} \, , \hspace{1cm} \Tr\, \bm{P} \equiv 0 \, ,
\eeq
and $\bm{\vep}_{rr'} := \vep_{rr'}$. Being the trace, $I$ is the total number density of photons in phase space, $\bm{P}$ captures the linear polarizations (``plus" and ``cross"), while $V$ captures the circular polarization. This leads to the decomposition of $f_{ab}$ into irreducible parts under LLTs
\beq \label{eq:fabpoldec}
f_{ab} \equiv \frac{1}{2} \( I \ep_{ab} + i V \vep_{ab} + P_{ab} \) \, ,
\eeq
where
\beq \label{eq:epvepdef}
\ep^{ab} := \ep_r^a \ep_r^b  \, ,  \hspace{1cm} \vep^{ab} := \vep_{rr'} \ep_r^a \ep_{r'}^b \, ,  \hspace{1cm} P^{ab} := P_{rr'} \ep_r^a \ep_{r'}^b \, ,
\eeq
satisfy the following identities 
\beq \label{eq:epiden}
k^a \ep_{ab} \equiv 0 \, , \hspace{1cm} \ep_{ab} \equiv \ep_{ba} \, , \hspace{1cm} \ep_a^c \ep_{cb} \equiv \ep_{ab} \, , \hspace{1cm} \ep_a^a \equiv 2 \, ,
\eeq
\beq   \label{eq:vepiden}
k^a \vep_{ab} \equiv 0 \, , \hspace{1cm} \vep_{ab} \equiv - \vep_{ba} \, , \hspace{1cm} \vep_{ab} \vep_{cd} \equiv 2\ep_{a[c} \ep_{d]b} \, , \hspace{1cm} \ep_a^c \vep_{cb} \equiv \vep_{ab} \, , \nn 
\eeq
\beq 
\vep_{abcd} k^d \equiv - \[ \vep_{ab} k_c + \vep_{bc} k_a + \vep_{ca} k_b \] \, , 
\eeq
and
\beq \label{eq:Piden}
k^a P_{ab} \equiv 0 \, , \hspace{1cm} P_a^a \equiv 0 \, , \hspace{1cm} P_{ab} \equiv P_{ba} \hspace{1cm} \vep_a^{\,\,\,c} P_{cb} \equiv \vep_b^{\,\,\,c} P_{ca} \, , \hspace{1cm} \ep_a^c P_{cb} \equiv P_{ab} \, .
\eeq
In particular, it will also be convenient to define
\beq
\ti{P}_{ab} := \vep_a^{\,\,\,c} P_{cb} \, ,
\eeq
which obeys \eqref{eq:Piden} as well. Finally, the inverse relation reads
\beq
I \equiv f_a^a \, , \hspace{1cm} P_{ab} \equiv 2 f_{(ab)} - \ep_{ab} f_c^c  \, , \hspace{1cm} V \equiv -i \vep^{ab} f_{ab} \, ,
\eeq 
and the BUU equation \eqref{eq:BUUfab} turns into
\bea
{\rm L} I = C_a^a \, , \hspace{1cm} {\rm L} V = -i \vep^{ab} C_{ab} \, , \hspace{1cm} {\rm L} P_{ab} = 2 C_{(ab)} - \ep_{ab} C_c^c \, .
\eea

\subsubsection{Dirac fermions}

We now consider the case of a Dirac particle of mass $m$ and charge $q$, along with its anti-particle. These correspond to two $2 \times 2$ blocks in $\bm{f}$, which we denote by $f_{ss'}^{\pm}(x,\vec{p})$, $s \in \{1,2\}$, with $\pm$ distinguishing the particle and anti-particle ones, respectively. The on-shell 4-momentum therefore obeys $p_a p^a \equiv -m^2$ and we will focus exclusively on the massive case $m \neq 0$ for simplicity.\footnote{See \cite{Fidler:2017pkg} for a treatment of the massless case.}

The wave-functions are the four Dirac spinor distributions $\{ u^{\pm}_s(x,\vec{p}) \}_{s=1,2}$ obeying the standard orthonormality relation 
\beq \label{eq:usornor}
\bar{u}^{\pm}_s u_{s'}^{\pm} \equiv \pm\, 2m\de_{ss'} \, , \hspace{1cm} \bar{u}_s^{\pm} u_{s'}^{\mp} \equiv 0 \, ,  \hspace{1cm} \bar{u} := u^{\dagger} \ga^0 \, ,
\eeq
the completeness relation
\beq \label{eq:uuclosure}
u^{\pm}_s(\vec{p})\, \bar{u}^{\pm}_s(\vec{p}) \equiv {\not} p \pm m \, ,
\eeq 
and the Dirac equation
\beq \label{eq:Diraceq}
\( {\not} p \mp m \) u^{\pm}_s(\vec{p}) \equiv 0 \, , \hspace{1cm} {\not} p := \ga^a p_a \, ,
\eeq
and relating the quantum field to the ladder operators as follows
\beq
\psi(X) = \sum_{s = 1,2} \int \frac{\ed^3 p}{(2\pi)^3 \sqrt{2E_p}} \[ u^+_s(\vec{p}) \, a^+_{\vec{p},s}\, e^{i p_a X^a} + u^-_s(\vec{p})\, (a_{\vec{p},s}^-)^{\dagger}\, e^{-i p_a X^a} \] \, .
\eeq
With these wave-functions, one can now construct the Dirac-indexed matrix distribution of particles and anti-particles, respectively, out of the $2 \times 2$ hermitian matrices $f^{\pm}_{ss'}$
\beq \label{eq:fpmoffpmss}
f^{\pm}(x,\vec{p}) := \pm \frac{1}{2m}\, f^{\pm}_{ss'}(x,\vec{p}) \, u^{\pm}_s(x,\vec{p})\, \bar{u}^{\pm}_{s'}(x,\vec{p}) \, ,  
\eeq
which therefore obey
\beq \label{eq:fDiracalgconstr}
\bar{f}^{\pm}(x,\vec{p}) \equiv f^{\pm}(x,\vec{p}) \, , \hspace{1cm} \( {\not} p \mp m \) f^{\pm}(x,\vec{p}) \equiv f^{\pm}(x,\vec{p}) \( {\not} p \mp m \) \equiv 0 \, .
\eeq
Their sum is the Fourier transform of the correlation function of free quantum fields $\bra \psi^{\frak{a}}(X)\, \bar{\psi}^{\frak{b}}(Y) \ket_{\ro}$, so they are invariant under U(1)GTs in particular
\beq
\ti{f}^{\pm}(x,\vec{p}) = e^{-iq\te(x)} f^{\pm}(x,\vec{p}) \, e^{i q \te(x)} \equiv f^{\pm}(x,\vec{p}) \, .
\eeq
As in the photon case, we demand that the wave-functions $u_s^{\pm}$ be Liouville-transported 
\beq
{\rm L}  u^{\pm}_s := p^a \[ \pa_a - \( \Si_{iba} p^b + q F_{ia} \) \frac{\pa}{\pa p^i} + \frac{1}{4}\, \Si_{bca} \ga^b \ga^c \] u^{\pm}_s = 0 \, ,
\eeq
for the corresponding BUU equations to become
\beq \label{eq:BUUfDir}
{\rm L}  f^{\pm} = \pm \frac{1}{2m}\, C^{\pm}_{ss'}  u^{\pm}_s \bar{u}^{\pm}_{s'} \, ,
\eeq
where now L is a straightforward generalization of \eqref{eq:Ltens} to two Dirac indices (the second one in the conjugate representation)
\beq
{\rm L} f^{\pm} := p^a \[ \pa_a - \( \Si_{iba} p^b + q F_{ia} \) \frac{\pa}{\pa p^i} \] f^{\pm} + \frac{1}{4}\, p^a \Si_{bca} \[ \ga^b \ga^c, f^{\pm} \] \, .
\eeq
However, as already noted in section \ref{sec:Liou}, any pair of Dirac indices can be turned into a Lorentz index. More precisely, a bar-hermitian Dirac matrix such as $f^{\pm}$ can be decomposed in the Clifford algebra basis $\{ 1, \ga^a, i\ga^{[a} \ga^{b]}, \ga^a \ga^5, i \ga^5 \}$. The general solution to the algebraic conditions \eqref{eq:fDiracalgconstr} takes the form \cite{Fidler:2017pkg}
\beq \label{eq:fpmofIS}
f^{\pm}(x,\vec{p}) = \frac{1}{4} \[ I^{\pm}(x,\vec{p}) + \ga^a \ga^5 S^{\pm}_a(x,\vec{p}) \] \( 1 \pm \frac{{\not } p}{m} \) \, ,
\eeq
where
\beq \label{eq:fanormalp}
p^a S^{\pm}_a(x,\vec{p}) \equiv 0 \, .
\eeq
Indeed, this condition makes $f^{\pm}$ bar-hermitian and it also implies that the $\( 1 \pm {\not } p/m \)$ factor in \eqref{eq:fpmofIS} can be put on either side of the square bracket, thus satisfying both of the last two equations in \eqref{eq:fDiracalgconstr}. That this is the general solution is then due to the fact that $I^{\pm}$ and $S_a^{\pm}$ have four independent components, for each sign, just like the original matrices $f^{\pm}_{ss'}$. The inverse relation is then simply
\beq
I^{\pm} \equiv \Tr\, f^{\pm} \, , \hspace{1cm} S^{\pm}_a \equiv \Tr \[ \ga_a \ga^5 f^{\pm} \]  \, .
\eeq
Since $\Tr\, f^{\pm} \equiv f^{\pm}_{ss}$, the $I^{\pm}$ are the total number density in phase space of particles and anti-particles, respectively, while the space-like pseudo-vectors $S^{\pm}_a$ correspond to the intrinsic spin direction when the fermion fluid is polarized (or in this case ``magnetized"). Indeed, $\Tr \[ \ga^a \ga^5 f^{\pm} \]$ is the statistical analogue of the spin pseudo-current of the Dirac field given in Eq. \eqref{eq:JJtcurrents}. In terms of these variables, the BUU equation \eqref{eq:BUUfDir} reads
\beq
{\rm L} I^{\pm} = \Tr\, C^{\pm} \, , \hspace{1cm} {\rm L} S_a^{\pm} = \Tr \[ \ga_a \ga^5 C^{\pm} \] \, ,
\eeq
where in the latter L is the Liouville operator acting on Lorentz vectors \eqref{eq:Ltens}. As in the case of the photon distribution, here too both sides are consistently normal to $p^a$, because the latter commutes with L.

\subsection{Intrinsic moment sources}

Until now we have considered only one way in which the matter distribution $f_{ss'}(x,\vec{p})$ affects the space-time fields $e_{\mu}^a(x)$ and $A_{\mu}(x)$ $-$ through the total number of particles in phase space. Indeed, in both the energy momentum tensor $T^{ab}$ and the electric current $J^a$ it is the trace of each block of $f_{ss'}$ that is involved (the $I$ components in the language of the previous section) and this is nothing but the statistical expectation value of the QFT number operators $N_{\vec{p},s} := a^{\dagger}_{\vec{p},s} a_{\vec{p},s}$. For example, a Dirac particle of mass $m$ and charge $q$ corresponds to the following sources for the space-time fields
\beq \label{eq:TJofI}
T^{ab}(x) = \int \frac{\ed^3 p}{(2\pi)^3 E_p} \, p^a p^b I(x,\vec{p}) \, , \hspace{1cm} J^a(x) = q \int \frac{\ed^3 p}{(2\pi)^3 E_p} \, p^a I(x,\vec{p}) \, .
\eeq
The question therefore arises of whether, and if so how, the polarization components affect the space-time fields as well. To get some intuition about this issue, note that in the case of electromagnetism, the above current will generate both electric and magnetic fields. However, the latter will only be due to the motion of the total charge, not to the {\it intrinsic} magnetic moment that comes from spin, and which lies in $S^a$, not $I$. 

In the examples we considered, we saw that the polarization components arise as Lorentz tensors $V$, $S^a$, $P^{ab}$, which means that they should be treated as ``intrinsic" (unresolved) multipole moments from the viewpoint of the space-time fields. We must therefore look for effective couplings of the space-time fields to such moments at the action level in order to derive the corresponding contributions at the level of the equations of motion. As in any multipole expansion, these moments couple to derivatives of $e_{\mu}^a$ and $A_{\mu}$ such as the spin connection $\Si^{ab}_{\mu}$, the field strengths $F_{ab}$ and $R_{abcd}$ and derivatives thereof. Consequently, the corresponding terms in the sources $T^{ab}$ and $J^a$ come with more space-time derivatives than the $\sim I$ contributions of Eq. \eqref{eq:TJofI}. Since we are working with long wave-length modes, these new terms would therefore appear as ``small" corrections to \eqref{eq:TJofI}. However, this is really the case only if there are no ``cancellations" in $I$. In the case of $T_{ab}$, the energy is a positive-definite quantity and therefore all contributions add up constructively. In $J^a$, however, the fact that $q$ can have either sign may indeed lead to a total $J^a$ that is many orders of magnitude smaller that the individual contributions. This is particularly the case in cosmology, since the universe is electrically neutral on large scales. For this reason, one could a priori have that the intrinsic magnetization contribution $\sim S^a$ is of the same order of magnitude as the $\sim I$ contribution. This is especially relevant, given that there exist large scale magnetic fields whose origin still remains a mystery (see for instance \cite{Durrer:2013pga} and references therein). Here we will therefore work out the leading order $\sim S^a$ part of $J^a$, but we will ignore the polarization corrections to $T^{ab}$. 

The coupling of interest in the action is $e F_{ab} M^{ab}/2$, where $M^{ab} \equiv - M^{ba}$ is a magnetic moment, because this is relativistic generalization of the usual magnetic dipole coupling $\vec{B} \cdot \vec{M}$, with $M^i := \vep^{ijk} M^{jk}/2$. Varying with respect to $A_{\mu}$, such a term modifies the current of a Dirac particle as follows
\beq \label{eq:Jamod}
J^a \to q \int \frac{\ed^3 p}{(2\pi)^3 E_p} \, p^a I + \na_b M^{ba} \, ,
\eeq 
which remains conserved thanks to the antisymmetry of $M^{ab}$. We must next build $M^{ab}$ out of $S^a$. To that end, we note that the spin charge is $\sim \vep_{abcd} \,p^c S^d$, since $S^a$ is the spin pseudo-vector and $p^a$ selects the ``time-component" in phase space. The ``spin moment" is therefore given by
\beq
S_{ab}(x) := \vep_{abcd} \int \frac{\ed^3 p}{(2\pi)^3 E_p}\, p^c S^d(x,\vec{p}) \, .
\eeq
To obtain a more familiar relation, we can use $p_a S^a(x,\vec{p}) \equiv 0$ to find
\beq
\frac{1}{2}\,\vep^{ijk} S^{jk}(x) = \int \frac{\ed^3 p}{(2\pi)^3} \[ \de^{ij} - \frac{p^i p^j}{E_p^2} \] S^j(x,\vec{p}) \approx \int \frac{\ed^3 p}{(2\pi)^3} \, S^i(x,\vec{p}) \, ,
\eeq
where the approximation holds if $S^i(x,\vec{p})$ is supported on non-relativistic momenta $p \ll E_p$. The magnetic moment is then given by the standard relation to the angular momentum
\beq
M^{ab} := \frac{gq}{2m}\, S^{ab} \, , 
\eeq
where $g$ is the ``gyromagnetic ratio" ($g \approx 2$). Thus, the total source current for the electromagnetic field receives a contribution \eqref{eq:Jamod} for each Dirac particle species.

\subsection{Example: the photon-electron-proton fluid} \label{sec:Compton}

As our particle content here we consider photons $(\ga)$, electrons $(e)$ and protons $(p)$, neglecting the internal structure of the latter, thus effectively treating it as a fundamental Dirac particle. We also focus on typical energies way below the electron mass $m_e \approx 0.51 \, {\rm MeV}$ and way above the Rydberg energy ${\rm Ry} \approx 14\, {\rm eV}$, so we can neglect positrons, anti-protons and the hydrogen bound states without spoiling energy-momentum conservation. The matrix distribution therefore takes the form
\beq
\bm{f} = \( \begin{array}{ccc} \bm{f}_{\ga} & \bm{0} & \bm{0} \\ \bm{0} & \bm{f}_e & \bm{0} \\ \bm{0} & \bm{0} & \bm{f}_p  \end{array} \) \, ,
\eeq
where the $\bm{f}_{\ga,e,p}$ blocks are $2 \times 2$ hermitian matrices, since these particles have two spin states. Since here we only care about the collision term, we will ignore the $x^{\mu}$ dependencies for notational simplicity. We will use the letters $k$, $p$ and $q$ for the momenta and $r$, $s$, $t$ for the discrete indices of $\bm{f}_{\ga,e,p}$, respectively. This way we can leave the particle label and momentum dependence implicit, i.e. recognizing them by the discrete indices
\beq
\bm{f}_{\ga}(\vec{k}) \to f_{rr'} \, , \hspace{1cm} \bm{f}_e(\vec{p}) \to f_{ss'} \, , \hspace{1cm} \bm{f}_p(\vec{q}) \to f_{tt'}  \, . 
\eeq
We will also need dummy momenta to perform the collision integrals, in which case we will simply use numbers and reflect the momentum dependence on the discrete indices again, e.g.
\beq
f_{r_n r'_n} \equiv f_{r_n r'_n}(\vec{k}_n) \, , \hspace{1cm} {\rm etc}.
\eeq
The on-shell 4-momenta thus obey 
\beq
k_a k^a \equiv 0 \, , \hspace{1cm} p_a p^a \equiv -m_e^2 \, , \hspace{1cm} q_a q^a \equiv -m_p^2 \, , 
\eeq 
and we use
\beq
k := |\vec{k}| \, , \hspace{1cm} E_p := \sqrt{m_e^2 + \vec{p}^2} \, , \hspace{1cm} E_q := \sqrt{m_p^2 + \vec{q}^2} \, .
\eeq
There are no unstable particles, so the interactions are dominated by the $2 \leftrightarrow 2$ processes
\beq
\ga + e \to \ga + e \, , \hspace{0.5cm} \ga + p \to \ga + p \, , \hspace{0.5cm} e + e \to e + e \, , \hspace{1cm} p + p \to p + p \, , \hspace{0.5cm} e + p \to e + p \, ,
\eeq
and the polarization-dependent BUU equation \eqref{eq:BUUgen} is thus given by
\bea
{\rm L} \bm{f}_{\ga} & = & \bm{C}_{\ga}^e + \bm{C}_{\ga}^p \, , \\
{\rm L} \bm{f}_e & = & \bm{C}_e^{\ga} + \bm{C}_e^e + \bm{C}_e^p \, , \\
{\rm L} \bm{f}_p & = & \bm{C}_p^{\ga} + \bm{C}_p^p + \bm{C}_p^e \, ,
\eea
where
\bea
C_{\ga,rr'}^e & := & \frac{1}{2} \int \frac{\ed^3 k_2}{(2\pi)^3 2 k_2} \frac{\ed^3 p_1}{(2\pi)^3 2 E_{p_1}} \frac{\ed^3 p_2}{(2\pi)^3 2 E_{p_2}} \, (2\pi)^4 \de^{(4)} ( k + p_1 - k_2 - p_2 )   \label{eq:Comptoner} \\
 & & \times \[  f_{s_2 s'_2} f_{r_2 r'_2} f_{s'_1 s_1}^{\circ}  f^{\circ}_{r'' r'}  \cA ( \ga_{r_2}, e_{s_2} \to \ga_r, e_{s_1}) \, \cA^* ( \ga_{r'_2}, e_{s'_2} \to \ga_{r''}, e_{s'_1})  \right. \nn \\
 & & \left. \hspace{0.5cm} - \, f_{s_1 s'_1} f_{r r''} f^{\circ}_{s'_2 s_2} f^{\circ}_{r'_2 r_2} \, \cA ( \ga_{r'}, e_{s_1} \to \ga_{r_2}, e_{s_2} ) \, \cA^* ( \ga_{r''}, e_{s'_1} \to \ga_{r'_2}, e_{s'_2} ) \] \, , \nn \\
\nn \\
C_{e,ss'}^{\ga} & := & \frac{1}{2} \int \frac{\ed^3 p_2}{(2\pi)^3 2 E_{p_2}} \frac{\ed^3 k_1}{(2\pi)^3 2 k_1} \frac{\ed^3 k_2}{(2\pi)^3 2 k_2} \, (2\pi)^4 \de^{(4)} ( p + k_1 - p_2 - k_2 )   \label{eq:Comptons} \\
 & & \times \[ f_{r_2 r'_2} f_{s_2 s'_2} f^{\circ}_{r'_1 r_1} f^{\circ}_{s'' s'} \, \cA ( \ga_{r_2}, e_{s_2} \to \ga_{r_1}, e_{s}) \, \cA^* ( \ga_{r'_2}, e_{s'_2} \to \ga_{r'_1}, e_{s''})  \right. \nn \\
 & & \left. \hspace{0.5cm} - \, f_{r_1 r'_1} f_{s s''} f^{\circ}_{r'_2 r_2} f^{\circ}_{s'_2 s_2}\, \cA ( \ga_{r_1}, e_{s'} \to \ga_{r_2}, e_{s_2} ) \, \cA^* ( \ga_{r'_1}, e_{s''} \to \ga_{r'_2}, e_{s'_2} ) \]  \, ,  \nn \\
\nn \\
C_{e,ss'}^e & := & \frac{1}{4} \int \frac{\ed^3 p_2}{(2\pi)^3 2 E_{p_2}} \frac{\ed^3 p_3}{(2\pi)^3 2 E_{p_3}} \frac{\ed^3 p_4}{(2\pi)^3 2 E_{p_4}} \, (2\pi)^4 \de^{(4)} ( p + p_2 - p_3 - p_4 )   \label{eq:Mollers} \\
 & & \times \[ f_{s_3 s'_3} f_{s_4 s'_4} f^{\circ}_{s'_2 s_2} f^{\circ}_{s'' s'} \, \cA ( e_{s_3}, e_{s_4} \to e_{s}, e_{s_2}) \, \cA^* ( e_{s'_3}, e_{s'_4} \to e_{s''}, e_{s'_2})  \right. \nn \\
 & & \left. \hspace{0.5cm} - \, f_{s_2 s'_2} f_{s s''} f^{\circ}_{s'_3 s_3} f^{\circ}_{s'_4 s_4}\, \cA ( e_{s'}, e_{s_2} \to e_{s_3}, e_{s_4} ) \, \cA^* ( e_{s''}, e_{s'_2} \to e_{s'_3}, e_{s'_4} ) \]   \, , \nn \\
\nn \\
C_{e,ss'}^p & := & \frac{1}{2} \int \frac{\ed^3 p_2}{(2\pi)^3 2 E_{p_2}} \frac{\ed^3 q_1}{(2\pi)^3 2 E_{q_1}} \frac{\ed^3 q_2}{(2\pi)^3 2 E_{q_2}} \, (2\pi)^4 \de^{(4)} ( p + q_1 - p_2 - q_2 )   \label{eq:Motts} \\
 & & \times \[ f_{t_2 t'_2} f_{s_2 s'_2} f^{\circ}_{t'_1 t_1} f^{\circ}_{s'' s'} \, \cA ( e_{s_2}, p_{t_2} \to e_{s}, p_{t_1}) \, \cA^* ( e_{s'_2}, p_{t'_2} \to e_{s''}, p_{t'_1})  \right. \nn \\
 & & \left. \hspace{0.5cm} - \, f_{t_1 t'_1} f_{s s''} f^{\circ}_{t'_2 t_2} f^{\circ}_{s'_2 s_2}\, \cA ( e_{s'}, p_{t_1} \to e_{s_2}, p_{t_2} ) \, \cA^* ( e_{s''}, p_{t'_1} \to e_{s'_2}, p_{t'_2} ) \]  \, , \nn
\eea
while $\bm{C}_{\ga}^p$, $\bm{C}_p^{\ga}$, $\bm{C}_p^p$ and $\bm{C}_p^e$ are the same as $\bm{C}_{\ga}^e$, $\bm{C}_e^{\ga}$, $\bm{C}_e^e$ and $\bm{C}_e^p$, respectively, but with the electrons and protons interchanged. We also remind that
\beq
f^{\circ}_{rr'} := \de_{rr'} + f_{rr'} \, , \hspace{1cm} f^{\circ}_{ss'} := \de_{ss'} - f_{ss'} \, , \hspace{1cm} f^{\circ}_{tt'} := \de_{tt'} - f_{tt'} \, .
\eeq
Invoking the polarization vectors $\ep_r^a \equiv \ep_r^a(\vec{k})$ for photons and the wave-functions $u_s \equiv u_s(\vec{p})$ and $u_t \equiv u_t(\vec{q})$ for electrons and protons, respectively, that were introduced in subsection \ref{sec:wavefunctions}, the involved scattering amplitudes to lowest order read
\beq
\begin{tikzpicture}[line width=1 pt, scale=2]
\node at (-4,0) {$i \cA( \ga_{r_2}, e_{s_2} \to \ga_{r_1}, e_{s_1}) :=$};
\begin{scope}[shift={(-2.5,0)}]
\draw[photon] (0,0.3) -- (0.6,-0.2);
\draw[electron] (0,-0.2) -- (1.8,-0.2);
\draw[photon] (1.2,-0.2) -- (1.8,0.3);
\end{scope}
\node at (-0.35,0) {$+$}; 
\begin{scope}[shift={(0,0)}]
\draw[photon] (0,0.3) -- (1.4,-0.2);
\draw[electron] (0,-0.2) -- (1.8,-0.2);
\draw[photon] (0.4,-0.2) -- (1.8,0.3);
\end{scope}
\end{tikzpicture}  \nn
\eeq
\bea
 & = & \frac{i}{2}\, e^2 \ep_a^{r_1} \ep_b^{r_2} \, \bar{u}_{s_1} \[ \frac{\ga^a \( {\not} p_1 + {\not} k_1 + m_e \) \ga^b}{k_1 \cdot p_1} - \frac{\ga^b \( {\not} p_2 - {\not} k_1 + m_e \) \ga^a}{k_1 \cdot p_2} \] u_{s_2} \nn \\
 & \equiv & \frac{i}{2}\, e^2 \ep_a^{r_1} \ep_b^{r_2} \, \bar{u}_{s_1} \[ \frac{\( \ga^a {\not} k_1 - 2 p_1^a \) \ga^b}{k_1 \cdot p_1} + \frac{\ga^b \( {\not} k_1 \ga^a + 2 p_2^a \)}{k_1 \cdot p_2} \] u_{s_2} \nn \\
 & \equiv & i e^2 \ep_a^{r_1} \ep_b^{r_2} \, \bar{u}_{s_1} A^{ab}(k_1,k_2,p_1,p_2)\, u_{s_2} \, , \label{eq:ACompe} 
\eea
\\
\beq
\begin{tikzpicture}[line width=1 pt, scale=2]
\node at (-4,0) {$i \cA( e_{s_3}, e_{s_4} \to e_{s_1}, e_{s_2}) :=$};
\begin{scope}[shift={(-2.5,0)}]
\draw[electron] (0,0.3) -- (1.8,0.3);
\draw[electron] (0,-0.3) -- (1.8,-0.3);
\draw[photon] (0.9,0.3) -- (0.9,-0.3);
\end{scope}
\node at (-0.35,0) {$+$}; 
\begin{scope}[shift={(0,0)}]
\draw[electron] (0,0.3) -- (1.8,-0.3);
\draw[electron] (0,-0.3) -- (1.8,0.3);
\draw[photon] (0.4,0.17) -- (0.4,-0.17);
\end{scope}
\end{tikzpicture}  \nn
\eeq
\\
\beq
 = - \frac{i}{2}\, e^2 \[  \frac{\bar{u}_{s_1} \ga_a u_{s_3} \bar{u}_{s_2} \ga^a u_{s_4}}{m_e^2 + p_1 \cdot p_3} -  \frac{\bar{u}_{s_1} \ga_a u_{s_4} \bar{u}_{s_2} \ga^a u_{s_3}}{m_e^2 + p_1 \cdot p_4} \] \, , \label{eq:AMolle} 
\eeq
\\
\beq  \label{eq:AMott} 
\begin{tikzpicture}[line width=1 pt, scale=2]
\node at (-4,0) {$i \cA( e_{s_2}, p_{t_2} \to e_{s_1}, p_{t_1}) :=$};
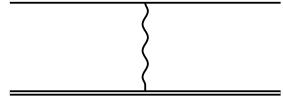
\begin{scope}[shift={(-2.5,0)}]
\draw[electron] (0,0.3) -- (1.8,0.3);
\draw[proton] (0,-0.3) -- (1.8,-0.3);
\draw[photon] (0.9,0.3) -- (0.9,-0.29);
\end{scope}
\end{tikzpicture} \nn
\eeq
\\
\beq
= \frac{i}{2}\, e^2 \frac{\bar{u}_{s_1} \ga_a u_{s_2} \bar{u}_{t_1} \ga^a u_{t_2}}{m_e^2 + p_1 \cdot p_2} \, ,
\eeq
\\
and similarly for the ones where protons and electrons are interchanged. In each case the energy-momentum conservation is understood.\footnote{Note also that the Feynman $i \ep$ regularization in the propagators is irrelevant here because the virtual particles cannot become real (on-shell) in the momentum region of interest.} It is also understood that the involved propagators are the full 2-point functions, so that the masses are the renormalized ones. We can check the Ward identities, i.e. that replacing $\ep^a_{r_1} \to k_1^a$ or $\ep^b_{r_2} \to k_2^b$ or both in Eq. \eqref{eq:ACompe} gives zero. The corresponding amplitude is therefore invariant under the transformations \eqref{eq:epraGT} and is thus consistently independent of the choice of basis $\ep_r^a$. We can then use the Dirac equation \eqref{eq:Diraceq}, the identity \eqref{eq:3gammaiden} and $k_a \ep_r^a \equiv 0$ to simplify the matrix in Eq. \eqref{eq:ACompe} 
\bea 
A^{ab}(k_1,k_2,p_1,p_2) & \to & \( \frac{p_2^a}{k_1 \cdot p_2} - \frac{p_1^a}{k_1 \cdot p_1} \) \ga^b + \frac{1}{2} \( \frac{1}{k_1 \cdot p_1} + \frac{1}{k_1 \cdot p_2} \) \( \et^{ab} {\not} k_1 - \ga^a k_1^b \) \nn \\
 & & - \,\frac{i}{2} \( \frac{1}{k_1 \cdot p_1} - \frac{1}{k_1 \cdot p_2} \) \vep^{ab}_{\,\,\,\,\,\,cd}\, k_1^c \ga^d \ga^5 \, . \label{eq:Aabdef}
\eea
Now once we plug the amplitudes inside the collision terms, we note that all the wave-functions $\ep_r^a$, $u_s$ and $u_t$ consistently contract with the corresponding matrix distributions $\bm{f}_{\ga,e,p}$ to form the Lorentz-indexed one $f_{ab}$ of Eq. \eqref{eq:fabdef} for the photons and the Dirac-indexed ones $f_{e,p}$ of Eq. \eqref{eq:fpmoffpmss} for the fermions, with the Dirac indexes kept again implicit. We can then use Eqs. \eqref{eq:fabpoldec} and \eqref{eq:fpmofIS} to express the result in terms of the desired quantities. As a concrete example, consider the creation term of $C^e_{\ga,ab}(\vec{k}_1) := C^e_{\ga,r_1 r'_1} \ep_a^{r_1} \ep_b^{r'_1}$ 
\bea
& \sim & \ep^a_{r_1} \ep^b_{r'_1} f_{s_2 s'_2} f_{r_2 r'_2} f_{s'_1 s_1}^{\circ}  f^{\circ}_{r''_1 r'_1} \, \cA ( \ga_{r_2}, e_{s_2} \to \ga_{r_1}, e_{s_1}) \, \cA^* ( \ga_{r'_2}, e_{s'_2} \to \ga_{r''_1}, e_{s'_1}) \nn  \\
 & = & \ep^a_{r_1} \ep^b_{r'_1} f_{s_2 s'_2} f_{r_2 r'_2} f_{s'_1 s_1}^{\circ} f^{\circ}_{r''_1 r'_1} \[ \ep_{r_1}^c \ep_{r_2}^d \bar{u}_{s_1} A_{cd}(k_1,k_2,p_1,p_2) \, u_{s_2}  \] \[ \ep_{r''_1}^e \ep_{r'_2}^f \bar{u}_{s'_2} \bar{A}_{ef}(k_1,k_2,p_1,p_2)\, u_{s'_1} \]  \nn \\
 & \equiv & (\ep_{r_1}^a \ep_{r_1}^c) \, ( f_{r_2 r'_2} \ep_{r_2}^d \ep_{r'_2}^f )\, ( f^{\circ}_{r''_1 r'_1} \ep_{r''_1}^e \ep_{r'_1}^b ) \nn \\
 & & \times\, \Tr \[ A_{cd}(k_1,k_2,p_1,p_2) \, ( u_{s_2} f_{s_2 s'_2} \bar{u}_{s'_2} )\, \bar{A}_{ef}(k_1,k_2,p_1,p_2) \, ( u_{s'_1} f_{s'_1 s_1}^{\circ} \bar{u}_{s_1} ) \] \nn \\
 & \equiv & 4 m_e^2 \ep_1^{af} f_2^{cd} f_1^{\circ,eb} \, \Tr \[ A_{fc}(k_1,k_2,p_1,p_2) \, f_{e,2}\, \bar{A}_{ed}(k_1,k_2,p_1,p_2) \, f^{\circ}_{e,1} \] \, . 
\eea
From there on one uses the gamma matrix trace technology, or the product table of the basis $\{ 1, \ga^a, i\ga^{[a} \ga^{b]}, \ga^a \ga^5, i \ga^5 \}$. Finally, following subsection \ref{sec:wavefunctions}, we can express the BUU equations in terms of the Lorentz scalars
\beq
I_{\ga} := f_a^a \, , \hspace{1cm} I_{e,p} := \Tr \, f_{e,p}  \, , \hspace{1cm} V := -i \vep^{ab} f_{ab} \, , 
\eeq
and vectors and tensor
\beq
S_{e,p}^a := \Tr \[ \ga^a \ga^5 f_{e,p} \]  \, , \hspace{1cm} P_{ab} := 2 f_{(ab)} - \ep_{ab} f_c^c \, ,
\eeq
keeping our notation convention, e.g.
\beq
I_{e,n} \equiv I_e(\vec{p}_n) \, , \hspace{1cm} {\rm etc}.
\eeq
We then use the identities Eqs. \eqref{eq:epiden}, \eqref{eq:vepiden} and \eqref{eq:Piden} to simplify the equations. In practice, it is sometimes also useful to use the following expressions
\beq 
\ep^{ab} = \et^{ab} - k^a l^b - l^a k^b \, ,  \hspace{1cm} \vep^{ab} = \vep^{abcd} \, k_c l_d \, , 
\eeq
where $l^a(\vec{k})$ satisfies 
\beq \label{eq:lalgcond}
l_a l^a \equiv 0 \, , \hspace{1cm} l_a k^a \equiv 1 \, , \hspace{1cm} l_a \ep_r^a \equiv 0 \, , \hspace{1cm} l^a P_{ab} \equiv 0 \, , \hspace{1cm} {\rm L} l^a \equiv 0 \, ,
\eeq
and thus completes the set $\{ k^a, \ep_1^a(\vec{k}), \ep_2^a(\vec{k}) \}$ into a normalized ``light-light-space-space" basis. Since the result cannot depend on the choice of $\ep_r^a$ basis, by gauge invariance, it cannot depend on $l^a$ either. We can now state the result for the collision terms\footnote{For these calculations we acknowledge the use of the symbolic tensor computation Mathematica package xAct \cite{xAct}.} 
\beq
{\rm L} I_{\ga} = C_{\ga}^e + C_{\ga}^p \, , \hspace{1cm} {\rm L} V = \ti{C}^e + \ti{C}^p \, , \hspace{1cm}  {\rm L} P_{ab} = C_{ab}^e + C_{ab}^p \, , 
\eeq
for the photons,
\beq
{\rm L} I_e = C^{\ga}_e + C_e^e + C^p_e \, , \hspace{1cm}  {\rm L} S_{e,a} = C_{e,a}^{\ga} + C_{e,a}^e + C_{e,a}^p \, , 
\eeq
for the electrons and
\beq
{\rm L} I_p = C^{\ga}_p + C_p^e + C^p_p \, , \hspace{1cm}  {\rm L} S_{p,a} = C_{p,a}^{\ga} + C_{p,a}^e + C_{p,a}^p \, , 
\eeq
for the protons, where (to linear order in the polarization variables)
\bea
C_{\ga}^e(\vec{k}_1) & := & e^4 \int \frac{\ed^3 k_2}{(2\pi)^3 2 k_2} \frac{\ed^3 p_1}{(2\pi)^3 2 E_{p_1}} \frac{\ed^3 p_2}{(2\pi)^3 2 E_{p_2}} \, (2\pi)^4 \de^{(4)} \( k_1 + p_1 - k_2 - p_2 \)   \label{eq:Cgae} \\
 & & \hspace{1cm} \times \[ \Pi_{\ga e} X  - {\rm I}^{\ga}_{e,2} P_{ab,2} X_2^a X_2^b \] + \Ord({\rm pol}^2) \, , \nn \\
\nn \\
\ti{C}^e(\vec{k}_1) & := & e^4 \int \frac{\ed^3 k_2}{(2\pi)^3 2 k_2} \frac{\ed^3 p_1}{(2\pi)^3 2 E_{p_1}} \frac{\ed^3 p_2}{(2\pi)^3 2 E_{p_2}} \, (2\pi)^4 \de^{(4)} \( k_1 + p_1 - k_2 - p_2 \)   \label{eq:Cte} \\
 & & \hspace{1cm} \times \[  {\rm I}^{\ga}_{e,2} V_2 \ti{X} - {\rm I}^{\ga}_{e,1} V_1 X + \( {\rm I}^e_{\ga,1} S_{e,1,a} - {\rm I}^e_{\ga,2} S_{e,2,a} \) \ti{X}_1^a \] + \Ord({\rm pol}^2) \, , \nn \\
\nn \\
C_{ab}^e(\vec{k}_1) & := & e^4 \[ \int \frac{\ed^3 k_2}{(2\pi)^3 2 k_2} \frac{\ed^3 p_1}{(2\pi)^3 2 E_{p_1}} \frac{\ed^3 p_2}{(2\pi)^3 2 E_{p_2}} \, (2\pi)^4 \de^{(4)} \( k_1 + p_1 - k_2 - p_2 \)  \right. \label{eq:Cabe} \\
 & & \hspace{1cm} \left. \times \( - 2 \Pi_{\ga e} X_{1,a} X_{1,b} + 2 {\rm I}^{\ga}_{e,2} P^{cd}_2 X_{ac} X_{bd} - {\rm I}^{\ga}_{e,1} P^{cd}_1 X_{abcd} \) \]^{\rm TT} + \Ord({\rm pol}^2) \, , \nn \\
\nn \\
C_e^{\ga}(\vec{p}_1) & := & e^4 \int \frac{\ed^3 p_2}{(2\pi)^3 2 E_{p_2}} \frac{\ed^3 k_1}{(2\pi)^3 2 k_1} \frac{\ed^3 k_2}{(2\pi)^3 2 k_2} \, (2\pi)^4 \de^{(4)} \( p_1 + k_1 - p_2 - k_2 \)   \label{eq:Cega} \\
 & &  \hspace{1cm} \times \[ \Pi_{\ga e} X + {\rm I}^{\ga}_{e,1} P_{ab,1} X_1^a X_1^b - {\rm I}^{\ga}_{e,2} P_{ab,2} X_2^a X_2^b  \] + \Ord({\rm pol}^2) \, , \nn \\
\nn \\
C_e^e(\vec{p}_1) & := & e^4 \int \frac{\ed^3 p_2}{(2\pi)^3 2 E_{p_2}} \frac{\ed^3 p_3}{(2\pi)^3 2 E_{p_3}} \frac{\ed^3 p_4}{(2\pi)^3 2 E_{p_4}} \, (2\pi)^4 \de^{(4)} \( p_1 + p_2 - p_3 - p_4 \) \Pi_{ee} Y + \Ord({\rm pol}^2) \, , \nn \\
\label{eq:Cee} \\
\nn \\
C_e^p(\vec{p}_1) & := & e^4 \int \frac{\ed^3 p_2}{(2\pi)^3 2 E_{p_2}} \frac{\ed^3 q_1}{(2\pi)^3 2 q_1} \frac{\ed^3 q_2}{(2\pi)^3 2 q_2} \, (2\pi)^4 \de^{(4)} \( p_1 + q_1 - p_2 - q_2 \) \Pi_{ep} Z + \Ord({\rm pol}^2) \, , \nn \\
 \label{eq:Cep} \\
\nn \\
C_{e,a}^{\ga}(\vec{p}_1) & := & e^4 \int \frac{\ed^3 p_2}{(2\pi)^3 2 E_{p_2}} \frac{\ed^3 k_1}{(2\pi)^3 2 k_1} \frac{\ed^3 k_2}{(2\pi)^3 2 k_2} \, (2\pi)^4 \de^{(4)} \( p_1 + k_1 - p_2 - k_2 \)   \label{eq:Caega} \\
 & & \times \[ {\rm I}_{\ga,2}^e S_{e,2,a} \ti{\ti{X}} - {\rm I}_{\ga,1}^e S_{e,1,a} X - {\rm I}_{\ga,2}^e S_{e,2}^b W_{ab} + {\rm I}_{e,2}^{\ga} V_2 W_a - {\rm I}_{e,1}^{\ga} V_1 [W_a]_{k_1 \leftrightarrow k_2, p_{1,2} \to -p_{1,2}} \] \nn \\
 & & +\, \Ord({\rm pol}^2) \, , \nn \\
\nn \\
C_{e,a}^e(\vec{p}_1) & := & e^4 \int \frac{\ed^3 p_2}{(2\pi)^3 2 E_{p_2}} \frac{\ed^3 p_3}{(2\pi)^3 2 E_{p_3}} \frac{\ed^3 p_4}{(2\pi)^3 2 E_{p_4}} \, (2\pi)^4 \de^{(4)} \( p_1 + p_2 - p_3 - p_4 \)   \label{eq:Caee} \\
 & & \hspace{1cm} \times \[ {\rm I}^e_{e,4} S_{e,3}^b \ti{Y}_{ab} + {\rm I}^e_{e,3} S_{e,4}^b [\ti{Y}_{ab}]_{p_3 \leftrightarrow p_4} - {\rm I}_{e,2}^e S_{e,1,a} Y - {\rm I}^e_{e,1} S_{e,2}^b Y_{ab}  \] + \Ord({\rm pol}^2) \, , \nn \\
\nn \\
C_{e,a}^p(\vec{p}_1) & := & e^4 \int \frac{\ed^3 p_2}{(2\pi)^3 2 E_{p_2}} \frac{\ed^3 q_1}{(2\pi)^3 2 q_1} \frac{\ed^3 q_2}{(2\pi)^3 2 q_2} \, (2\pi)^4 \de^{(4)} \( p_1 + q_1 - p_2 - q_2 \)   \label{eq:Caep} \\
 & & \times \[ \( {\rm I}_{p,2}^e S_{e,2,a} - {\rm I}_{p,1}^e S_{e,1,a} \) Z + \( {\rm I}_{e,2}^p S_{p,2,a} - {\rm I}_{e,1}^p S_{p,1,a}  \) \ti{Z} + {\rm I}^e_{p,2} S_{e,2}^b Z_{ab} - {\rm I}^p_{e,1} S_{p,1}^b \ti{Z}_{ab} \] \nn \\
 & & +\, \Ord({\rm pol}^2) \, , \nn 
\eea
and where we have defined the following dimensionless combinations
\bea
{\rm I}^e_{\ga,1} & := & I_{\ga,1} \( 1 + \frac{1}{2}\, I_{\ga,2} \) - \frac{1}{2}\, I_{e,2} \( I_{\ga,1} - I_{\ga,2} \) \, , \\
{\rm I}^e_{\ga,2} & := & I_{\ga,2} \( 1 + \frac{1}{2}\, I_{\ga,1} \) - \frac{1}{2}\, I_{e,1} \( I_{\ga,2} - I_{\ga,1} \) \, , \\
{\rm I}^{\ga}_{e,1} & := & I_{e,1} \( 1 - \frac{1}{2} \, I_{e,2} \) + \frac{1}{2}\, I_{\ga,2} \( I_{e,1} - I_{e,2} \) \, , \\
{\rm I}^{\ga}_{e,2} & := & I_{e,2} \( 1 - \frac{1}{2} \, I_{e,1} \) + \frac{1}{2}\, I_{\ga,1} \( I_{e,2} - I_{e,1} \) \, , \\
{\rm I}^e_{e,1} & := & I_{e,1} - \frac{1}{2} \( I_{e,1} I_{e,3} + I_{e,1} I_{e,4} - I_{e,3} I_{e,4} \) \, , \\
{\rm I}^e_{e,2} & := & I_{e,2} - \frac{1}{2} \( I_{e,2} I_{e,3} + I_{e,2} I_{e,4} - I_{e,3} I_{e,4} \) \, , \\
{\rm I}^e_{e,3} & := & I_{e,3} - \frac{1}{2} \( I_{e,3} I_{e,1} + I_{e,3} I_{e,2} - I_{e,1} I_{e,2} \) \, , \\
{\rm I}^e_{e,4} & := & I_{e,4} - \frac{1}{2} \( I_{e,4} I_{e,1} + I_{e,4} I_{e,2} - I_{e,1} I_{e,2} \) \, , \\
{\rm I}^p_{e,1} & := & I_{e,1} \( 1 - \frac{1}{2}\, I_{e,2} \) - \frac{1}{2}\, I_{p,2} \( I_{e,1} - I_{e,2} \) \, , \\
{\rm I}^p_{e,2} & := & I_{e,2} \( 1 - \frac{1}{2}\, I_{e,1} \) - \frac{1}{2}\, I_{p,1} \( I_{e,2} - I_{e,1} \) \, , \\
{\rm I}^e_{p,1} & := & I_{p,1} \( 1 - \frac{1}{2}\, I_{p,2} \) - \frac{1}{2}\, I_{e,2} \( I_{p,1} - I_{p,2} \) \, , \\
{\rm I}^e_{p,2} & := & I_{p,2} \( 1 - \frac{1}{2}\, I_{p,1} \) - \frac{1}{2}\, I_{e,1} \( I_{p,2} - I_{p,1} \) \, , \\
\Pi_{\ga e} & := & I_{\ga,2} I_{e,2} - I_{\ga,1} I_{e,1} + \frac{1}{2} \[ I_{e,1} I_{e,2} \( I_{\ga,1} - I_{\ga,2} \) - I_{\ga,1} I_{\ga,2} \( I_{e,1} - I_{e,2} \) \] \, , \\
\Pi_{ee} & := & I_{e,3} I_{e,4} - I_{e,1} I_{e,2} + \frac{1}{2} \[ I_{e,1} I_{e,3} \( I_{e,2} - I_{e,4} \) + I_{e,2} I_{e,4} \( I_{e,1} - I_{e,3} \) \] \, , \\
\Pi_{ep} & := & I_{e,2} I_{p,2} - I_{e,1} I_{p,1} + \frac{1}{2} \[ I_{e,1} I_{e,2} \( I_{p,1} - I_{p,2} \) + I_{p,1} I_{p,2} \( I_{e,1} - I_{e,2} \) \] \, , \\
A & := & \frac{k_1 \cdot p_1}{k_1 \cdot p_2} + \frac{k_1 \cdot p_2}{k_1 \cdot p_1} \, , \\
B & := & 1 - m_e^2 \( \frac{1}{k_1 \cdot p_1} - \frac{1}{k_1 \cdot p_2} \)  \, , \\
X & := & A + B^2 - 1  \, ,  \\
\ti{X} & := & A B  \, ,  \\
\ti{\ti{X}} & := & 1 + B^2 \, , \\
X_1^a & := & m_e \( \frac{p_1^a}{k_1 \cdot p_1} - \frac{p_2^a}{k_1 \cdot p_2} \) \, , \\
X_2^a & := & m_e \( \frac{p_1^a}{k_1 \cdot p_2} - \frac{p_2^a}{k_1 \cdot p_1} \) \, , \\
\ti{X}_1^a & := & \frac{1 - B}{m_e} \[ B k_1^a + k_2^a \] \, , \\
\ti{X}_2^a & := & \frac{1 - B}{m_e} \[ B k_2^a + k_1^a \] \, , \\
\ti{\ti{X}}^a & := & \frac{1 - B}{m_e}\, p_1^a + \frac{m_e}{k_1 \cdot p_1}\, k_1^a \, , \\
X^{ab} & := & \et^{ab} - \frac{p_1^a p_2^b}{k_1 \cdot p_1} + \frac{p_2^a p_1^b}{k_1 \cdot p_2} \, , \\
X^{abcd} & := & A \et^{ac} \et^{bd} - \et^{ac} X_1^b X_1^d - \et^{bc} X_1^a X_1^d \, , \\
Y & := & \( 1 + \frac{t}{u} + \frac{u}{t} \)^2 + 4 m_e^2 \( \frac{t}{u^2} + \frac{u}{t^2} \) + 4 m_e^4 \( \frac{1}{t^2} + \frac{1}{u^2} - \frac{1}{t u} \) \, , \\
Y^{ab} & := & \et^{ab} + 2 \( \frac{p_3^a p_4^b}{t} + \frac{p_4^a p_3^b}{u} \) + 2 m_e^2 \( \frac{p_1^a - p_3^a}{t} + \frac{p_1^a - p_4^a}{u} \) \( \frac{p_4^b}{t} + \frac{p_3^b}{u} \) \nn \\
 & & -\, 4 \( m_e^2 \et^{ab} + p_1^a p_1^b \) \( \frac{1}{t} + \frac{1}{u} + \frac{m_e^2}{tu} \)  \, , \\ 
\ti{Y}^{ab} & := & \[ \( 1 + \frac{u}{t} \)^2 + \frac{4m_e^2}{t} \( \frac{u}{t} - \frac{t}{u} \) + \frac{4m_e^4}{t} \( \frac{1}{t} - \frac{1}{u} \) \] \et^{ab} + \frac{2u}{t^2} \( p_3^a p_4^b - 2 p_1^{[a} p_4^{b]} \) \nn \\
 & & - \, \frac{2}{t} \[ 2 \( p_1^a - p_4^a \) p_1^b + \( p_1^a - p_3^a + p_4^a \) p_4^b \] + \frac{2}{u} \[ p_4^a \( p_1^b - p_4^b \) - 2 p_1^a p_1^b \] \nn \\
 & & +\, 2m_e^2 \[ \frac{1}{t^2} \[ \( p_1^a - p_3^a \) p_1^b - 4 p_1^{[a} p_4^{b]} + 2 p_3^a p_4^b \] + \frac{1}{u^2} \( p_1^a - p_4^a \) \( p_1^b - p_4^b \)  \right. \nn \\
 & & \left. \hspace{1.4cm} -\, \frac{1}{tu} \[ 2 p_1^a p_1^b + p_3^a \( p_1^b - p_4^b \) + 2 p_1^{[a} p_4^{b]} \] \] \, , \\
Z & := & 1 + \frac{m_p^2 + m_e^2 + 2 p_1 \cdot q_2}{m_e^2 + p_1 \cdot p_2} + \frac{2 (p_1 \cdot q_2)^2}{\( m_e^2 + p_1 \cdot p_2 \)^2} \, , \\
\ti{Z} & := & \frac{2 m_p m_e}{m_e^2 + p_1 \cdot p_2} \, , \\
Z^{ab} & := & \frac{p_1^a p_1^b + 2 q_2^a q_1^b}{m_e^2 + p_1 \cdot p_2} + \frac{m_p^2 \( p_1^a - p_2^b \) p_1^b + 2 \[ \( p_1^a - p_2^a \) q_2^b - q_2^a p_1^b \] p_1 \cdot p_2}{\( m_e^2 + p_1 \cdot p_2 \)^2} \, , \\
\ti{Z}^{ab} & := & \frac{m_p p_1^a q_2^b}{m_e \( m_e^2 + p_1 \cdot p_2 \)} + \frac{m_p m_e \( p_1^a - p_2^a \) q_2^b}{\( m_e^2 + p_1 \cdot p_2 \)^2} \, , \\
W^a & := & m_e^3 \( \frac{1}{k_1 \cdot p_1} - \frac{1}{k_1 \cdot p_2} \)^2 k_2^a - m_e \( \frac{1}{k_1 \cdot p_1} - \frac{1}{k_1 \cdot p_2} \) \( p_2^a + 2 k_1^a \) \\ 
 & & + \[ \frac{m_e}{k_1 \cdot p_1} \( \frac{k_1 \cdot p_2}{k_1 \cdot p_1} - 1 \) + \frac{1}{m_e} \( \frac{k_1 \cdot p_1}{k_1 \cdot p_2} - \frac{k_1 \cdot p_2}{k_1 \cdot p_1} \) \] p_1^a \, , \\
W^{ab} & := & \frac{1 - B}{m_e^2} \[ k_1^a k_1^b + B p_1^a p_1^b \] + \( 1 + B \) \[ \frac{k_1^a p_1^b}{k_1 \cdot p_1} - \frac{\( p_1^a - p_2^a \) k_1^b}{k_1 \cdot p_2} \]  \nn \, ,
\eea
we have made use of the Mandelstam variables
\beq
t := \( p_1 - p_3 \)^2 \equiv -2 \( m_e^2 + p_1 \cdot p_3 \) \, , \hspace{1cm} u := \( p_1 - p_4 \)^2 \equiv -2 \( m_e^2 + p_1 \cdot p_4 \)  \, , 
\eeq
in the $e + e \to e + e$ case and again the remaining terms can be obtained by interchanging electrons and protons. The ``TT" superscript in Eq. \eqref{eq:Cabe} stands for ``transverse-traceless part" 
\beq
X^{\rm TT}_{ab} := X_{ab} - \frac{1}{2}\, \ep_{ab} X_c^c \, ,
\eeq
so the right-hand side expression is consistently transverse with respect to $k_1^a$. Similarly, we also check that the integrands of \eqref{eq:Caega}, \eqref{eq:Caee} and \eqref{eq:Caep} are consistently transverse with respect to $p_1^a$. As mentioned at the end of subsection \ref{sec:collision}, the integrands of the collision matrices are not necessarily hermitian, only their integrals are, which for the real variables employed here translates into imaginary contributions. For instance, $\ti{C}^e$ contains an imaginary term of the form
\beq
\ti{C}^e \supset i \ti{P}_{ab,1} \int \frac{\ed^3 k_2}{(2\pi)^3 2 k_2} \frac{\ed^3 p_1}{(2\pi)^3 2 E_{p_1}} \frac{\ed^3 p_2}{(2\pi)^3 2 E_{p_2}} \, K^{ab}(\vec{k}_1,\vec{k}_2, \vec{p}_1, \vec{p}_2) \, .
\eeq 
To see that this is zero, we note that, whatever the $K^{ab}$ function, the result of the integral must be a tensor distribution by Lorentz invariance. Consequently, given the available dependencies, the integral can only be of the form
\beq
\int \frac{\ed^3 k_2}{(2\pi)^3 2 k_2} \frac{\ed^3 p_1}{(2\pi)^3 2 E_{p_1}} \frac{\ed^3 p_2}{(2\pi)^3 2 E_{p_2}} \, K^{ab}(\vec{k}_1, \vec{k}_2, \vec{p}_1, \vec{p}_2) = f_1(k_1)\, \et^{ab} + f_2(k_1)\, k_1^a k_1^b + f_3(k_1)\, \ep_1^{ab} \, ,
\eeq
for some ``form factors" $f_{1,2,3}$, which depend on the precise $K^{ab}$ function. Contracting this quantity with $\ti{P}_{ab,1}$ then gives trivially zero. Such simplifications occur with the spin polarizations $S_{e,p}^a$ as well and in real contributions to the collision terms too.

To conclude this example, observe that, as anticipated in section \ref{sec:advtet}, the matrix elements entering the above collision terms, e.g. $X, X_{1,2}^a, X^{ab}$, etc., are explicitly independent of $x^{\mu}$, i.e. they solely depend on the momenta $k^a$, $p^a$ and $q^a$. Importantly, this is true for all observer families $e_a^{\mu}(x)$, because the $S$-matrix is Lorentz invariant, so the $x^{\mu}$-dependence of LLTs is irrelevant. Note that the amplitudes \eqref{eq:ACompe}, \eqref{eq:AMolle} and \eqref{eq:AMott} do depend on $x^{\mu}$, but only because of the wave-functions $\ep_r^a$, $u_s$ and $u_t$. The latter must depend on $x^{\mu}$, because they must transform as vector distributions under LLTs for our equations to be covariant. Therefore, by expressing the Boltzmann matrix distribution $\bm{f}$ through Lorentz tensor distributions using the wave-functions, we precisely cancel out all occurrences of the latter in the BUU equation and the resulting tensor amplitudes are then explicitly $x^{\mu}$-independent. Remember also that for this last step to be possible we need the wave-functions to commute with L, so they must be Liouville-transported.

\subsection{Cosmic microwave background observables} \label{sec:CMB}

We can now consider the observables corresponding to the CMB photons. The quantity of interest is the photon distribution matrix at the observer position
\beq
\hat{f}_{ab}(\vec{k}) := f_{ab}(\hat{x}, \vec{k}) \, ,
\eeq
in which case $\vec{k}$ and the $ab$ indices are with respect to the actual observer frame $\hat{e}_a$, by construction. In practice, however, what one really measures is not directly $\hat{f}_{ab}$, but rather the intensity matrix distribution at the observer position $\hat{I}_{ab,cd}(\vec{k})$, which is defined as follows. First one considers the complexified (microscopic) electromagnetic field operator
\beq 
A_a(X) = \sum_{r = 1,2} \int \frac{\ed^3 k}{(2\pi)^3 \sqrt{2k}}\, a_{\vec{k},r}\, \ep^r_a(\vec{k}) \, e^{i k_b X^b} \, ,
\eeq
i.e. the operator given in Eq. \eqref{eq:microAaX} before taking the real part. The implicit definition of the (covariant) intensity matrix at $\hat{P}$ being
\beq \label{eq:Iabcddef}
\bra : F_{ab}(X)\, F^{\dagger}_{cd}(X) : \ket_{\ro(\hat{x})} \equiv \int \frac{\ed^3 k}{(2\pi)^3 2k} \, \hat{I}_{ab,cd}(\vec{k}) \, ,
\eeq 
where $F_{ab}(X)$ is the field strength \eqref{eq:FabX}, we find
\beq
\hat{I}_{ab,cd}(\vec{k}) \equiv k_a k_c \hat{f}_{bd}(\vec{k}) - k_a k_d \hat{f}_{bc}(\vec{k}) - k_b k_c \hat{f}_{ad}(\vec{k}) + k_b k_d \hat{f}_{ac}(\vec{k})  \, ,
\eeq
where the 4-momenta here are null $k_a k^a \equiv 0$. Let us next observe that, for a phase space distribution evaluated at the observer position $\hat{P}$, the $\vec{k}$ parameters play exactly the same role as $\hat{\om}$ and $\hat{n}^i$ in the case of the geodesic associated with localized sources, i.e. they parametrize the subspace of $T_{\hat{P}} \cM$ with respect to the spatial frame $\hat{e}_i$. We can therefore decompose the above $\vec{k}$-dependencies into the norm $\hat{\om} := k$ and direction $\hat{n} := \vec{k}/k$ and we subsequently express the latter using the standard angular parameterization of Eq. \eqref{eq:tevphobs}. More generally, we can make use of the geometrical machinery developed in section \ref{sec:geo} to describe the observer sky. First, note that what is observed in practice is the two-point function of the complexified electric field operator tangent to the observer sky 
\beq \label{eq:IABdef}
\hat{I}_{AB}(\vte, \hat{\om}) := \hat{k}^a_A \hat{k}^b_B \hat{I}_{0a,0b}(\vec{k}) \equiv \hat{\om}^2 \hat{k}^a_A \hat{k}^b_B \hat{f}_{ab}(\vec{k}) \equiv \hat{\om}^2 \hat{n}^i_A(\vte)\, \hat{n}^j_B(\vte)\, \hat{f}_{ij}(\vec{k}) \, , 
\eeq 
where we have used Eqs. \eqref{eq:kAadecomp}, \eqref{eq:Sshiftgf} and \eqref{eq:fabalgprop}. Thus, we basically observe the projection of the photon distribution $f_{ab}$ on the observed sky, up to a $\hat{\om}^2$ factor. From the third expression in the series of equalities \eqref{eq:IABdef}, we infer that the transformation rule under LLTs of this hermitian $2 \times 2$ matrix is 
\beq
\ti{\hat{I}}_{AB}(\ti{\vte}, \ti{\hat{\om}}) = \hat{\La}_{\para}^2(\vte)\, R_A^{\,\,\,C}(\vte)\, R_B^{\,\,\,D}(\vte) \, \hat{I}_{CD}(\vte, \hat{\om}) \, ,
\eeq
with $R^{AB}$ being the LLT-compensating Sachs rotations (see again section \ref{sec:geo}) and the $\hat{\La}_{\para}^2$ factor being due to the presence of the $\hat{\om}^2$ one in Eq. \eqref{eq:IABdef}. These are now clearly the components of a tensor field on the spectral observer sky $\Ss_{\rm spec}$ in the Sachs dyad basis. Expressing $f_{ab}$ in terms of $I$, $V$ and $P_{ab}$ through Eq. \eqref{eq:fabpoldec}, we then find
\beq
\hat{I}_{AB}(\vte,\hat{\om}) \equiv \frac{1}{2}\, \hat{\om}^2 \[ \de_{AB} \hat{I} + i \vep_{AB} \hat{V} + \hat{P}_{AB}  \](\vte,\hat{\om}) \, ,
\eeq
where
\beq \label{eq:PABdef}
\hat{P}_{AB}(\vte,\hat{\om}) := \hat{k}^a_A(\vte)\, \hat{k}^b_B(\vte)\, \hat{P}_{ab}(\vec{k}) \equiv \hat{n}^i_A(\vte)\, \hat{n}^i_B(\vte)\, \hat{P}_{ij}(\vec{k}) \, , \hspace{1cm} \hat{P}_{AA} \equiv 0 \, , \hspace{1cm} \hat{P}_{AB} \equiv \hat{P}_{BA} \, .
\eeq
Thus, the dimensionless CMB observables, that are the intensity spectrum map $\hat{I}(\vte,\hat{\om})$, circular polarization spectrum map $\hat{V}(\vte,\hat{\om})$ and linear polarization spectrum map $\hat{P}_{ab}(\vte,\hat{\om})$, are simply related to the phase space fields $I(x,\vec{k})$, $V(x,\vec{k})$ and $P_{ab}(x,\vec{k})$ by evaluation at the observer position $\hat{P}$ and projection on the observer sky for the $P_{ab}$. The dimensionful (spectral radiance) observables are then simply obtained by multiplying by $\hat{\om}^3$. The advantage of the dimensionless ones is that they transform tensorially on $\Ss_{\rm spec}$ under LLTs
\bea
\ti{\hat{I}}(\ti{\vte}, \ti{\hat{\om}}) & = & \hat{I}(\vte, \hat{\om}) \, , \label{eq:Ihtrans} \\
\ti{\hat{V}}(\ti{\vte}, \ti{\hat{\om}}) & = & \hat{V}(\vte, \hat{\om}) \, , \label{eq:Vhtrans} \\
\ti{\hat{P}}^{AB}(\ti{\vte}, \ti{\hat{\om}}) & = & R^A_{\,\,\,C}(\vte)\, R^B_{\,\,\,D}(\vte) \, \hat{P}^{CD}(\vte, \hat{\om}) \, , \label{eq:PhABtrans}
\eea
while the dimensionful ones have an extra $\hat{\La}^3_{\para}(\vte)$ factor. One can also express $\hat{P}_{AB}$ in terms of the Stokes parameters
\beq
\hat{P}^{AB} \equiv \hat{Q} \si_+^{AB} + \hat{U} \si_{\times}^{AB} \, , \hspace{1cm} \hat{Q} := \frac{1}{2}\, \si_+^{AB} \hat{P}_{AB} \, , \hspace{1cm} \hat{U} := \frac{1}{2}\, \si_{\times}^{AB} \hat{P}_{AB} \, ,
\eeq
where the $\bm{\si}_{+,\times}$ matrices have been defined in Eq. \eqref{eq:Paulipcdef}. In terms of the complex combination
\beq
\hat{P} := \hat{Q} + i \hat{U} \, ,
\eeq
the transformation \eqref{eq:PhABtrans} reads
\beq \label{eq:Phtrans}
\ti{\hat{P}}(\ti{\vte}, \ti{\hat{\om}}) = e^{2i\al(\vte)} \hat{P}(\vte, \hat{\om}) \, ,
\eeq 
where $\al$ is the angle in LLT-compensating local rotation $R^A_{\,\,\,B}(\vte)$ (see Eq. \eqref{eq:locrotS}). Finally, it is also conventional to express $\hat{I}$ in terms of some effective ``temperature" distribution $\hat{T}$ through the photon Bose-Einstein distribution
\beq \label{eq:IofT}
\hat{I}(\vte,\hat{\om}) \equiv \frac{2}{\exp \[ \frac{\hat{\om}}{\hat{T}(\vte,\hat{\om})} \] - 1} \, ,
\eeq
even though temperature is a macroscopic variable which therefore cannot depend on the microscopic momenta $\vec{k}$. 

We next decompose these fields in the basis of (spin-weighted) spherical harmonics (see for instance \cite{Durrer:2008eom} for a description). In the presence of tensors, such as $\hat{P}^{AB}$, this construction involves the covariant derivative on $\Ss$ with respect to the local rotations appearing in \eqref{eq:PhABtrans}, so we invoke the spin connection $w^{AB}_{\hat{A}}(\vte)$ on that space, i.e. the one for which the dyad $S_{\hat{A}}^A(\vte)$ is torsion-free. Given the dimensionality of $\Ss$, we can write
\beq
w^{AB}_{\hat{A}} \equiv \vep^{AB} w_{\hat{A}} \, ,
\eeq
so the covariant derivative reads
\beq
\na_{\hat{A}} \hat{X}^A := \pa_{\hat{A}} \hat{X}^A + w_{\hat{A}} \vep^A_{\,\,\,C} \hat{X}^C  \, ,
\eeq
and we will actually use the Sachs-indexed one $\na_A := S_A^{\hat{A}} \na_{\hat{A}}$. With Eq. \eqref{eq:dyad} and the 2-dimensional analogue of Eq. \eqref{eq:LVsc}, we then find
\beq
w_{\hat{A}} = - \de_{\hat{A}}^{\vph} \cos \vte \, , \hspace{1cm} w_A := S_A^{\hat{A}} w_{\hat{A}} = - \de_A^2 \cot \vte \, .
\eeq
Pay attention to the fact that, under a generic LLT at the observer, the frequency and angles mix, leading in particular to the rule \eqref{eq:paC3LLT} and thus
\beq \label{eq:naALLT3}
\ti{\na}_A = R_A^{\,\,\,B} \[ \hat{\La}_{\para} \na_B + \hat{\La}^0_{\,\,\,i} \hat{n}_B^i \hat{\om} \pa_{\om} \] \, ,
\eeq
where we have used Eqs. \eqref{eq:sbAAtrans} and \eqref{eq:sAOidef}. Therefore, this a covariant derivative only for fields on $\Ss$, i.e. with no $\hat{\om}$ dependence, or only under purely rotational LLTs for fields on $\Ss_{\rm spec}$. In what follows we will work with a real fully symmetric traceless tensor of arbitrary rank $\hat{T}_{A_1 \dots A_s}(\ti{\vte}, \ti{\hat{\om}})$, thus transforming as the generalization of \eqref{eq:PhABtrans} under LLTs
\beq  \label{eq:TA1Astrans}
\ti{\hat{T}}^{A_1 \dots A_s}(\ti{\vte}, \ti{\hat{\om}}) = R^{A_1}_{\,\,\,\,\,\,B_1}(\vte) \dots R^{A_s}_{\,\,\,\,\,\,B_s}(\vte) \, \hat{T}^{B_1 \dots B_s}(\vte, \hat{\om}) \, .
\eeq
This way we will be treating the three cases $\hat{I}$, $\hat{V}$ and $\hat{P}^{AB}$ simultaneously. In two dimensions such a tensor has only two independent components and these can be expressed in terms of two scalars under local rotations: the ``electric" and ``magnetic" fields\footnote{This expression can be obtained straightforwardly by performing a harmonic decomposition $\hat{T}_{A_1 \dots A_s} \equiv \sum_{k=0}^s \na_{\bra A_1 \dots A_k} h_{A_{k+1} \dots A_s \ket}$, where all the $h_{A_1 \dots A_k}$ are totally symmetric, traceless and transverse. In two dimensions these conditions are more than the number of independent components for $h_{A_1 \dots A_{k>1}}$, so these fields are zero, while the condition $\na_{A} h^{A} \equiv 0$ implies $h^{A} \equiv \ep^{AB} \na_{B} \ti{h}$, thus yielding the form \eqref{eq:ThdecompEhBh}.}
\beq \label{eq:ThdecompEhBh}
\hat{T}_{A_1 \dots A_s} \equiv \na_{A_1 \dots A_s} \hat{E} + \vep_{(A_1}^{\,\,\,\,\,\,B} \na_{A_2\dots A_s)B} \hat{B} \, , \hspace{1cm} \na_{A_1 \dots A_s} := \na_{\bra A_1} \dots \na_{A_s \ket}  \, ,
\eeq
where $\bra \dots \ket$ denotes full symmetrization and removal of traces. In terms of the complex operator
\beq
D_A := \[ \de_A^B + i \vep_A^{\,\,\,B} \] \na_B \, , 
\eeq
which satisfies the convenient identities
\beq
D_A D^A \equiv 0 \, , \hspace{1cm} \[ D_A, D_B \] \equiv 0 \, ,
\eeq
we obtain a much simpler result
\beq
\hat{T}_{A_1 \dots A_s} \equiv \hat{T}_{A_1 \dots A_s}^{\Cs} + {\rm c.c.} \, , \hspace{1cm} \hat{T}_{A_1 \dots A_s}^{\Cs} := D_{A_1} \dots D_{A_s} \( \hat{E} + i \hat{B} \)  \, ,
\eeq
as one can show by induction. The full information of the tensor can now be stored in the first component of the complexified field
\beq \label{eq:ThofEhBh}
\hat{T} := \hat{T}^{\Cs}_{1 \dots 1} \equiv D_1^s \( \hat{E} + i \hat{B} \) \, .
\eeq
Under LLTs at the observer, Eqs. \eqref{eq:TA1Astrans} and \eqref{eq:ThofEhBh} lead to the generalization of Eq. \eqref{eq:Phtrans}, i.e. the helicity $s$ representation of the local rotations on $\Ss$ 
\beq \label{eq:Thtrans}
\ti{\hat{T}}(\ti{\vte}, \ti{\hat{\om}}) = e^{i s\al(\vte)} \hat{T}(\vte, \hat{\om}) \, .
\eeq 
In the case of local boosts, however, the non-trivial transformation of the derivative \eqref{eq:naALLT3} and the simple result \eqref{eq:Thtrans}, imply that the electric and magnetic fields must mix and their transformation is non-local on $\Ss_{\rm spec}$ (inverse derivative operators). 

Now by decomposing $\hat{E}$ and $\hat{B}$ in the basis of spherical harmonics $Y_{lm}(\vte)$ in Eq. \eqref{eq:ThofEhBh}, we obtain a decomposition of $\hat{T}$ in the basis $\sim D_1^s Y_{lm}$, which are the spin-weighted spherical harmonics, i.e. the basis for helicity $s$ representations. The orthonormal elements are
\beq \label{eq:sYlmdef}
{}_s Y_{lm}(\vte) := (-1)^s \sqrt{\frac{(l-s)!}{(l+s)!}} \times \left\{ \begin{array}{cc} D_1^s Y_{lm}(\vte) & {\rm if} \,\,\, s \geq 0 \\ \bar{D}_1^{-s} Y_{lm}(\vte) & {\rm if} \,\,\, s \leq 0 \end{array} \right.  \, ,
\eeq
and in this context $D_1$ and $\bar{D}_1$ are the ``raising" and ``lowering" operators, respectively. In terms of the harmonic components
\bea
\hat{T}_{lm}(\hat{\om}) & := & \int \ed \Om \, {}_s Y^*_{lm}(\vte)\, \hat{T}(\vte, \hat{\om}) \, , \\
\hat{E}_{lm}(\hat{\om}) & := & \int \ed \Om \, Y^*_{lm}(\vte)\, \hat{E}(\vte, \hat{\om}) \, , \\
\hat{B}_{lm}(\hat{\om}) & := & \int \ed \Om \, Y^*_{lm}(\vte)\, \hat{B}(\vte, \hat{\om}) \, ,
\eea
Eq. \eqref{eq:ThofEhBh} reads 
\beq
\hat{T}_{lm}(\hat{\om}) \equiv \sqrt{\frac{(l-s)!}{(l+s)!}} \[ \hat{E}_{lm}(\hat{\om}) + i \hat{B}_{lm}(\hat{\om}) \] \, .
\eeq
Using the fact that $\hat{E}(\vte, \hat{\om})$ and $\hat{B}(\vte, \hat{\om})$ are real
\beq
\bar{\hat{E}}_{lm}(\hat{\om}) \equiv (-1)^m \hat{E}_{l,-m}(\hat{\om}) \, , \hspace{1cm} \bar{\hat{B}}_{lm}(\hat{\om}) \equiv (-1)^m \hat{B}_{l,-m}(\hat{\om}) \, ,
\eeq
we can now easily obtain them out of $\hat{T}_{lm}(\hat{\om})$
\bea
\hat{E}_{lm}(\hat{\om}) & \equiv & \frac{1}{2} \sqrt{\frac{(l+s)!}{(l-s)!}} \[ \hat{T}_{lm} + (-1)^m \bar{\hat{T}}_{l,-m} \] \, , \\
\hat{B}_{lm}(\hat{\om}) & \equiv & \frac{1}{2i} \sqrt{\frac{(l+s)!}{(l-s)!}} \[ \hat{T}_{lm} - (-1)^m \bar{\hat{T}}_{l,-m} \] \, .
\eea
Let us finally compute the transformation of $\hat{T}_{lm}$ under LLTs at the observer for completeness and compare with the results of \cite{Challinor:2002zh} as a consistency check. Under a local rotation at the observer, we have the relation
\beq \label{eq:sYlmrot}
{}_s Y_{lm}(\ti{\vte}) = e^{i s \al(\vte)} \sum_{m'=-l}^l \bar{D}^l_{mm'}(\hat{\te})\, {}_s Y_{lm'}(\vte) \, ,
\eeq
where $\bm{D}^l$ is the Wigner matrix associated to the Euler angles in $\vep_{ijk} \hat{\te}^{jk}$, yielding with Eq. \eqref{eq:Thtrans} the transformation rule
\bea
\ti{\hat{T}}_{lm}(\hat{\om}) & := & \int \ed \ti{\Om} \, {}_s Y^*_{lm}(\ti{\vte})\, \ti{\hat{T}}(\ti{\vte}, \hat{\om}) = \sum_{m'=-l}^l \bar{D}^l_{mm'}(\hat{\te}) \int \ed \Om\, {}_s Y^*_{lm'}(\vte)\, \hat{T}(\vte, \hat{\om}) \nn \\
 & \equiv & \sum_{m'=-l}^l \bar{D}^l_{mm'}(\hat{\te})\, \hat{T}_{lm'}(\hat{\om}) \, .
\eea
Under local boosts, however, the transformation is more complicated, even for the scalar quantities $\hat{I}_{lm}(\hat{\om})$ and $\hat{V}_{lm}(\hat{\om})$. First, we express the map \eqref{eq:Thtrans} in an active form, i.e. in terms of a unique $\hat{\om}$ parametrization
\beq \label{eq:Thtrans2}
\ti{\hat{T}}(\ti{\vte}, \hat{\om}) = e^{i s\al(\vte)} \hat{T}(\vte, \hat{\La}^{-1}_{\para}(\vte)\,\hat{\om})  \, ,
\eeq
because the two frequencies $\hat{\om}$ and $\ti{\hat{\om}}$ are related by an angle-dependent factor and here we are in harmonic space. Since we can rotate at will using Wigner matrices, without loss of generality, we can focus on the case where the boost is along the $i=3$ direction $\hat{\te}^{0i} = \de^i_3 \et$. Moreover, we will work at linear order in $\et$ for simplicity. The LLT-induced coordinate transformation given in Eqs. \eqref{eq:SOLLT} and \eqref{eq:hatomLLT} becomes
\beq
\ti{\vte} = \vte - \et \sin \vte + \Ord(\et^2) \, , \hspace{1cm}  \ti{\vph} = \vph \, ,  \hspace{1cm} \ti{\hat{\om}} = \( 1 + \et \cos \vte \) \hat{\om} + \Ord(\et^2) \, .
\eeq
In particular, with this choice of direction we have a diagonal Jacobian matrix, meaning that there is no compensating local rotation in the transformation of the dyad \eqref{eq:sbAAtrans}, i.e. $\al(\vte) = 0$ in \eqref{eq:Thtrans2}. Thus, expanding to linear order in $\et$ we find
\beq
\ti{\hat{T}}(\ti{\vte}, \hat{\om}) = \[ 1 - \et \cos \vte\, \hat{\om} \pa_{\hat{\om}} + \Ord(\et^2) \] \hat{T}(\vte,\hat{\om}) \, , 
\eeq 
and
\beq \label{eq:dOmYlmtrans}
\ed \ti{\Om} \, {}_s Y^*_{lm}(\ti{\vte}) = \ed \Om \[ 1 - \et \( 2 \cos \vte + \sin \vte \, \pa_{\vte} \) + \Ord(\et^2) \] {}_s Y^*_{lm}(\vte) \, .
\eeq
Using the following identities \cite{Varshalovich:1988ye}
\bea
\sin \vte\, \pa_{\vte}\, {}_s Y_{lm} & \equiv & l \, {}_s C_{l+1,m} \, {}_s Y_{l+1,m} + \frac{sm}{l(l+1)} \, {}_s Y_{lm} - \( l + 1 \) {}_s C_{lm} \, {}_s Y_{l-1,m} \, ,  \label{eq:Ylmid} \\
\cos \vte\, {}_s Y_{lm} & \equiv & {}_s C_{l+1,m} \, {}_s Y_{l+1,m} - \frac{sm}{l(l+1)} \, {}_s Y_{lm} + {}_s C_{lm} \, {}_s Y_{l-1,m} \, ,  \nn
\eea
where
\beq
{}_s C_{lm} := \sqrt{\frac{\( l^2 - m^2 \) \( l^2 - s^2 \)}{l^2 \( 4 l^2 - 1 \)}} \, ,
\eeq
we then obtain the variation
\bea
\de_{\et} T_{lm}(\hat{\om}) & := & \ti{\hat{T}}_{lm}(\hat{\om}) -  T_{lm}(\hat{\om}) \nn \\
 & \equiv & \int \ed \ti{\Om} \, {}_s Y^*_{lm}(\ti{\vte})\, \ti{\hat{T}}(\ti{\vte},\hat{\om}) - \int \ed \Om \, {}_s Y^*_{lm}(\vte)\, \hat{T}(\vte,\hat{\om}) \\
 & = & - \et \int \ed \Om \[ {}_s C_{l+1,m} \, {}_s Y^*_{l+1,m}(\vte) \( \hat{\om} \pa_{\hat{\om}} + l + 2 \) - \frac{sm}{l(l+1)}\, {}_s Y^*_{lm}(\vte) \( \hat{\om} \pa_{\hat{\om}} + 1 \) \right. \nn \\
 & & \hspace{2cm} \left. +\,  {}_s C_{lm} \, {}_s Y^*_{l-1,m}(\vte) \( \hat{\om} \pa_{\hat{\om}} - l + 1 \)  \] \hat{T}(\vte, \hat{\om})  + \Ord(\et^2) \nn \\
 & \equiv & - \et \[ {}_s C_{l+1,m} \( \hat{\om} \pa_{\hat{\om}} + l + 2 \) \hat{T}_{l+1,m}(\hat{\om}) - \frac{sm}{l(l+1)} \( \hat{\om} \pa_{\hat{\om}} + 1 \) \hat{T}_{lm}(\hat{\om}) \right. \nn \\
 & & \hspace{1cm}  \left. +\,  {}_s C_{lm} \( \hat{\om} \pa_{\hat{\om}} - l + 1 \) \hat{T}_{l-1,m}(\hat{\om}) \] + \Ord(\et^2) \, . \nn
\eea
Finally, it is straightforward to compute the variation of the total brightness 
\beq
\hat{T}_{lm} := \int_0^{\infty} \ed \hat{\om} \, \hat{\om}^3 \hat{T}_{lm}(\hat{\om}) \, , 
\eeq
by integrating by parts the $\sim \pa_{\hat{\om}}$ terms
\beq
\de_{\et} \hat{T}_{lm} = - \et \[ {}_2 C_{l+1,m} \( l - 2 \) \hat{T}_{l+1,m} + \frac{3sm}{l(l+1)}\, \hat{T}_{lm} - {}_2 C_{lm} \( l + 3 \) \hat{T}_{l-1,m} \] + \Ord(\et^2)  \, ,
\eeq
which is in agreement with \cite{Challinor:2002zh}.

\clearpage

\acknowledgments

We are grateful to Enea Di Dio, Ruth Durrer, Giuseppe Fanizza, Pierre Fleury and Fulvio Scaccabarozzi for useful discussions. This work is supported by a Consolidator Grant of the European Research Council (ERC-2015-CoG grant 680886).

\appendix

\section{Observer frames are orthonormal}  \label{app:obsfrortho}

Consider an observer with 4-velocity $\hat{u}$ performing measurements in an infinitesimal neighborhood $\hat{\cal I}$ of her position $\hat{P}$. She uses a local coordinate system $x^{\mu}$ to parametrize $\hat{\cal I}$ and therefore the basis $\pa_{\mu}$ to decompose tensors at $\hat{P}$. In particular, she parametrizes evolution with her proper time, meaning that the norm of the time-coordinate intervals $\ed t \equiv \ed x^0$ is unity
\beq \label{eq:proptime}
\hat{g}^{-1} \( \ed t, \ed t \) \equiv \hat{g}^{tt} = - 1 \, .
\eeq
Secondly, she parametrizes the measuring apparatus with spatial coordinates $x^{\al}$, where $\al \in \{ x,y,z \}$, and associates a physical distance to some interval $\ed x^{\al}$ using the Euclidean scalar product
\beq \label{eq:Eucscal}
\hat{g}^{-1} \( \ed x^{\al}, \ed x^{\be} \) \equiv \hat{g}^{\al\be} = \de^{\al\be} \, . 
\eeq
Finally, she is at rest with respect to these coordinates $\hat{u}^{\al} = 0$ and, since $t$ is her proper time, we have
\beq \label{eq:uobs}
\hat{u}^{\mu} = \de^{\mu}_t \, .
\eeq
Now conditions \eqref{eq:proptime} and \eqref{eq:Eucscal} on the inverse metric at $\hat{P}$ imply for the metric
\beq
\hat{g}_{tt} = - \frac{1}{1 + \hat{g}^{t\al} \hat{g}^{t\al}} \, , \hspace{1cm} \hat{g}_{t\al} = \frac{\hat{g}^{t\al}}{1 + \hat{g}^{t\be} \hat{g}^{t\be}} \, , \hspace{1cm} \hat{g}_{\al\be} = \de_{\al\be} - \frac{\hat{g}^{t\al} \hat{g}^{t\be}}{1 + \hat{g}^{t\ga} \hat{g}^{t\ga}}  \, .
\eeq
On the other hand, condition \eqref{eq:uobs} implies
\beq
-1 \equiv \hat{g}\( \hat{u}, \hat{u} \) = \hat{g}_{tt} \, ,
\eeq
and this leads to $\hat{g}^{t\al} = 0$. We thus have
\beq \label{eq:obsget}
\hat{g} \( \pa_{\mu}, \pa_{\nu} \) \equiv \hat{g}_{\mu\nu} = \et_{\mu\nu} \, , \hspace{1cm} \hat{u} = \hat{\pa}_t \, ,
\eeq
i.e. the $\pa_{\mu}$ basis is indeed orthonormal at $\hat{P}$.

\section{Active and passive diffeomorphisms} \label{app:diffsym}

The diffeomorphism symmetry of a differentiable manifold comes in two distinct manifestations, differing both at the conceptual and practical levels. Here we consider the space-time manifold $\cM$ as our working example, but everything we discuss in this section is applicable in general. 

Let us consider a tensor field $T$ on $\cal M$, which we can express in a coordinate-induced basis through a set of functions $T^{\nu_1 \dots \nu_n}_{\mu_1 \dots \mu_m}(x)$
\beq \label{eq:TcC}
T(x) = T^{\nu_1 \dots \nu_n}_{\mu_1 \dots \mu_m}(x) \, \pa_{\nu_1} \otimes \dots \otimes \pa_{\nu_n} \otimes \ed x^{\mu_1} \otimes \dots \otimes \ed x^{\mu_m} \, .
\eeq
The first manifestation of the diffeomorphism group are coordinate transformations, i.e. expressing the $x^{\mu}$ coordinates in terms of new ones $\ti{x}^{\mu}$
\beq
x^{\mu} = \vph^{\mu}(\ti{x}) \, ,
\eeq
through some diffeomorphism $\vph$. In that instance, the coordinate-independent expression of the tensor is invariant, by definition, 
\beq \label{eq:PMDcind}
T \to T \, ,
\eeq
so the transformation of the elementary bases 
\beq
\pa_{\mu} = \frac{\pa \ti{x}^{\mu}}{\pa x^{\nu}} \, \ti{\pa}_{\nu} \, , \hspace{1cm} \ed x^{\mu} = \frac{\pa x^{\mu}}{\pa \ti{x}^{\nu}}\, \ed \ti{x}^{\nu} \, ,
\eeq 
implies the following transformation of the components
\beq \label{eq:PMD}
\ti{T}^{\nu_1 \dots \nu_n}_{\mu_1 \dots \mu_m}(\ti{x}) = \frac{\pa \ti{x}^{\nu_1}}{\pa x^{\si_1}}(x) \dots \frac{\pa \ti{x}^{\nu_n}}{\pa x^{\si_n}}(x) \, \frac{\pa x^{\ro_1}}{\pa \ti{x}^{\mu_1}}(\ti{x}(x)) \dots \frac{\pa x^{\ro_m}}{\pa \ti{x}^{\mu_m}}(\ti{x}(x)) \, T^{\si_1 \dots \si_n}_{\ro_1 \dots \ro_m}(x) \, .
\eeq
Note that here both sides correspond to the same point $P \in {\cal M}$, which has coordinates $\ti{x}^{\mu}$ in the new system (lhs) and $x^{\mu}$ is the old one (rhs). Another perspective, however, is to interpret $x^{\mu}$ and $\ti{x}^{\mu}$ as the coordinates of two different points $P, \ti{P} \in \cM$ in a single coordinate system. From this viewpoint, expressing equation (\ref{eq:PMD}) at the same point on both sides would mean expressing it as the same coordinate, say $\ti{x}^{\mu}$,
\beq \label{eq:AMD}
\ti{T}^{\nu_1 \dots \nu_n}_{\mu_1 \dots \mu_m}(\ti{x}) = \frac{\pa \ti{x}^{\nu_1}}{\pa \vph^{\si_1}}(\vph(\ti{x})) \dots \frac{\pa \ti{x}^{\nu_n}}{\pa \vph^{\si_n}}(\vph(\ti{x})) \, \frac{\pa \vph^{\ro_1}}{\pa \ti{x}^{\mu_1}}(\ti{x}) \dots \frac{\pa \vph^{\ro_m}}{\pa \ti{x}^{\mu_m}}(\ti{x}) \, T^{\si_1 \dots \si_n}_{\ro_1 \dots \ro_m}(\vph(\ti{x})) \, .
\eeq
The reason we chose to express (\ref{eq:AMD}) in terms of $\vph^{\mu}(\ti{x})$, instead of $x^{\mu}(\ti{x})$ as in (\ref{eq:PMD}), is to stress that this is {\it not} a coordinate transformation. There is only a single coordinate system $\ti{x}^{\mu}$ and we are transforming the tensor $T$ using some function $\vph^{\mu}(\ti{x})$. Another way to see this is by noting that (\ref{eq:AMD}) takes the form of some operator acting on $T$
\beq \label{eq:AMDcind}
T \to \ti{T} = \vph_*(T) \, , 
\eeq 
i.e. we do modify the configuration, contrary to the coordinate transformation case (\ref{eq:PMDcind}) which is only a reparametrization of the same configuration. From (\ref{eq:AMD}) we see that the $\vph_*$ operator is the generalization of the composition operation to the tensor case
\beq
\vph_*(T) \equiv T \circ \vph \, ,
\eeq
i.e. now it also mixes the tensor indices with the Jacobian of $\vph(\ti{x})$, and it is known as the ``pull-back" operation induced by the $\vph$ map. 

To get more insight into the $\vph_*$ operator, we first consider the inverse map $\ti{x} = \vph^{-1}(x)$ and express it as a one-parameter family of diffeomorphisms $\vph^{-1}_s$ generated by some vector field $\xi^{\mu}$
\beq
\vph^{-1}_0 \equiv {\rm id} \, , \hspace{1cm} \vph^{-1}_1 \equiv \vph^{-1} \, , \hspace{1cm} \left. \pa_s \vph^{-1}_s \right|_{s = 0} \equiv \xi \, , 
\eeq 
so one reaches the new point $\ti{P}$ from the old one $P$ by flowing along an integral line of $\xi$ for a unit $s$-time. In particular, the solution takes the form
\beq
\vph^{\mu}(\ti{x}) = \ti{x}^{\mu} - \xi^{\mu}(\ti{x}) + \Ord(\xi^2) \, .
\eeq
With this and (\ref{eq:AMD}) we can now compute the variation induced by $\vph_*$ to linear order in $\xi^{\mu}$
\bea 
\de T^{\nu_1 \dots \nu_n}_{\mu_1 \dots \mu_m} & := & [\vph_*(T)]^{\nu_1 \dots \nu_n}_{\mu_1 \dots \mu_m} - T^{\nu_1 \dots \nu_n}_{\mu_1 \dots \mu_m} \nn \\
 & = &  - \xi^{\ro} \pa_{\ro} T^{\nu_1 \dots \nu_n}_{\mu_1 \dots \mu_m} + \sum_{k=1}^n T^{\nu_1 \dots \nu_{k-1} \ro \nu_{k+1} \dots \nu_n}_{\mu_1 \dots \mu_m} \pa_{\ro} \xi^{\nu_k} - \sum_{k=1}^m T^{\nu_1 \dots \nu_n}_{\mu_1 \dots \mu_{k-1} \ro \mu_{k+1} \dots \mu_m} \pa_{\mu_k} \xi^{\ro} + \Ord(\xi^2) \nn \\
 & \equiv & - \Lie_{\xi} T^{\nu_1 \dots \nu_n}_{\mu_1 \dots \mu_m} + \Ord(\xi^2)  \, , \label{eq:deT}
\eea
where $\Lie_{\xi}$ is the Lie derivative with respect to $\xi^{\mu}$. Since $\Lie_{\xi}$ is a covariant operator under coordinate transformations, this variation can be expressed independently of the choice of coordinate system
\beq
\de T := \ti{T} - T = - \Lie_{\xi} T + \Ord(\xi^2) \, .
\eeq
The operators $\Lie_{\xi}$ thus appear as the generators of the pull-back operation. They form an infinite-dimensional Lie algebra 
\beq
\[ \Lie_{\xi}, \Lie_{\xi'} \] \equiv \Lie_{(\Lie_{\xi} \xi')} \, ,
\eeq
and obey a Jacobi identity
\beq
\[ \Lie_{\xi}, \[ \Lie_{\xi'}, \Lie_{\xi''} \] \] + \[ \Lie_{\xi'}, \[ \Lie_{\xi''}, \Lie_{\xi} \] \] + \[ \Lie_{\xi''}, \[ \Lie_{\xi}, \Lie_{\xi'} \] \] \equiv 0 \, .
\eeq
These properties imply that the pull-back transformation is nothing but the exponential map of the generator\footnote{Of course, this description holds only for the diffeomorphisms that are connected to the identity.}
\beq 
\vph_*(T) = e^{-\Lie_{\xi}} T \, .
\eeq
In this form we obtain a generalization of the Taylor expansion in the generally-covariant context. For instance, for constant $\xi^{\mu}$ in some coordinate system we retrieve the translations of field theory in flat space-time, where the generators are $\Lie_{\xi} = \xi^{\mu} \pa_{\mu}$
\beq
f(x - \xi) = e^{- \xi^{\mu} \pa_{\mu}} f(x) \, , \hspace{1cm} \xi^{\mu} = {\rm constant} \, .
\eeq
In general $\xi^{\mu}$ is an arbitrary vector field and, depending on its shape, it can generate translations, rotations, etc., so (\ref{eq:AMD}) corresponds to translating $T$ along the integral lines of $\xi^{\mu}$. 

In conclusion, on the one hand we have coordinate transformations, while on the other hand we have pull-back transformations of tensor fields. In the former case, the coordinate-independent expression of the tensor $T$ is invariant, by definition, i.e. we do not change the configuration, only the way it is parametrized (\ref{eq:PMD}), hence the name ``passive'' diffeomorphism. In the latter case, the tensor field $T$ is genuinely transformed (\ref{eq:AMDcind}), hence the name ``active'' diffeomorphism. The latter are therefore the ``true'' transformations since they do modify the physical configuration and can be defined without requiring a coordinate system. 

Nevertheless, these two transformations are different manifestations of the same symmetry, since they are derived from the same relation (\ref{eq:PMD}) through two different interpretations. In particular, if an equation is invariant under passive diffeomorphisms, it is also invariant under the active ones, and vice-versa. Therefore, for questions of covariance, the distinction is irrelevant. In contrast, one of the cases where the distinction between the two transformations {\it is} relevant are the transformation properties of integrals over $\cal M$. So let us consider, as an example, a four-dimensional submanifold $U \subset {\cal M}$ and the integral of some scalar density of weight one $L$ over $U$
\beq
S_U := \int_U \ed^4 x\, L(x) \, .
\eeq
Under a passive diffeomorphism this quantity is invariant
\beq
\ti{S}_U = \int_U \ed^4 \ti{x}\, \ti{L}(\ti{x})  = \int_U \( \ed^4 x\, \det \frac{\pa \ti{x}}{\pa x} \) \( L(x) \det \frac{\pa x}{\pa \ti{x}} \) \equiv \int_U \ed^4 x\, L(x) \equiv S_U \, .
\eeq
In the active case, however, this quantity varies by a boundary term. Indeed, under an active diffeomorphism generated by $\xi^{\mu}$, we can write (\ref{eq:AMD}) as
\beq
\de_{\xi} T = - \Lie_{\xi} \[ T + \Ord(\xi) \] \, ,
\eeq
which for a scalar density of weight one gives a total derivative
\beq
\de_{\xi} L = - \Lie_{\xi} \[ L + \Ord(\xi) \] \equiv - \xi^{\mu} \pa_{\mu} \[ L + \Ord(\xi) \] - \[ L + \Ord(\xi) \] \pa_{\mu} \xi^{\mu} \equiv - \pa_{\mu} \[ \xi^{\mu} \[ L + \Ord(\xi) \] \] \, .
\eeq
Thus,
\beq
\de_{\xi} S_U = - \int_U \ed^4 x\, \pa_{\mu} \[ \xi^{\mu} \[ L + \Ord(\xi) \] \] = - \int_{\pa U} \[ L + \Ord(\xi) \] \xi^{\mu} \ed_{\mu}^3 x \, ,
\eeq
where
\beq
\ed_{\mu}^3 x := \frac{1}{3!} \, \vep_{\mu\nu\ro\si} \, \ed x^{\nu} \we \ed x^{\ro} \we \ed x^{\si} \, .
\eeq
The fact that the variation depends on the values of $L$ at the boundary $\pa U$, and is non-zero only when $\xi^{\mu}$ has a normal component to it, clearly shows that the field $L$ has been translated on $\cal M$ along $\xi$ and thus went through the boundary of $U$. In conclusion, it is important to know which type of diffeomorphism is being used when integrals are involved.

\section{Compact matrix formulation of the generalized BUU equation} \label{sec:compactBUU}

Let us define the tensor products of matrices
\bea
\bm{f}(x,\vec{p}_1, \dots, \vec{p}_n) & := & \bm{f}(x,\vec{p}_1) \otimes \dots \otimes \bm{f}(x,\vec{p}_n) \, , \\
\bm{f}_{\circ}(x,\vec{p}_1, \dots, \vec{p}_n) & := & \bm{f}_{\circ}(x,\vec{p}_1) \otimes \dots \otimes \bm{f}_{\circ}(x,\vec{p}_n) \, ,
\eea
whose index structure is set as follows
\beq
f_{s_1, \dots s_n; s'_1, \dots, s'_n}(x,\vec{p}_1, \dots, \vec{p}_n) \equiv f_{s_1 s'_1}(x,\vec{p}_1) \dots f_{s_n s'_n}(x,\vec{p}_n) \, .
\eeq
and also the notation
\beq
[\bm{\cA}_c]_{s_1, \dots s_n; r_1, \dots, r_m}(\vec{p}_1, \dots, \vec{p}_n; \vec{q}_1, \dots, \vec{q}_m) := \cA_c ( \vec{q}_1, r_1, \dots, \vec{q}_m, r_m \to \vec{p}_1, s_1, \dots, \vec{p}_n, s_n ) \, , 
\eeq
\beq
[\bm{\cA}_c^{\dagger}]_{s_1, \dots s_n; r_1, \dots, r_m}(\vec{p}_1, \dots, \vec{p}_n; \vec{q}_1, \dots, \vec{q}_m) := \cA^*_c ( \vec{p}_1, s_1, \dots, \vec{p}_n, s_n \to \vec{q}_1, r_1, \dots, \vec{q}_m, r_m ) \, .
\eeq
The latter is natural, because the ``in" states appear as kets in $\cA_c$, and therefore on its ``right side", whereas the ``out" states appear as bras on the left. The situation is reversed for $\cA_c^*$ because complex conjugation interchanges kets and bras. 

Considering the sets of indices on each side of the semi-column as a generalized index, we have that $\bm{f}$ and $\bm{A}_c$ appear as matrices in these generalized indices. We can therefore define the matrix product notation
\bea
& & [\bm{\cA}_c ( \vec{p}_1, \dots, \vec{p}_n; \vec{q}_1, \dots, \vec{q}_m)\, \bm{f}(x,\vec{q}_1, \dots, \vec{q}_m)]_{s_1, \dots s_n; r'_1, \dots r'_m} \nn \\
& & :=  [\bm{\cA}_c]_{s_1, \dots, s_n;r_1, \dots r_m} ( \vec{p}_1, \dots, \vec{p}_n; \vec{q}_1, \dots, \vec{q}_m)\, f_{r_1, \dots,r_m; r'_1, \dots, r'_m}(x,\vec{q}_1, \dots, \vec{q}_m) \, .   
\eea
With these conventions the creation and annihilation terms in Eqs. \eqref{eq:C+} and \eqref{eq:C-} now simply read
\bea
\bm{C}^+(x,\vec{p}) & \equiv & \frac{1}{2} \sum_{n, m = 0}^{\infty} \frac{1}{n! m!} \label{eq:C+n} \\
 & & \times \int \( \prod_{k=1}^n \frac{\ed^3 p_k}{(2\pi)^3 2 E_{p_k, s_k}} \prod_{l=1}^m \frac{\ed^3 q_l }{(2\pi)^3 2 E_{q_l, r_l} } \)  (2\pi)^4 \de^{(4)} \( p + \sum_{k=1}^n p_k - \sum_{l=1}^m q_l \) \nn   \\
 & & \times \, \Tr_+ \[ \frac{}{} \bm{\cA}_c ( \vec{p}_1, \dots, \vec{p}_n, \vec{p}; \vec{q}_1, \dots, \vec{q}_m) \, \bm{f}(x,\vec{q}_1, \dots, \vec{q}_m) \right. \nn \\
 & & \hspace{1.2cm} \left. \times\, \bm{\cA}_c^{\dagger}( \vec{q}_1, \dots, \vec{q}_m; \vec{p}_1, \dots, \vec{p}_n, \vec{p}) \, \bm{f}_{\circ}(x,\vec{p}_1, \dots, \vec{p}_n, \vec{p}) \] \nn \\ 
 & & + \, \bm{F}^+(x,\vec{p}) \, , \nn  
\eea
and
\bea
\bm{C}^-(x,\vec{p}) & \equiv & \frac{1}{2}  \sum_{n, m = 0}^{\infty} \frac{1}{n! m!} \label{eq:C-n} \\
 & & \times \int \( \prod_{k=1}^n \frac{\ed^3 p_k}{(2\pi)^3 2 E_{p_k, s_k}} \prod_{l=1}^m \frac{\ed^3 q_l }{(2\pi)^3 2 E_{q_l, r_l} } \)  (2\pi)^4 \de^{(4)} \( p + \sum_{k=1}^n p_k - \sum_{l=1}^m q_l \) \nn   \\
 & & \times \, \Tr_- \[ \bm{f}(x,\vec{p},\vec{p}_1, \dots, \vec{p}_n) \, \bm{\cA}_c^{\dagger}(\vec{p}, \vec{p}_1, \dots, \vec{p}_n; \vec{q}_1, \dots, \vec{q}_m) \right. \nn \\
 & & \hspace{1.2cm} \left. \times \, \bm{f}_{\circ}(x,\vec{q}_1, \dots, \vec{q}_m) \, \bm{\cA}_c ( \vec{q}_1, \dots, \vec{q}_m; \vec{p}, \vec{p}_1, \dots, \vec{p}_n) \]  \nn \\
 & & + \, \bm{F}^-(x,\vec{p})  \, . \nn
\eea
where $\Tr_{\pm}$ is the trace over all indices but the last/first pair, e.g.
\beq
[\Tr_+\, \bm{f}(x,\vec{p}_1, \dots, \vec{p}_n)]_{s_n s'_n} := f_{s_1, \dots, s_{n-1}, s_n; s_1, \dots, s_{n-1}, s'_n}(x,\vec{p}_1, \dots, \vec{p}_n) \, , 
\eeq 
\beq
[\Tr_-\, \bm{f}(x,\vec{p}_1, \dots, \vec{p}_n)]_{s_1 s'_1} := f_{s_1, s_2 \dots, s_n; s'_1, s_2 \dots, s_n}(x,\vec{p}_1, \dots, \vec{p}_n) \, .
\eeq

\bibliographystyle{JHEP}
\bibliography{Main}

\end{document}